\documentclass[11pt]{iopart}

\usepackage{iopams}
\usepackage{overpic,bbm,bm,epsfig}
\usepackage{multirow}
\usepackage{color}

\renewcommand{\theequation}{\thesection.\arabic{equation}}
\renewcommand{\thetable}{\thesection.\arabic{table}}
\renewcommand{\thefigure}{\thesection.\arabic{figure}}

\makeatletter

\newcommand{\Rmnum}[1]{\expandafter\@slowromancap\romannumeral #1@}
\makeatother

\begin{document}

\title[Zhi-zhong Xing $\&$ Zhen-hua Zhao]{
A review of $\mu$-$\tau$ flavor symmetry in neutrino physics}

\author{Zhi-zhong Xing$^{1}$ {\rm and} Zhen-hua Zhao$^{1,2}$}

\address{\footnotesize $^1$Institute of High Energy Physics and
School of Physical Sciences, University of Chinese Academy of
Sciences, Beijing 100049, China \\
$^2$Department of Physics, Liaoning Normal University, Dalian
116029, China} \ead{xingzz@ihep.ac.cn {\rm and}
zhaozhenhua@ihep.ac.cn} \vspace{10pt}
\begin{indented}
\item[] First version: December 2015; Modified version: March 2016
\end{indented}

\begin{abstract}
Behind the observed pattern of lepton flavor mixing is a partial or
approximate $\mu$-$\tau$ flavor symmetry --- a milestone on our road
to the true origin of neutrino masses and flavor structures. In this
review article we first describe the  features of $\mu$-$\tau$
permutation and reflection symmetries, and then explore their
various consequences on model building and neutrino phenomenology.
We pay particular attention to soft $\mu$-$\tau$ symmetry breaking,
which is crucial for our deeper understanding of the fine effects of
flavor mixing and CP violation.
\end{abstract}

%

\tableofcontents

\def\thefootnote{\arabic{footnote}}
\setcounter{footnote}{0}
\setcounter{table}{0}
\setcounter{equation}{0}
\setcounter{figure}{0}


\section{Introduction}

\subsection{A brief history of the neutrino families}

Soon after the French physicist Henri Becquerel first discovered the
radioactivity of uranium in 1896 \cite{Becquerel}, some nuclear
physicists began to focus their attention on the beta decays $(A, Z)
\to (A, Z+1) + e^-$, where the energy spectrum of the outgoing
electrons was expected to be {\it discrete} as constrained by the
energy and momentum conservations. Surprisingly, the British
physicist James Chadwick observed a {\it continuous} electron energy
spectrum of the beta decay in 1914 \cite{Chadwick1}, and such a
result was firmly established in the 1920s \cite{Ellis}. At that
time there were two typical points of view towards resolving this
discrepancy between the {\it observed} and {\it expected} energy
spectra of electrons: (a) the Danish theorist Niels Bohr intended to
give up the energy conservation law, and later his idea turned out
to be wrong; (b) the Austrian theorist Wolfgang Pauli preferred to
add in a new particle, which marked the birth of a new science. In
1930 Pauli pointed out that a light, spin-1/2 and neutral particle
--- known as the electron antineutrino today --- appeared in the
beta decay and carried away some energy and momentum in an invisible
way, and thus the energy spectrum of electrons in the process $(A,
Z) \to (A, Z+1) + e^- + \overline{\nu}^{}_e$ was continuous. Three
years later the Italian theorist Enrico Fermi took this hypothesis
seriously and developed an effective field theory of the beta decay
\cite{Fermi}, which made it possible to calculate the reaction rates
of nucleons and electrons or positrons interacting with neutrinos or
antineutrinos.

In 1936 the German physicist Hans Bethe pointed out that an inverse
beta decay mode of the form $\overline{\nu}^{}_e + (A, Z) \to (A,
Z-1) + e^+$ should be a feasible way to verify the existence of
electron antineutrinos produced from fission bombs or reactors
\cite{Bethe0}. This bright idea was elaborated by the Italian
theorist Bruno Pontecorvo in 1946 \cite{Pon46}, and it became more
realistic with the development of the liquid scintillation counting
techniques in the 1950s. For example, the invisible incident
$\overline{\nu}^{}_e$ triggers the reaction $\overline{\nu}^{}_e + p
\to n + e^+$, in which the emitted positron annihilates with an
electron and the daughter nucleus is captured in the detector. Both
events can be observed since they emit gamma rays, and the
corresponding flashes in the liquid scintillator are separated by
some microseconds. The American experimentalists Frederick Reines
and Clyde Cowan did the first reactor antineutrino experiment and
confirmed Pauli's hypothesis in 1956 \cite{RC}. Such a discovery
motivated Pontecorvo to speculate on the possibility of lepton
number violation and neutrino-antineutrino transitions in 1957
\cite{Pontecorvo}. His viewpoint was based on a striking conjecture
made by Fermi's doctoral student Ettore Majorana in 1937: a massive
neutrino could be its own antiparticle \cite{Majorana}. Whether
Majorana's hypothesis is true or not remains an open question in
particle physics.

In 1962 the muon neutrino --- a puzzling sister of the electron
neutrino --- was first observed by the American experimentalists
Leon Lederman, Melvin Schwartz and Jack Steinberger in a new
accelerator-based experiment \cite{Danby}. Their discovery more or
less motivated the Japanese theorists Ziro Maki, Masami Nakagawa and
Shoichi Sakata to think about lepton flavor mixing and $\nu^{}_e
\leftrightarrow \nu^{}_\mu$ transitions \cite{MNS}. The tau
neutrino, another sister of the electron neutrino, was finally
observed at the Fermilab in 2001 \cite{Kodama}. Today we are left
with three lepton families consisting of the charged members ($e$,
$\mu$, $\tau$) and the neutral members ($\nu^{}_e$, $\nu^{}_\mu$,
$\nu^{}_\tau$), together with their antiparticles. Table 1.1 shows
the total lepton number ($L$) and individual flavor numbers
($L^{}_e$, $L^{}_\mu$, $L^{}_\tau$) assigned to all the known
leptons and antileptons in the standard theory of electroweak
interactions. So far the nonconservation of $L^{}_e$, $L^{}_\mu$ and
$L^{}_\tau$ has been observed in a number of neutrino oscillation
experiments.
\begin{table}[t]
\caption{The total lepton number ($L$) and individual flavor numbers
($L^{}_e, L^{}_\mu, L^{}_\tau$) of three families of leptons and
antileptons in the standard theory of electroweak interactions.}
\vspace{0.1cm}
\begin{indented}
\item[]\begin{tabular}{lcccccc} \br
& \multicolumn{2}{c}{{\bf 1}st family} & \multicolumn{2}{c}{{\bf
2}nd family} & \multicolumn{2}{c}{{\bf 3}rd family} \\ \mr & $e^-$ &
$\nu^{}_e$ & $\mu^-$ & $\nu^{}_\mu$ & $\tau^-$ & $\nu^{}_\tau$ \\
\mr
$L$ & $+1$ & $+1$ & $+1$ & $+1$ & $+1$ & $+1$ \\
$L^{}_e$ & $+1$ & $+1$ & $0$ & $0$ & $0$ & $0$ \\
$L^{}_\mu$ & $0$ & $0$ & $+1$ & $+1$ & $0$ & $0$ \\
$L^{}_\tau$ & $0$ & $0$ & $0$ & $0$ & $+1$ & $+1$ \\ \br
& $e^+$ & $\overline{\nu}^{}_e$ & $\mu^+$ & $\overline{\nu}^{}_\mu$
& $\tau^+$ & $\overline{\nu}^{}_\tau$ \\ \mr
$L$ & $-1$ & $-1$ & $-1$ & $-1$ & $-1$ & $-1$ \\
$L^{}_e$ & $-1$ & $-1$ & $0$ & $0$ & $0$ & $0$ \\
$L^{}_\mu$ & $0$ & $0$ & $-1$ & $-1$ & $0$ & $0$ \\
$L^{}_\tau$ & $0$ & $0$ & $0$ & $0$ & $-1$ & $-1$ \\ \br
\end{tabular}
\end{indented}
\end{table}

The standard electroweak theory about charged leptons and neutrinos
was first formulated by the American theorist Steven Weinberg in
1967 \cite{W1967}. In this seminal paper the neutrinos were assumed
to be massless, and hence there should be no lepton flavor
conversion. Just one year later, a preliminary experimental evidence
for finite neutrino masses and lepton flavor mixing appeared thanks
to the first observation of solar neutrinos and their deficit as
compared with the prediction of the standard solar model
\cite{Davis}. The point was that the observed deficit of solar
electron neutrinos could easily be attributed to $\nu^{}_e \to
\nu^{}_\mu$ and $\nu^{}_e \to \nu^{}_\tau$ oscillations
\cite{GP1969} --- a pure quantum phenomenon which would not take
place if every neutrino were massless and the lepton flavor were
conserved. Hitherto the flavor oscillations of solar, atmospheric,
accelerator and reactor neutrinos (or antineutrinos) have all been
established \cite{PDG}, and thus a nontrivial extension of the
standard theory of electroweak interactions is unavoidable in order
to explain the origin of nonzero but tiny neutrino masses and the
dynamics of significant lepton flavor mixing.

At present the most popular and intriguing idea for neutrino mass
generation is the {\it seesaw} mechanism, which attributes the tiny
masses of three neutrinos to the existence of a few heavy degrees of
freedom and lepton number violation. There are three typical seesaw
mechanisms on the market:
\begin{itemize}
\item     Type-I seesaw --- two or three heavy right-handed
neutrinos are added into the standard theory and the lepton number
is violated by their Majorana mass term \cite{SS1,SS2,SS3,SS4,SS5};

\item     Type-II seesaw --- one heavy Higgs triplet is added into
the standard theory and the lepton number is violated by its
interactions with both the lepton doublet and the Higgs doublet
\cite{SS2a,SS2b,SS2c,SS2d,SS2e};

\item     Type-III seesaw --- three heavy triplet fermions are introduced
into the standard theory and the lepton number is violated by their
Majorana mass term \cite{He,Ma}.
\end{itemize}
After the heavy degrees of freedom are integrated out, all the
three seesaw mechanisms are convergent to a unique effective Majorana
neutrino mass operator \cite{Weinberg1979}.
Although a given seesaw scenario can qualitatively explain why the
neutrinos may have tiny masses, it is not powerful enough to
determine the flavor structures of charged leptons and neutrinos.
Hence a combination of the seesaw idea and possible flavor
symmetries is desirable so as to achieve some testable quantitative
predictions in the lepton sector.

In comparison with three lepton families, there exist three quark
families consisting of the up-type quarks ($u$, $c$, $t$) and the
down-type quarks ($d$, $s$, $b$), together with their antiparticles.
All these leptons and quarks constitute the flavor part of particle
physics, and their mass spectra, flavor mixing properties and CP
violation are the central issues of flavor dynamics. In this review
article we shall focus on the lepton flavors, especially the
$\mu$-$\tau$ flavor symmetry in the neutrino sector and its striking
impacts on the phenomenology of neutrino physics.

\subsection{The $\mu$-$\tau$ flavor symmetry stands out}

Since Noether's theorem was first published in 1918 \cite{N},
symmetries have been playing a very crucial role in understanding
the fundamental laws of Nature. In fact, symmetries are so powerful
that they can help simplify the complicated problems, classify the
intricate systems, pin down the conservation laws and even determine
the dynamics of interactions. In elementary particle physics there
are many successful examples of this kind, such as the continuous
space-time translation symmetries, the $\rm SU(3)^{}_{\rm q}$ quark
flavor symmetry and the $\rm SU(3)^{}_{\rm c} \times SU(2)^{}_{\rm
L} \times U(1)^{}_{\rm Y}$ gauge symmetries. Historically these
examples led us to the momentum and energy conservation laws, the
quark model and the standard model (SM) of electroweak and strong
interactions, respectively \cite{PDG}. The original symmetries in a
given theory may either keep exact or be broken, so as to make our
description of the relevant phenomena consistent with the
experimental observations. For instance, the electromagnetic $\rm
U(1)^{}_{\rm em}$ gauge symmetry, the strong $\rm SU(3)^{}_{\rm c}$
gauge symmetry and the continuous space-time translation symmetries
are all exact; but the $\rm SU(3)^{}_{\rm q}$ quark flavor symmetry,
the electroweak $\rm SU(2)^{}_{\rm L} \times U(1)^{}_{\rm Y}$ gauge
symmetry and the P and CP symmetries in weak interactions must be
broken. That is why exploring new symmetries and studying possible
symmetry-breaking effects have been one of the main streams in
particle physics, normally from lower energies to higher energies.

Although the SM has proved to be very successful in describing the
phenomena of strong, weak and electromagnetic interactions, it is by
no means a complete theory. The flavor part of this theory is
particularly unsatisfactory, because it involves many free
parameters but still cannot provide any solution to a number of
burning problems, such as the origin of neutrino masses and lepton
flavor mixing, the baryon number asymmetry of the Universe and the
nature of cold or warm dark matter. To go beyond the SM in this
connection, flavor symmetries are expected to serve for a powerful
guideline.

Flavor symmetries can be either Abelian or non-Abelian, either local
or global, either continuous or discrete, and either spontaneously
broken or explicitly broken. All these possibilities have been
extensively explored in the past few decades, so as to explain the
observed lepton and quark mass spectra and the observed flavor
mixing patterns \cite{Review1,Review2,Review3,Review4}. Given the
peculiar pattern of lepton flavor mixing which is suggestive of a
constant unitary matrix with some special entities (e.g.,
$1/\sqrt{2}$, $1/\sqrt{3}$ or $1/\sqrt{6}$), a lot of attention has
been paid to the global and discrete flavor symmetry groups in the
model building exercises. The advantages of such a choice are
obvious at least in the following aspects: (a) the model does not
involve any Goldstone bosons or additional gauge bosons which may
mediate harmful flavor-changing-neutral-current processes; (b) the
discrete group may come from some string compactifications or be
embedded in a continuous symmetry group; (c) the model contains no
family-dependent D-terms contributing to the sfermion masses if it
is built in a supersymmetric framework. Although many discrete
flavor symmetry groups have been taken into account in building
viable neutrino mass models, it remains unclear which one can
finally stand out as the unique basis of the true flavor dynamics of
leptons and quarks.

But it turns out to be clear that any promising discrete flavor
symmetry group in the neutrino sector has to accommodate the
simplest $\mu$-$\tau$ flavor symmetry --- a sort of $\rm Z^{}_2$
transformation symmetry with respect to the $\nu^{}_\mu$ and
$\nu^{}_\tau$ neutrinos, because the latter has convincingly
revealed itself through the currently available neutrino oscillation
data (as one can see in the next section). In other words, a partial
or approximate $\mu$-$\tau$ flavor symmetry must be behind the
observed pattern of the $3\times 3$ Pontecorvo-Maki-Nakagawa-Sakata
(PMNS) lepton flavor mixing matrix $U$ \cite{XZ2014}, and thus it
may serve as an especially useful low-energy window to look into the
underlying structures of lepton flavors at either the electroweak
scale or superhigh-energy scales. It is just this observation that
motivates us to review what we have learnt from
\begin{itemize}
\item     the $\mu$-$\tau$ {\it permutation} symmetry --- the
neutrino mass term is unchanged under the transformations
\begin{eqnarray}
\nu^{}_e \to \nu^{}_e \; , ~~~ \nu^{}_\mu \to \nu^{}_\tau \; , ~~~
\nu^{}_\tau \to \nu^{}_\mu \; ;
\end{eqnarray}

\item     the $\mu$-$\tau$ {\it reflection} symmetry --- the
neutrino mass term keeps unchanged under the transformations
\begin{eqnarray}
\nu^{}_e \to \nu^c_e \; , ~~~ \nu^{}_\mu \to \nu^c_\tau \; , ~~~
\nu^{}_\tau \to \nu^c_\mu \; ,
\end{eqnarray}
\end{itemize}
where the superscript ``$c$" denotes the charge conjugation of the
relevant neutrino field, and to explore their interesting
implications and consequences on various aspects of neutrino
phenomenology.

In particular, we stress that slight or soft breaking of the
$\mu$-$\tau$ flavor symmetry is expected to help resolve the octant
of the largest neutrino mixing angle $\theta^{}_{23}$ and even the
quadrant of the CP-violating phase $\delta$ in the standard
parametrization of the PMNS matrix $U$ \cite{PDG}, which consists of
three rotation angles ($\theta^{}_{12}$, $\theta^{}_{13}$,
$\theta^{}_{23}$) and one ($\delta$ in the Dirac case) or three
($\delta$, $\rho$, $\sigma$ in the Majorana case) CP-violating
phases. It may also offer a straightforward link between the
neutrino mass spectrum and the lepton flavor mixing pattern. All
these issues are very important on the theoretical side and highly
concerned in the ongoing and upcoming neutrino experiments.

The remaining parts of this review paper are organized in the
following way. In section 2 we begin with a brief introduction to
the phenomenology of lepton flavor mixing and neutrino oscillations,
followed by a short description of the main outcomes of various
solar, atmospheric, reactor and accelerator neutrino oscillation
experiments. In the three-flavor scheme a global analysis of current
experimental data leads us to a preliminary pattern of lepton flavor
mixing, which exhibits an approximate $\mu$-$\tau$ flavor symmetry.
Section 3 is devoted to an overview of the $\mu$-$\tau$ flavor
symmetry of the Majorana neutrino mass matrix and its connection
with the flavor mixing parameters. Two kinds of symmetries, the
$\mu$-$\tau$ {\it permutation} symmetry and the $\mu$-$\tau$ {\it
reflection} symmetry, will be classified and discussed. The typical
and instructive ways to slightly break the $\mu$-$\tau$ flavor
symmetry are introduced. In particular, the effects of $\mu$-$\tau$
symmetry breaking induced by radiative corrections are described in
some detail. The contribution of the charged-lepton sector to lepton
flavor mixing is also discussed. In section 4 we turn to some larger
discrete flavor symmetry groups to illustrate how the $\mu$-$\tau$
symmetry can naturally arise as a residual symmetry. Both the
bottom-up approach and the top-down approach will be described in
this connection. In section 5 we concentrate on the strategies of
model building with the help of the $\mu$-$\tau$ permutation or
reflection symmetry. A combination of the seesaw mechanism and the
$\mu$-$\tau$ flavor symmetry is taken into account, and the
relationship between the Friedberg-Lee symmetry and the $\mu$-$\tau$
flavor symmetry is explored. We also comment on the model-building
exercises associated with the light sterile neutrinos. Section 6 is
devoted to some phenomenological consequences of the $\mu$-$\tau$
flavor symmetry on some interesting topics in neutrino physics,
including neutrino oscillations in terrestrial matter, flavor
distributions of the Ultrahigh-energy (UHE) cosmic neutrinos at
neutrino telescopes, a possible connection between the leptogenesis
and low-energy CP violation, and a likely unified flavor texture of
leptons and quarks. The concluding remarks and an outlook are
finally made in section 7.

\def\thefootnote{\arabic{footnote}}
\setcounter{footnote}{0}
\setcounter{equation}{0}
\setcounter{table}{0}
\setcounter{figure}{0}

\section{Behind the lepton flavor mixing pattern}

To see why an approximate $\mu$-$\tau$ flavor symmetry is behind the
observed pattern of the PMNS lepton flavor mixing matrix $U$, let us
first introduce some basics about neutrino mixing and flavor
oscillations and then discuss current experimental constraints on
the structure of $U$.

\subsection{Lepton flavor mixing and neutrino oscillations}

Just similar to quark flavor mixing, lepton flavor mixing can also
take place provided leptonic weak charged-current interactions and
Yukawa interactions coexist in a simple extension of the SM. The
standard form of weak charged-current interactions of the charged
leptons and neutrinos reads
\begin{eqnarray}
-{\cal L}^{}_{\rm cc} = \frac{g}{\sqrt{2}} \sum_\alpha
\left[ \overline{\alpha^\prime_{\rm L}} \ \gamma^\mu
\nu^{}_{\alpha \rm L} W^-_\mu + {\rm h.c.} \right] \; ,
\end{eqnarray}
where $\alpha$ runs over $e$, $\mu$ and $\tau$, and the superscript
``$\prime$" denotes the flavor eigenstate of a charged lepton. The
leptonic Yukawa interactions are expected to be responsible for the
mass generation of both charged leptons and neutrinos after
spontaneous electroweak symmetry breaking, although the origin of
neutrino masses is very likely to involve some new degrees of
freedom and lepton number violation \cite{Xing09}. Without going
into details of a specific neutrino mass model, here we assume
massive neutrinos to be the Majorana particles and write out the
effective lepton mass terms as follows:
\begin{eqnarray}
-{\cal L}^{}_{\rm m} = \frac{1}{2} \ \overline{\nu^{}_{\alpha {\rm
L}}} \left( M^{}_\nu \right)^{}_{\alpha \beta} \nu^{c}_{\beta {\rm
R}} + \overline{\alpha^\prime_{\rm L}} \left( M^{}_l
\right)^{}_{\alpha\beta} \beta^\prime_{\rm R} + {\rm h.c.} \;
\end{eqnarray}
with $M^{}_\nu$ being symmetric and $M^{}_l$ being arbitrary.
Given the unitary matrices $O^{}_l$ and $O^{}_\nu$, $M^{}_\nu$
and $M^{}_l M^\dagger_l$ can be diagonalized through
the transformations
\begin{eqnarray}
O^\dagger_\nu M^{}_\nu O^*_\nu = D_\nu \equiv
\pmatrix{m^{}_1 & 0 & 0 \cr 0 & m^{}_2 & 0 \cr 0 & 0 & m^{}_3 \cr} \; ,
\nonumber \\
O^\dagger_l M^{}_l M^\dagger_l O^{}_l = D^{2}_l \equiv
\pmatrix{m^{2}_e & 0 & 0 \cr 0 & m^{2}_\mu & 0 \cr 0 & 0 &
m^{2}_\tau \cr} \; .
\end{eqnarray}
Then it is possible to reexpress ${\cal L}^{}_{\rm m}$ in terms of
the mass eigenstates of charged leptons and neutrinos:
\begin{eqnarray}
-{\cal L}^{}_{\rm m} = \frac{1}{2} \ \overline{\nu^{}_{i {\rm L}}}
\left( D^{}_\nu \right)^{}_{ij} \nu^{c}_{j {\rm R}} +
\overline{\alpha^{}_{\rm L}} \left( D^{}_l \right)^{}_{\alpha\beta}
\beta^{}_{\rm R} + {\rm h.c.} \; .
\end{eqnarray}
In doing so, one must consistently reexpress ${\cal L}^{}_{\rm cc}$
in terms of the relevant mass eigenstates:
\begin{eqnarray}
-{\cal L}^{}_{\rm cc} = \frac{g}{\sqrt{2}} \
\overline{\left( e \hspace{0.15cm} \mu \hspace{0.15cm}
\tau\right)^{}_{\rm L}} \ \gamma^\mu \ U
\pmatrix{ \nu^{}_1 \cr \nu^{}_2 \cr \nu^{}_3
\cr}^{}_{\rm L} W^-_\mu + {\rm h.c.} \; ,
\end{eqnarray}
where $U = O^\dagger_l O^{}_\nu$ is just the unitary PMNS matrix
which describes the strength of lepton flavor mixing in weak
interactions
\footnote{Whether $U$ is really unitary or not actually depends on
the mechanism of neutrino mass generation. In the canonical seesaw
mechanism \cite{SS1,SS2,SS3,SS4,SS5}, for instance, the mixing
between light and heavy Majorana neutrinos leads to tiny
unitarity-violating effects for $U$ itself. However, the unitarity
of $U$ has been tested at the percent level
\cite{Antusch1,Antusch2}, and thus it makes sense to assume $U$ to
be exactly unitary for the time being.}.

In a commonly chosen basis where the flavor eigenstates of three
charged leptons are identified with their mass eigenstates, the
flavor eigenstates of three neutrinos can be expressed as
\begin{eqnarray}
\pmatrix{ \nu^{}_e \cr \nu^{}_\mu \cr \nu^{}_\tau \cr} =
\pmatrix{
U^{}_{e 1} & U^{}_{e 2} & U^{}_{e 3} \cr U^{}_{\mu 1} & U^{}_{\mu 2}
& U^{}_{\mu 3} \cr U^{}_{\tau 1} & U^{}_{\tau 2} & U^{}_{\tau 3} \cr}
\pmatrix{ \nu^{}_1 \cr \nu^{}_2 \cr
\nu^{}_3 \cr} .
\label{2.6}
\end{eqnarray}
The nine elements of $U$ can be parameterized in terms of three
rotation angles and three CP-violating phases. For example, $U = V
P^{}_\nu$ with $V = O^{}_{23} O^{}_\delta O^{}_{13} O^\dagger_\delta
O^{}_{12}$ and $P^{}_\nu = {\rm Diag}\left\{e^{{\rm i}\rho}, e^{{\rm
i}\sigma}, 1\right\}$, where
\begin{eqnarray}
O^{}_{12} = \pmatrix{ c^{}_{12} & s^{}_{12} & 0 \cr
-s^{}_{12} & c^{}_{12} & 0 \cr 0 & 0 & 1 \cr} ,
\nonumber \\
O^{}_{13} = \pmatrix{ c^{}_{13} & 0 & s^{}_{13} \cr 0 & 1 & 0
\cr -s^{}_{13} & 0 & c^{}_{13} \cr} ,
\nonumber \\
O^{}_{23} = \pmatrix{ 1 & 0 & 0 \cr 0 & c^{}_{23} & s^{}_{23} \cr
0 & -s^{}_{23} & c^{}_{23} \cr} ,
\end{eqnarray}
and $O^{}_\delta = {\rm Diag}\{1, 1, e^{{\rm i}\delta}\}$
with $c^{}_{ij} \equiv \cos\theta^{}_{ij}$ and $s^{}_{ij} \equiv
\sin\theta^{}_{ij}$ (for $ij = 12, 13, 23$). To be more explicit,
\begin{eqnarray}
V = \pmatrix{ c^{}_{12} c^{}_{13} & s^{}_{12} c^{}_{13} & s^{}_{13}
e^{-{\rm i} \delta} \cr V^{}_{\mu 1} & V^{}_{\mu 2} & c^{}_{13}
s^{}_{23} \cr V^{}_{\tau 1} & V^{}_{\tau 2} & c^{}_{13} c^{}_{23}
\cr} ,
\end{eqnarray}
in which
\begin{eqnarray}
V^{}_{\mu 1} = -s^{}_{12} c^{}_{23} - c^{}_{12} s^{}_{13} s^{}_{23}
e^{{\rm i} \delta} \; ,
\nonumber \\
V^{}_{\mu 2} = c^{}_{12} c^{}_{23} - s^{}_{12} s^{}_{13} s^{}_{23}
e^{{\rm i} \delta} \; ,
\nonumber \\
V^{}_{\tau 1} = s^{}_{12} s^{}_{23} - c^{}_{12} s^{}_{13} c^{}_{23}
e^{{\rm i} \delta} \; ,
\nonumber \\
V^{}_{\tau 2} = -c^{}_{12} s^{}_{23} - s^{}_{12} s^{}_{13} c^{}_{23}
e^{{\rm i} \delta} \; .
\end{eqnarray}
Without loss of generality, the three mixing angles are all arranged
to lie in the first quadrant, while $\delta$ may vary from $0$ to
$2\pi$. The fact that the elements of $V$ in its first row and third
column are very simple functions of the flavor mixing angles
makes the latter have straightforward relations with the amplitudes
of solar ($\theta^{}_{12}$), reactor ($\theta^{}_{13}$) and
atmospheric ($\theta^{}_{23}$) neutrino oscillations, respectively
\cite{FX01}. In this parametrization $\delta$ is usually referred to
as the ``Dirac" phase, while $\rho$ and $\sigma$ are the Majorana
phases which have nothing to do with neutrino oscillations. If
massive neutrinos were the Dirac particles, one would simply forget
the Majorana phase matrix $P^{}_\nu$. Throughout this review we
mainly concentrate on the Majorana neutrinos, because they are well
motivated from a theoretical point of view. Then the symmetric
Majorana neutrino mass matrix can be reconstructed in terms of the
neutrino masses $m^{}_i$ (for $i=1,2,3$) and the PMNS matrix $U$ in
the chosen flavor basis:
\begin{eqnarray}
M^{}_\nu = \pmatrix{ M^{}_{ee} & M^{}_{e \mu} & M^{}_{e \tau} \cr
M^{}_{e \mu} & M^{}_{\mu\mu} & M^{}_{\mu \tau} \cr M^{}_{e \tau} &
M^{}_{\mu \tau} & M^{}_{\tau \tau} \cr} = U D^{}_\nu U^T \; .
\label{2.10}
\end{eqnarray}
In a specific neutrino mass model which is able to fix the texture
of $M^{}_\nu$, one may reversely obtain some testable predictions
for the neutrino masses and flavor mixing parameters.

${\cal L}^{}_{\rm cc}$ in
Eq. (2.5) tells us that a $\nu^{}_\alpha$ neutrino flavor can be
produced from the $W^+ + \alpha^- \to \nu^{}_\alpha$ interaction,
and a $\nu^{}_\beta$ neutrino flavor can be detected through the
$\nu^{}_\beta + W^- \to \beta^-$ interaction (for $\alpha, \beta =
e, \mu, \tau$). The effective Hamiltonian responsible for the
propagation of $\nu^{}_i$ in vacuum is expressed as
\begin{eqnarray}
{\cal H}^{}_{\rm eff} = \frac{1}{2E} M^{}_\nu M^\dagger_\nu
= \frac{1}{2E} V D^2_\nu V^\dagger \; ,
\end{eqnarray}
where $E \gg m^{}_i$ is the neutrino beam energy, and the Majorana
phase matrix $P^{}_\nu$ has been cancelled. Thanks to a
quintessentially quantum-mechanical effect, the $\nu^{}_\alpha \to
\nu^{}_\beta$ oscillation happens if the $\nu^{}_i$ beam travels a
proper distance $L$. The probability of such a flavor oscillation is
given by \cite{XZ}
\begin{eqnarray}
P(\nu^{}_\alpha \to \nu^{}_\beta) = \delta^{}_{\alpha\beta} - 4
\sum_{i<j} {\rm Re} \lozenge^{ij}_{\alpha\beta} \sin^2\Delta^{}_{ji}
\nonumber \\
\hspace{2.26cm} + 8 {\rm Im} \lozenge^{ij}_{\alpha\beta} \prod_{i<j}
\sin\Delta^{}_{ji} \; ,
\end{eqnarray}
where $\Delta^{}_{ji} \equiv \Delta m^2_{ji} L/\left(4 E\right)$
and $\lozenge^{ij}_{\alpha\beta} \equiv U^{}_{\alpha i} U^{}_{\beta
j} U^*_{\alpha j} U^*_{\beta i}$ (for $i,j = 1,2,3$ and $\alpha,
\beta = e, \mu, \tau$). The unitarity of the PMNS matrix $U$ leads
us to
\begin{eqnarray}
{\rm Im} \lozenge^{ij}_{\alpha\beta} = {\cal J}
\sum_\gamma \epsilon^{}_{\alpha \beta \gamma} \sum_k
\epsilon^{}_{ijk} \; ,
\end{eqnarray}
with ${\cal J} = c^{}_{12} s^{}_{12} c^2_{13} s^{}_{13} c^{}_{23}
s^{}_{23} \sin\delta$ being the so-called Jarlskog invariant \cite{J},
which is a universal measure of leptonic CP and T violation in neutrino
oscillations. The probability of $\overline{\nu}^{}_\alpha
\to \overline{\nu}^{}_\beta$ oscillations can be directly read off
from Eq. (2.12) by making the replacement $U \to U^*$. There are in
general two categories of neutrino oscillation experiments: {\it
appearance} ($\alpha \neq \beta$) and {\it disappearance} ($\alpha =
\beta$). Both the solar neutrino oscillations ($\nu^{}_e \to \nu^{}_e$)
and the reactor antineutrino oscillations ($\overline{\nu}^{}_e \to
\overline{\nu}^{}_e$) are of the disappearance type. In comparison,
the atmospheric muon-neutrino (or muon-antineutrino) oscillations
are essentially of the disappearance type, and the accelerator
neutrino (or antineutrino) oscillations can be of either type.

Given Eq. (2.5), the reactions $\nu^{}_e + e^- \to \nu^{}_e + e^-$
and $\overline{\nu}^{}_e + e^- \to \overline{\nu}^{}_e + e^-$ can
take place via the charged-current interactions. That is why the
behavior of neutrino (or antineutrino) flavor conversion in a dense
medium may be modified by the coherent forward $\nu^{}_e e^-$ or
$\overline{\nu}^{}_e e^-$ scattering. This matter effect is also
referred to as the Mikheyev-Smirnov-Wolfenstein (MSW) effect
\cite{MSW1,MSW2}. In this case Eqs. (2.11)---(2.13) have
to be replaced by their counterparts in matter \cite{Xing2004}.

\subsection{Current neutrino oscillation experiments}

The fact that neutrinos are massive and lepton flavors are mixed has
been firmly established in the past two decades, thanks to a number
of solar, atmospheric, reactor and accelerator neutrino (or
antineutrino) oscillation experiments \cite{PDG}. Let us briefly go
over some of them in the following, before we discuss what is behind
the observed pattern of lepton flavor mixing.

\begin{flushleft}
{\it A. Solar neutrino oscillations}
\end{flushleft}

The solar $^8{\rm B}$ neutrinos were first observed in the Homestake
experiment in 1968, but the measured flux was only about one third
of the value predicted by the standard solar model (SSM)
\cite{Davis,Bahcall}. This anomaly was later confirmed by other
experiments, such as GALLEX/GNO \cite{G1,G2}, SAGE \cite{Sage},
Super-Kamiokande \cite{SK} and SNO \cite{SNO}. The SNO experiment
was particularly crucial because it provided the first
model-independent evidence for the flavor conversion of solar
$\nu^{}_e$ neutrinos into $\nu^{}_\mu$ and $\nu^{}_\tau$ neutrinos.

The target material of the SNO detector is heavy water, which allows
the solar $^8{\rm B}$ neutrinos to be observed via the
charged-current (CC) reaction $\nu^{}_e + {\rm D} \to e^- + p + p$,
the neutral-current (NC) reaction $\nu^{}_\alpha + {\rm D} \to
\nu^{}_\alpha + p + n$ and the elastic-scattering process
$\nu^{}_\alpha + e^- \to \nu^{}_\alpha + e^-$ (for $\alpha = e, \mu,
\tau$) \cite{SNO}. The neutrino fluxes extracted from these three
channels can be expressed as $\phi^{}_{\rm CC} = \phi^{}_e$,
$\phi^{}_{\rm NC} = \phi^{}_e + \phi^{}_{\mu\tau}$ and $\phi^{}_{\rm
ES} = \phi^{}_e + 0.155 \phi^{}_{\mu \tau}$, where
$\phi^{}_{\mu\tau}$ denotes a sum of the fluxes of $\nu^{}_\mu$ and
$\nu^{}_\tau$ neutrinos. If there were no flavor conversion,
$\phi^{}_{\mu\tau} =0$ and $\phi^{}_{\rm CC} = \phi^{}_{\rm NC} =
\phi^{}_{\rm ES}$ would hold. The SNO data yielded $\phi^{}_{\rm CC}
= 1.68^{+0.06}_{-0.06} ({\rm stat})^{+0.08}_{-0.09} ({\rm syst})$,
$\phi^{}_{\rm ES} = 2.35^{+0.22}_{-0.22} ({\rm
stat})^{+0.15}_{-0.15} ({\rm syst}) > \phi^{}_{\rm CC}$ and
$\phi^{}_{\rm NC} = 4.94^{+0.21}_{-0.21} ({\rm
stat})^{+0.38}_{-0.34} ({\rm syst}) > \phi^{}_{\rm ES}$ \cite{SNO2},
demonstrating the existence of flavor conversion (i.e.,
$\phi^{}_{\mu\tau} \neq 0$) and supporting the SSM prediction for
$\phi^{}_{\rm NC}$ in a convincing way. In fact, the observed
deficit of solar $^8{\rm B}$ neutrinos is attributed to $\nu^{}_e
\to \nu^{}_\mu$ and $\nu^{}_e \to \nu^{}_\tau$ transitions modified
by significant matter effects in the core of the Sun. The survival
probability of $^8{\rm B}$ neutrinos may roughly approximate to
$P(\nu^{}_e \to \nu^{}_e) \simeq \sin^2\theta^{}_{12} \simeq 0.32$
\cite{Kayser}, leading us to $\theta^{}_{12} \simeq 34^\circ$.

Note that the recent Borexino experiment has done a real-time
measurement of the mono-energetic solar $^7{\rm Be}$ neutrinos and
observed a remarkable deficit corresponding to $P(\nu^{}_e \to
\nu^{}_e) =0.56 \pm 0.1$ \cite{B}. This result can approximately be
interpreted as a vacuum oscillation effect, since the low-energy
$^7{\rm Be}$ neutrino oscillation is not very sensitive to matter
effects in the Sun \cite{Kayser}. So one is left with the averaged
survival probability $P(\nu^{}_e \to \nu^{}_e) \simeq 1 - 0.5\sin^2
2\theta^{}_{12} \simeq 0.56$ for solar $^7{\rm Be}$ neutrinos, from
which $\theta^{}_{12} \simeq 35^\circ$ can be obtained. Such a
result is apparently consistent with the aforementioned value of
$\theta^{}_{12}$ extracted from the data of solar $^8{\rm B}$
neutrinos.
\begin{figure*}
\centering
\includegraphics[width=1\textwidth]{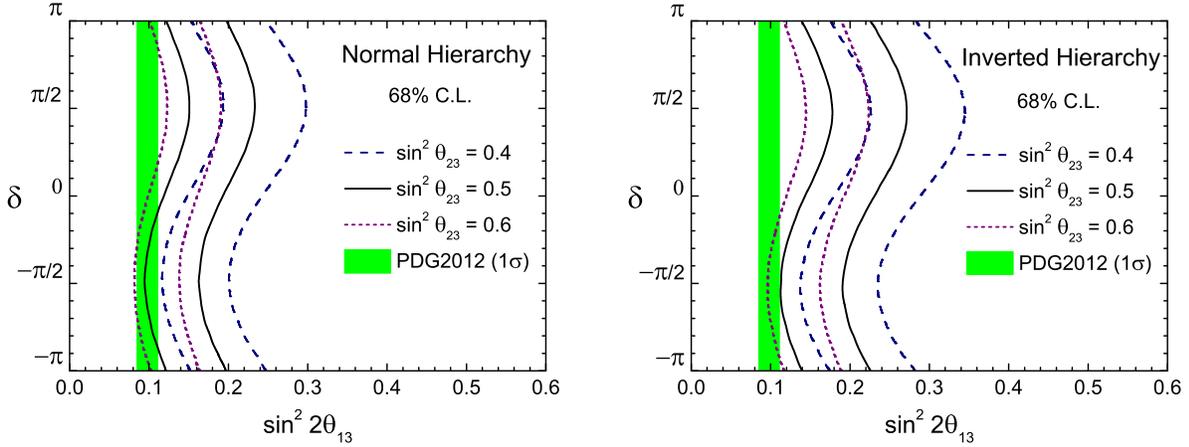}
\caption{The allowed region of $\sin^2 2\theta^{}_{13}$ changing
with the Dirac phase $\delta$ as constrained by the present T2K
neutrino oscillation data (curves) and the reactor antineutrino
oscillation data (green band). A preliminary hint for $\delta \sim
-\pi/2$ can therefore be observed.}
\end{figure*}

\begin{flushleft}
{\it B. Atmospheric neutrino oscillations}
\end{flushleft}

It is well known that the atmospheric $\nu^{}_\mu$,
$\overline{\nu}^{}_\mu$, $\nu^{}_e$ and $\overline{\nu}^{}_e$ events
are produced in the Earth's atmosphere by cosmic rays, mainly
through the decay modes $\pi^+ \to \mu^+ + \nu^{}_\mu$ with $\mu^+
\to e^+ + \nu^{}_e + \overline{\nu}^{}_\mu$ and $\pi^- \to \mu^- +
\overline{\nu}^{}_\mu$ with $\mu^- \to e^- + \overline{\nu}^{}_e +
\nu^{}_\mu$. If there were nothing wrong with the atmospheric
neutrinos that enter and excite an underground detector, they would
have an almost perfect spherical symmetry (i.e., the downward- and
upward-going neutrino fluxes should be equal, $\Phi^{}_e
(\theta^{}_z) = \Phi^{}_e (\pi -\theta^{}_z)$ and $\Phi^{}_\mu
(\theta^{}_z) = \Phi^{}_\mu (\pi -\theta^{}_z)$ with respect to the
zenith angle $\theta^{}_z$). In 1998 the Super-Kamiokande
Collaboration observed an approximate up-down flux symmetry for
atmospheric $\nu^{}_e$ and $\overline{\nu}^{}_e$ events and a
significant up-down flux asymmetry for atmospheric $\nu^{}_\mu$ and
$\overline{\nu}^{}_\mu$ events \cite{SK2}. Such a striking anomaly
could naturally be attributed to $\nu^{}_\mu \to \nu^{}_\tau$ and
$\overline{\nu}^{}_\mu \to \overline{\nu}^{}_\tau$ oscillations for
those upward-going $\nu^{}_\mu$ and $\overline{\nu}^{}_\mu$ events,
since the detector itself was insensitive to $\nu^{}_\tau$ and
$\overline{\nu}^{}_\tau$ events. This observation was actually the
first {\it model-independent} discovery of neutrino flavor
oscillations, and it marked an important turning point in
experimental neutrino physics.

In 2004 the Super-Kamiokande Collaboration did a careful
analysis of the disappearance probability of $\nu^{}_\mu$ and
$\overline{\nu}^{}_\mu$ events as a function of the neutrino flight
length $L$ over the neutrino energy $E$, and observed a clear dip
in the $L/E$ distribution as the first {\it direct} evidence for
atmospheric neutrino oscillations \cite{SK04}. Such a dip is
consistent with the sinusoidal probability of neutrino flavor
oscillations but incompatible with exotic new physics, such as the
neutrino decay and neutrino decoherence scenarios.

It is a great challenge to directly observe the atmospheric
$\nu^{}_\mu \to \nu^{}_\tau$ oscillation because this requires the
neutrino beam energy greater than a threshold of 3.5 GeV, such that
a tau lepton can be produced via the charged-current interaction of
incident $\nu^{}_\tau$ with the target nuclei in the detector. The
Super-Kamiokande data are found to be best described by neutrino
oscillations that include the $\nu^{}_\tau$ appearance in addition
to the overwhelming signature of the $\nu^{}_\mu$ disappearance. In
particular, a neural network analysis of the zenith-angle
distribution of multi-GeV contained events has recently demonstrated
the $\nu^{}_\tau$ appearance effect at the $3.8 \sigma$ level
\cite{SK13}.

\begin{flushleft}
{\it C. Accelerator neutrino oscillations}
\end{flushleft}

If the atmospheric $\nu^{}_\mu$ and $\overline{\nu}^{}_\mu$ flavors
oscillate, a fraction of the accelerator-produced $\nu^{}_\mu$ and
$\overline{\nu}^{}_\mu$ events may also disappear on their way to a
remote detector. This expectation has been confirmed by two
long-baseline neutrino oscillation experiments: K2K \cite{K2K} and
MINOS \cite{MINOS}. Both of them observed a reduction of the
$\nu^{}_\mu$ flux and a distortion of the $\nu^{}_\mu$ energy
spectrum, implying $\nu^{}_\mu \to \nu^{}_\mu$ oscillations. The
most amazing result obtained from the atmospheric and accelerator
neutrino oscillation experiments is $\sin^2 2\theta^{}_{23} \simeq
1$ or equivalently $\theta^{}_{23} \simeq 45^\circ$, which hints at
a likely $\mu$-$\tau$ flavor symmetry in the lepton sector.

Today the most important accelerator neutrino oscillation experiment
is the T2K experiment, which has discovered the $\nu^{}_\mu \to
\nu^{}_e$ appearance oscillations and carred out a precision
measurement of the $\nu^{}_\mu \to \nu^{}_\mu$ disappearance
oscillations. Since its preliminary data were first released in
2011, the T2K experiment has proved to be very successful in
establishing the $\nu^{}_e$ appearance out of a $\nu^{}_\mu$ beam at
the $7.3 \sigma$ level and constraining the neutrino mixing
parameters $\theta^{}_{13}$, $\theta^{}_{23}$ and $\delta$
\cite{T2K1,T2K2,T2K3}. The point is that the leading term of
$P(\nu^{}_\mu \to \nu^{}_e)$ is sensitive to $\sin^2 2\theta^{}_{13}
\sin^2\theta^{}_{23}$, and its sub-leading term is sensitive to
$\delta$ and terrestrial matter effects \cite{Freund}. Fig. 2.1 is
an illustration of the allowed region of $\sin^2 2\theta^{}_{13}$ as
a function of the CP-violating phase $\delta$, as constrained by the
present T2K data \cite{T2K3}. One can observe an unsuppressed value
of $\theta^{}_{13}$ in this plot, together with a preliminary hint
of $\delta$ around $-\pi/2$. The latter is also suggestive of a
possible $\mu$-$\tau$ reflection symmetry, as one will see in
section 3.

In comparison with K2K, MINOS and T2K, the accelerator-based OPERA
experiment was designed to search for the $\nu^{}_\tau$ appearance
in a $\nu^{}_\mu$ beam. After several years of data taking, the
OPERA Collaboration reported five $\nu^{}_\tau$ events in 2015.
These events have marked a discovery of the $\nu^{}_\mu \to
\nu^{}_\tau$ appearance oscillations with the $5.1 \sigma$
significance \cite{O}.

\begin{flushleft}
{\it D. Reactor antineutrino oscillations}
\end{flushleft}

Since the first observation of the $\overline{\nu}^{}_e$ events
emitted from the Savannah River reactor in 1956 \cite{RC}, nuclear
reactors have been playing a special role in neutrino physics. In
particular, $\theta^{}_{12}$ and $\theta^{}_{13}$ have been measured
in the KamLAND \cite{KM} and Daya Bay \cite{DYB0,DYB1} reactor
antineutrino experiments, respectively.

Given the average baseline length $L =180$ km, the KamLAND
experiment was sensitive to the $\Delta m^2_{21}$-driven
$\overline{\nu}^{}_e \to \overline{\nu}^{}_e$ oscillations and could
accomplish a terrestrial test of the large-mixing-angle MSW solution
to the long-standing solar neutrino problem. In fact, it succeeded
in doing so in 2003 \cite{KM}, with an impressive determination of
$\theta^{}_{12} \simeq 34^\circ$. A very striking sinusoidal
behavior of $P(\overline{\nu}^{}_e \to \overline{\nu}^{}_e)$ against
$L/E$ was also observed by the KamLAND Collaboration later on
\cite{KM2}.

The Daya Bay experiment was designed to probe the smallest lepton
flavor mixing angle $\theta^{}_{13}$ with an unprecedented
sensitivity $\sin^2 2\theta^{}_{13} \sim 1\%$ by measuring the
$\Delta m^2_{31}$-driven $\overline{\nu}^{}_e \to
\overline{\nu}^{}_e$ oscillation with a baseline length $L \simeq 2$
km. In 2012 the Daya Bay Collaboration announced a $5.2\sigma$
discovery of $\theta^{}_{13} \neq 0$ and obtained $\sin^2
2\theta^{}_{13} = 0.092 \pm 0.016 ({\rm stat}) \pm 0.005 ({\rm
syst})$ \cite{DYB0}. Some similar but less significant results were
also achieved in the RENO \cite{RENO} and Double Chooz \cite{DC}
reactor antineutrino oscillation experiments.

The Daya Bay experiment has also measured the energy dependence of
the $\overline{\nu}^{}_e$ disappearance and seen a nearly full
oscillation cycle against $L/E$ \cite{DYB2}. The updated result
$\sin^2 2\theta^{}_{13} = 0.090^{+0.008}_{-0.009}$ is
obtained in the three-flavor framework. A combination of the Daya
Bay measurement of $\theta^{}_{13}$ and the T2K measurement of a
relatively strong $\nu^{}_\mu \to \nu^{}_e$ appearance signal
\cite{T2K3} drives a slight but intriguing preference for $\delta
\sim -\pi/2$, as shown in Fig. 2.1. In addition, the relatively large
$\theta^{}_{13}$ is so encouraging that the next-generation
precision experiments should be able to determine the neutrino mass
ordering and the CP-violating phase $\delta$ in the foreseeable
future \cite{Wang}.

\subsection{The observed pattern of the PMNS matrix}

In the three-flavor scheme there are six independent parameters
which govern the behaviors of neutrino oscillations: two neutrino
mass-squared differences $\Delta m^2_{21}$ and $\Delta m^2_{31}$,
three flavor mixing angles $\theta^{}_{12}$, $\theta^{}_{13}$ and
$\theta^{}_{23}$, and the Dirac CP-violating phase $\delta$. Those
successful atmospheric, solar, accelerator and reactor neutrino
oscillation experiments discussed above allow us to determine
$\Delta m^2_{21}$, $|\Delta m^2_{31}|$, $\theta^{}_{12}$,
$\theta^{}_{13}$ and $\theta^{}_{23}$ to a good degree of accuracy.
The ongoing and future neutrino oscillation experiments are expected
to fix the sign of $\Delta m^2_{31}$ and pin down the value of
$\delta$.
\begin{table}[t]
\caption{The best-fit values, together with the 1$\sigma$
and 3$\sigma$ intervals, for the six three-flavor neutrino
oscillation parameters from a global analysis of current
experimental data \cite{Fogli}.} \vspace{0.1cm}
\begin{indented}
\item[]\begin{tabular}{cccc} \br Parameter & Best fit &
1$\sigma$ range & 3$\sigma$ range \\ \mr \multicolumn{4}{c}{Normal
mass ordering $(m^{}_1 < m^{}_2 < m^{}_3$)} \\ \mr \vspace{0.1cm}
$\Delta m^2_{21}/10^{-5} ~{\rm eV}^2$ & $7.54$  & 7.32 --- 7.80 &
6.99 --- 8.18 \\ \vspace{0.1cm}
$\Delta m^2_{31}/10^{-3} ~ {\rm eV}^2$~ & $2.47$ & 2.41 --- 2.53 &
2.26 --- 2.65 \\ \vspace{0.1cm}
$\sin^2\theta_{12}/10^{-1}$ & $3.08$ & 2.91 --- 3.25 & 2.59 --- 3.59
\\ \vspace{0.1cm}
$\sin^2\theta_{13}/10^{-2}$ & $2.34$ & 2.15 --- 2.54 & 1.76 --- 2.95
\\ \vspace{0.1cm}
$\sin^2\theta_{23}/10^{-1}$ & $4.37$  & 4.14 --- 4.70 & 3.74 ---
6.26 \\ \vspace{0.1cm}
$\delta/\pi$ &  $1.39$ & 1.12 --- 1.77 & 0.00
--- 2.00 \\ \br
\multicolumn{4}{c}{Inverted mass ordering $(m^{}_3 < m^{}_1 <
m^{}_2$)} \\ \mr \vspace{0.1cm}
$\Delta m^2_{21}/10^{-5} ~{\rm eV}^2$ & $7.54$  & 7.32 --- 7.80 &
6.99 --- 8.18 \\ \vspace{0.1cm}
$\Delta m^2_{13}/10^{-3} ~ {\rm eV}^2$~ & $2.42$ & 2.36 --- 2.48 &
2.22 --- 2.60 \\ \vspace{0.1cm}
$\sin^2\theta_{12}/10^{-1}$ & $3.08$ & 2.91 --- 3.25 & 2.59 --- 3.59
\\ \vspace{0.1cm}
$\sin^2\theta_{13}/10^{-2}$ & $2.40$ & 2.18 --- 2.59 & 1.78 --- 2.98
\\ \vspace{0.1cm}
$\sin^2\theta_{23}/10^{-1}$ & $4.55$  & 4.24 --- 5.94 & 3.80 ---
6.41 \\ \vspace{0.1cm}
$\delta/\pi$ &  $1.31$ & 0.98 --- 1.60 & 0.00
--- 2.00 \\ \br
\end{tabular}
\end{indented}
\end{table}
\begin{figure*}
\centering
\includegraphics[width=.9\textwidth]{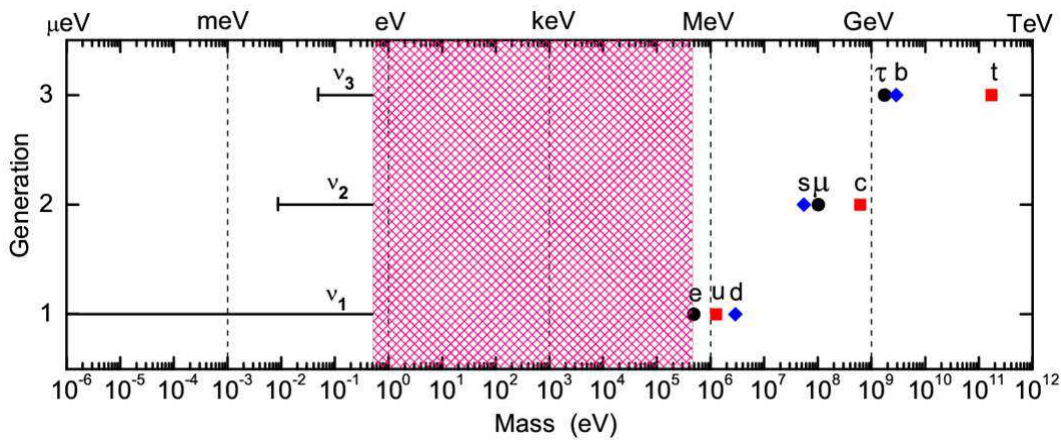}
\caption{A schematic illustration of the fermion mass spectrum of
the SM at the electroweak scale, where the three neutrino masses are
assumed to have a normal ordering.}
\end{figure*}

A global analysis of the available data on solar (SNO,
Super-Kamiokande, Borexino), atmospheric (Super-Kamiokande),
accelerator (MINOS, T2K) and reactor (KamLAND, Daya Bay, RENO)
neutrino (or antineutrino) oscillations has been done by several
groups \cite{Fogli,Valle,GG}. Here we quote the main results
obtained by Capozzi {\it et al} \cite{Fogli} in Table 2.1
\footnote{The notations $\delta m^2 \equiv m^2_2 - m^2_1$ and
$\Delta m^2 \equiv m^2_3 - (m^2_1 + m^2_2)/2$ have been used in Ref.
\cite{Fogli}. Their relations with $\Delta m^2_{21}$ and $\Delta
m^2_{31}$ are $\Delta m^2_{21} = \delta m^2$ and $\Delta m^2_{31} =
\Delta m^2 + \delta m^2/2$.}.
Some immediate comments are in order.
\begin{itemize}
\item     The unfixed sign of $\Delta m^2_{31}$ implies two
possible neutrino mass orderings: normal ($m^{}_1 < m^{}_2 <
m^{}_3$) or inverted ($m^{}_3 < m^{}_1 < m^{}_2$). Here ``normal"
means that the mass ordering of the neutrinos is parallel to that of
the charged leptons or the quarks of the same charge (i.e., $m^{}_e
\ll m^{}_\mu \ll m^{}_\tau$, $m^{}_u \ll m^{}_c \ll m^{}_t$ and
$m^{}_d \ll m^{}_s \ll m^{}_b$, as shown in Fig. 2.2 \cite{keV}). A
good theoretical reason for $\Delta m^2_{31} >
0$ or $\Delta m^2_{13} > 0$ has been lacking
\footnote{If the neutrino mass ordering is finally found to be
inverted, one may always reorder it to $m^\prime_1 < m^\prime_2 <
m^\prime_3$ by setting $m^\prime_1 = m^{}_3$, $m^\prime_2 = m^{}_1$
and $m^\prime_3 = m^{}_2$, equivalent to a transformation
$(\nu^{}_1, \nu^{}_2, \nu^{}_3) \to (\nu^\prime_2, \nu^\prime_3,
\nu^\prime_1)$. In this case the elements of $U$ must be reordered
in a self-consistent way: $U \to U^\prime$, in which
$U^\prime_{\alpha 1} = U^{}_{\alpha 3}$, $U^\prime_{\alpha 2} =
U^{}_{\alpha 1}$ and $U^\prime_{\alpha 3} = U^{}_{\alpha 2}$ (for
$\alpha = e, \mu, \tau$) \cite{XingMO}. Of course, such a reordering
treatment does not change any physical content of massive
neutrinos.}.

\item     The output values of $\theta^{}_{13}$, $\theta^{}_{23}$
and $\delta$ in such a global fit are more or less sensitive to the
sign of $\Delta m^2_{31}$. That is why it is crucial to determine
the neutrino mass hierarchy in the ongoing and upcoming reactor
(JUNO \cite{JUNO}), atmospheric (PINGU \cite{PINGU}) and accelerator
(NO$\nu$A \cite{NOVA} and LBNE \cite{LBNE}) neutrino oscillation
experiments.

\item     The hint $\delta \neq 0$ (or $\pi$) at the
$1\sigma$ level is preliminary but encouraging, simply because it
implies a potential effect of leptonic CP violation which is likely
to show up in some long-baseline neutrino oscillation experiments in
the foreseeable future. In particular, the best-fit value of
$\delta$ is quite close to $-\pi/2$, implying an approximate
$\mu$-$\tau$ reflection symmetry as one can see later on.

\item     The possibility of $\theta^{}_{23} = \pi/4$ cannot be
excluded at the $1\sigma$ or $2\sigma$ level, and hence a more
precise determination of $\theta^{}_{23}$ is desirable so as to
resolve its octant. Since $\theta^{}_{23} = \pi/4$ is a natural
consequence of the $\mu$-$\tau$ flavor symmetry in the neutrino
sector, the positive or negative deviation of $\theta^{}_{23}$ from
$\pi/4$ may have profound implications on the structure of the PMNS
matrix $U$ and the building of viable neutrino mass models.
\end{itemize}
In short, the sign of $\Delta m^2_{31}$, the octant of
$\theta^{}_{23}$ and the value of $\delta$ remain unknown. Whether
these three open issues are potentially correlated with one another
is an intriguing question.

Combining Eqs. (2.8) and (2.9) with Table 2.1, we obtain the
remarkable result
\begin{eqnarray}
|U^{}_{\mu 1}| \simeq |U^{}_{\tau 1}| \; , ~~ |U^{}_{\mu 2}|
\simeq |U^{}_{\tau 2}| \; , ~~ |U^{}_{\mu 3}| \simeq |U^{}_{\tau
3}| \;
\end{eqnarray}
to a reasonably good degree of accuracy. This result becomes
more transparent when the allowed ranges of the nine PMNS matrix
elements are explicitly given at the $3\sigma$ level:
\begin{eqnarray}
\left| U\right| \simeq \pmatrix{ 0.79 - 0.85 & 0.50
- 0.59 & 0.13 - 0.17 \cr 0.19 - 0.56 & 0.41
- 0.74 & 0.60 - 0.78 \cr 0.19 - 0.56 & 0.41
- 0.74 & 0.60 - 0.78 \cr} \;
\end{eqnarray}
in the $\Delta m^2_{31} >0$ case; or
\begin{eqnarray}
\left| U\right| \simeq \pmatrix{ 0.89 - 0.85 & 0.50
- 0.59 & 0.13 - 0.17 \cr 0.19 - 0.56 & 0.40
- 0.73 & 0.61 - 0.79 \cr 0.20 - 0.56 & 0.41
- 0.74 & 0.59 - 0.78 \cr} \;
\end{eqnarray}
in the $\Delta m^2_{31} <0$ case. In either case the pattern of $U$
is significantly different from that of quark flavor mixing
described by the Cabibbo-Kobayashi-Maskawa (CKM) matrix $V^{}_{\rm
CKM}$ \cite{C,Kobayashi}. The latter is close to the identity matrix
because its maximal flavor mixing angle is the Cabibbo angle
$\theta^{}_{\rm C} \simeq 13^\circ$ \cite{PDG}.

In fact, the equality $|U^{}_{\mu i}| = |U^{}_{\tau i}|$ (for
$i=1,2,3$) exactly holds if either of the following two sets of
conditions can be satisfied \cite{XZ08}:
\begin{eqnarray}
|U^{}_{\mu i}| = |U^{}_{\tau i}| \;\;\; \Longleftrightarrow \;\;
\left\{
\begin{array}{l}
\theta^{}_{23} = \displaystyle
\frac{\pi}{4} \; , ~~ \theta^{}_{13} = 0 \; ; \\
\hspace{0cm} {\rm or} \\
\theta^{}_{23} = \displaystyle \frac{\pi}{4} \; , ~~ \delta = \pm
\frac{\pi}{2} \; .
\end{array}
\right.
\label{2.17}
\end{eqnarray}
The possibility of $\theta^{}_{13} = 0$ has been ruled out, but
$\theta^{}_{23} =\pi/4$ and $\delta = -\pi/2$ are both allowed at
the $1\sigma$ or $2\sigma$ level (and $\delta = \pi/2$ is allowed at
the $3\sigma$ level) as shown in Table 2.1. That is why we claim
that there must be an approximate $\mu$-$\tau$ flavor symmetry
behind the observed pattern of $U$. In this case the $\mu$-$\tau$
symmetry is expected to be a good starting point for model building,
no matter what larger flavor groups it belongs to. On the
experimental side, it is imperative to measure $\theta^{}_{23}$ and
$\delta$ as accurately as possible, so as to fix the strength of
$\mu$-$\tau$ symmetry breaking.

At this point it is also worth mentioning that three of the six
unitarity triangles of $U$ (the so-called Majorana triangles)
in the complex plane, defined by the three orthogonality relations
\begin{eqnarray}
\triangle^{}_1 : & \hspace{0.2cm} & U^{}_{e 2} U^*_{e 3} + U^{}_{\mu
2} U^*_{\mu 3} + U^{}_{\tau 2} U^*_{\tau 3} = 0 \; ,
\nonumber \\
\triangle^{}_2 : & \hspace{0.2cm} & U^{}_{e 3} U^*_{e 1} + U^{}_{\mu
3} U^*_{\mu 1} + U^{}_{\tau 3} U^*_{\tau 1} = 0 \; ,
\nonumber \\
\triangle^{}_3 : & \hspace{0.2cm} & U^{}_{e 1} U^*_{e 2} + U^{}_{\mu
1} U^*_{\mu 2} + U^{}_{\tau 1} U^*_{\tau 2} = 0 \; ,
\end{eqnarray}
will become the {\it isosceles} triangles provided $\theta^{}_{23} =
\pi/4$ and $\delta = \pm\pi/2$ are satisfied. This interesting
property, which can be intuitively seen in Fig. 2.3 \cite{Zhu}, is
another reflection of the $\mu$-$\tau$ flavor symmetry as an
instructive guideline for the study of leptonic CP violation. We
shall go into details of this flavor symmetry and its breaking
mechanisms in the subsequent sections.
\begin{figure*}
\centering
\includegraphics[width=.86\textwidth]{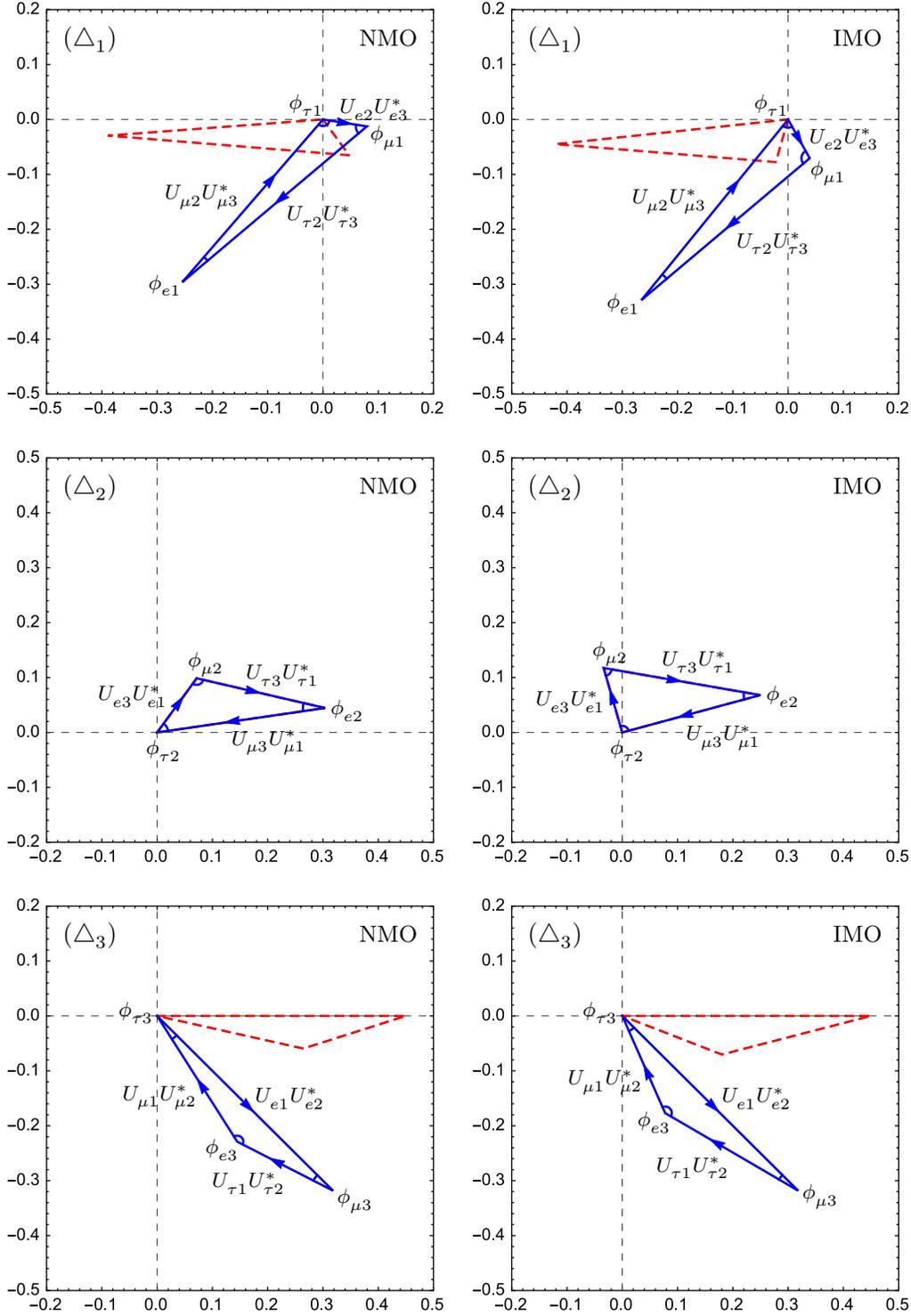}
\vspace{0.2cm} \caption{The real shapes and orientations of three
Majorana unitarity triangles in the complex plane, plotted by
assuming $\left(\rho, \sigma\right) = \left(0, \pi/4\right)$ and
inputting the best-fit values of $\theta^{}_{12}$, $\theta^{}_{13}$,
$\theta^{}_{23}$ and $\delta$ \cite{GG} in the normal mass ordering
(NMO: left panel) or inverted mass ordering (IMO: right panel) case.
The dashed triangles correspond to $\left(\rho, \sigma\right) =
\left(0, 0\right)$ for comparison.}
\end{figure*}

\def\thefootnote{\arabic{footnote}}
\setcounter{footnote}{0}
\setcounter{equation}{0}
\setcounter{table}{0}
\setcounter{figure}{0}

\section{An overview of the $\mu$-$\tau$ flavor symmetry}

The approximate $\mu$-$\tau$ flavor symmetry displayed in the PMNS
matrix $U$ is also expected to show up in the mass matrix of either
charged leptons or neutrinos. From a model-building point of view,
one needs to know what forms of $M^{}_{l}$ and $M^{}_{\nu}$ can
give rise to the observed pattern of lepton flavor mixing as well as
the correct mass spectra of $(e, \mu, \tau)$ and $(\nu^{}_1, \nu^{}_2,
\nu^{}_3)$. Since $U = O^\dagger_l O^{}_\nu$ is a measure of
the mismatch between $O^{}_{l}$ and $O^{}_{\nu}$ which have been
used to diagonalize $M^{}_l M^\dagger_l$ and $M^{}_\nu$ in Eq.
(2.3), we prefer to work in the basis where $O^{}_{l}$ equals the
identity matrix (i.e., $M^{}_{l}$ itself is simply a diagonal
matrix). In this basis the neutrino mixing effects are completely
determined by the structure of $M^{}_{\nu}$. The latter can be
reconstructed in terms of $U$ and the neutrino masses as shown
in Eq. (\ref{2.10}). The six independent elements of $M^{}_\nu$
turn out to be
\begin{eqnarray}
M^{}_{\alpha \beta} \equiv \langle m\rangle^{}_{\alpha \beta} =
\sum^{3}_{i=1} m^{}_i U^{}_{\alpha i} U^{}_{\beta i} \; ,
\end{eqnarray}
where $\alpha$ and $\beta$ run over $e, \mu, \tau$. Of course,
$M^{}_\nu$ totally consists of nine independent physical parameters.

With the help of the standard parametrization of $U$ in Eqs.
(2.7)---(2.9), we express the six effective neutrino mass terms as
follows:
\begin{eqnarray}
\langle m\rangle^{}_{ee} & = & \overline m^{}_1 c^2_{12} c^2_{13}+
\overline m^{}_2 s^2_{12} c^2_{13} + m^{}_3 \tilde s^{*2}_{13} \; ,
\nonumber \\
\langle m\rangle^{}_{\mu\mu} & = & \overline m^{}_1 \left(s^{}_{12}
c^{}_{23} + c^{}_{12} \tilde s^{}_{13} s^{}_{23} \right)^2
\nonumber \\
&& + \overline m^{}_2 \left(c^{}_{12} c^{}_{23} - s^{}_{12} \tilde
s^{}_{13} s^{}_{23} \right)^2 + m^{}_3 c^2_{13} s^2_{23} \; ,
\nonumber \\
\langle m\rangle^{}_{\tau\tau} & = & \overline m^{}_1
\left(s^{}_{12} s^{}_{23} - c^{}_{12} \tilde s^{}_{13}
c^{}_{23}\right)^2
\nonumber \\
&& + \overline m^{}_2 \left(c^{}_{12} s^{}_{23} + s^{}_{12} \tilde
s^{}_{13} c^{}_{23} \right)^2 + m^{}_3 c_{13}^2 c_{23}^2 \; ,
\nonumber \\
\langle m\rangle^{}_{e\mu} & = & -\overline m^{}_1 c^{}_{12}
c^{}_{13} \left(s^{}_{12} c^{}_{23} + c^{}_{12} \tilde s^{}_{13}
s^{}_{23} \right)
\nonumber \\
&& + \overline m^{}_2 s^{}_{12} c^{}_{13} \left( c^{}_{12} c^{}_{23} -
s^{}_{12} \tilde s^{}_{13} s^{}_{23} \right)
\nonumber \\
&& + m^{}_3 c^{}_{13} \tilde s^{*}_{13} s^{}_{23} \; ,
\nonumber \\
\langle m\rangle^{}_{e\tau} & = & \overline m^{}_1 c^{}_{12} c^{}_{13}
\left(s^{}_{12} s^{}_{23} - c^{}_{12} \tilde s^{}_{13} c^{}_{23}\right)
\nonumber \\
&& -\overline m^{}_2 s^{}_{12} c^{}_{13} \left(c^{}_{12} s^{}_{23} +
s^{}_{12}\tilde s^{}_{13} c^{}_{23} \right)
\nonumber \\
&& + m^{}_3 c^{}_{13} \tilde s^{*}_{13}c^{}_{23} \; ,
\nonumber \\
\langle m\rangle^{}_{\mu\tau} & = & -\overline m^{}_1 \left(s^{}_{12}
s^{}_{23} - c^{}_{12} \tilde s^{}_{13} c^{}_{23} \right) \left(
s^{}_{12} c^{}_{23}+ c^{}_{12} \tilde s^{}_{13} s^{} _{23} \right)
\nonumber \\
&& -\overline m_2^{} \left(c^{}_{12} s^{}_{23} +
s^{}_{12}\tilde  s^{}_{13} c^{}_{23} \right) \left(c^{}_{12}
c^{}_{23} - s^{}_{12} \tilde s^{}_{13} s^{}_{23} \right)
\nonumber\\
&& + m^{}_3 c^2_{13} c^{}_{23} s^{}_{23} \; ,
\end{eqnarray}
where $\overline{m}^{}_1 \equiv m^{}_1 e^{2{\rm i} \rho}$,
$\overline{m}^{}_2 \equiv m^{}_2 e^{2{\rm i} \sigma}$ and
$\tilde{s}^{}_{13} \equiv s^{}_{13} e^{{\rm i} \delta}$ are defined
for the sake of simplicity. The numerical profiles of $|\langle
m\rangle^{}_{\alpha\beta}|$ versus the lightest neutrino mass (i.e.,
$m^{}_{1}$ in the normal hierarchy (NH) case or $m^{}_{3}$ in the
inverted hierarchy (IH) case) are illustrated in Fig. 3.1 \cite{YL},
where the best-fit values of two neutrino mass-squared differences
and three flavor mixing angles presented in Table 2.1 are input
while the Dirac and Majorana phases are allowed to vary in the
intervals $[0,2\pi)$ and $[0,\pi)$, respectively. It is obvious that
the relations $|\langle m\rangle^{}_{\mu\mu}| \simeq |\langle
m\rangle^{}_{\tau\tau}|$ and $|\langle m\rangle^{}_{e\mu}|\simeq
|\langle m\rangle^{}_{e\tau}|$ hold in most of the parameter space.
Instead of regarding this observation as the reflection of a kind of
flavor ``anarchy" \cite{anarchy}, we treat it seriously as an
indication of the $\mu$-$\tau$ symmetry which can always be embedded
in a much larger flavor symmetry group.

The $\mu$-$\tau$ flavor symmetry itself is so powerful that it
constrains the texture of $M^{}_\nu$ by establishing some equalities
or linear relations among its six independent elements. In other
words, the $\mu$-$\tau$ symmetry has been identified as the {\it
minimal} (and thus most convincing) flavor symmetry in the neutrino
sector to give rise to $|\langle m\rangle^{}_{\mu\mu}| = |\langle
m\rangle^{}_{\tau\tau}|$ and $|\langle m\rangle^{}_{e\mu}|= |\langle
m\rangle^{}_{e\tau}|$. It is therefore expected to be a successful
bridge between neutrino phenomenology and model building. In this
section we shall mainly describe the salient features of
$\mu$-$\tau$ permutation and reflection symmetries and outline some
typical ways to softly break them. The larger flavor symmetry groups
and model-building exercises will be discussed in sections 4 and 5,
respectively.
\begin{figure*}
\centering
\includegraphics[width=6.1in]{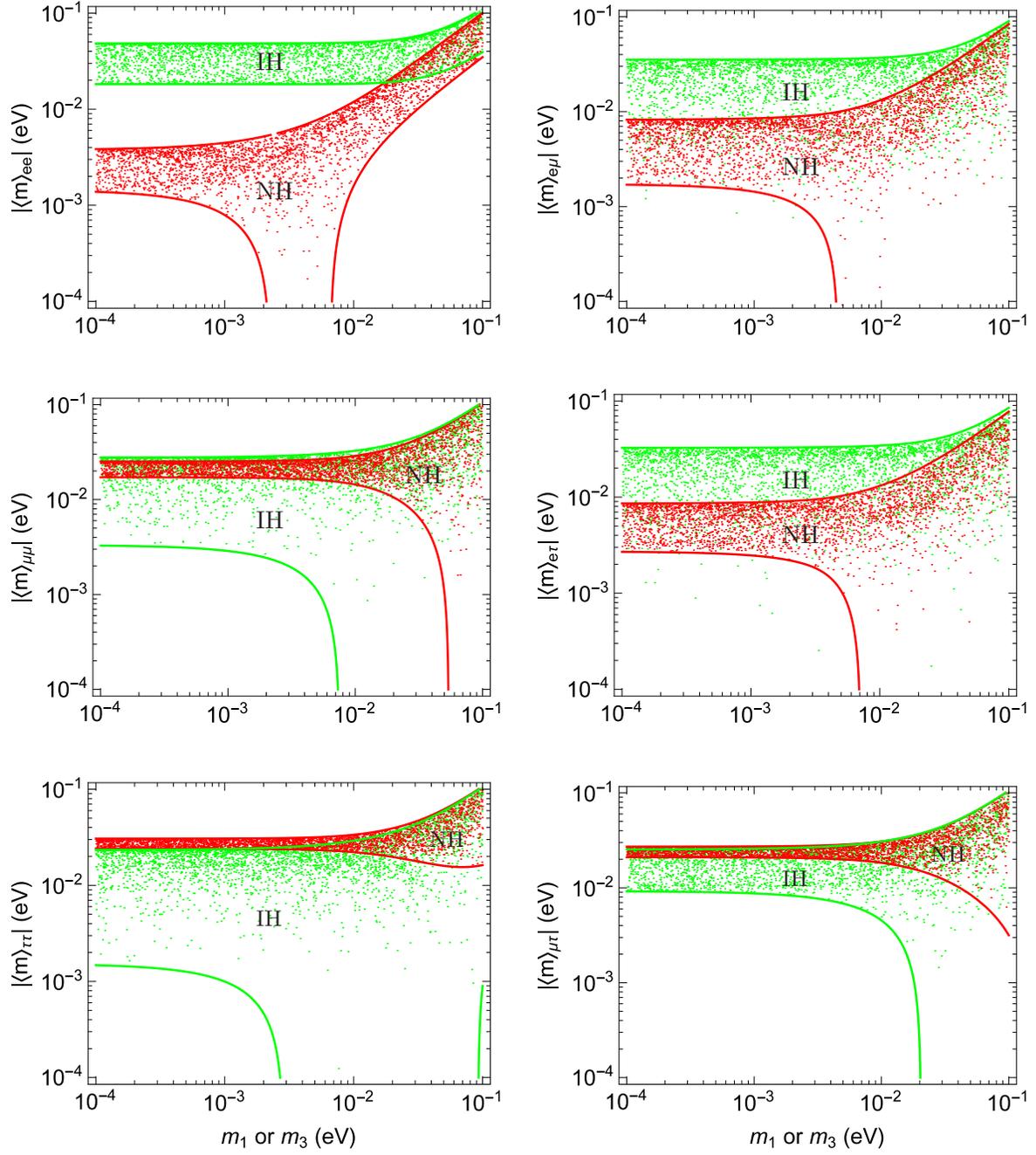}
\caption{The profiles of $|\langle m\rangle^{}_{\alpha \beta}|$
versus $m^{}_{1}$ (normal hierarchy or NH: red region) or $m^{}_{3}$
(inverted hierarchy or IH: green region), where $\delta \in [0,
2\pi)$, $\rho \in [0, \pi)$, $\sigma \in [0, \pi)$, and the best-fit
values of $\Delta m^2_{21}$, $\Delta m^2_{31}$, $\theta^{}_{12}$,
$\theta^{}_{13}$ and $\theta^{}_{23}$ listed in Table 2.1 are
input.}
\end{figure*}

\subsection{The $\mu$-$\tau$ permutation symmetry}

Let us begin with the $\mu$-$\tau$ permutation symmetry which has
been defined in Eq. (1.1) \cite{MT-P,MT-P2,MT-P3,MT-P4}.
Historically, the discussion about this simple flavor symmetry was
motivated by the experimental facts $\theta^{}_{23} \sim \pi/4$ and
$\sin^2_{}{2\theta^{}_{13}}<0.18$ \cite{Chooz}, which point to
$\theta^{}_{23} = \pi/4$ and $\theta^{}_{13} =0$ as an ideal
possibility. In this case the PMNS matrix $U$ is  greatly simplified
and may take the following form in a chosen phase convention:
\begin{eqnarray}
U = \frac{1}{\sqrt2}\pmatrix{\sqrt2 \ c^{}_{12} & \sqrt2 \ s^{}_{12}
& 0 \cr -s^{}_{12} & c^{}_{12} & -\kappa \cr -\kappa s^{}_{12} &
\kappa c^{}_{12} & 1} \; ,
\end{eqnarray}
where $\kappa=\pm 1$ will manifest itself in the corresponding
neutrino mass matrix $M^{}_\nu$. The latter can be easily
reconstructed as follows:
\begin{eqnarray}
M^{}_{\nu} = \frac{1}{2}\pmatrix{ 2m^{}_{11} & -\sqrt{2} \ m^{}_{12}
& -\kappa\sqrt{2} \ m^{}_{12} \cr \cdots & m^{}_{22} +m^{}_{3} &
\kappa\left(m^{}_{22}-m^{}_{3}\right) \cr \cdots & \cdots &
m^{}_{22}+m^{}_{3}} \; ,
\end{eqnarray}
in which
\begin{eqnarray}
m^{}_{11} = \overline m^{}_{1}c^{2}_{12} +
\overline m^{}_{2}s^{2}_{12} \; ,
\nonumber \\
m^{}_{12} = \left(\overline m^{}_{1} - \overline m^{}_{2}\right)
c^{}_{12} s^{}_{12} \; ,
\nonumber \\
m^{}_{22} = \overline m^{}_{1} s^{2}_{12} + \overline
m^{}_{2}c^{2}_{12} \; .
\end{eqnarray}
It is clear that the elements of $M^{}_{\nu}$ satisfy the relations
$M^{}_{e \mu} = \kappa M^{}_{e\tau}$ and $M^{}_{\mu \mu}= M^{}_{\tau
\tau}$. So $M_{\nu}^{}$ is invariant under the $\mu$-$\tau$
permutation operation $\nu^{}_{\mu} \leftrightarrow \kappa
\nu^{}_{\tau}$, which is represented by the transformation matrix
\begin{eqnarray}
S^{\pm} = \pmatrix{ 1 & 0 & 0 \cr 0 & 0 & \kappa \cr 0 & \kappa & 0} \; ,
\end{eqnarray}
where $S^\pm$ corresponds to $\kappa = \pm 1$. So the most general
neutrino mass matrix respecting the $\mu$-$\tau$ permutation
symmetry can be parameterized as
\begin{eqnarray}
M^{}_{\nu} = \pmatrix{
A & B &\kappa B \cr B & C & D \cr \kappa B & D & C } \; ,
\end{eqnarray}
where $A$, $B$, $C$ and $D$ are in general complex. For simplicity,
here these four parameters are all taken to be real \cite{Review2}.
Both $\kappa = \pm 1$ lead to $\theta^{}_{23}=\pi/4$,
as the phases or signs of the elements of $U$ can always be
rearranged to assure its mixing angles to lie in the first quadrant.
This kind of rephasing is implemented by redefining the phases of
relevant lepton fields. In the following we shall focus on the
$\kappa = +1$ case. Note that the $\mu$-$\tau$ permutation symmetry
has no definite prediction for $\theta^{}_{12}$. In fact,
\begin{eqnarray}
\tan{2\theta^{}_{12}} = \frac{2\sqrt{2} \ B}{C+ D-A} \; ,
\end{eqnarray}
and the three neutrino mass eigenvalues of $M^{}_\nu$ are given by
\begin{eqnarray}
m^{}_{1} = \frac{1}{2}\left[A+C+ D
-\sqrt{\left(A-C-D\right)^2+8B^2}\right] \; ,
\nonumber \\
m^{}_{2} = \frac{1}{2}\left[A+C+ D
+\sqrt{\left(A-C-D\right)^2+8B^2}\right] \; ,
\nonumber \\
m^{}_{3} = C-D \; .
\end{eqnarray}
Depending on the specific values of the relevant free parameters,
the neutrino mass spectrum can be either normal ($\Delta m^{2}_{31} >0$)
or inverted ($\Delta m^{2}_{31} <0$).

The neutrino mass matrix $M^{}_\nu$ in Eq. (3.7) will become more
predictive if its free parameters are further constrained. Let us
consider two simple but instructive cases for illustration:
\begin{itemize}
\item     If $C+D-A=B$ is assumed, then Eq. (3.8) leads us to the
prediction $\sin{\theta^{}_{12}}=1/\sqrt{3}$ or equivalently
$\theta^{}_{12} \simeq 35.3^\circ$. In this case we are left with
the well-known tri-bimaximal (TB) mixing pattern of massive
neutrinos \cite{TB1,TB2}:
\begin{eqnarray}
U^{}_{\rm TB} = \frac{1}{\sqrt6}\pmatrix{ 2& \sqrt{2} &0 \cr -1 &
\sqrt2 & -\sqrt3 \cr -1 & \sqrt2 & \sqrt3} \; ,
\end{eqnarray}
together with the mass eigenvalues $m^{}_{1}=C+ D-2B$, $m^{}_{2}=C+
D+B$ and $m^{}_{3}=C- D$.

\item     If $C+ D-A=0$ is taken, then Eq. (3.8)
yields $\theta^{}_{12} = \theta^{}_{23}=\pi/4$. In this case one
arrives at another special pattern of the neutrino mixing matrix ---
the so-called bi-maximal (BM) flavor mixing pattern \cite{BM1,BM2}:
\begin{eqnarray}
U^{}_{\rm BM} = \frac{1}{2}\pmatrix{ \sqrt{2} & \sqrt{2} & 0 \cr -1
& 1 & -\sqrt{2}\cr -1 & 1 & \sqrt{2}} \; ,
\end{eqnarray}
as well as the mass eigenvalues $m^{}_{1}=C+ D-\sqrt{2} B$,
$m^{}_{2}=C+ D+ \sqrt{2} B$ and $m^{}_{3}=C- D$.
\end{itemize}
Considering other constraints on the free parameters of $M^{}_\nu$
in Eq. (3.7), one may similarly derive other neutrino mixing
patterns which maintain $\theta^{}_{23}=\pi/4$ and
$\theta^{}_{13}=0$ but predict different values of $\theta^{}_{12}$
(e.g., the golden-ratio \cite{GR} and hexagonal
\cite{Hexagonal1,Hexagonal2} patterns). In such examples all the
three flavor mixing angles are constants, which have nothing to do
with the neutrino masses. It is possible to link $\theta^{}_{12}$
with the ratio of $m^{}_1$ to $m^{}_2$ in the following way
\cite{FX2006,SDP}:
\begin{eqnarray}
\tan\theta^{}_{12} = \sqrt{\frac{m^{}_1}{m^{}_2}} \; ,
\end{eqnarray}
if $A =0$ (i.e., $\langle m\rangle^{}_{ee} =0$) holds. On the other
hand, taking $C=D$ will lead to $m^{}_{3}=0$, as one can see from
Eq. (3.9). Such a special neutrino mass spectrum is not in conflict
with the present experimental data, and it may have some interesting
consequences in neutrino phenomenology \cite{Zhu}.

At this point let us point out that the canonical $\mu$-$\tau$
permutation symmetry can be generalized in a way described by the
transformation matrix \cite{GJKLST}
\begin{eqnarray}
S(\theta,\phi) = \pmatrix{ 1 & 0 & 0 \cr 0 & \cos{2 \theta} & \sin{2
\theta} \ e^{-2 {\rm i} \phi} \cr 0 & \sin{2 \theta} \ e^{2 {\rm i}
\phi} & - \cos{ 2 \theta}} \; .
\end{eqnarray}
Obviously, this symmetry will be reduced to the canonical one if
$\theta = \pi/4$ and $\phi = 0$ are taken. The requirement of
$M^{}_\nu$ being invariant under $S(\theta, \phi)$ imposes the
following constraints on its elements:
\begin{eqnarray}
M^{}_{e \tau} - e^{2 {\rm i} \phi} M^{}_{e \mu} \tan{ \theta } = 0
\; ,
\nonumber \\
\left(e^{-2 {\rm i} \phi} M^{}_{\tau \tau} - e^{2 {\rm i} \phi}
M^{}_{\mu \mu}\right) + 2 M^{}_{\mu \tau} \cos{ 2 \theta } = 0 \; .
\end{eqnarray}
Such an $M^{}_\nu$ will result in a flavor mixing matrix $U$
consisting of $\theta^{}_{13}=0$ and $\theta^{}_{23} = \theta$.
The point is that $S(\theta, \phi)$ has an eigenvector
\begin{eqnarray}
u=\pmatrix{0 \cr e^{-{\rm i} \phi} \sin{\theta} \cr -e^{{\rm i} \phi}
\cos{\theta}} \;
\end{eqnarray}
with the eigenvalue $-1$. Hence the relation $S(\theta, \phi) u= -u$
together with $[S(\theta, \phi)]^\dagger M^{}_{\nu} [S(\theta,
\phi)]^* = M^{}_\nu$ will lead us to $S(\theta, \phi) (M^{}_\nu
u^*)= - (M^{}_\nu u^*)$, indicating that $M^{}_\nu u^*$ is
proportional to $u$. One may therefore draw the conclusion that $u$
constitutes one column of $U$. On the other hand, it is
straightforward to show that the reconstructed $M^{}_\nu$ via Eq.
(2.10) in terms of $U$ featuring $\theta^{}_{13}=0$ satisfies Eq.
(3.14). In other words, $M^{}_\nu$ always assumes an $S(\theta,
\phi)$ symmetry of the form given in Eq. (3.13) for the case of
$\theta^{}_{13}=0$. But in what follows we shall focus on the
canonical $\mu$-$\tau$ permutation symmetry, simply because it is
much simpler and favored by the experimental result $\theta^{}_{23}
\simeq \pi/4$.

Finally, it is worth pointing out that there is a remarkable
variant of the $\mu$-$\tau$ permutation symmetry --- the so-called
$\mu$-$\tau$ permutation {\it antisymmetry} \cite{AS,AS2}. When the
latter is concerned, the neutrino mass matrix satisfies the
transformation
\begin{eqnarray}
S^{+} M^{\rm AS}_{\nu} S^{+}=- M^{\rm AS}_{\nu}\;,
\end{eqnarray}
and thus it has a special form
\begin{eqnarray}
M^{\rm AS}_{\nu} = \pmatrix{
0 & B & -B \cr B & C & 0 \cr -B & 0 & -C} \; .
\end{eqnarray}
The unitary matrix used to diagonalize this special mass matrix is
given by
\begin{eqnarray}
U^{}_{\rm AS} = \frac{1}{2N} \pmatrix{
2 B^* & 2 B^* & -2C \cr
C^* + {\rm i}N &  C^* - {\rm i}N &  2B \cr
C^* - {\rm i}N &  C^* + {\rm i}N &  2B} \; ,
\end{eqnarray}
where $N = \sqrt{2|B|^2+|C|^2}$. In this case $\theta^{}_{12}$ and
$\theta^{}_{23}$ are both equal to $\pi/4$, and a finite value of
$\theta^{}_{13}$ is allowed. The three mass eigenvalues of $M^{\rm
AS}_\nu$ are ${\rm i} N$, $-{\rm i} N$ and 0, respectively. To fit
current neutrino oscillation data, one has to introduce proper
perturbations to $M^{\rm AS}_\nu$ so as to break its $\mu$-$\tau$
permutation antisymmetry and arrive at an acceptable neutrino mass
spectrum \cite{AS2}. Note that an arbitrary Majorana neutrino mass
matrix can always be decomposed into two parts: the first part
respects the $\mu$-$\tau$ permutation symmetry, and the second part
possesses the $\mu$-$\tau$ permutation antisymmetry. In such a
treatment the second part may serve as a perturbation term to
characterize the $\mu$-$\tau$ symmetry breaking effects. This point
will become clearer in section 3.3.

To summarize, the $\mu$-$\tau$ permutation symmetry was motivated by
the early experimental data of neutrino oscillations and has played
an important role in understanding the lepton flavor structures. The
discovery of a relatively large value of $\theta^{}_{13}$
\cite{DYB1} requires one to go beyond the original model-building
scope in this respect, either by taking into account much larger
$\mu$-$\tau$ symmetry breaking effects \cite{GHY,HY} or by paying
particular attention to the $\mu$-$\tau$ {\it reflection} symmetry.
The latter approach is more attractive because it can both produce a
nonzero $\theta^{}_{13}$ and predict $\delta =\pm \pi/2$ in the
first place (i.e., in the symmetry limit).

\subsection{The $\mu$-$\tau$ reflection symmetry}

The $\mu$-$\tau$ reflection symmetry was originally put forward by
Harrison and Scott \cite{MT-R}, who were inspired by the approximate
$\mu$-$\tau$ universality in the neutrino oscillation data. They
started from the possibility of $|U^{}_{\mu i}| = |U^{}_{\tau i}|$
(for $i=1,2,3$), switched off two Majorana phases and arrived at the
following parametrization of $U$ in a proper phase convention:
\begin{eqnarray}
U = \pmatrix{ u^{}_{1} & u^{}_{2} & u^{}_{3} \cr v^{}_{1} & v^{}_{2}
& v^{}_{3} \cr v^{*}_{1}& v^{*}_{2}& v^{*}_{3} \cr} \; ,
\end{eqnarray}
where the elements $u^{}_{i}$ have been arranged to be real. Thanks
to its unitarity, $U$ contains four independent real parameters
which can be chosen as $u^{}_{1}$, $u^{}_2$, $|v^{}_1|$ and
$\gamma^{}_1 \equiv \arg\left(v^{}_1\right)$. For instance, the
phase difference $\gamma^{}_{21} \equiv \gamma^{}_{2}-\gamma^{}_{1}$
is determined via
\begin{eqnarray}
\cos{\gamma^{}_{21}} = \frac{-u^{}_2 u^{}_1}{2|v^{}_2 v^{}_1|} =
\frac{-u^{}_2 u^{}_1}{\displaystyle \sqrt{\left(1-u^2_2\right)
\left(1-u^2_1 \right)}} \; .
\end{eqnarray}
The Jarlskog invariant \cite{J} of this neutrino mixing pattern
turns out to be
\begin{eqnarray}
|\mathcal{J}| & = & \left|{\rm Im}\left(u^{}_1 u^{*}_2 v^*_1
v^{}_2\right)\right| = u^{}_1 u^{}_2 |v^{}_1
v^{}_{2}\sin{\gamma^{}_{21}}|
\nonumber \\
& = & \frac{1}{2}u^{}_1 u^{}_2 \sqrt{1-u^2_1-u^2_2}
 =\frac{1}{2}\left|U^{}_{e1}U^{}_{e2}U^{}_{e3}\right| \; .
\end{eqnarray}
On the other hand, $\mathcal{J}$ is given by
\begin{eqnarray}
|\mathcal{J}| & = &
c^{}_{12}s^{}_{12}c^2_{13}s_{13}c^{}_{23}s^{}_{23}
\left|\sin{\delta}\right|
\nonumber \\
& = & \frac{1}{2}\left|U^{}_{e1} U^{}_{e2} U^{}_{e3} \sin{\delta}
\right| \sin 2 \theta^{}_{23} \;
\end{eqnarray}
in the standard parametrization of $U$, as pointed out below Eq.
(2.13). If $\theta^{}_{13}\neq 0$ or equivalently $U^{}_{e 3} \neq
0$, one will immediately obtain $\sin{2\theta^{}_{23}}
\left|\sin{\delta}\right| = 1$ from Eqs. (3.21) and (3.22). This
interesting result implies that the condition $|U^{}_{\mu i}| =
|U^{}_{\tau i}|$ must yield $\theta^{}_{23}=\pi/4$ and $\delta=\pm
\pi/2$ --- the same conclusion has been drawn in Eq. (2.17). Current
neutrino oscillation data have given $\theta^{}_{13} \simeq 9^\circ$
\cite{DYB1} and provided a preliminary but noteworthy hint $\delta
\sim -\pi/2$ \cite{T2K3}. That is why the special mixing pattern in
Eq. (3.19), which is invariant under the $\mu$-$\tau$ reflection
symmetry transformation (namely, $\left(S^+ U\right)^* =U$), is
phenomenologically appealing.

Now let us turn to the neutrino mass matrix. It is easy to show that
an $M^{}_\nu$ obeying the condition
\begin{eqnarray}
\left(S^+ M^{}_{\nu}S^+\right)^* = M^{}_{\nu} \;
\end{eqnarray}
can lead us to the neutrino mixing matrix in Eq. (3.19). In other
words, $M^{}_\nu$ is required to be invariant with respect to the
$\mu$-$\tau$ reflection transformations defined by Eq. (1.2)
\footnote{This is actually a kind of generalized CP symmetry
\cite{GCP1,GCP2} --- a framework which enables us to combine the
flavor symmetry with the CP symmetry in a consistent way. Such a
scenario will be discussed in section 4.3.}.
To be explicit, one may parametrize this kind of $M^{}_\nu$ as
follows:
\begin{eqnarray}
M^{}_{\nu} = \pmatrix{
A & B & B^{*} \cr B & C & D \cr B^* & D & C^*\cr } \; ,
\end{eqnarray}
where $A$ and $D$ are two real parameters.

Note that it is possible to obtain a neutrino mass matrix of
the above form without invoking the $\mu$-$\tau$ reflection
symmetry. In Refs. \cite{BMV1,BMV2} the authors started from a
particular neutrino mass matrix at a seesaw scale,
\begin{eqnarray}
M^{0}_{\nu} = m\pmatrix{
1 & 0 & 0 \cr 0 & 0 & 1 \cr 0 & 1 & 0 \cr  } \; ,
\end{eqnarray}
and allowed it to run down to a much lower scale via the one-loop
renormalization-group equations (RGEs) in the supersymmetry
framework \cite{CIPV1,CIPV2}. Then the resulting neutrino mass
matrix at the electroweak scale reads as
\begin{eqnarray}
M^{}_{\nu} & = & \left(I+\Delta^\dagger\right) M^{0}_{\nu}
\left(I+\Delta^*\right)
\nonumber \\
& = & m \pmatrix{ 1+2\delta^{}_{ee} &
\delta^{*}_{e\mu}+\delta^{}_{e\tau} &
\delta^{}_{e\mu}+\delta^{*}_{e\tau}  \cr \cdots &
2\delta^{}_{\mu\tau} & 1+\delta^{}_{\mu\mu}+\delta^{}_{\tau\tau} \cr
\cdots & \cdots & 2\delta^{*}_{\mu\tau}\cr }
\end{eqnarray}
with $I$ being the identity matrix and $\Delta$ representing the
Hermitian radiative correction matrix,
\begin{eqnarray}
\Delta = \pmatrix{ \delta^{}_{ee} & \delta^{}_{e\mu} &
\delta^{}_{e\tau} \cr \delta^{*}_{e\mu} & \delta^{}_{\mu\mu} &
\delta^{}_{\mu\tau} \cr \delta^{*}_{e\tau} & \delta^{*}_{\mu\tau} &
\delta^{}_{\tau\tau}\cr } \; .
\end{eqnarray}
It is obvious that the texture of $M^{}_\nu$ in Eq. (3.26) is the
same as that in Eq. (3.24), but the former as a model-building
example crucially depends on the special form of $M^0_\nu$ and the
relevant RGEs. We shall subsequently concentrate on the pattern of
$M^{}_{\nu}$ in Eq. (3.24) and assume it to result from the
$\mu$-$\tau$ reflection symmetry.

The correspondence between the flavor mixing matrix in Eq. (3.19)
and the neutrino mass matrix in Eq. (3.24) (i.e., the former as a
consequence of the latter) was first noticed in Ref. \cite{GL}. It
can be verified by taking $M^{}_{\nu} = U D^{}_\nu U^{\rm T}$ in Eq.
(3.23),
\begin{eqnarray}
S^+U^* D^{}_{\nu}U^\dagger S^+ = U D^{}_\nu U^{\rm T} \; ,
\end{eqnarray}
where $D^{}_\nu$ has been defined in Eq. (2.3). Hence $S^+U^*$ is
identical to $U$ up to a diagonal phase matrix $X$
--- namely, $S^+U^*  =U X$, in which $X^{}_{ii}$ is either an
arbitrary phase factor for $m^{}_i=0$ or $\pm 1$ for
$m^{}_{i}\neq0$. We are therefore led to a particular neutrino
mixing matrix of the form
\begin{eqnarray}
U = \pmatrix{ \sqrt{X^{}_{11}}u^{}_{1} & \sqrt{X^{}_{22}}u^{}_{2} &
\sqrt{X^{}_{33}}u^{}_{3} \cr v^{}_{1} & v^{}_{2} & v^{}_{3} \cr
X^{}_{11}v^{*}_{1} & X^{}_{22}v^{*}_{2} & X^{}_{33}v^{*}_{3} \cr }
\; .
\end{eqnarray}
It is clear that the relation $|U^{}_{\mu i}| = |U^{}_{\tau i}|$
holds, and this pattern of $U$ will have the same form as the one
given in Eq. (3.19) if $X^{}_{ii} =1$ (for $i=1,2,3$) is taken.
Provided $X^{}_{ii}$ happens to be $-1$, one may redefine the
corresponding neutrino mass eigenstate $\nu^{}_{i}$ as
$\nu^{\prime}_{i}={\rm i}\nu^{}_{i}$ to absorb the imaginary factor
in the first row of $U$ and assure the second and third rows of $U$
to satisfy $U^{}_{\mu i} = U^*_{\tau i}$. In the meantime the mass
eigenvalue $m^{}_i$ should be replaced by $m^\prime_i = -m^{}_i$,
implying that $\nu^\prime_i$ has a Majorana phase equal to $\pi/2$.

In short, as for the Majorana neutrinos, the flavor mixing matrix
resulting from the mass matrix $M^{}_\nu$ in Eq. (3.24) can also be
parameterized as that given by Eq. (3.19), but the relevant Majorana
phases are fixed to be 0 or $\pi/2$. One may see this interesting
conclusion from another angle. Let us convert $M^{}_{\nu}$ in Eq.
(3.24) into a real matrix by a unitary transformation in the (2,3)
plane \cite{YL2}:
\begin{eqnarray}
U^\dagger_{23} M^{}_{\nu}U^*_{23} = \pmatrix{ A & \sqrt{2} \ {\rm
Im} B & \sqrt{2} \ {\rm Re} B \cr \cdots & D-{\rm Re} C & {\rm Im} C
\cr \cdots & \cdots & D+{\rm Re} C \cr} \;
\end{eqnarray}
with
\begin{eqnarray}
U^{}_{23}=\frac{1}{\sqrt 2}\pmatrix{ \sqrt 2&  0& 0\cr  0& {\rm i} &
1\cr 0& -{\rm i}& 1\cr }.
\end{eqnarray}
The mass matrix in Eq. (3.30) can subsequently be diagonalized by a
real orthogonal matrix $O$ to get real mass eigenvalues, and the
resulting neutrino mixing matrix $U = U^{}_{23}O$ takes the form
given by Eq. (3.19) with special Majorana phases. To conclude, the
$\mu$-$\tau$ reflection symmetry can not only predict
$\theta^{}_{23} =\pi/4$ and $\delta =\pm \pi/2$ but also constrain
the corresponding Majorana phases to be 0 or $\pi/2$.

It is worth stressing that Eq. (3.24) is not the only form of
$M^{}_\nu$ which is able to generate both $\theta^{}_{23} =\pi/4$
and $\delta =\pm \pi/2$. In fact, the elements of a neutrino mass
matrix just need to satisfy the following relation to achieve this
goal \cite{AS,Yasue1,Yasue2,Yasue3,Yasue4}:
\begin{eqnarray}
\sum_{\alpha = e}^{\tau} M^{*}_{e\alpha} M^{}_{\alpha\tau} =
\sum_{\alpha = e}^{\tau} M^{}_{e\alpha} M^{*}_{\alpha\mu} \; .
\end{eqnarray}
There are some other solutions to this equation, besides that given
by Eq. (3.24). For example,
\begin{eqnarray}
M^{}_{ee} = M^{}_{\mu\tau} \; , \hspace{0.4cm} M^{}_{e\mu} =
M^{*}_{e\tau} \; , \hspace{0.4cm} M^{}_{\mu\mu} = M^{*}_{\tau\tau}
\; ,
\end{eqnarray}
leading to a special neutrino mass matrix of the form
\begin{eqnarray}
M^{}_{\nu} = \pmatrix{
A& B & B^{*} \cr B& C & A \cr B^* & A &C^* \cr } \; .
\end{eqnarray}
Note that this matrix is not equivalent to that given by Eq. (3.24)
even if $D = A$ is taken, because here $A$ is a complex parameter
instead of a real one. Note also that the present matrix does not
possess a transparent flavor symmetry (e.g., the $\mu$-$\tau$
reflection symmetry), and thus it is difficult to be derived from an
underlying flavor model. That is why we pay particular attention to
the pattern of $M^{}_\nu$ in Eq. (3.24), which is much more favored
from the model-building point of view.
\begin{figure*}
\vspace{-0.4cm} \centering
\includegraphics[width=6.8in]{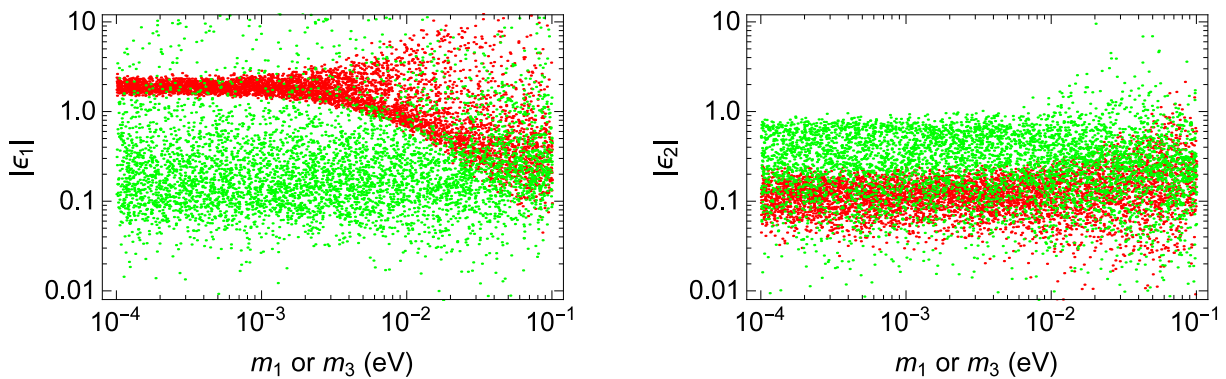}
\caption{The profiles of $|\epsilon^{}_{1}|$ and $|\epsilon^{}_{2}|$
versus the lightest neutrino mass $m^{}_{1}$ (normal hierarchy:
red region) or $m^{}_{3}$ (inverted hierarchy: green region),
where the $1\sigma$ ranges of relevant neutrino oscillation parameters
listed in Table 2.1 have been input.}
\end{figure*}

\subsection{Breaking of the $\mu$-$\tau$ permutation symmetry}

Now that the $\mu$-$\tau$ permutation symmetry gives rise to
$\theta^{}_{13} =0$ and $\theta^{}_{23} =\pi/4$, the observed result
of $\theta^{}_{13}$ ($\simeq 9^\circ$) and a possible deviation of
$\theta^{}_{23}$ from $\pi/4$ definitely signify the breaking of
this interesting flavor symmetry. One nevertheless should discuss
the symmetry-breaking effects at the mass matrix level, in light of
the fact that the symmetry is realized for the mass matrix rather
than for the flavor mixing matrix. To quantify the strength of
$\mu$-$\tau$ symmetry breaking, it is necessary to introduce a
characteristic measure of such effects. In fact, the most general
perturbation to a Majorana neutrino mass matrix $M^{(0)}_\nu$ with
the $\mu$-$\tau$ permutation symmetry can be decomposed into a
symmetry-conserving part and a symmetry-violating part:
\begin{eqnarray}
M^{(1)}_{\nu} & = & \frac{1}{2}\pmatrix{ 2 \delta^{}_{ee}&
\delta^{}_{e\mu}+\delta^{}_{e\tau} &
\delta^{}_{e\mu}+\delta^{}_{e\tau} \cr
\delta^{}_{e\mu}+\delta^{}_{e\tau} &
\delta^{}_{\mu\mu}+\delta^{}_{\tau\tau} & 2 \delta^{}_{\mu\tau} \cr
\delta^{}_{e\mu}+\delta^{}_{e\tau} & 2 \delta^{}_{\mu\tau} &
\delta^{}_{\mu\mu}+\delta^{}_{\tau\tau} }
\nonumber \\
& + & \frac{1}{2} \pmatrix{ 0 & \delta^{}_{e\mu}-\delta^{}_{e\tau} &
\delta^{}_{e\tau}-\delta^{}_{e\mu} \cr
\delta^{}_{e\mu}-\delta^{}_{e\tau}&
\delta^{}_{\mu\mu}-\delta^{}_{\tau\tau} & 0 \cr
\delta^{}_{e\tau}-\delta^{}_{e\mu} & 0 & \delta^{}_{\tau\tau}
-\delta^{}_{\mu\mu}} \;
\end{eqnarray}
whose parameters are small in magnitude. Because the
symmetry-conserving part can be absorbed via a redefinition of the
original elements of $M^{(0)}_\nu$, we are left with the full
neutrino mass matrix $M^{}_{\nu} = M^{(0)}_{\nu} + M^{(1)}_{\nu}$ in
the following form \cite{GJKLST}:
\begin{eqnarray}
M^{}_{\nu}=\pmatrix{ A^\prime & B^\prime\left(1+\epsilon^{}_1\right)
& B^\prime\left(1-\epsilon^{}_1\right) \cr
B^\prime\left(1+\epsilon^{}_1\right) &
C^\prime\left(1+\epsilon^{}_2\right) &  D^\prime \cr
B^\prime\left(1-\epsilon^{}_1\right) &  D^\prime &
C^\prime\left(1-\epsilon^{}_2\right) }
\end{eqnarray}
with
\begin{eqnarray}
A^\prime=A+\delta^{}_{ee}\;, \hspace{0.5cm} &
B^\prime=B+\displaystyle
\frac{\delta^{}_{e\mu}+\delta^{}_{e\tau}}{2}\; ,
\nonumber \\
D^\prime=D+\delta^{}_{\mu\tau}\; ,\hspace{0.5cm} &
C^\prime=C+\displaystyle\frac{\delta^{}_{\mu\mu}+
\delta^{}_{\tau\tau}}{2}\; ,
\nonumber \\
\epsilon^{}_1=\displaystyle\frac{\delta^{}_{e\mu}-
\delta^{}_{e\tau}}{2B^\prime} \; , \hspace{0.55cm} & \epsilon^{}_2=
\displaystyle\frac{\delta^{}_{\mu\mu}-\delta^{}_{\tau\tau}}
{2C^\prime} \; .
\end{eqnarray}
So it is convenient to use the dimensionless parameters
\begin{eqnarray}
\epsilon^{}_{1}=\frac{M^{}_{e \mu}- M^{}_{e\tau}}{M^{}_{e \mu}+
M^{}_{e\tau}} \ ,\hspace{0.7cm} \epsilon^{}_{2}=\frac{M^{}_{\mu
\mu}-M^{}_{\tau \tau}}{M^{}_{\mu \mu}+ M^{}_{\tau\tau}} \;
\end{eqnarray}
to measure the effects of $\mu$-$\tau$ symmetry breaking. If
$|\epsilon^{}_{1}|$ and $|\epsilon^{}_{2}|$ are both small enough
(e.g., $\lesssim 0.2$), then one may argue that the neutrino mass
matrix $M^{}_\nu$ possesses an approximate $\mu$-$\tau$ permutation
symmetry.

The small parameters $\epsilon^{}_{1}$ and $\epsilon^{}_{2}$ can be
linked to the observable quantities through a reconstruction of
$M^{}_{\alpha \beta}$ in terms of the neutrino mass and flavor
mixing parameters. Note that $|\epsilon^{}_{1,2}|$ are not
rephasing-invariant --- namely, they are sensitive to the phase
transformations
\begin{eqnarray}
\nu^{}_{e}\to e^{{\rm i}\phi^{}_1}\nu^{}_{e}\ ,
\hspace{0.4cm} \nu^{}_{\mu}\to e^{{\rm
i}\phi^{}_2}\nu^{}_{\mu}\ , \hspace{0.4cm} \nu^{}_{\tau}\to
e^{{\rm i}\phi^{}_3}\nu^{}_{\tau} \; .
\end{eqnarray}
In this case the values of $|\epsilon^{}_{1,2}|$ will accordingly
change:
\begin{eqnarray}
\left|\epsilon^{}_1\right| \to \frac{\left|M^{}_{e
\mu}-M^{}_{e\tau}e^{{\rm i}\phi^{}_{32}}\right|} {\left|M^{}_{e
\mu}+ M^{}_{e\tau}e^{{\rm i}\phi^{}_{32}}\right|} \; ,
\nonumber \\
\left|\epsilon^{}_2\right| \to \frac{\left|M^{}_{\mu\mu}
-M^{}_{\tau\tau}e^{2{\rm i}\phi^{}_{32}}\right|} {\left|M^{}_{\mu
\mu} + M^{}_{\tau\tau}e^{2{\rm i}\phi^{}_{32}}\right|} \; ,
\end{eqnarray}
where $\phi^{}_{32} \equiv \phi^{}_{3}-\phi^{}_{2}$. Hence the
values of $|\epsilon^{}_{1,2}|$ are not fully physical.
Nevertheless, $\phi^{}_{32}$ is supposed to be strongly suppressed
in the presence of an approximate $\mu$-$\tau$ permutation symmetry.
So one expects that this small phase difference should not change
the main features of $|\epsilon^{}_{1,2}|$. Based on this reasonable
argument, we neglect $\phi^{}_{32}$ for the time being and shall
take account of it when necessary. Fig. 3.2 illustrates the possible
values of $|\epsilon^{}_{1,2}|$ against the lightest neutrino mass,
where the relevant neutrino oscillation parameters used to
reconstruct $M^{}_{\alpha\beta}$ take values in their $1\sigma$
ranges given by Table 2.1. We see that $|\epsilon^{}_{1}|$ and
$|\epsilon^{}_{2}|$ can be simultaneously small when the neutrino
mass spectrum has an inverted hierarchy, and $|\epsilon^{}_1|$ is
unacceptably large when $m^{}_1$ is very small (i.e., when the
neutrino mass hierarchy is normal). The numerical results shown in
Fig. 3.2 can be understood by using the approximate expressions of
$|\epsilon^{}_{1,2}|$ to be given below. Thanks to the smallness of
$\theta^{}_{13} \sim 0.15$ and $\Delta\theta^{}_{23} \equiv
\theta^{}_{23}-\pi/4 \in [-0.09, +0.09]$, it is possible to obtain
\begin{eqnarray}
\epsilon^{}_{1} &\sim & -\Delta\theta^{}_{23}-\frac{m^{}_{11} \tilde
\theta^{}_{13} -m^{}_{3} \tilde \theta^{*}_{13} }{m^{}_{12}} \; ,
\nonumber \\
\epsilon^{}_{2} &\sim &
2\Delta\theta^{}_{23}-\frac{4m^{}_{22}\Delta\theta^{}_{23}
+2m^{}_{12}\tilde \theta^{}_{13} } {m^{}_{22}+m^{}_{3}} \; ,
\end{eqnarray}
where the notations introduced in Eqs. (3.2) and (3.5) have been
used (in particular, here $\tilde{\theta}^{}_{13} \equiv \theta^{}_{13}
e^{{\rm i} \delta}$). For the sake of simplicity, we consider the
properties of $|\epsilon^{}_{1,2}|$ in three typical cases \cite{GJP}.
\begin{itemize}
\item    \underline{$m^{}_{1}\ll m^{}_{2}\ll m^{}_{3}$}. Given this
highly hierarchical neutrino mass spectrum, we obtain
\begin{eqnarray}
|\epsilon^{}_{1}| \sim
\frac{m^{}_{3}\theta^{}_{13}}{m^{}_{2}c^{}_{12} s^{}_{12}}\simeq 1.9
\; , \hspace{0.5cm} |\epsilon^{}_{2}| \sim  2|\Delta\theta^{}_{23}|
\ ,
\end{eqnarray}
implying that $M^{}_{\nu}$ itself does not really exhibit an
approximate $\mu$-$\tau$ permutation symmetry.

\item    \underline{$m^{}_{1}\simeq m^{}_2 \gg m^{}_3$}.
Because of the near equality of $m^{}_{1}$ and $m^{}_{2}$ in this
case, the Majorana phases $\rho$ and $\sigma$ become relevant:
\begin{eqnarray}
\epsilon^{}_{1} &\sim &  -\Delta\theta^{}_{23}-\frac{\displaystyle
c^2_{12}+ s^2_{12}e^{2{\rm i}\left(\sigma-\rho\right)}}
{\displaystyle\left[1-e^{2{\rm i}\left(\sigma-\rho\right)}\right]
c^{}_{12}s^{}_{12}} \tilde \theta^{}_{13} \; ,
\nonumber \\
\epsilon^{}_{2} & \sim & -2\Delta\theta^{}_{23}-
2\frac{\displaystyle \left[1-e^{2{\rm
i}\left(\sigma-\rho\right)}\right] c^{}_{12}s^{}_{12}}{\displaystyle
c^2_{12}+s^2_{12}e^{2{\rm i} \left(\sigma-\rho\right)}}   \tilde
\theta^{}_{13} \; .
\end{eqnarray}
Note that the coefficient of $\tilde \theta^{}_{13}$ in
$\epsilon^{}_{1}$ is inversely proportional to that in
$\epsilon^{}_{2}$. So $|\epsilon^{}_{1}|$ will be much larger than
$1$ for $\sigma-\rho\sim 0$, and $|\epsilon^{}_{2}|$ will have a too
large value for $\sigma-\rho\sim \pi/2$. A careful analysis indicates
that $|\epsilon^{}_{1}|$ and $|\epsilon^{}_{2}|$ have no chance to
be small enough at the same time.

\item    \underline{$m^{}_{1}\simeq m^{}_{2} \simeq m^{}_{3}$}.
In this case let us consider four sets of special but typical values
of $\rho$ and $\sigma$. Given $(\rho,\sigma)=(0,0)$ or
$(\pi/2,\pi/2)$, for example, $|\epsilon^{}_{1}|$ will be much
larger than $1$ owing to a nearly complete cancellation between the
two components of $m^{}_{12}$  as one can see in Eq. (3.5). If
$(\rho,\sigma)=(0, \pi/2)$ or $(\pi/2, 0)$, then $\epsilon^{}_{1,2}$
can approximate to
\begin{eqnarray}
\epsilon^{}_{1} &\sim &  -\Delta\theta^{}_{23} + \frac{\displaystyle
e^{-{\rm i}\delta}\mp\left(c^2_{12}-s^2_{12} \right)e^{{\rm i}
\delta}}{\displaystyle 2c^{}_{12}s^{}_{12}}\theta^{}_{13} \; ,
\nonumber \\
\epsilon^{}_{2}& \sim &\frac{\displaystyle 2\left(1\mp c^2_{12} \pm
s^2_{12}\right)\Delta \theta^{}_{23} \mp 4c^{}_{12}s^{}_{12} \tilde
\theta^{}_{13} } {\displaystyle 1\pm c^2_{12}\mp s^2_{12}} \; ,
\end{eqnarray}
which are allowed to be simultaneously small.
\end{itemize}
We conclude that a quasi-degenerate neutrino mass spectrum allows
$M^{}_{\nu}$ to show an approximate $\mu$-$\tau$ permutation
symmetry. This conclusion can also be put in another way: given
$m^{}_{1} \simeq m^{}_2 \simeq m^{}_3$, the Majorana neutrino mass
matrix with an approximate $\mu$-$\tau$ permutation symmetry is able
to generate a phenomenologically viable pattern of neutrino mixing
--- a pattern compatible with current experimental data, as one will
see later on. A similar approach has been discussed in Ref.
\cite{smirnov} to explore the gap between $U^{}_{\rm TB}$ and the
experimentally-favored pattern as well as what such a gap implies
on the underlying texture of $M^{}_\nu$. Because $U^{}_{\rm TB}$
can be obtained by invoking the condition
$M^{}_{ee} + M^{}_{e\mu} = M^{}_{\mu \mu} + M^{}_{\mu
\tau}$ for $M^{}_\nu$ on the basis of the
$\mu$-$\tau$ permutation symmetry, one needs $\epsilon^{}_{1}$,
$\epsilon^{}_{2}$ and an extra small parameter to describe the
actual departure of $M^{}_\nu$ from
$M^{\rm TB}_\nu \equiv U^{}_{\rm TB} D^{}_\nu U^T_{\rm TB}$.
Hence a similar conclusion has been reached \cite{smirnov}:
only in the $m^{}_{1}\simeq m^{}_{2}
\simeq m^{}_{3}$ case can $M^{}_\nu$ possess an approximate
$\mu$-$\tau$ symmetry and give rise to a viable neutrino
mixing pattern which is very close to the form of $U^{}_{\rm TB}$.
In this sense $U^{}_{\rm TB}$ itself is accidental, at least
from a point of view of model building, although its
predictions for $\theta^{}_{12}$ and $\theta^{}_{23}$ are quite
close to the experimental values.

Instead of reconstructing $M^{}_\nu$ in terms of neutrino masses and
flavor mixing parameters to constrain the sizes of
$\epsilon^{}_{1,2}$, now we follow a more straightforward way to
study the texture of $M^{}_{\nu}$ with an approximate $\mu$-$\tau$
permutation symmetry (i.e., both $\epsilon^{}_1$ and $\epsilon^{}_2$
are assumed to be small from the beginning) and explore the direct
dependence of $\tilde \theta^{}_{13}$ and $\Delta \theta^{}_{23} $
on $\epsilon^{}_{1,2}$. There are two good reasons for doing so: (a)
it is important to identify what kind of symmetry breaking can give
rise to the viable phenomenological consequences in a model-building
exercise; (b) some specific symmetry-breaking patterns are able to
predict a few interesting correlations among the physical parameters
and thus may bridge the gap between the unknown and known
parameters. Let us first derive the expressions of $\tilde
\theta^{}_{13}$ and $\Delta \theta^{}_{23} $ arising from the most
general symmetry breaking pattern. In the standard parametrization
the unitary matrix used for diagonalizing a Majorana neutrino mass
matrix takes the form $U=P^{}_{\phi}VP^{}_{\nu}$, where $V$ and
$P^{}_{\nu}$ have been given in Eqs. (2.6)---(2.9), and
$P^{}_{\phi}={\rm{Diag}}\{e^{{\rm i}\phi^{}_{1}},e^{{\rm
i}\phi^{}_{2}},e^{{\rm i}\phi^{}_{3}}\}$. Although $\phi^{}_{1}$,
$\phi^{}_2$ and $\phi^{}_3$ have no physical meaning, they are
necessary in the diagonalization process and thus should not be
ignored from here. When the $\mu$-$\tau$ permutation symmetry is
exact, $\phi^{}_{2}$ is automatically equal to $\phi^{}_{3}$. In the
presence of small symmetry-breaking effects, $\phi^{}_{32}$ becomes
a finite but small quantity. By taking $M^{}_{\nu} = U D^{}_\nu U^T$
in Eq. (3.36) and doing perturbation expansions for those small
quantities, one can obtain the following relations which connect
$\phi^{}_{32}$, $\Delta \theta^{}_{23} $ and $\tilde \theta^{}_{13}$
with the perturbation parameters $\epsilon^{}_{1,2}$:
\begin{eqnarray}
m^{}_{12}\Delta\theta^{}_{23}+m^{}_{11}\tilde \theta^{}_{13}-
m^{}_{3}\tilde \theta^{*}_{13}
+ \frac{m^{}_{12}}{2} \left(2\epsilon^{}_{1}+{\rm i}
\phi^{}_{32}\right ) = 0 \; ,
\nonumber \\
m^{}_{22-3}\Delta\theta^{}_{23}+m^{}_{12}\tilde \theta^{}_{13}
+\frac{m^{}_{22+3}}{2} \left(\epsilon^{}_{2}+{\rm i}\phi^{}_{32}
\right) = 0 \; ,
\end{eqnarray}
in which $m^{}_{22\pm 3} \equiv m^{}_{22}\pm m^{}_{3}$ are defined.
After solving these equations in a straightforward way, we obtain
$\tilde \theta^{*}_{13}$ and $\Delta \theta^{}_{23} $ as functions
of $\epsilon^{}_{1,2}$ \cite{GJKLST}:
\begin{eqnarray}
&& \tilde \theta^{*}_{13} = \left(2\Delta m^{2}_{31}\right)^{-1}
\left( 2m^{}_3 m^{}_{12} c^2_{12}\epsilon^{}_1 +2\overline m^{}_1
m^*_{12}c^2_{12}\epsilon^*_1 \right.
\nonumber \\
&& \hspace{0.85cm} \left. + m^{}_3
m^{}_{22+3}c^{}_{12}s^{}_{12}\epsilon^{}_2+ \overline
m^{}_1m^*_{22+3}c^{}_{12}s^{}_{12}\epsilon^*_2\right)
\nonumber \\
&& \hspace{0.85cm} + \left(2\Delta m^{2}_{32}\right)^{-1}
\left(2 m^{}_3 m^{}_{12}s^2_{12}\epsilon^{}_1 +2 \overline
m^{}_2 m^*_{12}s^2_{12}\epsilon^*_1 \right.
\nonumber \\
&& \hspace{0.82cm} \left. -m^{}_3
m^{}_{22+3}c^{}_{12}s^{}_{12}\epsilon^{}_2 -\overline m^{}_2
m^*_{22+3}c^{}_{12}s^{}_{12}\epsilon^*_2\right) \; ,
\nonumber \\
&& \Delta\theta^{}_{23} = {\rm Re} \left\{ \left(2\Delta
m^{2}_{31}\right)^{-1}\left[2m^{}_{12} c^{}_{12}s^{}_{12}
\left(\overline m^{*}_{1}\epsilon^{}_1+m^{}_3
\epsilon^{*}_{1}\right) \right. \right.
\nonumber \\
&& \hspace{1.1cm} \left. + m^{}_{22+3}s^2_{12} \left(\overline
m^{*}_{1} \epsilon^{}_2+ m^{}_3 \epsilon^{*}_{2}\right) \right]
\nonumber \\
&& \hspace{1.1cm} -\left(2\Delta m^{2}_{32}\right)^{-1}
\left[2m^{}_{12} c^{}_{12}s^{}_{12} \left(\overline m^{*}_{2}
\epsilon^{}_1+m^{}_3 \epsilon^{*}_{1}\right) \right.
\nonumber \\
&& \hspace{1.03cm} \left.\left. -m^{}_{22+3} c^2_{12}
\left(\overline m^{*}_{2}\epsilon^{}_2+m^{}_3
\epsilon^{*}_{2}\right)\right]\right\} \; .
\end{eqnarray}
With the help of these results, one may get a ball-park feeling of
the dependence of $\tilde \theta^{}_{13}$ and $\Delta\theta^{}_{23}$
on $\epsilon^{}_{1,2}$ in some particular cases to be discussed
below. One will see that the values of $\theta^{}_{13}$ and
$\Delta\theta^{}_{23}$ are strongly correlated with the neutrino
mass spectrum when the strength of $\mu$-$\tau$ permutation symmetry
breaking (i.e., the size of $\epsilon^{}_{1,2}$) is specified.

In the assumption of CP conservation, Eq. (3.46) is simplified to
\begin{eqnarray}
&& \theta^{}_{13} = \frac{2 m^{}_{12} c^2_{12}\epsilon^{}_1
+m^{}_{22+3}c^{}_{12}s^{}_{12}\epsilon^{}_2}
{2 \left(m^{}_3 \mp m^{}_1\right)}
\nonumber \\
&& \hspace{0.85cm} + \frac{2m^{}_{12} s^2_{12}
\epsilon^{}_1 -m^{}_{22+3}c^{}_{12}s^{}_{12}\epsilon^{}_2}
{2  \left(m^{}_3 \mp m^{}_2\right)} \; ,
\nonumber \\
&& \Delta\theta^{}_{23} = \frac{2m^{}_{12} c^{}_{12}
s^{}_{12}\epsilon^{}_1 +m^{}_{22+3}s^2_{12}\epsilon^{}_2}
{2 \left(m^{}_3 \mp m^{}_1\right)}
\nonumber \\
&& \hspace{1.15cm} - \frac{2m^{}_{12}
c^{}_{12}s^{}_{12}\epsilon^{}_1 -m^{}_{22+3}c^2_{12}\epsilon^{}_2}
{2  \left(m^{}_3 \mp m^{}_2\right)} \; ,
\end{eqnarray}
in which the signs ``$\mp$" correspond to $\overline{m}^{}_{1,2} =
\pm m^{}_{1,2}$ for $\rho, \sigma = 0$ or $\pi/2$. Some more
specific discussions and comments are in order.

(1) As for the neutrino mass spectrum with a vanishingly small
$m^{}_{1}$, the expression of $\theta^{}_{13}$ can be further
simplified to
\begin{eqnarray}
\theta^{}_{13} \sim \frac{1}{2}\sqrt{r} \
c^{}_{12}s^{}_{12}\left(2\epsilon^{}_{1}-\epsilon^{}_{2}\right)
\simeq 0.04 \left(2\epsilon^{}_{1}-\epsilon^{}_{2}\right ) \ ,
\end{eqnarray}
where $r \equiv \Delta m^2_{21}/\Delta m^2_{31} \simeq 0.03$. Given
$|\epsilon^{}_{1,2}| \lesssim 0.2$, it is impossible to get the
observed value of $\theta^{}_{13}$ from the above expression.

(2) In the $m^{}_{1}\simeq m^{}_2 \gg m^{}_3$ case, $\theta^{}_{13}$
and $\Delta\theta^{}_{23}$ are sensitive to the combination of
$\rho$ and $\sigma$
\footnote{In the CP-conserving case under consideration, $\rho$ and
$\sigma$ can only take a value of 0 or $\pi/2$.}.
If $\rho$ and $\sigma$ are equal, $\theta^{}_{13}$ will be highly
suppressed:
\begin{eqnarray}
\theta^{}_{13}\sim \frac{1}{4}r c^{}_{12}s^{}_{12}\left(2\epsilon^{}_1
-\epsilon^{}_2\right) \simeq 0.004 \left(2\epsilon^{}_1-\epsilon^{}_2
\right) \; .
\end{eqnarray}
Otherwise, $\theta^{}_{13}$ turns out to be
\begin{eqnarray}
\theta^{}_{13}\sim \frac{1}{2} \cos{2\theta^{}_{12}}
\sin{2\theta^{}_{12}} \left(2{\epsilon}^{}_1-\epsilon^{}_{2}\right)
\; ,
\end{eqnarray}
which is still unable to fit its experimental value.

(3) When the neutrino mass spectrum is nearly degenerate, one may
consider the following four special but typical cases.
\begin{itemize}
\item \underline{$(\rho, \sigma) = (0,0)$}. In this case
$\theta^{}_{13}$ approximates to
\begin{eqnarray}
\theta^{}_{13}\sim \frac{2m^{2}_0}{\Delta m^2_{31}}r c^{}_{12}
s^{}_{12}\epsilon^{}_2 \; ,
\end{eqnarray}
where $m^{}_{0}$ denotes the overall neutrino mass scale. The factor
$m^2_0/\Delta m^2_{31}$ can enhance the value of $\theta^{}_{13}$ as
$m^{}_0$ goes up, but $m^{}_0\leq 0.1$ eV is expected to hold in
light of the present cosmological upper bound on the sum of three
neutrino masses (e.g., $m^{}_1 + m^{}_2 + m^{}_3 \simeq 3 m^{}_0
<0.23$ eV at the $95\%$ confidence level \cite{Planck}). So
$\theta^{}_{13}$ is at most 0.03. Such a result is certainly
unacceptable.

\item \underline{$(\rho, \sigma) = (0,\pi/2)$}. In this case we
obtain
\begin{eqnarray}
\theta^{}_{13}\sim \frac{2m^2_0}{\Delta m^2_{31}}c^{}_{12}s^{}_{12}
\left(2c^2_{12}\epsilon^{}_{1}+s^2_{12}\epsilon^{}_{2}\right) \; ,
\nonumber\\
\Delta\theta^{}_{23}\sim \frac{2m^2_0}{\Delta m^2_{31}}s^{2}_{12}
\left(2c^2_{12}\epsilon^{}_{1}+s^2_{12}\epsilon^{}_{2}\right) \; .
\end{eqnarray}
Thanks to the enhancement factor $m^2_0/\Delta m^2_{31}$,
$\theta^{}_{13}$ is easy to reach the measured value. Moreover,
there is a correlation between $|\Delta\theta^{}_{23}|$ and
$\theta^{}_{13}$, i.e., $|\Delta\theta^{}_{23}|\sim
\theta^{}_{13}s^{}_{12}/c^{}_{12} \simeq6^\circ$. This relatively
large $|\Delta \theta^{}_{23} |$ will be tested in the near future.

\item \underline{$(\rho, \sigma) = (\pi/2,0)$}. In this case the
results are
\begin{eqnarray}
\theta^{}_{13}\sim \frac{2m^2_0}{\Delta m^2_{31}}c^{}_{12}s^{}_{12}
\left(2s^2_{12}\epsilon^{}_{1}+c^2_{12}\epsilon^{}_{2}\right) \; ,
\nonumber\\
\Delta\theta^{}_{23}\sim \frac{2m^2_0}{\Delta m^2_{31}}c^{2}_{12}
\left(2s^2_{12}\epsilon^{}_{1}+c^2_{12}\epsilon^{}_{2}\right) \; .
\end{eqnarray}
The correlation between $|\Delta\theta^{}_{23}|$ and $\theta^{}_{13}$
predicts $|\Delta \theta^{}_{23}| \sim \theta^{}_{13}c^{}_{12}/s^{}_{12}
\simeq 13^\circ$, which is too large to be acceptable.

\item \underline{$(\rho, \sigma) = (\pi/2,\pi/2)$}. This special case
is disfavored because $\theta^{}_{13}$ is extremely suppressed by
the factor $\Delta m^2_{21}/m^2_0 $, as one can see from
\begin{eqnarray}
\theta^{}_{13}\sim \frac{\Delta m^2_{21}}{4m^2_0}c^{}_{12}s^{}_{12}
\epsilon^{}_1 \; .
\end{eqnarray}
\end{itemize}
To summarize, in the CP-conserving case considered above only the
example of a quasi-degenerate neutrino mass spectrum with
$(\rho,\sigma)=(0,\pi/2)$ is still allowed by current experimental
data.

When CP violation is taken into account, one has to deal with some more
free parameters. But this possibility is certainly more realistic
and more interesting, because it is related to the asymmetry between
matter and antimatter in a variety of weak interaction processes
including neutrino oscillations. Here let us take two
typical examples for illustration.

(A) In the first example we assume the Majorana phases to take
trivial values (i.e., $0$ or $\pi/2$) and the symmetry-breaking
parameters to be purely imaginary (i.e., $\epsilon^{}_{1,2}={\rm
i}|\epsilon^{}_{1,2}|$). Then Eq. (3.46) leads us to the results
$\Delta\theta^{}_{23} =0$ and
\begin{eqnarray}
\tilde \theta^{*}_{13} = {\rm i} \frac{2m^{}_{12} c^2_{12}
|\epsilon^{}_1| +m^{}_{22+3}c^{}_{12}s^{}_{12}|\epsilon^{}_2|} {2
\left(m^{}_3 \pm m^{}_1\right)}
\nonumber \\
\hspace{0.86cm} +
{\rm i} \frac{2m^{}_{12} s^2_{12}|\epsilon^{}_1|
-m^{}_{22+3}c^{}_{12}s^{}_{12}|\epsilon^{}_2|}
{2 \left(m^{}_3 \pm m^{}_2\right)} \; ,
\end{eqnarray}
which implies $\delta = \pm\pi/2$. These results are actually the
same as those predicted by the $\mu$-$\tau$ reflection symmetry,
simply because the perturbation under consideration is so special
that the overall neutrino mass matrix $M^{}_\nu$ as given in Eq.
(3.36) respects this flavor symmetry. Note that $\theta^{}_{13}$ has
the same expressions as those shown in Eqs. (3.48)---(3.50) when
$m^{}_{1}$ or $m^{}_{3}$ is vanishingly small. If the
quasi-degenerate neutrino mass spectrum is concerned, then
\begin{eqnarray}
\underline{(\rho, \sigma) =(0,0):} \hspace{0.2cm} &
\theta^{}_{13} \sim \frac{\Delta m^2_{21}}{8m^2_0}
c^{}_{12}s^{}_{12}\left(2|{\epsilon}^{}_1|-|{\epsilon}^{}_2|\right) \; ,
\nonumber \\
\underline{(\rho, \sigma) =(0,\pi/2):} \hspace{0.2cm} &
\theta^{}_{13} \sim \frac{2m^2_0}{\Delta m^2_{31}}
c^{}_{12}s^{3}_{12}
\left(2|{\epsilon}^{}_1|-|{\epsilon}^{}_2|\right) \; ,
\nonumber \\
\underline{(\rho, \sigma) =(\pi/2,0):} \hspace{0.2cm} &
\theta^{}_{13} \sim \frac{2m^2_0}{\Delta m^2_{31}}
c^{3}_{12}s^{}_{12}\left(2|{\epsilon}^{}_1|
-|{\epsilon}^{}_2|\right) \; ,
\nonumber \\
\underline{(\rho, \sigma) =(\pi/2,\pi/2):} \hspace{0.2cm} &
\theta^{}_{13} \sim \frac{r}{2}
c^{}_{12}s^{}_{12}\left(2|{\epsilon}^{}_1|-|{\epsilon}^{}_2| \right)
\;
\end{eqnarray}
can be obtained. Similar to the results achieved in the
CP-conserving case, an acceptable value of $\theta^{}_{13}$ is only
obtainable from Eq. (3.56) with $(\rho,\sigma)=(0,\pi/2)$ or
$(\pi/2,0)$. It is worth mentioning that $\theta^{}_{13}$ is always
proportional to $2|\epsilon^{}_{1}|-|{\epsilon}^{}_{2}|$ in this
case, because it is the combination
$2{\epsilon}^{}_{1}-{\epsilon}^{}_{2}$ that is invariant under the
phase transformations in Eq. (3.39) when ${\epsilon}^{}_{1,2}$ are
imaginary.

(B) In the second example we relax $\rho$ and $\sigma$ to see
whether $\theta^{}_{13}$ is possible to fit current experimental
data in the $m^{}_{1}\simeq m^{}_{2}\gg m^{}_{3}$ case, where
\begin{eqnarray}
\theta^{}_{13} =
\sqrt{\cos^2{2\theta^{}_{12}}\left[1-\cos{2\left(\sigma-\rho
\right)}\right]^2 +\sin^2{2\left(\sigma-\rho\right)}}
\nonumber \\
\hspace{1cm} \times \frac{1}{2} c^{}_{12}s^{}_{12} \left|2\epsilon^{}_1
-\epsilon^{}_2\right| \; .
\end{eqnarray}
The maximal value of $\theta^{}_{13}$ turns out to be $\theta^{\rm
max}_{13} \simeq 0.25\left|2\epsilon^{}_1-\epsilon^{}_2\right|$ at
$\sigma-\rho=\pi/4$. It is therefore hard to make $\theta^{}_{13}$
compatible with the data unless $\epsilon^{}_{1,2}$ have a relative
phase of $\pi$ and take their upper limit $\sim 0.2$ (for a valid
perturbation expansion). In this extreme case $\Delta \theta^{}_{23}
\simeq -0.03 \cos{[{\rm arg}(\epsilon^{}_1)]} \lesssim 2^\circ$. Of
course, the chance for this extreme case to happen is rather small.
So it is fair to say that the $m^{}_3 \to 0$ limit seems not to be
favored by $M^{}_\nu$ in Eq. (3.36) even when CP violation is taken
into account. Finally, it is worth emphasizing that $\delta$ may
take any value in the interval $[0,2\pi)$ regardless of the strength
of $\mu$-$\tau$ symmetry breaking. This point can be
interpreted as follows. The vanishing of $\theta^{}_{13}$ and
$\Delta\theta^{}_{23}$ is protected by the $\mu$-$\tau$ permutation
symmetry, so their realistic magnitudes are subject to the small
effects of symmetry breaking. In contrast, $\delta$ is not well
defined when the $\mu$-$\tau$ permutation symmetry is exact, and it
turns out to have physical meaning only after this symmetry is
broken and the resulting neutrino mass matrix $M^{}_\nu =
M^{(0)}_\nu + M^{(1)}_\nu$ contains the nontrivial phases. Hence
$\delta$ is sensitive to the phases of $\epsilon^{}_{1,2}$. As shown
in Eq. (3.55), $\delta =\pm\pi/2$ can emerge if $\epsilon^{}_{1,2}$
are purely imaginary. Note that a finite value of $\delta$ may arise
even when $\epsilon^{}_{1,2}$ are real, if $M^{(0)}_\nu$ itself
involves the nontrivial Majorana phases \cite{GJKLST}. Here let us take
$\epsilon^{}_2=2\epsilon^{}_1$ as an example which resembles the
symmetry breaking induced by the RGE running effects, as one can see
later. In this special case $\delta$ is given by
\begin{eqnarray}
\tan{\delta}= \frac{m^{}_2 \sin{2\sigma}-m^{}_1\sin{2\rho}} {m^{}_1
\cos{2\rho}-m^{}_2 \cos{2\sigma}-m^{}_3r} \; ,
\end{eqnarray}
which is not directly associated with ${\epsilon}^{}_{1,2}$
and may be large no matter whether the neutrino mass spectrum
exhibits a normal or inverted hierarchy.

The above discussions are subject to the smallness of $\mu$-$\tau$
symmetry breaking --- namely, the symmetry-breaking terms are
relatively small as compared with the entries of $M^{(0)}_\nu$. In a
given neutrino mass model the elements of $M^{(0)}_\nu$ may not be
at the same order of magnitude, and some of them are even possible
to vanish at the tree level. Provided the higher-order contributions
break the flavor symmetry, then they may play a dominant role in
those originally vanishing entries to make the predictions of the
resulting neutrino mass matrix $M^{}_\nu$ compatible with current
neutrino oscillation data. Of course, there is nothing wrong with
this situation, but it motivates us to reconsider the physical
implications of an approximate $\mu$-$\tau$ permutation symmetry.
Though a concrete model may more or less correlate the
symmetry-breaking terms appearing in different elements of
$M^{}_\nu$, we just treat them as independent parameters in our
subsequent analysis. A good example of this kind is a hierarchical
neutrino mass matrix which can lead us to the mass spectrum
$m^{}_{1}< m^{}_{2}\ll m^{}_{3}$. In this case the $\mu\mu$,
$\mu\tau$ and $\tau\tau$ entries of $M^{}_\nu$ are expected to be
much larger than the others in magnitude by a factor
$\sim\sqrt{1/r}=\sqrt{\Delta m^2_{31}/\Delta m^2_{21}}$. If these
large entries arise from a simple flavor symmetry and those smaller
ones come from the symmetry-breaking effects, a typical form of
$M^{}_{\nu}$ may be \cite{RNM04}
\begin{eqnarray}
M^{}_{\nu}= m^{}_{0}\pmatrix{ d \epsilon & c\epsilon & b\epsilon \cr
c\epsilon & 1+a\epsilon & -1\cr b\epsilon & -1 & 1+\epsilon\cr } \;
,
\end{eqnarray}
where $\epsilon$ denotes a small perturbation (or symmetry-breaking)
parameter, and $a$, $b$, $c$ and $d$ are all the real coefficients
of $\mathcal O(1)$. A straightforward calculation yields the flavor
mixing angles and neutrino mass eigenvalues as follows:
\begin{eqnarray}
\theta^{}_{13} \simeq \frac{1}{2\sqrt 2}\left(b-c\right) \epsilon \; ,
\nonumber \\
\Delta \theta^{}_{23} \simeq \frac{1}{4}\left(a-1\right) \epsilon \; ,
\nonumber \\
\tan{2\theta^{}_{12}} \simeq \frac{2\sqrt{2}\left(b+c\right)}{a+1-2d} \; ;
\end{eqnarray}
and
\begin{eqnarray}
m^{}_{1} \simeq \frac{1}{4}\epsilon m^{}_{0}\left(2d+a+1 - \Delta
\right) \; ,
\nonumber \\
m^{}_{2} \simeq \frac{1}{4}\epsilon m^{}_{0}\left(2d+a+1 + \Delta
\right) \; ,
\nonumber \\
m^{}_{3} \simeq 2m^{}_{0} \; ,
\end{eqnarray}
in which $\Delta=\sqrt{\left(2d-a-1\right)^2+8\left(b+c\right)^2}$.
The small perturbation parameter $\epsilon$ is found to be
\begin{eqnarray}
\epsilon \simeq \frac{4\sqrt{ r}}{\sqrt{\displaystyle
\Delta \left(2d+a+1\right)}} \; .
\end{eqnarray}
We see that $\theta^{}_{13}$ and $\Delta \theta^{}_{23} $ are
proportional to $b-c$ and $a-1$, respectively. Moreover,
$\theta^{}_{13}$ is about $\sqrt{r}=\sqrt{\Delta m^2_{21}/\Delta
m^2_{31}}$ up to an $\mathcal O(1)$ factor, in agreement with the
experimental result. So the smallness of $\theta^{}_{13}$ is
connected to the hierarchy between $\sqrt{\Delta m^2_{21}}$ and
$\sqrt{\Delta m^2_{31}}$ in this scenario.

If $m^{}_{1}\simeq m^{}_2 \gg m^{}_3$ holds, $M^{}_{\nu}$ is
expected to exhibit a different hierarchical structure, in which the
$ee$, $\mu\mu$, $\mu\tau$ and $\tau\tau$ entries should be much
larger than the others. To illustrate, a simple but instructive
example of this kind is
\begin{eqnarray}
M^{}_{\nu} = m^{}_{0}\pmatrix{ 2+d\epsilon & c\epsilon & b\epsilon
\cr c\epsilon & 1+a\epsilon & 1\cr b\epsilon & 1 & 1+\epsilon\cr} \; .
\end{eqnarray}
The results for three flavor mixing angles are similar to those
given in Eq. (3.60), but the neutrino mass eigenvalues turn out to
be
\begin{eqnarray}
m^{}_{1} \simeq 2m^{}_0 + \frac{1}{4}\epsilon m^{}_{0}\left(2d+a+1 -
\Delta\right) \; ,
\nonumber \\
m^{}_{2} \simeq 2m^{}_0 + \frac{1}{4}\epsilon m^{}_{0}\left(2d+a+1 +
\Delta\right) \; ,
\nonumber \\
m^{}_{3} \simeq \frac{1}{2} \epsilon m^{}_0 \left(a+1\right) \; .
\end{eqnarray}
In this case $\epsilon \simeq 2r/\Delta$, leading to a high suppression
of $\theta^{}_{13}$ and $\Delta \theta^{}_{23} $. The point is that
$m^{}_1$ is equal to $m^{}_2$ at the leading order, and their
degeneracy is lifted at the next-to-leading order. So $\epsilon$
should be at the order of $(m^{}_2-m^{}_1)/(m^{}_2+m^{}_1)\sim
0.01$. To get around this problem, one may assume $a=-1$, $b=-c$ and
$d=0$, such that $r = \Delta =0$ holds and $\epsilon$ may not be
strongly suppressed. In this situation it is possible to obtain an
appreciable value of $\theta^{}_{13}$, but lifting the degeneracy of
$m^{}_{1}$ and $m^{}_2$ requires a higher-order perturbation.

Last but not least, there is a well-known form of $M^{}_\nu$ which
yields an inverted neutrino mass spectrum with $m^{}_1=-m^{}_2$ at
the tree level:
\begin{eqnarray}
M^{}_{\nu}= m^{}_0\pmatrix{
e\epsilon & 1+d\epsilon &1+c\epsilon \cr
1+d\epsilon & b\epsilon & \epsilon\cr
1+c\epsilon & \epsilon & a\epsilon\cr } \; ,
\end{eqnarray}
in which the most general next-to-leading-order terms have been
included. This particular texture of $M^{}_\nu$ can be attributed to
an $L^{}_e-L^{}_\mu-L^{}_\tau$ flavor symmetry \cite{STP} --- namely,
the leading-order entries have a vanishing
$L^{}_e-L^{}_\mu-L^{}_\tau$ charge but the others do not. After a
straightforward calculation, the phenomenological consequences of
this neutrino mass matrix can be summarized as follows:
\begin{eqnarray}
\theta^{}_{13} \simeq \frac{\left(a-b\right) \epsilon}{2\sqrt 2} \; ,
\nonumber \\
\Delta \theta^{}_{23}  \simeq \frac{\left(c-d\right) \epsilon}{2} \; ,
\nonumber \\
\tan{2\theta^{}_{12}} \simeq \frac{4\sqrt{2}}{\left(2+a+b-2e\right)
\epsilon} \; ;
\end{eqnarray}
and
\begin{eqnarray}
m^{}_{1} \simeq \frac{1}{4}\left[\left(2e+2+a+b\right)\epsilon -
4\sqrt{2}\right] m^{}_0 \; ,
\nonumber \\
m^{}_{2} \simeq \frac{1}{4}\left[\left(2e+2+a+b\right)\epsilon +
4\sqrt{2}\right] m^{}_0 \; ,
\nonumber \\
m^{}_3 \simeq \frac{1}{2}\left(a+b-2\right) \epsilon m^{}_0 \; .
\end{eqnarray}
Similar to the previous case, the symmetry-breaking parameter
$\epsilon$ has to be exceedingly small unless $a=-b$ and $e=-1$ are
assumed. But such an assumption will lead to $\tan 2\theta^{}_{12}
\simeq \sqrt{2}/\epsilon$, implying an unacceptably large value of
$\theta^{}_{12}$ for $\epsilon \lesssim 0.2$. Namely, the pattern of
$M^{}_\nu$ in Eq. (3.65) is not favored by current experimental
data. We have certainly assumed CP conservation in the above
examples. When CP violation is taken into account, some of the model
parameters become complex \cite{MR} and the procedure for
diagonalizing a complex and hierarchical neutrino mass matrix can be
found in Ref. \cite{SFK}.

\subsection{Breaking of the $\mu$-$\tau$ reflection symmetry}

Let us proceed to discuss possible ways of breaking the $\mu$-$\tau$
reflection symmetry. Without involving any model details, the most
general correction to a neutrino mass matrix with the $\mu$-$\tau$
reflection symmetry is in the form as given by Eq. (3.35). It can be
divided into two parts: the part respecting the $\mu$-$\tau$
reflection symmetry and the part violating this interesting
symmetry. Namely,
\begin{eqnarray}
M^{(1)}_\nu & = & \frac{1}{2}\pmatrix{ 2{\rm Re} \delta^{}_{ee} &
\delta^{}_{e\mu}+\delta^{*}_{e\tau}
&\delta^{*}_{e\mu}+\delta^{}_{e\tau} \cr
\delta^{}_{e\mu}+\delta^{*}_{e\tau} &
\delta^{}_{\mu\mu}+\delta^{*}_{\tau\tau} & 2{\rm Re}
\delta^{}_{\mu\tau} \cr \delta^{*}_{e\mu}+\delta^{}_{e\tau} & 2{\rm
Re} \delta^{}_{\mu\tau} & \delta^{*}_{\mu\mu}+\delta^{}_{\tau\tau} }
\nonumber \\
& + & \frac{1}{2}\pmatrix{ 2{\rm i Im} \delta^{}_{ee} &
\delta^{}_{e\mu}-\delta^{*}_{e\tau} &
\delta^{}_{e\tau}-\delta^{*}_{e\mu} \cr
\delta^{}_{e\mu}-\delta^{*}_{e\tau} &
\delta^{}_{\mu\mu}-\delta^{*}_{\tau\tau} & 2{\rm i Im}
\delta^{}_{\mu\tau} \cr \delta^{}_{e\tau}-\delta^{*}_{e\mu} & 2{\rm
i Im} \delta^{}_{\mu\tau} &
\delta^{}_{\tau\tau}-\delta^{*}_{\mu\mu}}
\end{eqnarray}
in a way similar to Eq. (3.35). As the first part of $M^{(1)}_\nu$
can be absorbed into the original neutrino mass matrix given in Eq.
(3.24), which is now defined as $M^{(0)}_\nu$ in the $\mu$-$\tau$
reflection symmetry limit, the full neutrino mass matrix
$M^{}_\nu = M^{(0)}_\nu + M^{(1)}_\nu$ can be parametrized as
\begin{eqnarray}
M^{}_\nu = \pmatrix{ \widehat A \left(1+{\rm i}\epsilon^{}_3\right)
& \widehat B \left(1+\epsilon^{}_1\right) & \widehat B^{*}
\left(1-\epsilon^{*}_1\right) \cr \cdots & \widehat C
\left(1+\epsilon^{}_2\right) & \widehat D \left(1+{\rm i}
\epsilon^{}_4\right) \cr \cdots & \cdots & \widehat C^{*}
\left(1-\epsilon^{*}_2\right)}
\end{eqnarray}
with
\begin{eqnarray}
\widehat A = A+{\rm Re} \delta^{}_{ee} \; , \hspace{0.5cm} \widehat
B = B+\frac{\delta^{}_{e\mu}+\delta^{*}_{e\tau}}{2} \; ,
\nonumber \\
\widehat D = D+{\rm Re} \delta^{}_{\mu\tau} \; , \hspace{0.4cm}
\widehat C = C+\frac{\delta^{}_{\mu\mu}+\delta^{*}_{\tau\tau}}{2} \;
;
\end{eqnarray}
and
\begin{eqnarray}
\epsilon^{}_1 = \frac{M^{}_{e \mu}- M^{*}_{e\tau}}{M^{}_{e
\mu}+M^{*}_{e\tau}} = \frac{\delta^{}_{e\mu}-\delta^{*}_{e\tau}}{2
\widehat B} \; ,
\nonumber \\
\epsilon^{}_2 = \frac{M^{}_{\mu\mu}-M^{*}_{\tau \tau}}
{M^{}_{\mu \mu}+ M^{*}_{\tau\tau}}
= \frac{\delta^{}_{\mu\mu}-\delta^{*}_{\tau\tau}}{2 \widehat C} \; ,
\nonumber \\
\epsilon^{}_3 = \frac{{\rm Im} M^{}_{ee}}{{\rm Re} M^{}_{ee}} =
\frac{{\rm Im} \delta^{}_{ee}}{\widehat A} \; ,
\nonumber \\
\epsilon^{}_4 = \frac{{\rm Im} M^{}_{\mu \tau}}{{\rm Re} M^{}_{\mu
\tau}} = \frac{{\rm Im} \delta^{}_{\mu\tau}}{\widehat D} \; .
\end{eqnarray}
Hence these dimensionless quantities characterize the strength of
$\mu$-$\tau$ reflection symmetry breaking. They should be small
(e.g., $\lesssim 0.2$) so that $M^{}_\nu$ in Eq. (3.69) may possess
an approximate $\mu$-$\tau$ reflection symmetry. Note that rephasing
the neutrino fields actually allows us to remove
$\epsilon^{}_{3,4}$, so we shall omit $\epsilon^{}_{3,4}$ without
loss of generality in the following discussions. As pointed out in
section 3.2, the $\mu$-$\tau$ reflection symmetry of a neutrino mass
matrix $M^{(0)}_\nu$ requires the corresponding neutrino mixing
matrix $U^{(0)}=P^{(0)}_\phi V^{(0)} P^{(0)}_\nu$ to have a special
pattern in which $\theta^{(0)}_{23}=\pi/4$, $\delta^{(0)}=\pm
\pi/2$, $\rho^{(0)}=0$ or $\pi/2$ and $\sigma^{(0)} =0$ or $\pi/2$
hold. Moreover, the three phases of $P^{(0)}_\phi$ have to satisfy
$\phi^{(0)}_1= \phi^{(0)}_2 + \phi^{(0)}_3=0$ although they have no
physical meaning. When this flavor symmetry is slightly broken, the
resulting flavor mixing parameters will depart from the above
values. Let us define the relevant deviations as $\Delta \phi^{}_1
\equiv \phi^{}_1-0$, $\Delta \phi \equiv
\left(\phi^{}_2+\phi^{}_3\right)/2 -0$, $\Delta \theta^{}_{23}
\equiv \theta^{}_{23}- \pi/4$, $\Delta \delta \equiv
\delta-\delta^{(0)}$, $\Delta \rho \equiv \rho-\rho^{(0)}$ and
$\Delta \sigma \equiv \sigma-\sigma^{(0)}$, and suppose that all of
them are small quantities governed by $\epsilon^{}_{1,2}$. The
explicit relations of these quantities with $\epsilon^{}_{1,2}$ can
be established by doing a perturbation expansion for $M^{}_\nu = U
D^{}_\nu U^{T}$. After some calculations, one arrives at
\begin{eqnarray}
\overline m^{}_1 c^2_{12} \Delta \rho + \overline m^{}_2 s^2_{12}
\Delta \sigma = - m^{}_{11} \Delta \phi^{}_1 \; ,
\nonumber \\
\overline m^{}_1 s^2_{12} \Delta \rho +
\overline m^{}_2 c^2_{12} \Delta \sigma
-2 m^{}_{12} \tilde s^{}_{13} \Delta \theta^{}_{23}
=-m^{}_{22-3}\Delta \phi  \; ,
\nonumber \\
\overline m^{}_1 c^{}_{12} ({\rm i}s^{}_{12} +c^{}_{12}\tilde
s^{}_{13} ) \Delta \rho -  \overline m^{}_2 s^{}_{12} ({\rm
i}c^{}_{12}-s^{}_{12}\tilde s^{}_{13}) \Delta \sigma
\nonumber \\
\hspace{0.5cm} - \frac{1}{2} (m^{}_{12}+{\rm i}m^{}_{11+3}\tilde
s^{}_{13} ) \Delta \theta^{}_{23} + \frac{1}{2} m^{}_{11-3}\tilde
s^{}_{13} \Delta \delta
\nonumber \\
\hspace{0.5cm}
= \frac{1}{2} (m^{}_{12}-{\rm i}m^{}_{11+3}\tilde s^{}_{13} )
(\epsilon^{}_1-{\rm i}\Delta \phi^{}_1-{\rm i}\Delta \phi ) \; ,
\nonumber \\
\overline m^{}_1 s^{}_{12} ({\rm i}s^{}_{12}
+2c^{}_{12} \tilde s^{}_{13} )\Delta \rho
+ \overline m^{}_2 c^{}_{12} ({\rm i}c^{}_{12}-2 s^{}_{12}
\tilde s^{}_{13} )\Delta \sigma
\nonumber \\
\hspace{0.5cm} -m^{}_{22-3}\Delta \theta^{}_{23}
+  m^{}_{12} \tilde s^{}_{13} \Delta \delta
\nonumber \\
\hspace{0.5cm} = \frac{1}{2} (m^{}_{22+3}-2{\rm i}
m^{}_{12}\tilde s^{}_{13} )
( \epsilon^{}_2 - 2{\rm i}\Delta \phi ) \;,
\end{eqnarray}
where $m^{}_{11\pm 3}$ and $m^{}_{22\pm3}$ stand for $m^{}_{11}\pm
m^{}_3$ and $m^{}_{22}\pm m^{}_3$, respectively. Moreover,
$\overline m^{}_1$ (or $\overline m^{}_2$) and $\tilde s^{}_{13}$
are equal to $\pm m^{}_1$ (or $\pm m^{}_2$) and $\pm s^{}_{13}$,
respectively, for $\rho^{(0)} =0$ or $\pi/2$ (or $\sigma^{(0)} =0$
or $\pi/2$) and $\delta^{(0)}=\pm \pi/2$. Note that the above four
equations actually correspond to the conditions of $\mu$-$\tau$
reflection symmetry breaking defined in Eq. (3.71). After solving these
equations, one will see the clear dependence of $\Delta
\theta^{}_{23}$, $\Delta \delta$, $\Delta \rho$ and $\Delta \sigma$
on the perturbation parameters ${\rm R}^{}_{1,2} \equiv {\rm
Re}(\epsilon^{}_{1,2})$ and ${\rm I}^{}_{1,2} \equiv {\rm
Im}(\epsilon^{}_{1,2})$. For clarity, we express their results in
the following parametrizations:
\begin{eqnarray}
\Delta \theta^{}_{23} & = c^{\theta}_{r1} {\rm R}^{}_1 +
c^{\theta}_{i1} {\rm I}^{}_1 + c^{\theta}_{r2} {\rm R}^{}_2
+ c^{\theta}_{i2} {\rm I}^{}_2 \; ,
\nonumber \\
\Delta \delta & = c^{\delta}_{r1} {\rm R}^{}_1 +
c^{\delta}_{i1} {\rm I}^{}_1 + c^{\delta}_{r2} {\rm R}^{}_2
+ c^{\delta}_{i2} {\rm I}^{}_2 \; ,
\nonumber \\
\Delta \rho & = c^{\rho}_{r1} {\rm R}^{}_1 +
c^{\rho}_{i1} {\rm I}^{}_1 + c^{\rho}_{r2} {\rm R}^{}_2
+ c^{\rho}_{i2} {\rm I}^{}_2 \; ,
\nonumber \\
\Delta \sigma & = c^{\sigma}_{r1} {\rm R}^{}_1 +
c^{\sigma}_{i1} {\rm I}^{}_1 + c^{\sigma}_{r2} {\rm R}^{}_2
+ c^{\sigma}_{i2} {\rm I}^{}_2 \; .
\end{eqnarray}
The lengthy expressions of these 16 coefficients
are listed in the Appendix for reference.

Since the Majorana phases cannot be pinned down in a foreseeable
future, we only discuss the properties of $\Delta \theta^{}_{23}$
and $\Delta \delta$ in several typical schemes regarding the
neutrino mass spectrum.

(1) As for $m^{}_1 \ll m^{}_2 \ll m^{}_3$, the expressions of
$\Delta \theta^{}_{23}$ and $\Delta \delta$ can approximate to
\begin{eqnarray}
\Delta \theta^{}_{23} \simeq & \left(2m^2_3\right)^{-1}
\left(-2 m^2_{2} c^{2}_{12} s^{2}_{12} {\rm R}^{}_1
-4 m^{2}_{2} c^{}_{12} s^3_{12} \tilde s^{}_{13} {\rm I}^{}_1 \right.
\nonumber \\
& \left. + m^2_3 {\rm R}^{}_2 - \bar m^{}_{2} m^{}_3 c^{}_{12} s^{}_{12}
\tilde s^{}_{13} {\rm I}^{}_2\right) \; ,
\nonumber \\
\Delta \delta \simeq & \hspace{-0.2cm} \left( 2 \overline m^{}_2
m^{}_3 c^{}_{12} s^{}_{12} \tilde s^{}_{13}\right)^{-1} \left[
m^{2}_{2} c^{2}_{12} s^{2}_{12} \left(2 {\rm R}^{}_1 +
{\rm R}^{}_2 \right) \right.
\nonumber \\
& \left. \hspace{-0.2cm} + m^2_3 \left(c^2_{12} - s^2_{12} \right)
s^2_{13} \left( 2{\rm R}^{}_1-{\rm R}^{}_2\right) \right] \nonumber \\
& \hspace{-0.2cm} -2 s^2_{12} {\rm I}^{}_1 - \left(2 \overline
m^{}_2 \right)^{-1} m^{}_3 {\rm I}^{}_2 \;.
\end{eqnarray}
In this case the largest coefficient for $\Delta \theta^{}_{23}$ is
$|c^{\theta}_{r2}|\simeq 0.5$. On the other hand, $\Delta
\theta^{}_{23}$ is almost insensitive to ${\rm R}^{}_1$ and ${\rm
I}^{}_{1,2}$. In contrast, $\delta$ is more sensitive to the
symmetry breaking because of an enhancement factor $1/s^{}_{13}$.
Numerically, the coefficients $|c^{\delta}_{r1,i1,i2}|$ (or
$|c^{\delta}_{r2}|$) take values of $\mathcal O(1)$ (or $\mathcal
O(0.1)$) around $m^{}_1 \sim 0.001$ eV.

(2) The results in the $m^{}_1 \simeq m^{}_2 \gg m^{}_3$ case
strongly depend on the values of $\rho^{(0)}$ and $\sigma^{(0)}$.
When these two phases are equal, one will have
\begin{eqnarray}
\Delta \theta^{}_{23} \simeq -s^2_{13} {\rm R}^{}_1 - r c^{}_{12}
s^{}_{12} \tilde s^{}_{13} \left({\rm I}^{}_1- \frac{3}{4} {\rm
I}^{}_2 \right) -\frac{1}{2} {\rm R}^{}_2  \; ,
\nonumber \\
\Delta \delta \simeq \left(4 c^{}_{12} s^{}_{12} \tilde
s^{}_{13}\right)^{-1} \left[  \left(c^2_{12} - s^2_{12}\right) s^2_{13}
(2{\rm R}^{}_1 + {\rm R}^{}_2) \right. \nonumber \\
\left. \hspace{0.9cm} - r c^{2}_{12} s^{2}_{12} (2 {\rm R}^{}_1
-{\rm R}^{}_2) \right ] + {\rm I}^{}_1- \frac{3}{4} {\rm I}^{}_2 \;
,
\end{eqnarray}
where $r \equiv \Delta m^{2}_{21}/\Delta m^2_{31} \simeq 0.03$.
Among the relevant coefficients, $|c^\theta_{r2}|$ and
$|c^\delta_{i1,i2}|$ are of $\mathcal O(1)$, and the others are much
smaller. If $\rho^{(0)}$ and $\sigma^{(0)}$ take different values,
such as $\rho^{(0)} =0$ and $\sigma^{(0)} =\pi/2$, the expressions
of $\Delta \theta^{}_{23}$ and $\Delta \delta$ turn out to be
\begin{eqnarray}
\Delta \theta^{}_{23} \simeq & -4 c^{2}_{12} s^2_{12} {\rm R}^{}_1
-\left(c^{2}_{12} - s^{2}_{12}\right) c^{}_{12} s^{}_{12} \tilde
s^{}_{13} \left(4{\rm I}^{}_1 - {\rm I}^{}_2\right)
\nonumber \\
& - \frac{1}{2} \left(c^{2}_{12} - s^{2}_{12}\right)^2 {\rm R}^{}_2
\; ,
\nonumber \\
\Delta \delta \simeq & \left(r c^{}_{12} s^{}_{12}\right)^{-1}
\left[c^{}_{12} s^{}_{12} \left(c^2_{12} - s^2_{12}\right)
\left(4 {\rm I}^{}_1 - {\rm I}^{}_2\right) \right.
\nonumber \\
& \left. - \tilde s^{}_{13} \left(2{\rm R}^{}_1 + {\rm R}^{}_2\right)
\right] \; .
\end{eqnarray}
In this case $|c^\theta_{r1}|\sim 1$  is the largest coefficient for
$\Delta \theta^{}_{23}$ while the other three are smaller by at
least one order of magnitude. $\delta$ is very unstable against the
symmetry breaking as its coefficients can easily exceed 10 due to
the enhancement factor $1/r$.

(3) When the mass spectrum $m^{}_1 \simeq m^{}_2 \simeq m^{}_3$ is
concerned, the effects of $\mu$-$\tau$ reflection symmetry breaking
in $M^{}_\nu$ may be significantly magnified, as one can see from
the corresponding results of $\Delta \theta^{}_{23}$ and $\Delta
\delta$. In the $\rho^{(0)}=\sigma^{(0)}=0$ case one obtains
\begin{eqnarray}
\Delta \theta^{}_{23} \simeq 2 \frac{m^2_1} {\Delta m^2_{31}}
\left[2 s^2_{13} {\rm R}^{}_1 - r c^{}_{12} s^{}_{12} \tilde
s^{}_{13} \left(2 {\rm I}^{}_1 - {\rm I}^{}_2 \right) + {\rm R}^{}_2
\right] ,
\nonumber \\
\Delta \delta \simeq \left( c^{}_{12} s^{}_{12} \tilde
s^{}_{13}\right)^{-1} \left\{ \left[r c^{2}_{12} s^{2}_{12} +
\left(c^2_{12} - s^2_{12}\right) s^2_{13}\right] {\rm R}^{}_1
\right.
\nonumber \\
\hspace{0.9cm} + 2 \frac{m^2_1} {\Delta m^2_{31}} \left[ r
c^{2}_{12} s^{2}_{12} - \left(c^2_{12} - s^2_{12}\right) s^2_{13}
\right] {\rm R}^{}_2
\nonumber \\
\hspace{0.93cm} \left. - \frac {m^2_1} {\Delta m^2_{31}}
\left(c^2_{12} - s^2_{12}\right) \left(2{\rm I}^{}_1 - {\rm
I}^{}_2\right) \right\} .
\end{eqnarray}
Given $m^{}_1 \sim 0.1$ eV, for example, $|c^{\theta}_{r2}|$ can
reach 6 while $|c^{\theta}_{r1,i1,i2}|$ are much smaller. $\Delta
\delta$ is highly sensitive to ${\rm I}^{}_{1,2}$ whose relevant
coefficients are of $\mathcal O(10)$, but it is insensitive to ${\rm
R}^{}_{1,2}$ whose relevant coefficients are of $\mathcal O(0.1)$.
On the other hand, ($\rho^{(0)}$, $\sigma^{(0)}$) = (0, $\pi/2$)
will lead $\Delta \theta^{}_{23}$ and $\Delta \delta$ to
\begin{eqnarray}
\Delta \theta^{}_{23} \simeq 2\frac{m^2_1}{\Delta m^2_{31}}
\left[2c^2_{12} s^2_{12} {\rm R}^{}_1
+2\left(c^2_{12}-s^2_{12}\right) c^{}_{12} s^{}_{12} \tilde
s^{}_{13} {\rm I}^{}_1 \right.
\nonumber \\
\hspace{0.95cm} \left. +s^4_{12} {\rm R}^{}_2 + c^{}_{12}s^{3}_{12}
\tilde s^{}_{13} {\rm I}^{}_2 \right] \; ,
\nonumber \\
\Delta \delta \simeq \frac{m^2_1}{\Delta m^2_{21}} \left[
8\frac{m^2_1}{\Delta m^2_{31}}c^{}_{12}s^{}_{12} \tilde
s^{}_{13}\left(2c^2_{12} {\rm R}^{}_1-s^2_{12} {\rm R}^{}_2\right)
\right.
\nonumber \\
\hspace{0.95cm} \left. + 4\left(c^{2}_{12}-s^{2}_{12}\right)
{\rm I}^{}_1+2c^2_{12} {\rm I}^{}_2 \right] \; .
\end{eqnarray}
In this case $|c^{\theta}_{r1,r2}|$ and $|c^{\theta}_{i1,i2}|$ are
respectively of $\mathcal O(1)$ and $\mathcal O(0.1)$, and all the
coefficients associated with $\Delta \delta$ may be enhanced to 100.

In short, $\Delta \delta$ and $\Delta \theta^{}_{23}$ tend to be
enlarged when the degeneracy of the neutrino masses grows and
$\rho^{(0)}$ takes a different value from $\sigma^{(0)}$.
Furthermore, $\delta$ is in general more unstable than
$\theta^{}_{23}$ with respect to the effects of $\mu$-$\tau$
reflection symmetry breaking. In particular, the coefficients
associated with $\Delta \delta$ can reach a value about 100 provided
$m^{}_1 \simeq m^{}_2 \simeq m^{}_3$ and $\rho^{(0)} \neq
\sigma^{(0)}$. Furthermore, $\Delta \theta^{}_{23}$ is more
sensitive to ${\rm R}^{}_{1,2}$ while $\Delta \delta$ is more
sensitive to ${\rm I}^{}_{1,2}$. With the help of such observations,
one may break the $\mu$-$\tau$ reflection symmetry in a proper way
whenever necessary in a model-building exercise.

\subsection{RGE-induced $\mu$-$\tau$ symmetry breaking effects}

If a certain flavor symmetry is associated with the neutrino mass
generation and lepton flavor mixing at a superhigh energy scale
$\Lambda^{}_{\rm FS}$, it will be broken due to the RGE evolution of
$M^{}_\nu$ from $\Lambda^{}_{\rm FS}$ down to the electroweak scale
$\Lambda^{}_{\rm EW} \sim 10^2$ GeV. In this case the significant
difference between $m^{}_{\mu}$ and $m^{}_{\tau}$ is just a source
of symmetry breaking which can be transmitted to $M^{}_\nu$ via the
RGE running effects. From a model-building point of view, it is in
general natural to introduce a kind of flavor symmetry at an energy
scale much higher than $\Lambda^{}_{\rm EW}$ (e.g., the seesaw
scale). So the RGE-triggered symmetry breaking effects should be
taken into account when such a model is confronted with the
available experimental data at low energies \cite{OZ}. It is
therefore worthwhile to explore how the $\mu$-$\tau$ flavor symmetry
is broken via the RGEs. At the one-loop level the RGE running
behavior of $M^{}_{\nu}$ is described by \cite{RGE1,RGE2,RGE3,RGE4}
\begin{eqnarray}
\frac{{\rm d}M^{}_{\nu}}{{\rm d}t} = C \left(Y^{\dagger}_l Y^{}_l
\right)^{T} M^{}_\nu +C M^{}_\nu \left(Y^{\dagger}_l Y^{}_l
\right) + \alpha M^{}_{\nu} \;
\end{eqnarray}
with $t\equiv (1/16\pi^2)\ln (\mu/\Lambda^{}_{\rm FS})$, in which
$\mu$ denotes an energy scale between $\Lambda^{}_{\rm EW}$ and
$\Lambda^{}_{\rm FS}$, $C=1$ and
\begin{eqnarray}
\alpha \simeq -\frac{6}{5}g^2_1-6 g^2_2+6 y^2_t \;
\end{eqnarray}
within the minimal supersymmetry standard model (MSSM). In Eq.
(3.79) the $\alpha$-term just provides an overall rescaling factor
which will be referred to as $I^{}_{\alpha}$, while the other two
terms may change the structure of $M^{}_\nu$. In the basis chosen
for Eq. (3.79), the Yukawa coupling matrix of three charged leptons
is diagonal: $Y^{}_l = {\rm Diag}\{y^{}_e,y^{}_\mu, y^{}_\tau\}$.
Because of $y^{}_{e} \ll y^{}_\mu \ll y^{}_\tau$, it is reasonable
to neglect the contributions of both $y^{}_{e}$ and $y^{}_\mu$ in
the subsequent discussions. So we are left with $y^{2}_\tau =
\left(1 + \tan^2{\beta}\right) m^2_\tau/v^2$ in the MSSM with $v
\simeq 246$ GeV being the vacuum expectation value of the Higgs
fields. There are normally two ways to proceed: a) one may follow
the method described in Refs. \cite{KSB,CEIN,RGE5} to derive the
differential RGEs of neutrino masses and flavor mixing parameters
from Eq. (3.79); b) one may first integrate Eq. (3.79) to arrive at
the RGE-corrected neutrino mass matrix at $\Lambda^{}_{\rm EW}$ and
then diagonalize the latter to obtain neutrino masses and flavor
mixing parameters. Here we follow the second approach so as to see
how $M^{}_\nu$ is modified by the RGE running effects in a
transparent way. After integrating Eq. (3.79), we get
\cite{IRGE1,IRGE2}
\begin{eqnarray}
M^\prime_{\nu}= I^{}_{\alpha} I^{\dagger}_\tau M^{}_{\nu}
I^{*}_{\tau} \;
\end{eqnarray}
at $\Lambda^{}_{\rm EW}$, where $I^{}_\tau \simeq {\rm Diag} \{1, 1,
1-\Delta^{}_{\tau}\}$ and
\begin{eqnarray}
I^{}_{\alpha} = {\rm exp}\left( \int^{\Lambda^{}_{\rm
EW}}_{\Lambda^{}_{\rm FS}} \alpha {\rm d}t \right) , \hspace{0.4cm}
\Delta^{}_{\tau} = \int^{\Lambda^{}_{\rm FS}}_{\Lambda^{}_{\rm
EW}}y^2_{\tau} {\rm d}t \; .
\end{eqnarray}
Given $\Lambda^{}_{\rm FS}\sim 10^{14}$ GeV, for example,
$I^{}_{\alpha}$ changes from 0.9 to 0.8 and $\Delta^{}_\tau$ changes
from 0.002 to 0.044 when $\tan{\beta}$ varies from 10 to 50.

Now let us consider a specific neutrino mass matrix respecting the
$\mu$-$\tau$ permutation symmetry at $\Lambda^{}_{\rm FS}$, such as
the one given in Eq. (3.4) with $\kappa = +1$. With the help of Eq.
(3.81), one may calculate the RGE corrections to this matrix at
$\Lambda^{}_{\rm EW}$ and express the result in a way parallel to
Eq. (3.4):
\begin{eqnarray}
M^\prime_{\nu} \simeq I^{}_{\alpha} \left[ M^{}_\nu - \Delta^{}_\tau
\pmatrix{ 0 & 0 & M^{}_{e \tau} \cr 0 & 0 & M^{}_{\mu \tau} \cr
M^{}_{e \tau} & M^{}_{\mu \tau} & 2 M^{}_{\tau\tau}\cr} \right] \; .
\end{eqnarray}
It becomes obvious that the term proportional to $\Delta^{}_\tau$
measures the strength of $\mu$-$\tau$ symmetry breaking. One may
diagonalize $M^\prime_\nu$ via the unitary transformation $U^\prime
= P^{\prime}_{\phi} V^\prime P^{\prime}_{\nu}$, where $V^\prime$ is
an analogue of $V$ shown in Eq. (2.8), and $P^\prime_\phi$ (or
$P^\prime_\nu$) is an analogue of $P^{}_\phi$ (or $P^{}_\nu$) as
given above Eq. (3.45). Note again that $P^\prime_\phi$ should not be
ignored in this treatment so as to keep everything consistent,
although its phases have no physical meaning. After a lengthy but
straightforward calculation, we obtain the neutrino masses
\begin{eqnarray}
m^{\prime}_{1} & \simeq &
I^{}_{\alpha}\left(1-s^{2}_{12}\Delta^{}_{\tau}\right)m^{}_{1} \; ,
\nonumber \\
m^{\prime}_{2} & \simeq &
I^{}_{\alpha}\left(1-c^{2}_{12}\Delta^{}_{\tau}\right)m^{}_{2} \; ,
\nonumber \\
m^{\prime}_{3} & \simeq &
I^{}_{\alpha}\left(1-\Delta^{}_{\tau}\right) m^{}_3 \;
\end{eqnarray}
at $\Lambda^{}_{\rm EW}$; and the flavor mixing angles \cite{DGR}
\begin{eqnarray}
\theta^{\prime}_{12} & \simeq & \theta^{}_{12}+
\frac{1}{2}c^{}_{12}s^{}_{12} \frac{\left|\overline m^{}_1+\overline
m^{}_2\right|^2}{\Delta m^2_{21}} \Delta^{}_{\tau} \; ,
\nonumber \\
\theta^{\prime}_{13} & \simeq & c^{}_{12}s^{}_{12}
\frac{m^{}_{3}\left| \overline m^{}_1 - \overline
m^{}_2 \right|} {\Delta m^2_{31}} \Delta^{}_{\tau} \; ,
\nonumber \\
\theta^{\prime}_{23} & \simeq & \frac{\pi}{4}+
\frac{\left|\overline m^{}_1 + m^{}_3\right|^2s^{2}_{12} +
\left|\overline m^{}_2+m^{}_3\right|^2c^{2}_{12}}{2 \Delta m^2_{31}}
\Delta^{}_{\tau} \;
\end{eqnarray}
at $\Lambda^{}_{\rm EW}$. In addition, the two Majorana phases at
the electroweak scale are given by
\begin{eqnarray}
\rho^{\prime} & \simeq & \rho - \frac{m^{}_1 m^{}_2 \sin{2
\left(\rho-\sigma\right) c^{2}_{12}}}{\Delta
m^2_{21}}\Delta^{}_{\tau} \; ,
\nonumber \\
\sigma^{\prime} & \simeq & \sigma-\frac{m^{}_1 m^{}_2
\sin{2\left(\rho-\sigma\right)s^{2}_{12}}}{\Delta
m^2_{21}}\Delta^{}_{\tau} \; .
\end{eqnarray}
Note that the validity of these interesting analytical results
depends on the prerequisite that none of the physical parameters
evolves violently from $\Lambda^{}_{\rm FS}$ down to
$\Lambda^{}_{\rm EW}$ to make the perturbation expansions invalid.
In this sense it is more general to discuss the running effects of
relevant flavor parameters by deriving their differential RGEs from
Eq. (3.79). But Eqs. (3.84)---(3.86) remain very useful for us to
see some salient features of radiative corrections to the neutrino
masses and flavor mixing parameters. Some more comments and
discussions are in order.
\begin{itemize}
\item     If the neutrino mass ordering is inverted or
quasi-degenerate, the radiative correction to $\theta^{}_{12}$ tends
to be appreciable because it may be enhanced by a factor of
$\mathcal O(100)$ in either of these two cases, although $y^{}_\tau$
itself is strongly suppressed even in the MSSM. Nevertheless, a
proper choice of $\rho-\sigma \sim \pm \pi/2$ can give rise to a
significant cancellation in $|\overline m^{}_1 + \overline m^{}_2|$
and thus suppress the aforementioned enhancement and stabilize the
running behavior of $\theta^{}_{12}$ \cite{NYO,BDMP}. If the
neutrino mass ordering is normal, $\theta^\prime_{12} -
\theta^{}_{12} \sim c^{}_{12}s^{}_{12}\Delta^{}_\tau/2$ is expected
to be small. The higher the scale $\Lambda^{}_{\rm FS}$ is, the
smaller the corresponding value of $\theta^{}_{12}$ will be as
compared with its value at $\Lambda^{}_{\rm EW}$, because it always
increases when running down in the MSSM scheme \cite{MST}.

\item As for $\theta^{}_{13}$, the RGE running effects are small
no matter whether the neutrino mass ordering is normal or inverted.
To generate $\theta^{}_{13} \sim 9^\circ$ at $\Lambda^{}_{\rm EW}$
from $\theta^{}_{13} =0$ at $\Lambda^{}_{\rm FS}$, the value of
$\tan{\beta}$ has to be larger than 50 unless the absolute neutrino
mass scale is unreasonably higher than 0.1 eV \cite{GPRR}. In this
case the bottom-quark Yukawa coupling eigenvalue $y^{}_b$ would lie
in the non-perturbation regime at a superhigh energy scale --- a
result which is definitely unacceptable \cite{MSSM}. Namely, the
RGE-induced $\mu$-$\tau$ permutation symmetry breaking is hard to
fit the observed value of $\theta^{}_{13}$, and thus one has to
introduce a nonzero $\theta^{}_{13}$ at $\Lambda^{}_{\rm FS}$. In
the latter case the analytical approximations obtained in Eqs.
(3.84)---(3.86) remain valid
\footnote{In this case the expression of $\theta^\prime_{13}$ in Eq.
(3.85) should be modified as $\theta^\prime_{13} \simeq
\theta^{}_{13} + c^{}_{12} s^{}_{12} m^{}_3 \left| \overline{m}^{}_1
- \overline{m}^{}_2\right| \Delta^{}_\tau/\Delta m^2_{31}$.},
simply because $\theta^{}_{13}$ itself does not significantly affect
other parameters \cite{RGE5}.

\item An appreciable deviation of $\theta^{\prime}_{23}$ from $\pi/4$
can be produced to fit the present experimental data when the
neutrino masses are nearly degenerate. In particular, such a
deviation will be positive in the MSSM scheme if $\Delta m^2_{31}
>0$ holds, as one can see from Eq. (3.85). This interesting observation
offers a potential correlation between the octant of $\theta^{}_{23}$
and the neutrino mass hierarchy, and two numerical examples will be
presented below \cite{LZ}.
\end{itemize}
Finally it is worth pointing out that a finite and meaningful value
of $\delta$ will be generated along with a nonzero value of
$\theta^{}_{13}$ after the exact $\mu$-$\tau$ permutation symmetry
at $\Lambda^{}_{\rm FS}$ is broken via the RGE running effects. Of
course, the real seeds of such a radiative generation of $\delta$
must be the nontrivial Majorana phases $\rho$ and $\sigma$ at
$\Lambda^{}_{\rm FS}$ \cite{Mei,DGR2}.

The above approach also applies to the RGE-triggered $\mu$-$\tau$
reflection symmetry breaking. Consider that a neutrino mass matrix
$M^{}_\nu$ exhibiting the $\mu$-$\tau$ reflection symmetry is
generated at a high scale $\Lambda^{}_{\rm FS}$. Then the
corresponding neutrino mass matrix $M^\prime_\nu$ at the electroweak
scale $\Lambda^{}_{\rm EW}$ can be deduced through Eq. (3.81) when
the RGE running effects are taken into account. It is possible to
reparametrize $M^{\prime}_\nu$ in the form given by Eq. (3.69) with
$\epsilon^{}_2 = 2\epsilon^{}_1 = \Delta^{}_\tau$ as well as
$\epsilon^{}_{3,4} = 0$ \cite{YL2}. Therefore, we can derive the
RGE-induced $\Delta \theta^{}_{23}$, $\Delta \delta$, $\Delta \rho$
and $\Delta \sigma$ from Eq. (3.73) by taking ${\rm R^{}_2} = 2 {\rm
R}^{}_1 = \Delta^{}_\tau$ and ${\rm I}^{}_{1,2} = 0$. While the
result of $\Delta \theta^{}_{23}$ is the same as that shown in Eq.
(3.85), the result of $\Delta \delta$ appears as
\begin{eqnarray}
\frac{\Delta \delta}{\Delta^{}_\tau} = & \frac{m^{}_3
\left(\overline m^{}_2 - \overline m^{}_1\right)
c^{}_{12} s^{}_{12}} { \left(m^2_3 - m^2_1\right) \tilde s^{}_{13}} +
\left( \frac {\overline m^{}_1 s^2_{12}} {\overline m^{}_1 - m^{}_3} +
\frac{\overline m^{}_2 c^2_{12}} {\overline m^{}_2 - m^{}_3} \right)
\nonumber \\
& \cdot \frac{\left(\overline m^{}_1 + m^{}_3\right) c^2_{12}
- \left(\overline m^{}_2 + m^{}_3\right) s^2_{12}}
{\overline m^{}_1 + \overline m^{}_2} \cdot \frac{\tilde s^{}_{13}}
{c^{}_{12} s^{}_{12}} \; .
\end{eqnarray}
For illustration, we show the possible values of $\Delta
\theta^{}_{23}$, $\Delta \delta$, $\Delta \rho$ and $\Delta \sigma$
against the lightest neutrino mass in four typical cases in Fig.
3.3. Case (a): ($\rho^{(0)}$, $\sigma^{(0)}$) = (0, 0) and
$\tan{\beta}=50$ with a normal neutrino mass ordering; Case (b):
($\rho^{(0)}$, $\sigma^{(0)}$) = (0, 0) and $\tan{\beta}=30$ with an
inverted mass ordering; Case (c): ($\rho^{(0)}$,
$\sigma^{(0)}$) = (0, $\pi/2$) and $\tan{\beta}=10$ with a normal
neutrino mass ordering; and Case (d): ($\rho^{(0)}$, $\sigma^{(0)}$)
= (0, $\pi/2$) and $\tan{\beta}=10$ with an inverted mass
ordering. We have taken $\delta^{(0)}=-\pi/2$ in our numerical
calculations. As one can see, the RGE running effects may be
significant in the $m^{}_1 \simeq m^{}_2 \gg m^{}_3$ case, and
especially in the $m^{}_1 \simeq m^{}_2 \simeq m^{}_3$ case with
with $\sigma^{(0)} \neq \rho^{(0)}$. Moreover, $\Delta
\theta^{}_{23}$ tends to be larger than $\Delta \delta$ when
$\rho^{(0)}$ and $\sigma^{(0)}$ take the same values, or vice versa.
\begin{figure*}
\vspace{0.17cm}
\centering
\includegraphics[width=6.in]{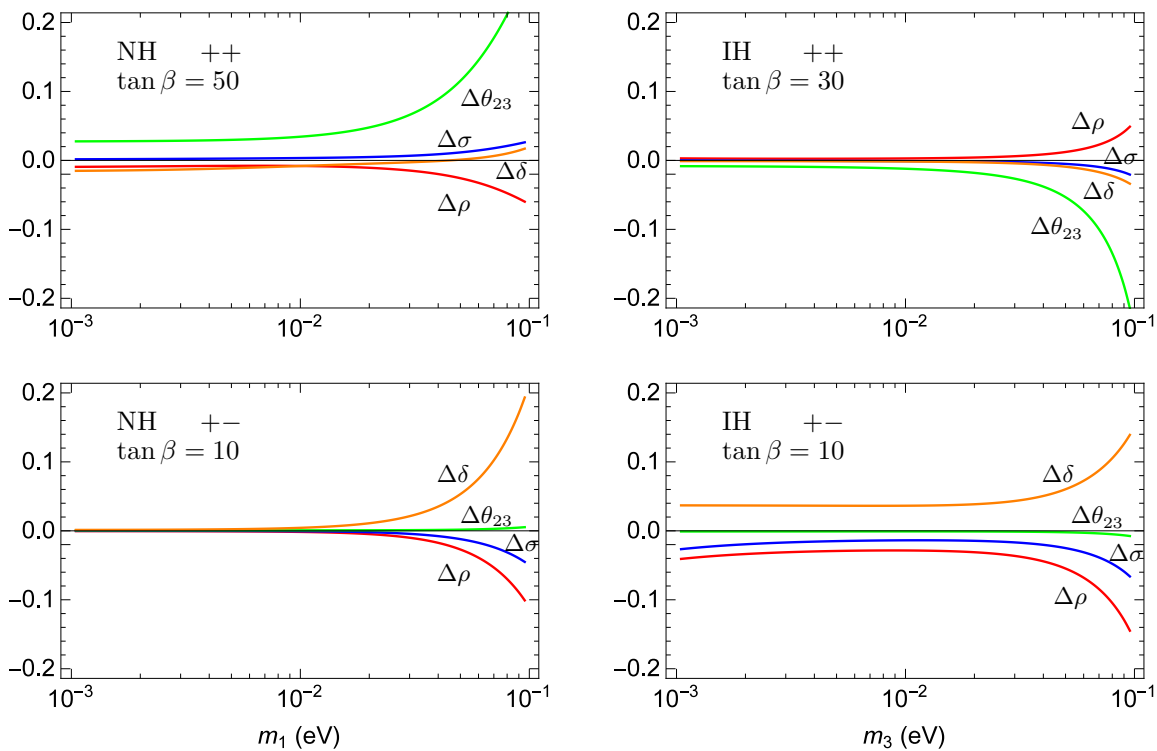}
\caption{The possible values of $\Delta \theta^{}_{23}$, $\Delta
\delta$, $\Delta \rho$ and $\Delta \sigma$ against the lightest
neutrino mass in four typical cases. We have taken $\Lambda^{}_{\rm
FS} \sim 10^{14}$ GeV for the flavor symmetry scale. The sign
``$+-$" or ``$++$" corresponds to $(\rho^{(0)}, \sigma^{(0)}) = (0,
0)$ or $(\rho^{(0)}, \sigma^{(0)}) = (0, \pi/2)$. }
\end{figure*}

We have discussed the RGE-induced corrections to a Majorana neutrino
mass matrix which initially possesses the $\mu$-$\tau$ permutation
or reflection symmetry. Now we turn to radiative corrections to
$|U^{}_{\mu i}| = |U^{}_{\tau i}|$, since these equalities are a
straightforward consequence of the $\mu$-$\tau$ symmetry. In other
words, the observed effects of $\mu$-$\tau$ symmetry breaking in the
PMNS matrix $U$ can be attributed to the RGE-induced corrections to
$|U^{}_{\mu i}| = |U^{}_{\tau i}|$ when the neutrino masses and
flavor mixing parameters run from $\Lambda^{}_{\rm FS}$ down to
$\Lambda^{}_{\rm EW}$. In this case it is possible to get the
correct octant of $\theta^{}_{23}$ and even the correct quadrant of
$\delta$ \cite{LZ}. Let us illustrate this striking point with the
help of the one-loop RGEs in the MSSM framework. Note that the
framework of the SM for the RGE running is less interesting in this
connection for two simple reasons: (a) it is difficult to make the
deviation of $\theta^{}_{23}$ from $45^\circ$ appreciable even if
the neutrinos have a nearly degenerate mass spectrum; (b) the SM
itself largely suffers from the vacuum-stability problem for the
measured value of the Higgs mass ($\simeq 125$ GeV) as the flavor
symmetry scale $\Lambda^{}_{\rm FS}$ is above $10^{10}$ GeV
\cite{XZZ}. For simplicity, here we start from $\theta^{}_{23}
=\pi/4$ and $\delta = -\pi/2$ at $\Lambda^{}_{\rm FS} \sim 10^{14}$
GeV to fit the observed pattern of the PMNS matrix $U$ at
$\Lambda^{}_{\rm EW} \sim 10^2$ GeV by taking account of the RGE
evolution. Hence we are not subject to the special values of the
Majorana phases as constrained by the $\mu$-$\tau$ reflection
symmetry imposed on the neutrino mass matrix $M^{}_\nu$. Namely, we
mainly concentrate on the radiative breaking of the equalities
$|U^{}_{\mu i}| = |U^{}_{\tau i}|$ without going into details of the
model-building issues.

We first define three $\mu$-$\tau$ ``asymmetries" of the PMNS matrix
$U$ and then figure out their expressions in the standard
parametrization as follows:
\begin{eqnarray}
\Delta^{}_1 & \equiv & |U^{}_{\tau 1}|^2 - |U^{}_{\mu 1}|^2
\nonumber \\
& = & \left( \cos^2\theta^{}_{12} \sin^2\theta^{}_{13} -
\sin^2\theta^{}_{12} \right) \cos 2\theta^{}_{23}
\nonumber \\
&& - \sin 2\theta^{}_{12} \sin\theta^{}_{13} \sin 2\theta^{}_{23}
\cos\delta \; ,
\nonumber \\
\Delta^{}_2 & \equiv & |U^{}_{\tau 2}|^2 - |U^{}_{\mu 2}|^2
\nonumber \\
& = & \left( \sin^2\theta^{}_{12} \sin^2\theta^{}_{13} -
\cos^2\theta^{}_{12} \right) \cos 2\theta^{}_{23}
\nonumber \\
&& + \sin 2\theta^{}_{12} \sin\theta^{}_{13} \sin 2\theta^{}_{23}
\cos\delta \; ,
\nonumber \\
\Delta^{}_3 & \equiv & |U^{}_{\tau 3}|^2 - |U^{}_{\mu 3}|^2 =
\cos^2\theta^{}_{13} \cos 2\theta^{}_{23} \; .
\end{eqnarray}
It is clear that the three asymmetries satisfy the sum rule
$\Delta^{}_1 + \Delta^{}_2 + \Delta^{}_3 =0$, and they vanish when
the exact $\mu$-$\tau$ flavor symmetry holds (i.e., when
$\theta^{}_{13} =0$ and $\theta^{}_{23} =\pi/4$, or when $\delta =
\pm\pi/2$ and $\theta^{}_{23} =\pi/4$). The present best-fit results
of neutrino mixing parameters listed in Table 2.1 indicate that the
possibility of $\delta = -\pi/2$ and $\theta^{}_{23} =\pi/4$ is
slightly more favored than either the possibility of $\delta =
\pi/2$ and $\theta^{}_{23} =\pi/4$ or that of $\theta^{}_{13} =0$
and $\theta^{}_{23} =\pi/4$. Hence we infer that the natural
condition for all the three $\Delta^{}_i$ to vanish should be
$\delta = -\pi/2$ and $\theta^{}_{23} = \pi/4$ at $\Lambda^{}_{\rm
FS} \sim 10^{14}$ GeV, and the observed pattern of lepton flavor
mixing at $\Lambda^{}_{\rm EW} \sim 10^2$ GeV should be a
consequence of the most ``economical" $\mu$-$\tau$ symmetry breaking
triggered by the RGE running effects from $\Lambda^{}_{\rm FS}$ to
$\Lambda^{}_{\rm EW}$ \cite{LZ}. Explicitly, the one-loop RGEs of
$\Delta^{}_i$ are given by
\begin{eqnarray}
\frac{{\rm d}\Delta^{}_1}{{\rm d}t} & = &
-y^2_\tau \left[ \xi^{}_{21} \left( |U^{}_{\tau
1}|^2 \Delta^{}_2 + |U^{}_{\tau 2}|^2 \Delta^{}_1 + |U^{}_{e 3}|^2
\right) \right.
\nonumber \\
& & + \xi^{}_{31} \left( |U^{}_{\tau 1}|^2 \Delta^{}_3 +
|U^{}_{\tau 3}|^2 \Delta^{}_1 + |U^{}_{e 2}|^2 \right)
\nonumber \\
& & + \zeta^{}_{21} \left(
|U^{}_{\tau 1}|^2 \Delta^{}_2 + |U^{}_{\tau 2}|^2 \Delta^{}_1 +
|U^{}_{e 3}|^2 \right) \cos\Phi^{}_{12}
\nonumber \\
& & + \zeta^{}_{31} \left( |U^{}_{\tau 1}|^2 \Delta^{}_3 +
|U^{}_{\tau 3}|^2 \Delta^{}_1 + |U^{}_{e 2}|^2 \right)
\cos\Phi^{}_{13}
\nonumber \\
& & \left. + {\cal J} \left(\zeta^{}_{21} \sin\Phi^{}_{12} -
\zeta^{}_{31} \sin\Phi^{}_{13} \right)\right] \; ,
\nonumber \\
\frac{{\rm d}\Delta^{}_2}{{\rm d}t} & = &
y^2_\tau \left[ \xi^{}_{21} \left( |U^{}_{\tau
1}|^2 \Delta^{}_2 + |U^{}_{\tau 2}|^2 \Delta^{}_1 + |U^{}_{e 3}|^2
\right) \right.
\nonumber \\
& & - \xi^{}_{32} \left( |U^{}_{\tau 2}|^2 \Delta^{}_3 +
|U^{}_{\tau 3}|^2 \Delta^{}_2 + |U^{}_{e 1}|^2 \right)
\nonumber \\
& & + \zeta^{}_{21} \left(
|U^{}_{\tau 1}|^2 \Delta^{}_2 + |U^{}_{\tau 2}|^2 \Delta^{}_1 +
|U^{}_{e 3}|^2 \right) \cos\Phi^{}_{12}
\nonumber \\
& & - \zeta^{}_{32}
\left( |U^{}_{\tau 2}|^2 \Delta^{}_3 + |U^{}_{\tau 3}|^2 \Delta^{}_2
+ |U^{}_{e 1}|^2 \right) \cos\Phi^{}_{23}
\nonumber \\
& & \left. + {\cal J} \left(\zeta^{}_{21} \sin\Phi^{}_{12}
- \zeta^{}_{32} \sin\Phi^{}_{23} \right)\right] \; ,
\nonumber \\
\frac{{\rm d}\Delta^{}_3}{{\rm d}t} & = &
y^2_\tau \left[ \xi^{}_{31} \left( |U^{}_{\tau
1}|^2 \Delta^{}_3 + |U^{}_{\tau 3}|^2 \Delta^{}_1 + |U^{}_{e 2}|^2
\right) \right.
\nonumber \\
& & + \xi^{}_{32} \left( |U^{}_{\tau 2}|^2 \Delta^{}_3 +
|U^{}_{\tau 3}|^2 \Delta^{}_2 + |U^{}_{e 1}|^2 \right)
\nonumber \\
& & + \zeta^{}_{31} \left(
|U^{}_{\tau 1}|^2 \Delta^{}_3 + |U^{}_{\tau 3}|^2 \Delta^{}_1 +
|U^{}_{e 2}|^2 \right) \cos\Phi^{}_{13}
\nonumber \\
& & + \zeta^{}_{32}
\left( |U^{}_{\tau 2}|^2 \Delta^{}_3 + |U^{}_{\tau 3}|^2 \Delta^{}_2
+ |U^{}_{e 1}|^2 \right) \cos\Phi^{}_{23}
\nonumber \\
& & \left. - {\cal J} \left(\zeta^{}_{31} \sin\Phi^{}_{13}
- \zeta^{}_{32} \sin\Phi^{}_{23} \right)\right] \; ,
\end{eqnarray}
where $\xi^{}_{ij} \equiv (m^2_i + m^2_j)/\Delta m^2_{ij}$,
$\zeta^{}_{ij} \equiv 2m^{}_i m^{}_j/\Delta m^2_{ij}$,
$\cos\Phi^{}_{ij} \equiv {\rm Re}(U^{}_{\tau i} U^*_{\tau
j})^2/|U^{}_{\tau i} U^*_{\tau j}|^2$ as well as $\sin\Phi^{}_{ij}
\equiv {\rm Im}(U^{}_{\tau i} U^*_{\tau j})^2/ |U^{}_{\tau i}
U^*_{\tau j}|^2$ \cite{LZ}. The two Majorana CP-violating phases
$\rho$ and $\sigma$ affect the evolution of $\Delta^{}_i$ via
$\cos\Phi^{}_{ij}$ and $\sin\Phi^{}_{ij}$. Given the fact $|\Delta
m^2_{31}| \simeq |\Delta m^2_{32}| \sim 30 \Delta m^2_{21}$ with
$\Delta m^2_{21} \simeq 7.5 \times 10^{-5} ~{\rm eV}^2$
\cite{Fogli}, we expect $\xi^{}_{21} \gg |\xi^{}_{31}| \simeq
|\xi^{}_{32}|$ to hold in most cases. But this does not necessarily
mean that $\Delta^{}_3$ should be more stable against radiative
corrections than $\Delta^{}_1$ and $\Delta^{}_2$, because their
running behaviors also depend on the initial inputs of
$|U^{}_{\alpha i}|^2$. An appreciable deviation of $\theta^{}_{23}$
from $\pi/4$, which is sensitive or equivalent to an appreciable
deviation of $\Delta^{}_3$ from zero, requires a sufficiently large
value of $\tan\beta$. Hence the resulting octant of $\theta^{}_{23}$
is controlled by the neutrino mass ordering (i.e., the sign of
$\Delta m^2_{31}$ or $\Delta m^2_{32}$, or equivalently the sign of
$\xi^{}_{31}$ or $\xi^{}_{32}$). Since the signs of $\xi^{}_{ij}$
and $\zeta^{}_{ij}$ are always the same, one may adjust the evolving
direction of $\Delta^{}_3$ without much fine-tuning of the other
relevant parameters.
\begin{table}[t]
\caption{The RGE-triggered corrections to $|U^{}_{\mu i}| =
|U^{}_{\tau i}|$ for Majorana neutrinos running from $\Delta^{}_i
=0$ at $\Lambda^{}_{\rm FS}$ down to $\Lambda^{}_{\rm EW}$ in the
MSSM with $\tan\beta = 31$.} \vspace{0.1cm}
\begin{indented}
\item[]\begin{tabular}{lll} \br
Parameter & $\Lambda^{}_{\rm FS}\sim 10^{14}$ GeV &
$\Lambda^{}_{\rm EW} \sim 10^2$ GeV \\ \mr
\multicolumn{3}{c}{Example I (Capozzi {\it et al} \cite{Fogli})} \\ \mr
\vspace{0.1cm}
$m^{}_{1} ~ ({\rm eV} )$ & 0.100 & 0.087 \\ \vspace{0.1cm}
$\Delta m^{2}_{21} ~ ({\rm eV}^2 )$ & $1.70 \times 10^{-4}$ &
$7.54 \times 10^{-5}$ \\ \vspace{0.1cm}
$\Delta m^{2}_{31} ~ ({\rm eV}^2 )$ & $-2.98 \times 10^{-3}$ &
$-2.34 \times 10^{-3}$ \\ \vspace{0.1cm}
$\theta^{}_{12}$ & $35.2^\circ$ & $33.7^\circ$ \\ \vspace{0.1cm}
$\theta^{}_{13}$ & $10.1^\circ$ & $8.9^\circ$ \\ \vspace{0.1cm}
$\theta^{}_{23}$ & $45.0^\circ$ & $42.4^\circ$ \\ \vspace{0.1cm}
$\delta$ & $270^\circ$ & $236^\circ$ \\ \vspace{0.1cm}
$\rho$ & $-82^\circ$ & $-66^\circ$ \\ \vspace{0.1cm}
$\sigma$ & $19^\circ$ & $27^\circ$ \\ \vspace{0.1cm}
${\cal J}$ & $-0.040$ & $-0.029$ \\ \vspace{0.1cm}
$\Delta^{}_{1}$ & 0 & 0.054 \\ \vspace{0.1cm}
$\Delta^{}_{2}$ & 0 & $-0.142$ \\ \vspace{0.1cm}
$\Delta^{}_{3}$ & 0 & 0.088 \\ \br
\multicolumn{3}{c}{Example II (Forero {\it et al} \cite{Valle})} \\ \mr
\vspace{0.1cm}
$m^{}_{1} ~ ({\rm eV} )$ & 0.100 & 0.087 \\ \vspace{0.1cm}
$\Delta m^{2}_{21} ~ ({\rm eV}^2 )$ & $2.12 \times 10^{-4}$
& $7.60 \times 10^{-5}$ \\ \vspace{0.1cm}
$\Delta m^{2}_{31} ~ ({\rm eV}^2 )$ & $3.50 \times 10^{-3}$
& $2.48 \times 10^{-3}$ \\ \vspace{0.1cm}
$\theta^{}_{12}$ & $32.1^\circ$ & $34.6^\circ$ \\ \vspace{0.1cm}
$\theta^{}_{13}$ & $6.9^\circ$ & $8.8^\circ$ \\ \vspace{0.1cm}
$\theta^{}_{23}$ & $45.0^\circ$ & $48.9^\circ$ \\ \vspace{0.1cm}
$\delta$ & $270^\circ$ & $241^\circ$ \\ \vspace{0.1cm}
$\rho$ & $-76^\circ$ & $-45^\circ$ \\ \vspace{0.1cm}
$\sigma$ & $17^\circ$ & $29^\circ$ \\ \vspace{0.1cm}
${\cal J}$ & $-0.027$ & $-0.030$ \\ \vspace{0.1cm}
$\Delta^{}_{1}$ & 0 & 0.111 \\ \vspace{0.1cm}
$\Delta^{}_{2}$ & 0 & 0.022 \\ \vspace{0.1cm}
$\Delta^{}_{3}$ & 0 & $-0.133$ \\ \br
\end{tabular}
\end{indented}
\end{table}

We proceed to present two numerical examples to illustrate radiative
corrections to the equalities $|U^{}_{\mu i}| = |U^{}_{\tau i}|$,
corresponding to the best-fit results of neutrino oscillation
parameters given in Refs. \cite{Fogli} and \cite{Valle}. We start
from $\Delta^{}_i =0$ (i.e., $\theta^{}_{23} = 45^\circ$ and $\delta
= 270^\circ$) at $\Lambda^{}_{\rm FS} \sim 10^{14}$ GeV and run them
down to $\Lambda^{}_{\rm EW} \sim 10^2$ GeV via the RGEs that have
been given in Eq. (3.89). Table 3.1 summarizes the typical inputs
and outputs of these two examples. Some discussions are in order.

(1) In Example I with the inverted neutrino mass ordering, the
best-fit results of $\Delta m^2_{21}$, $\Delta m^2_{31}$,
$\theta^{}_{12}$, $\theta^{}_{13}$, $\theta^{}_{23}$ and $\delta$ at
$\Lambda^{}_{\rm EW}$ \cite{Fogli} can be successfully reproduced
from the proper inputs at $\Lambda^{}_{\rm FS}$. In this case
$\theta^{}_{23} (\Lambda^{}_{\rm FS}) - \theta^{}_{23}
(\Lambda^{}_{\rm EW}) \simeq 2.6^\circ$ and $\delta
(\Lambda^{}_{\rm FS}) - \delta (\Lambda^{}_{\rm EW}) \simeq
34^\circ$ hold thanks to the RGE running effects, and thus
$\theta^{}_{23} (\Lambda^{}_{\rm EW})$ lies in the first octant and
$\delta (\Lambda^{}_{\rm EW})$ is located in the third quadrant.

(2) In Example II only the normal neutrino mass ordering allows us
to arrive at $\theta^{}_{23}(\Lambda^{}_{\rm EW}) \simeq 48.9^\circ$
from $\theta^{}_{23}(\Lambda^{}_{\rm FS}) = 45^\circ$. In this case
we obtain $\delta(\Lambda^{}_{\rm EW}) \simeq 241^\circ$ from
$\delta(\Lambda^{}_{\rm FS}) = 270^\circ$, a result consistent with
the best-fit value of $\delta$ \cite{Valle}. The future long- and
medium-baseline neutrino oscillation experiments are going to pin
down the octant of $\theta^{}_{23}$ and the quadrant of $\delta$,
and then it will be possible to test the expected correlation
between the neutrino mass ordering and the deviation of
$\theta^{}_{23}$ (or $\delta$) from $45^\circ$ (or $270^\circ$).

(3) The running behaviors of $\Delta^{}_i$ from $\Lambda^{}_{\rm
FS}$ down to $\Lambda^{}_{\rm EW}$ can similarly be understood. In
view of $\Delta^{}_3 = \cos^2\theta^{}_{13} \cos 2\theta^{}_{23}$
given in Eq. (3.88), one must have $\Delta^{}_3 (\Lambda^{}_{\rm
EW}) > 0$ for $\theta^{}_{23} (\Lambda^{}_{\rm EW}) < 45^\circ$ in
Example I, and $\Delta^{}_3 (\Lambda^{}_{\rm EW}) < 0$ for
$\theta^{}_{23} (\Lambda^{}_{\rm EW})
> 45^\circ$ in Example II. In comparison, the evolution of
$\Delta^{}_1$ or $\Delta^{}_2$ is not so obvious, but $\Delta^{}_1 +
\Delta^{}_2 + \Delta^{}_3 =0$ always holds at any energy scale
between $\Lambda^{}_{\rm EW}$ and $\Lambda^{}_{\rm FS}$ \cite{LZ}.

(4) One has to adjust the initial values of $\rho$ and $\sigma$ at
$\Lambda^{}_{\rm FS}$ in a careful way to control the running
behaviors of six neutrino oscillation parameters, so that their
best-fit results at $\Lambda^{}_{\rm EW}$ can be correctly
reproduced. Nevertheless, we find that it is really possible to
resolve the octant of $\theta^{}_{23}$ and the quadrant of $\delta$
through the radiative breaking of $|U^{}_{\mu i}| = |U^{}_{\tau i}|$
by inputting proper values of the absolute neutrino mass $m^{}_1$
and the MSSM parameter $\tan\beta$. Once such unknown parameters are
measured or constrained to a better degree of accuracy in the
future, it will be possible to examine whether the quantum
corrections can really accommodate the observed effects of
$\mu$-$\tau$ symmetry breaking.

\subsection{Flavor mixing from the charged-lepton sector}

So far we have taken the charged-lepton mass matrix $M^{}_{l}$ to be
diagonal. If a specific texture of the neutrino mass matrix
$M^{}_{\nu}$ fails to generate the observed pattern of lepton flavor
mixing, however, a non-diagonal form of $M^{}_{l}$ should be taken
into account. Of course, $M^{}_{l}$ is usually non-diagonal in a
realistic flavor-symmetry model. In a grand unified theory (GUT),
for example, $M^{}_{l}$ is always associated with the down-type
quark mass matrix $M^{}_{\rm d}$ and thus both of them should be
non-diagonal. In all these contexts it is meaningful to consider the
contribution of $M^{}_l$ to the overall lepton flavor mixing matrix
$U = O^\dagger_l O^{}_\nu$
\cite{charged1,charged2,charged3,charged4,
charged5,charged5,charged6,charged7,charged8,charged9,charged10,
charged11,charged12,charged13,charged14,charged15,charged16}, where
the phases of $O^{}_{l}$ and $O^{}_\nu$ deserve some particular
attention. To clarify the issue, we follow the formulas given in
Ref. \cite{AK} but adopt a slightly different parametrization $O=
U^{}_{23}U^{}_{13}U^{}_{12}P^{}_{\alpha}$ for $O^{}_l$ or
$O^{}_\nu$, where $P^{}_{\alpha} = {\rm Diag}\{e^{{\rm i}
\alpha^{}_{1}},e^{{\rm i}\alpha^{}_{2}},e^{{\rm i}\alpha^{}_{3}}\}$
and
\begin{eqnarray}
U^{}_{12} = \pmatrix{ c^{}_{12} & \tilde{s}^{*}_{12} & 0 \cr
-\tilde{s}^{}_{12} & c^{}_{12} & 0 \cr 0 & 0 & 1 \cr} ,
\nonumber \\
U^{}_{13} = \pmatrix{ c^{}_{13} & 0 & \tilde{s}^{*}_{13} \cr 0 & 1 &
0 \cr -\tilde{s}^{}_{13} & 0 & c^{}_{13} \cr} ,
\nonumber \\
U^{}_{23} = \pmatrix{ 1 & 0 & 0 \cr 0 & c^{}_{23} &
\tilde{s}^{*}_{23} \cr 0 & -\tilde{s}^{}_{23} & c^{}_{23} \cr} ,
\end{eqnarray}
with the definition $\tilde s^{}_{ij} = s^{}_{ij} e^{{\rm
i}\delta^{}_{ij}}$ (for $ij = 12, 13, 23$). In comparison, we recall
the standard parametrization
$O=P^{}_{\phi}O^{}_{23}U^{}_{13}O^{}_{12}P^{}_{\nu}$ with
$P^{}_{\phi}={\rm Diag}\{e^{{\rm i}\phi^{}_{1}}, e^{{\rm
i}\phi^{}_{2}}, e^{{\rm i}\phi^{}_{3}}\}$ and $P^{}_{\nu}={\rm
Diag}\{e^{{\rm i}\rho},e^{{\rm i}\sigma},1\}$. These two
descriptions are related to each other as follows:
\begin{eqnarray}
\delta^{}_{12}=\phi^{}_2-\phi^{}_1 \; , \hspace{0.3cm}
\delta^{}_{13}=\delta+\phi^{}_3-\phi^{}_1 \; ,
\hspace{0.3cm} \delta^{}_{23}=\phi^{}_3-\phi^{}_2\; ,
\nonumber \\
\alpha^{}_1=\phi^{}_1+\rho \; ,\hspace{0.3cm}
\alpha^{}_2=\phi^{}_2+\sigma \; , \hspace{0.3cm}
\alpha^{}_3=\phi^{}_3 \; .
\end{eqnarray}
To be more specific, the lepton flavor mixing matrix $U$ in the new
parametrization reads
\begin{eqnarray}
U & = & O^{\dagger}_{l}O^{}_{\nu} = P^{l\dagger}_{\alpha}
U^{l\dagger}_{12} U^{l\dagger}_{13} U^{l\dagger}_{23}
U^{\nu}_{23}U^{\nu}_{13}U^{\nu}_{12}P^{\nu}_{\alpha}
\nonumber \\
& = &
P^{l\dagger}_{\alpha}U^{}_{23}U^{}_{13}U^{}_{12}P^{\nu}_{\alpha} \;
,
\end{eqnarray}
where the phase matrix $P^{l}_{\alpha}$ is irrelevant in physics as
it can be rotated away via a redefinition of the phases of three
charged-lepton fields. In most cases of the model-building
exercises, the three angles of $O^{}_l$ and the $(1,3)$ angle of
$O^{}_\nu$ are significantly small
\footnote{Of course, a given $O^{}_l$ with one or two large angles
may correct $O^{}_\nu$ in a remarkable way \cite{FPR}. But this
possibility is beyond the scope of our interest in the present
article.}.
To the leading order of such small parameters, the three mixing
angles of $U$ can approximate to \cite{AK}
\begin{eqnarray}
\tilde s^{}_{12} \simeq & \tilde s^{\nu}_{12}-\tilde
\theta^{l}_{12}c^{\nu}_{12}c^{\nu}_{23} +\tilde
\theta^{l}_{13}c^{\nu}_{12}\tilde s^{\nu*}_{23} \; ,
\nonumber \\
\tilde s^{}_{13} \simeq &
\tilde \theta^{\nu}_{13} -\tilde \theta^{l}_{13}c^{\nu}_{23}
-\tilde \theta^{l}_{12}\tilde s^{\nu}_{23} \; ,
\nonumber \\
\tilde s^{}_{23} \simeq & \tilde s^{\nu}_{23}- \tilde
\theta^{l}_{23}c^{\nu}_{23} \; .
\end{eqnarray}
A particularly interesting case is that $\theta^{l}_{13}$ and
$\theta^{\nu}_{13}$ are both negligibly small as compared with
$\theta^{l}_{12}$, leading us to the approximate results
\begin{eqnarray}
\delta \simeq \delta^l_{12} - \delta^\nu_{12} - \pi \; ,
\nonumber \\
s^{}_{13} \simeq \theta^{l}_{12} s^{\nu}_{23} \; ,
\nonumber \\
s^{}_{12} \simeq s^{\nu}_{12} +
\theta^{l}_{12}c^{\nu}_{12}c^{\nu}_{23} \cos{\delta} \; .
\end{eqnarray}
In obtaining Eq. (3.94), we have used Eq. (3.91) as well as
$\delta^{}_{23} \simeq \delta^{\nu}_{23}$ and $\delta^{}_{12} \simeq
\delta^{\nu}_{12}$. Given the approximate $\mu$-$\tau$ flavor
symmetry, $\theta^{\nu}_{23}$ is supposed to be around $\pi/4$. If
$\theta^{l}_{12}$ approximates to the Cabibbo angle of quark flavor
mixing $\theta^{}_{\rm C} \simeq 0.22$
\footnote{This assumption makes sense in some GUT
models, where $M^{}_l$ and $M^{}_{\rm d}$ are usually related to
each other in a similar way.},
then Eq. (3.94) can give rise to an interesting relation
$\theta^{}_{13} \simeq \theta^{}_{\rm C}/\sqrt{2} \simeq 0.15$,
which is in good agreement with the present experimental data.
Moreover, Eq. (3.94) implies
\begin{eqnarray}
s^{}_{12} \simeq  s^{\nu}_{12} +
c^{\nu}_{12}\theta^{}_{13}\cos{\delta} \; .
\end{eqnarray}
If $O^{}_\nu \simeq U^{}_{\rm BM}$ as given in Eq. (3.11), one will
arrive at $s^{}_{12} \gtrsim \left(1-\theta^{}_{13}\right)/\sqrt{2}$
from Eq. (3.95). This lower bound is apparently inconsistent with
the experimental result of $\theta^{}_{12}$. If $O^{}_\nu \simeq
U^{}_{\rm TB}$ as given in Eq. (3.10), however, the situation will
be different because $\theta^{\nu}_{12}$ itself is already close to
the observed value of $\theta^{}_{12}$. In this case $\delta$ is
required to approach $\pm \pi/2$ in order to suppress the
contribution from the second term in Eq. (3.95).

Although the pattern of lepton flavor mixing appears strikingly
different from that of quark flavor mixing, they have been
speculated to have a potential link. In this respect a viable GUT
model may offer an ideal context where the relevant quarks and
leptons reside in the same representations such that their
respective Yukawa coupling matrices can be naturally correlated
\footnote{To see some realistic models with certain flavor symmetries
being considered in the GUT framework, we refer the readers to
Ref. \cite{Review3} and references therein.}.
For instance, the  aforementioned relation $\theta^{}_{13}
\simeq\theta^{}_{\rm C}/\sqrt{2}$ makes a possible unification of
quarks and leptons quite appealing. Such a relation will come out if
$M^{}_{\nu}$ respects the $\mu$-$\tau$ permutation symmetry and
$O^{}_{l}$ has a CKM-like structure (i.e., $\theta^{l}_{12}$ is very
close to the Cabibbo angle $\theta^{}_{\rm C}$ while
$\theta^{l}_{13}$ and $\theta^{l}_{23}$ are negligibly small).
Moreover, the well-known Gatto-Sartori-Tonin (GST) relation
$\theta^{}_{\rm C}\simeq \sqrt{m^{}_d/m^{}_s}$ \cite{GST} tempts us
to attribute the Cabibbo angle to the down-type quark mass matrix
$M^{}_{\rm d}$. The latter happens to be related to $M^{}_l$ in the
GUT framework \cite{Jue}. Let us take the SU(5) GUT model, which can
be embedded in the SO(10) GUT model, as an example. Accordingly,
each family of quarks and leptons are grouped into the SU(5)
representations ${\bf \bar 5}$ and ${\bf 10}$ in the manner
\begin{eqnarray}
{\bf{\bar 5}}&=\pmatrix {d^{c}_{\rm r} \cr d^{c}_{\rm b} \cr
d^{c}_{\rm g} \cr e \cr -\nu^{}_e} , \hspace{0.15cm} {\bf
10}&=\pmatrix {0& u^{c}_{\rm g} & -u^c_{\rm b} & u^{}_{\rm r} &
d^{}_{\rm r} \cr \cdot & 0 & u^{c}_{\rm r} & u^{}_{\rm b} &
d^{}_{\rm b} \cr \cdot & \cdot & 0 & u^{}_{\rm g} & d^{}_{\rm g} \cr
\cdot & \cdot & \cdot & 0 & e^c \cr \cdot & \cdot & \cdot & \cdot &
0 }
\end{eqnarray}
with the subscripts ``r", ``b" and ``g" being the color indices of
quarks. Since the right-handed neutrino fields $N^{}_{i}$ are the
SU(5) singlets, their masses can be much larger than the electroweak
symmetry breaking scale. In the minimal SU(5) GUT scenario whose
Higgs sector only contains the 5-dimensional representations
$H^{}_{\bf 5}$ and $H^{}_{\bf \bar 5}$, the Yukawa interaction terms
include $Y^{ij}_{\rm u} {\bf 10}^{}_i {\bf 10}^{}_j H^{}_{\bf 5} $,
$Y^{ij}_\nu  {\bf \bar 5}^{}_i N^c_j H^{}_{\bf 5}$ and $Y^{ij}_{\rm
d} {\bf 10}^{}_i {\bf \bar 5}^{}_j H^{}_{\bf \bar 5}$ with $i$ and
$j$ being the family indices \cite{Review3}. After the SU(5)
symmetry is broken down to the SM gauge symmetry, the Yukawa
coupling matrices of quarks and leptons take the following forms:
\begin{eqnarray}
-\mathcal L^{}_{\rm mass} & = & Y^{ij}_{\rm u}
Q^{}_i u^c_j H^{}_{\rm u}+ Y^{ij}_\nu L^{}_i N^c_j H^{}_{\rm u}
+ Y^{ij}_{\rm d} Q^{}_i  d^c_j  H^{}_{\rm d}
\nonumber \\
&& \hspace{-0.11cm} + Y^{ij}_{\rm d} e^c_i L^{}_j H^{}_{\rm d} \; ,
\end{eqnarray}
in which the relevant notations are self-explanatory. It is well
known that the last two terms in Eq. (3.97) imply $M^{}_{\rm d} =
M^{T}_l$ after the electroweak gauge symmetry breaking. Hence it
seems natural to expect $\theta^{l}_{12}\simeq \theta^{}_{\rm C}$ at
the GUT scale if $\theta^{}_{\rm C}$ mainly originates from
$M^{}_{\rm d}$. On the other hand, the mass relations
$m^{}_{d}=m^{}_{e}$, $m^{}_{s}=m^{}_{\mu}$ and
$m^{}_{b}=m^{}_{\tau}$ derived from $M^{}_{\rm d}=M^{T}_l$ at the
GUT scale are difficult to fit current experimental data even though
the RGE running effects of those masses are taken into account. Such
a problem forces us to extend the Higgs sector with higher
dimensional representations. A famous example of this kind is the
45-dimensional Higgs representation $H^{}_{\bf \overline{45} }$
introduced in the Georgi-Jarlskog (GJ) mechanism \cite{GJ}. This
Higgs field contributes to the down-type quark sector and the
charged-lepton sector via the Yukawa interaction term
\begin{eqnarray}
Y^{ij}_{\rm d} {\bf 10}^{}_i {\bf \bar 5}^{}_j H^{}_{\bf
\overline{45}} \hspace{0.1cm} \Longrightarrow \hspace{0.1cm}
Y^{ij}_{\rm d} \left(Q^{}_i  d^c_j- 3 e^c_i L^{}_j\right) H^{}_{\rm
d} \; ,
\end{eqnarray}
in which the factor $-3$ is a Clebsch-Gordan coefficient. Given such
a difference between $M^{}_{\rm d}$ and $M^{T}_l$, the GJ mechanism
may lead to a few more realistic mass relations. But a new problem
emerges in this case: $\theta^{l}_{12} \simeq \theta^{}_{\rm C}/3$,
which is too small to enhance $\theta^{}_{13}$ from 0 to the
observed value.

Following the model-building strategies outlined in Refs.
\cite{AGMS,SFK2012}, one may certainly consider some more options
for the Higgs representations in order to derive
$\theta^{l}_{12}\simeq \theta^{}_{\rm C}$ as well as the viable mass
relations. If only the dimension-four operators are allowed, then
only the 5- and 45-dimensional Higgs representations can contribute
to lepton and quark masses. When the dimension-five operators are
taken into account, more possibilities will arise. The contributions
of a given operator to the down-type quark and charged-lepton masses
are parametrized as $Y^{ij}_{\rm d} \left( Q^{}_i d^c_j+ c^{}_{ij}
e^c_i L^{}_j\right) H^{}_{\rm d}$ with $c^{}_{ij}$ being the
Clebsch-Gordan coefficient. A list of all the candidate operators
and the associated $c^{}_{ij}$ coefficients can be found in Ref.
\cite{NewGUT}.

Assuming the $(1,3)$ and $(2,3)$ rotation angles to be negligible,
here we focus on the $(1,2)$ submatrices of $M^{}_{\rm d}$ and
$M^{}_l$ which are mainly relevant to the first two families. In
order to have the GST relation as a natural outcome, let us consider
the texture \cite{W77,F77}
\begin{eqnarray}
M^{}_{\rm d} = \pmatrix{0 & b \cr c & a} \; , \hspace{0.5cm} M^{}_l
= \pmatrix{0 & c^{}_{c}c \cr c^{}_{b}b & c^{}_{a} a} \; ,
\end{eqnarray}
where $b \simeq c$ holds from an empirical observation, and
$c^{}_{a,b,c}$ denote the Clebsch-Gordan coefficients of the
relevant entries --- they can be different from one another,
indicating that different entries are generated by different Higgs
representations. Given the complex $(1,2)$ rotation submatrix of
$U^{}_{12}$ defined in Eq. (3.90), a diagonalization of $M^{}_{\rm
d}$ or $M^{}_l$ is straightforward. Since $a$, $b$ and $c$ are
related to $m^{}_d$, $m^{}_s$ and $\theta^{}_{\rm C}$, one just
needs to find out a proper combination of $c^{}_{a}$, $c^{}_b$ and
$c^{}_c$ which can yield the realistic electron and muon masses as
well as the desired relation $\theta^{l}_{12} \simeq \theta^{}_{\rm
C}$. It is found that only the combination of $c^{}_b=1/2$ and
$c^{}_a=c^{}_c=6$ can satisfy the above requirement
\cite{AM2011,MPRS}. In this case the relation $\theta^{}_{13}\simeq
\theta^{}_{\rm C}/\sqrt{2}$ can finally be achieved provided
$M^{}_\nu$ possesses the $\mu$-$\tau$ flavor symmetry. But
$\delta^{l}_{12}$ and $\delta^{\nu}_{12}$ remain undetermined,
preventing us from calculating $\delta$ and $s^{\nu}_{12}$ based on
Eq. (3.94). One may follow the idea of spontaneous CP breaking to
find a way out. To be specific, the CP symmetry is imposed on the
model in the beginning to forbid the presence of CP violation, and
it is spontaneously broken by some specific fields so that the
resulting CP-violating phases are under control. A good example of
this kind has been given in Ref. \cite{AGMS2}, where a scalar field
acquires an imaginary vacuum expectation value and thus breaks the
CP symmetry to the maximum level. This vacuum expectation value
leads to $\delta^{l}_{12}=\pm \pi/2$, while $\delta^{\nu}_{12}$
remains vanishing since the neutrino sector has no link to the
relevant scalar field. So we arrive at $\delta = \pm\pi/2$ for the
PMNS matrix $U$, and $s^{\nu}_{12}$ should take a value which is not
far away from the experimental value of $s^{}_{12}$. In such an
explicit model-building exercise one may construct $O^{}_\nu =
U^{}_{\rm TB}$ to make things easy.

\def\thefootnote{\arabic{footnote}}
\setcounter{footnote}{0}
\setcounter{equation}{0}
\setcounter{table}{0}
\setcounter{figure}{0}

\section{Larger flavor symmetry groups}

Before introducing some concrete models to show how to realize the
$\mu$-$\tau$ flavor symmetry, let us first formulate the connection
between neutrino mixing and flavor symmetries from a
group-theoretical angle and then outline the general strategy for
generating a particular neutrino mixing pattern. In addition, we
shall briefly describe some recent works on combining the flavor
symmetries and generalized CP (GCP) symmetries. Such a new treatment
can not only pave the way for realizing the $\mu$-$\tau$ flavor
symmetry but also lay the foundation for embedding it in a larger
flavor symmetry group. The latter will become relevant especially
when one is ambitious to nail down the PMNS matrix $U$ completely,
since the $\mu$-$\tau$ symmetry itself can only fix a part of the
texture of $U$.

\subsection{Neutrino mixing and flavor symmetries}

As shown in section 3.1, $\theta^{}_{13} =0$ and $\theta^{}_{23}
=\pi/4$ lead the third column of $U$ to the form $(0,
-1,1)^{T}/\sqrt{2}$, implying a $\rm Z^{}_2$ or $\mu$-$\tau$
permutation symmetry of $M^{}_\nu$. And the reverse is also true. In
fact, a link between the pattern of neutrino mixing and a certain
flavor symmetry of $M^{}_\nu$ like this always exists, irrespective
of any particular form of $U$ in a sense as follows. In the basis
where $M^{}_l$ is diagonal, $U$ is identified as the unitary matrix
used to diagonalize $M^{}_\nu$ (i.e., $U^\dagger M^{}_\nu U^* =
D^{}_\nu$). Hence the relation $M^{}_\nu u^{*}_i= m^{}_i u^{}_i$ is
implied, where $u^{}_i$ denotes the $i$-th column of $U$ and
$m^{}_i$ is the $i$-th neutrino mass eigenvalue corresponding to
$u^{}_i$. Then it is easy to check that $M^{}_\nu$ must be invariant
under the transformations $\mathcal S^{}_i$ (for $i=1,2,3$):
\begin{eqnarray}
\mathcal S^\dagger_i M^{}_\nu \mathcal S^*_i & = \mathcal
S^\dagger_i (m^{}_i u^{}_i u^{T}_i-m^{}_j u^{}_j u^{T}_j -m^{}_k
u^{}_k u^{T}_k)
\nonumber \\
& = m^{}_i u^{}_i u^{T}_i + m^{}_j u^{}_j u^{T}_j + m^{}_k u^{}_k
u^{T}_k
\nonumber \\
& = U D_\nu U^{T}= M^{}_\nu \; ,
\end{eqnarray}
where $\mathcal S^{}_i$ are defined in terms of $u^{}_i$ via
\begin{eqnarray}
\mathcal S^{}_i = u^{}_i u^\dagger_i-(u^{}_j u^\dagger_j
+u^{}_k u^\dagger_k) \; .
\end{eqnarray}
All the three $\mathcal S^{}_i$ are unitary and commute with one
another. Thanks to the relation $\mathcal S^{}_i \mathcal S^{}_j
=\mathcal S^{}_k$, only two of them are independent. Furthermore,
each of $\mathcal S^{}_i$ has order 2 as indicated by $\mathcal
S^2_i=1$. Summarizing these features, we reach the conclusion that
$M^{}_\nu$ always has the Klein symmetry $\rm K^{}_4= Z^{}_2 \times
Z^{}_2$ \cite{Lam1,Lam2,Lam3}. The symmetry operators can be
explicitly constructed in terms of $u^{}_i$ in an approach described
by Eq. (4.2). For instance,
\begin{eqnarray}
\mathcal S^{\rm TB}_1 = \frac{1}{3} \pmatrix{1 & -2 & -2 \cr -2 & -2
& 1 \cr -2 & 1 & -2 } \; ,
\nonumber \\
\mathcal S^{\rm TB}_2 = \frac{1}{3} \pmatrix{-1 & 2 & 2 \cr 2 & -1 &
2 \cr 2 & 2 & -1 } \; ,
\nonumber \\
\mathcal S^{\rm TB}_3 = \pmatrix{-1 & 0 & 0 \cr
0 & 0 & -1 \cr 0 & -1 & 0 } \; ,
\end{eqnarray}
when $U$ takes the well-known flavor mixing pattern $U^{}_{\rm TB}$
as given in Eq. (3.10). Note that $\mathcal S^{\rm TB}_3$ is nothing
but the $\mu$-$\tau$ permutation operation in Eq. (3.6), up to a
sign rearrangement.

The above conclusion can be put another way. If $M^{}_\nu$ assumes a
flavor symmetry described by $\mathcal S^{}_i$, then the normalized
invariant eigenvector of $\mathcal S^{}_i$ (defined by $\mathcal
S^{}_i u^{}_i=u^{}_i$) will constitute one column of $U$
\cite{Lam1,Lam2,Lam3}
\footnote{But which column it will occupy is a matter of convention
and can be determined upon some phenomenological considerations.}.
One may verify this observation as follows. The invariance of
$M^{}_\nu$ with respect to $\mathcal S^{}_i$ yields
\begin{eqnarray}
M^{}_\nu u^{*}_i = \mathcal S^\dagger_i M^{}_\nu \mathcal S^* u^*_i
\hspace{0.2cm} \Longrightarrow \hspace{0.2cm} \mathcal S^{}_i
M^{}_\nu u^*_i = M^{}_\nu u^*_i \; ,
\end{eqnarray}
which in turn implies that $M^{}_\nu u^*_i$ is identical with
$u^{}_i$ up to a coefficient (i.e., the neutrino mass eigenvalue
$m^{}_i$):
\begin{eqnarray}
M^{}_\nu u^*_i = m^{}_i u^{}_i \hspace{0.2cm} \Longrightarrow
\hspace{0.2cm} u^\dagger_i M^{}_\nu u^*_i = m^{}_i \; .
\end{eqnarray}
It should be noted that the resulting flavor mixing pattern is
independent of $m^{}_i$. This point means that one may obtain a
constant pattern of neutrino mixing directly from the flavor
symmetry by bypassing a concrete form of $M^{}_\nu$. To do so,
however, one has to know all the three symmetry operators $\mathcal
S^{}_i$. If only a single $\mathcal S^{}_i$ is known, then $U$ can
be partially determined. For example, when $M^{}_\nu$ respects a
symmetry defined by $\mathcal S^{\rm TB}_1$ or $\mathcal S^{\rm
TB}_2$ given in Eq. (4.3), one column of $U$ will be
$(2,-1,-1)^{T}/\sqrt{6}$ or $(1,1,1)^{T}/\sqrt{3}$, respectively. We
are therefore led to the so-called TM1 or TM2 mixing pattern
\cite{Lam1,MMTB1,TMa,TMb,TMc,MMTB2,TMd}:
\begin{eqnarray}
U^{}_{\rm TM1} = \frac{1}{\sqrt{6}}\pmatrix{ 2  & \sqrt{2} c &
\sqrt{2} \tilde s^*  \cr -1  & \sqrt{2}c+ \sqrt{3} \tilde s &
\sqrt{2} \tilde s^*  -\sqrt{3} c \cr -1  & \sqrt{2}c-\sqrt{3} \tilde
s & \sqrt{2} \tilde s^* + \sqrt{3} c } \; ,
\nonumber  \\
U^{}_{\rm TM2} = \frac{1}{\sqrt 6}\pmatrix{ 2 c & \sqrt{2} &2 \tilde
s^* \cr - c + \sqrt{3} \tilde s  & \sqrt{2} & - \tilde s^* -\sqrt{3}
c \cr - c - \sqrt{3} \tilde s  & \sqrt{2} &  -\tilde s^* +\sqrt{3} c
} \; ,
\end{eqnarray}
where $\tilde s \equiv \sin{\theta}e^{{\rm i}\varphi}$ has been
defined. In Eq. (4.6) the location of the invariant eigenvector of
$\mathcal S^{\rm TB}_1$ or $\mathcal S^{\rm TB}_2$ has been
specified. The model-building details and phenomenological
consequences of these two particular neutrino mixing patterns will
be elaborated in section 5.3. Note that an $\mathcal S^{}_i$ can
only limit one column of $U$ up to the associated Majorana phase.
The reason lies in the fact that when one $u^{}_i$ is rephased as a
whole, it remains an invariant eigenvector of $\mathcal S^{}_i$. As
for the charged leptons, the diagonal Hermitian matrix
$M^{}_lM^\dagger_l$ possesses the symmetry
\begin{eqnarray}
\mathcal T^\dagger M^{}_l M^\dagger_l \mathcal T= M^{}_l M^\dagger_l
\; ,
\end{eqnarray}
in which $\mathcal T \equiv {\rm Diag}\{e^{{\rm i}\phi^{}_1},
e^{{\rm i}\phi^{}_2}, e^{-{\rm i}(\phi^{}_1+\phi^{}_2)}\}$ is a
diagonal phase matrix. Conversely, the Hermitian matrix $M^{}_l
M^\dagger_l$ must be diagonal if it satisfies the phase
transformation in Eq. (4.7). It is worth pointing out that the above
discussions keep valid in the basis of non-diagonal $M^{}_l$. But in
this case one needs to make an appropriate transformation of those
basis-dependent quantities. For example, the symmetry operator which
is imposed on $M^{}_l M^\dagger_l$ will become $\mathcal
T^\prime=O^{}_l \mathcal T O^{\dagger}_l$ with $O^{}_l$ being the
unitary matrix used to diagonalize $M^{}_l M^\dagger_l$.

Although it is quite possible that the $\mathcal S^{}_i$ and
$\mathcal T$ symmetries just arise by accident, we choose to
identify them as the residual symmetries from a larger flavor
symmetry group $\rm G^{}_F$ in the following \cite{Lam1,Lam2,Lam3}.
In other words, the subgroup $\rm G^{}_\nu$ (or ${\rm G}^{}_l$)
generated by $\mathcal S^{}_i$ (or $\mathcal T$) remains intact in
the neutrino (or charged-lepton) sector when $\rm G^{}_F$ itself is
broken down:
\begin{eqnarray}
{\rm G^{}_F} \Longrightarrow \left\{ \begin{array}{ll}
\hspace{-0.15cm} {\rm G^{}_\nu}=\langle \mathcal S^{}_i \rangle &
{\rm for} \hspace{0.3cm} M^{}_\nu \; , \\ \hspace{-0.15cm} {\rm
G}^{}_l=\langle \mathcal T \rangle & {\rm for} \hspace{0.3cm} M^{}_l
M^\dagger_l \; .
\end{array} \right.
\end{eqnarray}
From the bottom-up point of view, one may get a hold of $\rm G^{}_F$
by demanding that ${\rm G}^{}_l$ and $\rm G^{}_\nu$ close to a group
\cite{bottom-up}. In order for $\rm G^{}_F$ to be finite, there must
exist an $n$ such that $\mathcal T^n=1$. As already remarked, we
need a non-degenerate diagonal $\mathcal T$ to guarantee the
diagonality of $M^{}_l M^\dagger_l$, implying $n\geq 3$. For the
sake of simplicity, let us tentatively set $n =3$
\footnote{As a matter of fact, another choice of $n$ does not
necessarily yield a finite group. For example, it has been checked
that no finite group can be obtained by using a $\mathcal T$ with
$3<n<30$ and $\mathcal S^{\rm TB}_{2,3}$ as the generators
\cite{Review3}.}.
This means that $\mathcal T$ has $1$, $\omega$ and $\omega^2$ as its
diagonal entries, where $\omega \equiv e^{{\rm i}2\pi/3}$. But the
ordering of these three entries has to be specified, because
different options may result in different $\rm G^{}_F$. Given ${\rm
G}^{}_l =\langle \mathcal T ={\rm Diag} \{1, \omega, \omega^2\}
\rangle$ and $\rm G^{}_\nu = \langle \mathcal S^{\rm
TB}_{2},\mathcal S^{\rm TB}_{3} \rangle$, for example, we can arrive
at $\rm G^{}_F=S^{}_4$ (the permutation group of four objects).
Furthermore, it has been shown that $\rm S^{}_4$ (or a group
containing $\rm S^{}_4$ as a subgroup) is the unique discrete group
that is able to naturally accommodate the flavor mixing pattern
$U^{}_{\rm TB}$ from the group-theoretical arguments
\cite{Lam3,Lam4}. Indeed, it is likely that merely one $\mathcal
S^{}_i$ symmetry is the surviving symmetry from $\rm G^{}_F$. In
this situation $\rm G^{}_F$ is found to be successively $\rm
S^{}_4$, $\rm A^{}_4$ (the alternating group of four objects) or
$\rm S^{}_3$ (the permutation group of three objects) when the
generator of $\rm G^{}_\nu$ is only one of $\mathcal S^{\rm TB}_1$,
$\mathcal S^{\rm TB}_2$ and $\mathcal S^{\rm TB}_3$. Interestingly,
the $\rm A^{}_4$ group that has been popularly considered for
realizing $U^{}_{\rm TB}$ \cite{A41,A42} actually does not contain a
subgroup for the last column of $U^{}_{\rm TB}$. But an accidental
symmetry in such models may elevate the $\rm A^{}_4$ symmetry to
$S^{}_4$ \cite{Lam5}, making $U^{}_{\rm TB}$ available. Roughly
speaking, the group generated by $\mathcal T$ and $\mathcal
S^{}_{i}$ tends to be small when $\mathcal S^{}_i$ have some regular
forms like those illustrated in Eq. (4.3). If the neutrino mixing
pattern and the relevant $\mathcal S^{}_i$ lack any regularity, the
resulting $\rm G^{}_F$ will become much larger and even not finite.
That is why the observed pattern of lepton flavor mixing is more
likely to originate from a reasonable flavor symmetry than that of
quark flavor mixing \cite{Lam3}. In order to accommodate the
experimental value of $\theta^{}_{13}$, which more or less lowers
the regularity of $U$, a much larger flavor symmetry group has to be
invoked (an example of this kind is $\Delta(6N^2)$ \cite{Group} with
$N$ being a big number) \cite{TFH1,TFH2,Lam6,KSLIM}.

The aforementioned procedure of constructing $\rm G^{}_F$ from
certain $\rm G^{}_\nu$ and ${\rm G}^{}_l$ can be reversed: assuming
a flavor symmetry group $\rm G^{}_F$ for the charged leptons and
neutrinos, one may take some of its subgroups as the residual
symmetries and derive the associated flavor mixing pattern
\cite{Lam5}. Such an exercise can be done in a purely
group-theoretical way with no need of building an explicit dynamic
model. First of all, we need to divide the elements of $\rm G^{}_F$
into two categories
--- one with the elements of order 2 and the other with the elements
of order $\geq 3$. Then we may identify three elements $S^{}_i$
which satisfy $\mathcal S^{}_i \mathcal S^{}_j = \mathcal S^{}_j
\mathcal S^{}_i = \mathcal S^{}_k$ (for $i\neq j \neq k$) in the
former category as the generators of $\rm G^{}_\nu$, and fix an
element $\mathcal T$ in the latter category as the generator of
${\rm G}^{}_l$. In the basis where $\mathcal T$ is diagonal, $U$
will be composed of the normalized invariant eigenvectors of
$\mathcal S^{}_{i}$. If there are not three such order-2 elements,
then only one order-2 element can be identified as the residual
symmetry for $M^{}_\nu$ (i.e., $\rm G^{}_\nu=Z^{}_2$) in which case
$U$ can be partly determined. In fact, some authors have considered
the possibility of reducing $\rm G^{}_\nu$ from $\rm Z^{}_2 \times
Z^{}_2$ to $\rm Z^{}_2$ so as to leave $\theta^{}_{13}$ as a free
parameter \cite{GSF1,GSF2,HS,GSF3}.

\subsection{Model building with discrete flavor symmetries}

Now let us recapitulate some key issues concerning an application of
the group-theoretical arguments to some model-building exercises.
The most important issue in building a flavor-symmetry model is how
to break $\rm G^{}_F$ while preserving the desired residual
symmetries. To serve this purpose, one may introduce a new type of
scalar fields
--- {\it flavons}, which do not carry any quantum number of the SM
gauge symmetry but can constitute some nontrivial representations of
$\rm G^{}_F$. They break a given flavor symmetry via acquiring their
appropriate vacuum expectation values (VEVs). In the canonical
(type-I) seesaw mechanism the Lagrangian responsible for generating the
lepton masses may take a general form as
\begin{eqnarray}
-\mathcal L^{}_{\rm mass} & = & \sum^{}_{\gamma} y^{\gamma}_l
\overline{L^{\rho}} \ l^{\sigma} H^{}_d
\frac{\phi^{\gamma}_l}{\Lambda} +\sum^{}_{\gamma} y^{\gamma}_\nu
\overline{L^{\rho}} N^{\sigma}  H^{}_u
\frac{\phi^{\gamma}_\nu}{\Lambda}
\nonumber \\
&& + \sum^{}_{\gamma} y^{\gamma}_N N^{\sigma} N^{\sigma}
\frac{\phi^{\gamma}_N}{\Lambda} + {\rm h.c.} \; ,
\end{eqnarray}
in which $l$ and $N$ stand respectively for the right-handed
charged-lepton and neutrino fields, and the superscripts $\rho$,
$\sigma$ and $\gamma$ denote the representations occupied by the
fermion and flavon fields. Since the flavor symmetry breaks in
different manners in the charged-lepton and neutrino sectors, the
flavon fields for these two sectors are distinct. But when
$\phi^{}_\nu$ and $\phi^{}_N$ have the same transformation
properties, they can be identified as the same field. In order to
separate $\phi^{}_l$ from $\phi^{}_{\nu,N}$, an extra symmetry is
usually needed (e.g., an auxiliary $\rm Z^{\rm aux}_2$ symmetry
under which $\phi^{}_l$ and $l$ are odd while the other fields are
even). Furthermore, there must be enough independent flavon fields
so that the Yukawa coupling parameters present in each sector can
fit the lepton masses. Note that $\Lambda$ represents a high energy
scale above which the new fields may become active, and the ratio of
a flavon's VEV to $\Lambda$ (denoted as $\epsilon=\langle \phi
\rangle/\Lambda$) is typically assumed to be a small quantity such
that the magnitude of a Yukawa coupling parameter can be regulated
by the power of $\epsilon$ --- the spirit of the Froggatt-Nielsen
mechanism for explaining the observed quark mass hierarchy
\cite{FN}. This point will become relevant when one needs to either
produce a hierarchical mass matrix like $M^{}_l$ or discuss the soft
breaking of a flavor symmetry by higher-order terms. In this
connection it is necessary to know what forms the flavon VEVs
$\langle \phi^{}_{l,\nu,N} \rangle$ should take in order to break
all the flavor symmetries apart from ${\rm G}^{}_l$ and $\rm
G^{}_\nu$. One can easily verify that $M^{}_l M^\dagger_l$ will be
invariant under $\mathcal T$ if $\langle \phi^\gamma_l \rangle$
satisfies the condition \cite{Lam5}
\begin{eqnarray}
\mathcal T^\gamma \langle \phi^\gamma_l \rangle= \langle
\phi^\gamma_l \rangle \; .
\end{eqnarray}
Similarly, the effective neutrino mass matrix resulting from the
seesaw mechanism will be unchanged with respect to $\mathcal S^{}_i$
provided \cite{Lam5}
\begin{eqnarray}
\mathcal S^{\gamma}_i \langle \phi^{\gamma}_\nu \rangle = \langle
\phi^{\gamma}_\nu \rangle \;, \hspace{0.5cm} \mathcal S^{\gamma}_i
\langle \phi^{\gamma}_N \rangle = \langle \phi^{\gamma}_N \rangle \;
.
\end{eqnarray}
If a flavon field has no way to offer the proper VEV required by Eq.
(4.10) or (4.11), it has to be forbidden to acquire a VEV and
therefore contributes nothing to the lepton masses. With these
observations in mind, we are well prepared to assign suitable
representations for the fermion and flavon fields in order to build
a phenomenologically viable model.

We proceed to illustrate the above points using a simple flavor
model based on $\rm G^{}_F = S^{}_4$ \cite{Lam5}. The 24 elements of
$\rm S^{}_4$ belong to five conjugacy classes in which the orders
are 1, 2, 2, 3 and 4, respectively. Undoubtedly, the generators
$\mathcal S^{}_i$ of $\rm G^{}_\nu$ are contained in the second or
third class while the generator $\mathcal T$ of ${\rm G}^{}_l$
resides in the fourth or fifth class. For simplicity, we choose an
order-3 element to be $\mathcal T$
\footnote{If an order-4 element is chosen as $\mathcal T$, then the
flavor mixing pattern $U^{}_{\rm BM}$ shown in Eq. (3.11) will arise.}.
There are five irreducible representations for $\rm S^{}_4$: {\bf
1}, ${\bf 1^\prime}$, {\bf 2}, {\bf 3} and ${\bf 3^\prime}$, whose
dimensions are self-explanatory. Once the flavor symmetry is
specified, a question immediately follows: which representation(s)
should the lepton doublets furnish? With an order-2 element of $\rm
G^{}_F$ as the residual symmetry for $M^{}_\nu$, only two flavors
can mix if all the lepton doublets reside in the one-dimensional
representations \cite{Lam2}. In the situation that one lepton
doublet is in a one-dimensional representation and the other two
form a two-dimensional representation, one column of $U$ will have a
vanishing entry \cite{Lam2}. That is why the three lepton doublets
are commonly organized into a three-dimensional representation. Here
we put them in the representation {\bf 3}, implying that the
residual symmetries will be defined in this representation. To be
specific, $\mathcal T^{\bf 3}$ may be represented by ${\rm Diag}
\{1, \omega, \omega^2 \}$ for a diagonal $M^{}_l M^\dagger_l$. As
for $\mathcal S^{\bf 3}_i$, they can be represented by the $\mathcal
S^{\rm TB}_i$ in Eq. (4.3). Moreover, the explicit forms of
$\mathcal T$ and $\mathcal S^{}_i$ in other irreducible
representations are listed in Table 4.1 \cite{Lam5}, where
$\sigma^{}_1$ is the first Pauli matrix, and the flavon VEV
alignments (with the identification $\phi^{\gamma}_\nu=
\phi^{\gamma}_N$) preserving the $\mathcal T$ and $\mathcal S^{}_i$
symmetries are also shown. Note that $\phi^{\bf 1}_l$ and $\phi^{\bf
1}_\nu$ are always allowed to have VEVs which will never break any
symmetry. Furthermore, $l$ and $N$ are also assigned to the
representation {\bf 3}. Because of the product rule ${\bf 3}\times
{\bf 3}= {\bf 1}+ {\bf 2} + {\bf 3}+ {\bf 3^\prime}$, only the {\bf
1}, {\bf 2}, {\bf 3} and ${\bf 3^\prime}$ flavon representations may
contribute to the lepton masses. Among them, $\phi^{\bf 2}_l$ and
$\phi^{\bf 3}_\nu$ cannot receive nonzero VEVs which would otherwise
break the desired residual symmetries. The VEVs $\langle \phi^{\bf
1}_l \rangle$, $\langle \phi^{\bf 3}_l \rangle$ and $\langle
\phi^{\bf 3^\prime}_l \rangle$ help give rise to a diagonal $M^{}_l$
in the following form:
\begin{eqnarray}
M^{}_l & = & \langle H^{}_d \rangle \pmatrix{ y^{}_e & 0 & 0 \cr 0 &
y^{}_\mu & 0 \cr 0 & 0 & y^{}_\tau \cr} \; ,
\end{eqnarray}
where $y^{}_e = y^{\bf 1}_l/\sqrt{3} + \sqrt{2} y^{\bf
3^\prime}_l/3$, $y^{}_\mu = y^{\bf 1}_l/\sqrt{3} - \sqrt{2}y^{\bf
3^\prime}_l/6 + y^{\bf 3}_l/\sqrt{6}$ and $y^{}_\tau = y^{\bf
1}_l/\sqrt{3} - \sqrt{2} y^{\bf 3^\prime}_l/6 - y^{\bf
3}_l/\sqrt{6}$ containing the proper Clebsch-Gordan coefficients.
Note that there are three free parameters in Eq. (4.12), exactly
enough to fit the three charged-lepton masses. On the other hand,
$\langle \phi^{\bf 1}_\nu \rangle$, $\langle \phi^{\bf 2}_\nu
\rangle$ and $\langle \phi^{\bf 3}_\nu \rangle$ will result in a
Dirac neutrino mass matrix of the form
\begin{eqnarray}
M^{}_{\rm D} = \langle H^{}_u \rangle \pmatrix{ y^{}_{11} &
y^{}_{12} & y^{}_{12} \cr y^{}_{12} & y^{}_{22} & y^{}_{23} \cr
y^{}_{12} & y^{}_{23} & y^{}_{22} \cr} \; ,
\end{eqnarray}
in which the four independent elements are given by $y^{}_{11} =
y^{\bf 1}_\nu/\sqrt{3}+ \sqrt{2}y^{\bf 3^\prime}_\nu/3$, $y^{}_{12}
= y^{\bf 2}_\nu /\sqrt{6} - \sqrt{2} y^{\bf 3^\prime}_\nu/6$,
$y^{}_{22} = y^{\bf 2}_\nu /\sqrt{6} + \sqrt{2}y^{\bf
3^\prime}_\nu/3$ and $y^{}_{23} = y^{\bf 1}_\nu/\sqrt{3} -
\sqrt{2}y^{\bf 3^\prime}_\nu/6$. One can see that the texture of
$M^{}_{\rm D}$ has the same $(2,3)$ permutation symmetry as the one
shown in Eq. (3.7), and $y^{}_{22} + y^{}_{23} = y^{}_{11} +
y^{}_{12}$ holds. In addition, the texture of the heavy Majorana
neutrino mass matrix $M^{}_N$ is identical to that of $M^{}_{\rm D}$
with the replacements $y^{\bf 1,2,3^\prime}_{\nu} \Longrightarrow
y^{\bf 1,2,3^\prime}_{N}$. Then it is straightforward to follow the
canonical seesaw formula $M^{}_\nu = M^{}_{\rm D} M^{-1}_N M^T_{\rm
D}$ to show that the neutrino mixing pattern $U^{}_{\rm TB}$ can be
achieved. Finally, let us stress that the allowed flavon VEVs and
the neutrino mixing pattern will change if just one order-2 element
is preserved as the residual symmetry.
\begin{table}[t]
\caption{The explicit forms of $\mathcal T$, $\mathcal S^{}_i$,
$\langle \phi^{}_{l} \rangle$ and $\langle \phi^{}_{\nu} \rangle$ in
each irreducible representation with $\mathcal T$ being diagonal
\cite{Lam5}.} \vspace{0.1cm}
\begin{indented}
\item[]\begin{tabular}{ccccc} \br
 & ${\bf 1^\prime}$ & {\bf 2} & {\bf 3} & ${\bf 3^\prime}$ \\ \mr
\vspace{0.1cm}
$\mathcal T$ &  1 & Diag$\{\omega,\omega^2\}$ &
Diag$\{1,\omega,\omega^2\}$ & Diag$\{1,\omega,\omega^2\}$\\
\vspace{0.1cm} $\langle \phi^{}_l \rangle$ &  1 & $(0,0)^{T}$ &
$(1,0,0)^{T}$ & $(1,0,0)^{T}$ \\ \vspace{0.1cm} $\mathcal S^{}_1$ &
$-1$ & $\sigma^{}_1$ & $\mathcal S^{\rm TB}_1$ & $-\mathcal S^{\rm
TB}_1$ \\ \vspace{0.1cm} $\mathcal S^{}_2$  & 1 & Diag$\{1,1\}$ &
$\mathcal S^{\rm TB}_2$ & $\mathcal S^{\rm TB}_2$ \\ \vspace{0.1cm}
$\mathcal S^{}_3$  & $-1$ & $\sigma^{}_1$ & $\mathcal S^{\rm TB}_3$
& $-\mathcal S^{\rm TB}_3$ \\ \vspace{0.1cm} $\langle \phi^{}_\nu
\rangle$  & 0 & $(1,1)^{T}$ & $(0,0,0)^{T}$ & $(1,1,1)^{T}$  \\  \br
\end{tabular}
\end{indented}
\end{table}

As illustrated in Table 4.1, the flavon VEVs have to align in some
particular directions to preserve the desired residual symmetries.
Whether such special alignments can be naturally achieved matters,
as it has something to do with whether the corresponding
flavor-symmetry model is convincing or not. In this regard the most
popular method of deriving special flavon VEVs is the so-called
$F$-term alignment mechanism \cite{VEV}. The latter invokes the
supersymmetry, allowing one to take advantage of the $\rm U(1)$
R-symmetry --- $\rm U(1)^{}_R$ (under which the superpotential terms
should carry a total charge of 2)
\footnote{This special symmetry originates from the fact that in the
supersymmetry algebra the relations $\{ Q, Q^\dagger \} = P$ and $\{
Q, Q \} = \{ Q^\dagger, Q^\dagger \}=[ Q, P] = [ Q^\dagger, P] = 0$
keep invariant with respect to the $\rm U(1)$ transformation $Q \to
Q \exp{({\rm i} \phi)}$ \cite{MSSM}, where $Q$ and $P$ stand
respectively for the generators of supersymmetry and space-time
translations, and the curly and square brackets stand respectively
for the anticommutation and commutation relations. Since $Q$ itself
carries a nonzero $\rm U(1)^{}_R$ charge, the distinct components of
a supermultiplet which are connected to one another via the action
of $Q$ always have different $\rm U(1)^{}_R$ charges. For a
superfield with the $\rm U(1)^{}_R$ charge $n$, for example, its
$\phi$, $\psi$ and $F$ components have the $n$, $n-1$ and $n-2$
charges, respectively \cite{MSSM}. The superpotential terms are
accordingly required to assume a total charge of 2 in order for
their $F$ components to conserve $\rm U(1)$ R-symmetry.}.
The charge assignments for this symmetry generally go as follows:
the superfields for the SM fermions carry a charge of 1, while the
superfields for Higgs and flavons are neutral. Hence the
superpotential for generating lepton masses is not affected by this
symmetry, but the flavon fields themselves cannot form the
superpotential terms any more. To constrain the flavon VEVs, we may
introduce the driving fields (denoted by $\psi$) \cite{VEV} which
transform trivially with respect to the SM gauge symmetry but
nontrivially under the relevant flavor symmetry. In particular, they
carry a charge of 2 of the $\rm U(1)^{}_R$ symmetry, implying that
they must appear linearly in the superpotential terms. To be more
explicit, the superpotential involving both $\psi$ and the flavon
fields takes the form
\begin{eqnarray}
W = \psi \left(M^2_1+ M^{}_2 \phi + c \phi^2+ \cdots \right) \; ,
\end{eqnarray}
where $M^{}_{1}$ and $M^{}_{2}$ are two dimension-one parameters,
and $c$ is dimensionless. If the supersymmetry remains intact when
the flavor symmetry suffers breakdown, the flavon potential will be
given by
\begin{eqnarray}
V(\phi) = \sum^{}_{a} \left|\frac{\partial W}{\partial
\psi^{}_a}\right|^2 \; ,
\end{eqnarray}
where $a$ is used to distinguish different components of the driving
fields in multi-dimensional representations. By definition, the
flavon VEVs are taken to minimize the potential $V(\phi)$:
\begin{eqnarray}
\left| -F^*_{\psi^{}_a} \right| = \left| \frac{\partial W} {\partial
\psi^{}_a} \right|=0 \; .
\end{eqnarray}
For illustration, let us explain how the VEV alignments of
$\phi^{\bf 3}_l$ and $\phi^{\bf 3^\prime}_l$ in Table 4.1
\footnote{A possible way of realizing the VEV alignments of
$\phi^{\bf 2}_\nu$ and $\phi^{\bf 3}_\nu$ in Table 4.1 can be found
in Ref. \cite{VEV2}.}
can be derived with the help of three driving fields $\psi^{\bf 1}$,
$\psi^{\bf 3^\prime}$ and $\psi^{\bf 2}$. The related superpotential
appears as
\begin{eqnarray}
W=& \psi^{\bf 1} \left( c^{}_1 \phi^{\bf 3}_l \phi^{\bf 3}_l+ c^{}_2
\phi^{\bf 3^\prime}_l \phi^{\bf 3^\prime}_l-M^2 \right)+ \psi^{\bf
3^\prime} \left( c^{}_3 \phi^{\bf 3}_l \phi^{\bf 3^\prime}_l \right)
\nonumber \\
& \hspace{-0.15cm} +\psi^{\bf 2} \left( c^{}_4 \phi^{\bf 3}_l
\phi^{\bf 3}_l + c^{}_5 \phi^{\bf 3^\prime}_l \phi^{\bf 3^\prime}_l
+ c^{}_6 \phi^{\bf 3}_l \phi^{\bf 3^\prime}_l \right) \; .
\end{eqnarray}
Note that a term like $\psi^{\bf 3^\prime}\left( M^\prime \phi^{\bf
3^\prime}_l \right)$ has been forbidden by the $\rm Z^{aux}_2$
symmetry mentioned below Eq. (4.9). The conditions for the flavon
VEVs $\langle \phi^{\bf 3}_l \rangle=(v^{}_1,v^{}_2,v^{}_3)^{T}$ and
$\langle \phi^{\bf 3^\prime}_l \rangle= (v^{\prime}_1, v^{\prime}_2,
v^{\prime}_3)^{T}$ turn out to be
\begin{eqnarray}
c^{}_1 \left(v^2_1 + 2v^{}_2 v^{}_3\right) + c^{}_2 \left(v^{\prime
2}_1 + 2v^{\prime}_2 v^{\prime}_3\right) - M^2 = 0 \; ,
\nonumber \\
c^{}_3 \pmatrix{ v^{}_2 v^{\prime}_3 - v^{}_3 v^{\prime}_2 \cr
v^{}_1 v^{\prime}_2 - v^{}_2 v^{\prime}_1  \cr
v^{}_3 v^{\prime}_1 - v^{}_1 v^{\prime}_3 } =
\pmatrix{ 0 \cr 0 \cr 0 } \; ,
\end{eqnarray}
as well as
\begin{eqnarray}
c^{}_4 \pmatrix { v^2_2+ 2v^{}_1 v^{}_3 \cr v^2_3+ 2v^{}_1 v^{}_2 } +
c^{}_5 \pmatrix { v^{\prime 2}_2+ 2v^{\prime}_1 v^{\prime}_3 \cr
v^{\prime 2}_3+ 2v^{\prime}_1 v^{\prime}_2 }
\nonumber \\
+ c^{}_6 \pmatrix { v^{}_2 v^{\prime}_2+ v^{}_1 v^{\prime}_3
+v^{\prime}_1 v^{}_3 \cr -v^{}_3 v^{\prime}_3-v^{}_1 v^{\prime}_2
-v^{\prime}_1 v^{}_2} =\pmatrix{ 0 \cr 0  } \; .
\end{eqnarray}
The VEV alignments illustrated in Table 4.1 (i.e., $v^{}_{2,3}=
v^{\prime}_{2,3}=0$) apparently satisfy these equations. In
particular, the scale of $v^{}_1$ and $v^\prime_1$ is determined by
$M$ --- the only dimensional parameter in Eq. (4.17), which is
supposed to be located at a high energy scale. However, a driving
field in the representation {\bf 1} may not be welcome in some
specific models, in which case the mass term like that in Eq. (4.17)
will be absent and the trivial vacuum (i.e., all the flavon VEVs are
vanishing) offers a solution to the conditions for minimizing
$V(\phi)$. To find a way out, the negative and
supersymmetry-breaking flavon mass terms $-m^2_{\phi} |\phi|^2$ are
usually invoked to drive the flavon VEVs from the trivial ones
\cite{VEV}.

Last but not least, let us point out that there exists a class of
models where no residual symmetries survive. Such models are named
the {\it indirect models} while those featuring residual symmetries
are referred to as the {\it direct models} \cite{KL09}. This
classification is based on how the symmetries for $M^{}_l
M^{\dagger}_l$ and $M^{}_\nu$ come about. In the direct models
$M^{}_l M^{\dagger}_l$ and $M^{}_\nu$ are constrained by ${\rm
G}^{}_l$ and $\rm G^{}_\nu$ --- the remnants of $\rm G^{}_F$,
respectively. The relevant flavon VEVs should preserve these
residual symmetries. In the indirect models the special forms of
$M^{}_l M^{\dagger}_l$ and $M^{}_\nu$ arise accidentally, and the
flavon VEVs do not necessarily keep invariant under any symmetry. In
a realistic indirect model $M^{}_l$ and $M^{}_N$ are typically taken
to be diagonal. On the other hand, the Yukawa interaction term of
the neutrinos takes the form
\begin{eqnarray}
y^{}_i \overline {L^{\bf 3}} N^{}_i   H^{}_u \frac{\phi^{\bf
3}_i}{\Lambda} \; ,
\end{eqnarray}
which can be expressed as
\begin{eqnarray}
y^{}_i \left(\overline {L^{}_e} v^{i}_1+ \overline {L^{}_\mu}
v^{i}_2 + \overline {L^{}_\tau} v^{i}_3 \right) N^{}_i H^{}_u
\frac{1 }{\Lambda} \;
\end{eqnarray}
after the flavon fields have developed their VEVs as $\langle
\phi^{\bf 3}_i \rangle = (v^{i}_1, v^{i}_2, v^{i}_3)^{T}$. The
effective Majorana neutrino mass matrix turns out to be
\cite{Review3}
\begin{eqnarray}
M^{}_\nu = \sum^{}_i \frac{ y^2_i \langle H^{}_u \rangle^2}{M^{}_i
\Lambda^2} \langle \phi^{\bf 3}_i \rangle \langle \phi^{\bf 3}_i
\rangle^{T} \; ,
\end{eqnarray}
where $M^{}_i$ (for $i=1,2,3$) denote the right-handed neutrino
masses. Interestingly, the resulting PMNS matrix $U$ has a special
structure: its three columns are proportional to $\langle \phi^{\bf
3}_1 \rangle$, $\langle \phi^{\bf 3}_2 \rangle$ and $\langle
\phi^{\bf 3}_3 \rangle$, provided the latter are orthogonal to one
another. One is therefore encouraged to choose these column vectors
to align with the columns of the desired PMNS matrix $U$ \cite{SD}.
In order to achieve the flavor mixing pattern $U^{}_{\rm TB}$, for
instance, the following alignments for $\langle \phi^{\bf 3}_i
\rangle$ are favored:
\begin{eqnarray}
\langle \phi^{\bf 3}_1 \rangle = v^{}_1
\pmatrix{2 \cr -1 \cr -1 \cr} \; , \hspace{0.4cm}
\langle \phi^{\bf 3}_2 \rangle = v^{}_2
\pmatrix{1 \cr 1 \cr 1 \cr} \; ,
\nonumber \\
\langle \phi^{\bf 3}_3 \rangle = v^{}_3
\pmatrix{0 \cr -1 \cr 1 \cr} \; .
\end{eqnarray}
These VEVs either stay invariant or change their signs under
$\mathcal S^{\rm TB}_i$, but none of $\mathcal S^{\rm TB}_i$ is
preserved by all the three VEVs. Since $\langle \phi^{\bf 3}_i
\rangle$ appear quadratically in $M^{}_\nu$, the sign changes are
actually irrelevant and thus $M^{}_\nu$ acquires the $\mathcal
S^{\rm TB}_i$ symmetries accidentally. To summarize, in the indirect
models the flavor symmetry is mainly responsible for constraining
the Yukawa coupling structure of the neutrinos to have the form
given in Eq. (4.20). It can also assist the generation of special
flavon VEVs like those shown in Eq. (4.23).

\subsection{Generalized CP and spontaneous CP violation}

As discussed above, the residual-symmetry approach lacks the
predictive power for the Majorana phases. In addition, obtaining a
reasonable value of $\theta^{}_{13}$ at the leading order either
needs a large $\rm G^{}_F$ which will complicate the theory or
reduces $\rm G^{}_\nu$ from $\rm Z^{}_2 \times Z^{}_2$ to $\rm
Z^{}_2$ which will be less predictive. It is found that the GCP
symmetry \cite{GCP1,GCP2} can constrain the Majorana phases and
allow a finite $\theta^{}_{13}$ to be predicted. That is why the GCP
has recently attracted a lot of attention in the model-building
exercises, especially since $\theta^{}_{13}$ was experimentally
determined \cite{Thesis}.

A theoretical reason for the introduction of the GCP concept lies in
the fact that the canonical CP symmetry is not always consistent
with the discrete non-Abelian flavor symmetry. To consistently
define the CP symmetry in the context of a discrete flavor symmetry,
a nontrivial condition has to be fulfilled as one can see later on.
For the sake of simplicity and without loss of generality, let us
consider a scalar multiplet $\Phi=(\phi,\phi^*)^{T}$ on which $\rm
G^{}_F$ acts as
\begin{eqnarray}
\Phi \to \rho(g) \Phi \; , \hspace{0.5cm} g \in {\rm G^{}_F} \; .
\end{eqnarray}
Here $\rho$ is used to denote the representation constituted by
$\Phi$ which is not necessarily irreducible. On the other hand, a CP
symmetry is expected to act on $\Phi$ in the form
\begin{eqnarray}
\Phi \to X \Phi^*  \; ,
\end{eqnarray}
in which $X$ is unitary to keep the kinetic term $|\partial \Phi|^2$
invariant. Accordingly, a successive implementation of the CP, $g
\in \rm G^{}_F$ and {\it inverse} CP symmetries will transform
$\Phi$ through the following path:
\begin{eqnarray}
\Phi \to X \Phi^* \to X \rho(g)^* \Phi^* \to X \rho(g)^* X^{-1} \Phi
\; .
\end{eqnarray}
The consistency requirement suggests that $X \rho(g)^* X^{-1}$
should be a transformation belonging to $\rm G^{}_F$. We are
therefore required to combine the CP symmetry with $\rm G^{}_F$ as
follows \cite{GCP1,GCP2}:
\begin{eqnarray}
X \rho(g)^* X^{-1}= \rho(g^\prime) \; , \hspace{0.5cm}
g^\prime \in {\rm G^{}_F} \; .
\end{eqnarray}
Thanks to the preservation of the multiplication rules, we just need
to assure the consistency of $X$ with the generators of $\rm
G^{}_F$. The canonical CP symmetry (i.e., $X=1$) always follows when
$\rho$ is a real representation. But in a theory involving the
complex representations $X=1$ itself may not satisfy the consistency
condition, in which case a GCP symmetry has to be invoked. Note that
the GCP transformation may interchange different representations
(especially those which are complex conjugate to each other). That
is why one chooses to define CP on a vector space spanned by $\phi$
together with its complex conjugate counterpart $\phi^*$. In
principle, different real representations can also be connected by a
GCP transformation \cite{GCP2}. Hence $\rho$ has to contain all the
representations connected by the GCP transformations. Moreover, all
the $X$ matrices allowed by Eq. (4.27) constitute a representation
of the automorphism group $\rm Aut(G^{}_F)$ of $\rm G^{}_F$. Let us
consider two elements $a^{}_1$ and $a^{}_2$ of $\rm Aut(G^{}_F)$
whose representation matrices are $X^{}_1$ and $X^{}_2$,
respectively. The representation matrix of $a^{}_2a^{}_1$ is $X^{}_2
W X^{}_1$, as determined by the reasoning \cite{GCP2}:
\begin{eqnarray}
\rho(g^{\prime\prime}) & = X^{}_2 \rho(g^\prime)^* X^{-1}_2
= X^{}_2 W \rho(g^\prime) W X^{-1}_2
\nonumber \\
& = X^{}_2 W X^{}_1 \rho(g)^* X^{-1}_1 W X^{-1}_2 \; ,
\end{eqnarray}
where $W$ is the matrix interchanging the complex conjugate
components of $\Phi$ (i.e., $\Phi^*=W\Phi$), and it has the
properties
\begin{eqnarray}
W^2=1 \; , \hspace{0.5cm} \rho(g)=W \rho(g)^* W \; .
\end{eqnarray}
Note that for an $X$ satisfying Eq. (4.27), $\rho(g)X$ is also a
solution but it does not supply any additional physical
implications. Hence the GCP transformations of physical interest are
given by $\rm Aut(G^{}_F)$ modding out those equivalent ones and
form a group called $\rm H^{}_{CP}$. Consequently, the group $\rm
G^{}_{CP}$ constituted by $\rm G^{}_F$ and $\rm H^{}_{CP}$ is
isomorphic to their semi product \cite{GCP1,GCP2}
\begin{eqnarray}
\rm G^{}_{CP} = G^{}_F \rtimes H^{}_{CP} \; .
\end{eqnarray}
And the multiplication rule for any two elements of $\rm G^{}_{CP}$
is given by
\begin{eqnarray}
\left(g^{}_1, h^{}_1\right) \left(g^{}_2, h^{}_2\right) =
\left(g^{}_1 g^\prime_2 , h^{}_1 h^{}_2\right) \; ,
\end{eqnarray}
where $g^\prime_2 = h^{}_1 g^{}_2 h^{-1}_1$.

Now let us look at some physical implications of the GCP. By
definition, the GCP symmetries constrain $M^{}_l M^\dagger_l$ and
$M^{}_\nu$ in the following way:
\begin{eqnarray}
X^\dagger M^{}_l M^\dagger_l X= (M^{}_l M^\dagger_l)^* \; ,
\hspace{0.5cm} X^\dagger M^{}_\nu X^*= M^*_\nu \; .
\end{eqnarray}
Since $M^{}_l M^\dagger_l$ and $M^{}_\nu$ are diagonalized by
$O^{}_l$ and $O^{}_\nu$, respectively, one may derive the following
relations from Eq. (4.32):
\begin{eqnarray}
X^\dagger_l D^2_l X^{}_l = D^2_l \; , \hspace{0.5cm}
X^\dagger_\nu D^{}_\nu X^{*}_\nu = D^{}_\nu \; ,
\end{eqnarray}
in which
\begin{eqnarray}
X^{}_l = O^\dagger_l X  O^*_l \; , \hspace{0.5cm}
X^{}_\nu = O^\dagger_\nu X O^*_\nu   \; .
\end{eqnarray}
It becomes clear that $X^{}_l$ is a diagonal phase matrix. In
comparison, $X^{}_\nu$ is also diagonal but its finite entries are
$\pm 1$. These results allow us to obtain a PMNS matrix with the
property
\begin{eqnarray}
X^\dagger_l U X^{}_\nu = U^* \; .
\end{eqnarray}
It is easy to check that this relation will lead us to the trivial
CP phases \cite{CP1,CP2}. Hence it is necessary to break the GCP
symmetry in order to accommodate CP violation. A phenomenologically
interesting and economical way of breaking $\rm G^{}_F \rtimes
H^{}_{CP}$ is to preserve the residual symmetries ${\rm G}^{}_l$ and
$\rm G^{}_\nu \rtimes H^\nu_{CP} = Z^{}_2 \times H^\nu_{CP}$ in the
charged-lepton and neutrino sectors \cite{GCP1}, respectively, as
illustrated in Fig. 4.1. This goal can be achieved by requiring the
related flavon VEVs to satisfy the condtions
\begin{eqnarray}
\mathcal{T} \langle \phi^{}_{l} \rangle = \langle \phi^{}_{l}
\rangle \; , \hspace{0.5cm} \mathcal{S} \langle \phi^{}_\nu \rangle=
X \langle \phi^{}_\nu \rangle^* = \langle \phi^{}_\nu \rangle \; ,
\end{eqnarray}
in which $\mathcal{T}$, $\mathcal{S}$ and $X$ are the representation
matrices for the generators of ${\rm G}^{}_l$, $\rm Z^{}_2$ and $\rm
H^\nu_{CP}$, respectively. Furthermore, the commutation between $\rm
Z^{}_2$ and $\rm H^\nu_{CP}$ requires
\begin{eqnarray}
X \mathcal{S}^* X^{-1} = \mathcal{S} \; .
\end{eqnarray}
In light of $\mathcal{S}^2=1$, we may diagonalize $\mathcal{S}$ by a
unitary matrix $O^{}_{\mathcal S}$ (i.e., $O^\dagger_{\cal S} {\cal
S} O^{}_{\cal S} = D^{}_{\mathcal S} = \pm {\rm Diag} \{-1, 1, -1
\}$). One can see that the first and third eigenvalues of $\cal S$
are degenerate. In this case it is possible to redefine
$O^{}_{\mathcal{S}}$ by carrying out a complex $(1,3)$ rotation. In
combination with the condition given in Eq. (4.37), this freedom
allows us to have $O^{}_{\mathcal S} O^{T}_{\mathcal S} = X$. Taking
account of the invariance of $M^{}_\nu$ under $\rm Z^{}_2 \times
H^\nu_{CP}$,
\begin{eqnarray}
\mathcal{S}^\dagger M^{}_\nu \mathcal{S}^* = M^{}_\nu \;,
\hspace{0.5cm} X^\dagger M^{}_\nu X^*= M^*_\nu \; ,
\end{eqnarray}
one finds
\begin{eqnarray}
D^{}_{\mathcal S} O^\dagger_{\mathcal S} M^{}_\nu O^*_{\mathcal S} =
O^\dagger_{\mathcal S} M^{}_\nu O^*_{\mathcal S} D^{}_{\mathcal S}
\; ,
\nonumber \\
O^\dagger_{\mathcal S} M^{}_\nu O^*_{\mathcal S} =
\left(O^\dagger_{\mathcal S} M^{}_\nu O^*_{\mathcal S}\right)^* \; .
\end{eqnarray}
This means that $O^\dagger_{\mathcal S} M^{}_\nu O^*_{\mathcal S}$
can be diagonalized by a real orthogonal matrix
\begin{eqnarray}
R(\theta) = \pmatrix{ \cos{\theta} & 0 & \sin{\theta} \cr
0 & 1 & 0 \cr -\sin{\theta} & 0 & \cos{\theta}} \; .
\end{eqnarray}
The PMNS matrix turns out to be \cite{GCP1}
\begin{eqnarray}
U = U^\dagger_l O^{}_{\mathcal S} R(\theta) P^{}_\nu \; ,
\end{eqnarray}
up to possible permutations of its rows or columns. Here $U^{}_l$ is
determined by ${\rm G}^{}_l$, whereas $P^{}_\nu$ is a diagonal matrix
(with its entries being $\pm 1$ or $\pm{\rm i}$) to keep the three
neutrino mass eigenvalues positive. In such an approach $\theta$ is
the only free parameter associated with the flavor mixing angles and
CP-violating phases \cite{GCP1}, and thus we are led to a few
testable correlations among them. Of course, the value of $\theta$
can be determined by confronting it with the experimental result of
$\theta^{}_{13}$.
\begin{figure}
\vspace{0.2cm}
\centerline{\includegraphics[width=0.4\textwidth]{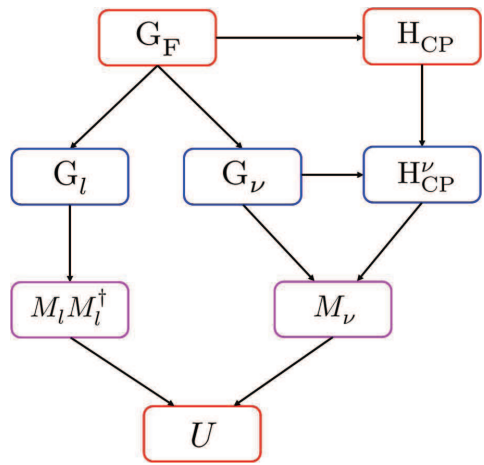}}
\caption{A schematic illustration for the combination of $\rm
G^{}_F$ and GCP as well as their breaking pattern which has direct
consequence on the PMNS flavor mixing matrix $U$. Here we focus on
the case of $\rm G^{}_\nu=Z^{}_2$.}
\end{figure}

In the above direct-model approach, the values of CP-violating
phases are purely determined by the group structure of $\rm G^{}_F
\rtimes H^{}_{CP}$. So we are left with little room for manoeuvre
once the flavor symmetry has been specified. Furthermore, this
approach typically leads the CP phases to some simple or special
numbers such as 0, $\pm \pi/2$ or $\pi$
\cite{example1,example2,example3,example4}, although nontrivial CP
phases may arise as the flavor symmetry becomes larger
\cite{DZ1,DZ2}. Alternatively, the indirect models can provide
various nontrivial CP phases which are easier to control. In such
models the canonical CP symmetry is consistent with $\rm G^{}_F$ and
all the parameters are real. A nontrivial CP phase is introduced by
the complex VEV of a flavon field $\phi$ which transforms trivially
with respect to $\rm G^{}_F$ \cite{SCP1,SCP2,ZZH14,SCP3}. In order
to control the phase of $\langle \phi \rangle$, one may introduce an
extra ${\rm Z}^{}_n$ symmetry under which $\phi$ carries a single
charge \cite{SCPKing}. When the $F$-term alignment mechanism
\cite{VEV} is used to shape the flavon VEVs, there will be a
superpotential term relevant to $\phi$: $\psi
\left(\phi^n/\Lambda^{n-2} \pm M^2\right)$, where $\psi$ denotes a
driving field which is neutral under ${\rm Z}^{}_n$ and $\rm
G^{}_F$. Therefore, $\langle \phi \rangle$ is required to satisfy
the condition
\begin{eqnarray}
\left| -F^{*}_\psi \right| = \left|\frac{\langle \phi \rangle^n}
{\Lambda^{n-2}} \pm M^2 \right|=0 \; ,
\end{eqnarray}
which may lead to any nontrivial phase for $\langle \phi \rangle$
with the help of an adjustable $n$
\cite{SCPKing2,SCPKing3,SCPKing4}.

\def\thefootnote{\arabic{footnote}}
\setcounter{footnote}{0}
\setcounter{equation}{0}
\setcounter{table}{0}
\setcounter{figure}{0}

\section{Realization of the $\mu$-$\tau$ flavor symmetry}

We proceed to discuss how to realize the $\mu$-$\tau$ flavor
symmetry in different physical contexts. First of all, let us make
some general remarks.

(1) $M^{}_l M^\dagger_l$ and $M^{}_\nu$ cannot simultaneously
possess the same $\mu$-$\tau$ flavor symmetry. Otherwise, the
resulting lepton flavor mixing pattern would only contain a finite
value of $\theta^{}_{12}$ and thus be unrealistic. That is why one
usually takes $M^{}_l$ to be diagonal and attributes the effects of
lepton flavor mixing purely to $M^{}_\nu$ in a specific
model-building exercise. Of course, one may impose a flavor symmetry
on $M^{}_{l}$ but keep $M^{}_{\nu}$ diagonal, or allow both of them
to be structurally non-diagonal and unparallel. Different basis
choices are equivalent in principle, but one basis is likely to be
advantageous over another in practice when building an explicit
model to determine or constrain the flavor structures of both
charged leptons and neutrinos. To illustrate this phenomenological
point, we assume $M^{}_{l} M^\dagger_l$ to have the $\mu$-$\tau$
permutation symmetry in a form like that given in Eq. (3.7), leading
to the eigenvalues $m^2_\alpha$ (for $\alpha = e, \mu, \tau$) whose
expressions are analogous to those given in Eq. (3.9). In this case
one has to tolerate some significant cancellations among the
parameters of $M^{}_{l} M^\dagger_l$ in order to generate a strongly
hierarchical charged-lepton mass spectrum (i.e., $m^{}_e \ll
m^{}_\mu \ll m^{}_\tau$ as shown in Fig. 2.2 \cite{PDG}). It is
certainly possible to conjecture that there exists a flavor symmetry
which guarantees the exact cancellations and thus $m^{}_{e} =
m^{}_{\mu} =0$ at the lowest order, and it is the effect of flavor
symmetry breaking that gives rise to nonzero $m^{}_{e}$ and
$m^{}_{\mu}$ together with the associated flavor mixing parameters.
Assuming the strength of symmetry breaking to be $\epsilon \sim
\mathcal O(0.1)$, one may expect that the finite values of
$m^{}_{\mu}$, $\theta^{}_{13}$ and $\Delta \theta^{}_{23} \equiv
\theta^{}_{23} -\pi/4$ emerge at the ${\cal O}(\epsilon)$ level
while nonzero $m^{}_{e}$ arises at the ${\cal O}(\epsilon^3)$ level.
To achieve such results, however, a delicate arrangement of the
relevant perturbation terms is indispensable \cite{GFM}. In
contrast, the neutrino mass spectrum should not exhibit a strong
hierarchy unless $m^{}_1$ is negligibly small. Another point is also
noteworthy: although the lepton flavor mixing matrix is a constant
matrix in the symmetry limit, it may be strongly correlated with the
mass spectrum of charged leptons or neutrinos (or both of them)
after the flavor symmetry is broken. When discussing the effects of
$\mu$-$\tau$ flavor symmetry breaking, one therefore prefers to work
in the basis of $M^{}_l$ being diagonal such that it is more
straightforward to infer some information about the neutrino mass
spectrum from the experimental observations of $\theta^{}_{13}$ and
$\theta^{}_{23}$. In other words, let us simply concentrate on the
$\mu$-$\tau$ permutation or reflection symmetry of $M^{}_\nu$ and
its possible breaking schemes in the basis where $M^{}_l$ itself is
diagonal.

(2) A ${\rm U(1)}^{}_e \times {\rm U(1)}^{}_\mu \times {\rm
U(1)}^{}_\tau$ flavor symmetry or its subgroup should be invoked in
the charged-lepton sector to make $M^{}_l M^\dagger_l$ diagonal.
Nevertheless, the left-handed charged-lepton and neutrino fields
reside in the same $\rm SU(2)^{}_{L}$ doublets, so the symmetry
placed on one sector should equally act on the other sector. Hence
the major challenge in model building is to preserve a symmetry in
one sector but break it properly in the other sector. To serve this
purpose, additional Higgs (or flavon) fields furnishing nontrivial
representations of the flavor symmetry are generally needed. In this
connection the Higgs (or flavon) fields, which are neutral with
respect to the symmetry for keeping the diagonality of $M^{}_l
M^\dagger_l$ but transform nontrivially under the $\mu$-$\tau$
flavor symmetry, can be employed to break the would-be degeneracy
between $m^{}_\mu$ and $m^{}_\tau$.

(3) A number of flavor-symmetry models, which are phenomenologically
viable, have typically used the canonical seesaw mechanism to
explain the smallness of three neutrino masses. Accordingly, at
least two right-handed neutrinos are introduced in addition to the
SM fields. No matter whether they assume the $\mu$-$\tau$ flavor
symmetry or not, the effective Majorana neutrino mass matrix
$M^{}_\nu$ will have this symmetry provided the left-handed
neutrinos respect the same symmetry. This interesting point can be
explicitly seen from the seesaw formula $M^{}_\nu = M^{}_{\rm D}
M^{-1}_N M^{T}_{\rm D}$ \cite{SS1,SS2,SS3,SS4,SS5} by taking the
form of $M^{}_{\rm D}$ as
\begin{eqnarray}
M^{}_{\rm D} = \langle H \rangle \pmatrix{
y^{}_{11} &  y^{}_{12} & y^{}_{13} \cr
y^{}_{21} & y^{}_{22} & y^{}_{23} \cr
y^{}_{21} & y^{}_{22} & y^{}_{23} } \; ,
\end{eqnarray}
where $M^{}_N$ is an arbitrary symmetric matrix. However, one might
wish that the right-handed neutrinos should also obey some flavor
symmetries in order to reduce the number of free parameters
\cite{Lam7}.

\subsection{Models with the $\mu$-$\tau$ permutation symmetry}

In this subsection we aim to use a series of models, proposed by
Grimus and Lavoura \cite{GL,GL1,GL2}, to illustrate the essential
ingredients for building a phenomenologically viable model based on
the $\mu$-$\tau$ flavor symmetry. The main idea of their models is
that the ${\rm U(1)}^{}_e \times {\rm U(1)}^{}_\mu \times {\rm
U(1)}^{}_\tau$ symmetry is conserved by the Yukawa interactions but
softly broken by the dimension-three Majorana mass terms of
right-handed neutrinos. Thus let us introduce three right-handed
neutrinos $N^{}_\alpha$ (for $\alpha=e,\mu,\tau$) and allow each of
them to carry an associated lepton number. To properly break the
flavor symmetry, three Higgs doublets $H^{}_i$ (for $i=1,2,3$) are
needed and each of them develops a VEV $v^{}_i$ (for
$\sqrt{v^{2}_{1} + v^{2}_{2} + v^{2}_{3}} \simeq 246$ GeV). In
addition to the lepton-number symmetries, two $\rm Z^{}_2$-type
flavor symmetries are also enforced \cite{GL1}:
\begin{eqnarray}
{\rm Z}^{\mu\tau}_2: \hspace{0.35cm} L^{}_\mu \leftrightarrow
L^{}_\tau, \hspace{0.15cm} \mu^{}_{\rm R} \leftrightarrow
\tau^{}_{\rm R}, \hspace{0.15cm} N^{}_\mu \leftrightarrow N^{}_\tau,
\hspace{0.15cm} H^{}_3 \to -H^{}_3 \; ;
\nonumber \\
{\rm Z}^{\rm aux}_2: \hspace{0.2cm} \mu^{}_{\rm R}, \hspace{0.1cm}
\tau^{}_{\rm R}, \hspace{0.1cm} H^{}_2, \hspace{0.1cm} H^{}_3
\hspace{0.15cm} {\rm change \hspace{0.15cm} sign} \; .
\end{eqnarray}
Here ${\rm Z}^{\mu\tau}_2$ is the $\mu$-$\tau$ permutation symmetry
interchanging all the fermion fields of the $\mu$ and $\tau$
flavors. Note that $H^{}_3$ is odd under the ${\rm Z}^{\mu\tau}_2$
symmetry, and thus it will break this symmetry when obtaining a VEV.
Such a requirement is actually necessary for obtaining $m^{}_\mu
\neq m^{}_\tau$, as one will see later on. In addition, there is an
auxiliary symmetry ${\rm Z}^{\rm aux}_2$ which separates $H^{}_1$
from $H^{}_{2}$ and $H^{}_3$. Because of this symmetry, $H^{}_2$ and
$H^{}_3$ are responsible for the mass generation of $\mu$ and $\tau$
while $H^{}_1$ only couples to $e^{}_{\rm R}$ as well as
$N^{}_\alpha$. Given the above setting, the Lagrangian for
generating lepton masses is expressed as \cite{GL3}
\begin{eqnarray}
-\mathcal L^{}_{\rm mass} = & y^{}_1 \overline{L^{}_e} e^{}_{\rm R}
H^{}_1 + y^{}_2 \left(\overline{L^{}_\mu} \mu^{}_{\rm R}
+\overline{L^{}_\tau} \tau^{}_{\rm R} \right) H^{}_2
\nonumber \\
& \hspace{-0.1cm} + y^{}_3 \left(\overline{L^{}_\mu} \mu^{}_{\rm R}
- \overline{L^{}_\tau} \tau^{}_{\rm R} \right) H^{}_3 +
y^{}_4 \overline{L^{}_e} N^{}_e \widetilde H^{}_1
\nonumber \\
& \hspace{-0.1cm} + y^{}_5 \left(\overline{L^{}_\mu} N^{}_\mu +
\overline{L^{}_\tau} N^{}_\tau \right) \widetilde H^{}_1 + {\rm
h.c.} \; ,
\end{eqnarray}
where $\widetilde H^{}_{1} \equiv {\rm i} \sigma^{}_2 H^{}_{1}$.
After the electroweak symmetry breaking, $M^{}_l$ and $M^{}_{\rm D}$
turn out to be diagonal:
\begin{eqnarray}
M^{}_l = \pmatrix{ y^{}_1 v^{}_1 & 0 & 0 \cr 0 & y^{}_2  v^{}_2
+y^{}_3  v^{}_3 & 0 \cr 0 & 0 & y^{}_2  v^{}_2 - y^{}_3  v^{}_3 \cr}
\; ,
\nonumber \\
M^{}_{\rm D} = \pmatrix{ y^{}_4 v^{}_1 & 0 & 0 \cr 0 & y^{}_5
v^{}_1 & 0 \cr 0 & 0 & y^{}_5 v^{}_1 \cr} \; .
\end{eqnarray}
Note that it is the coexistence of $H^{}_2$ and $H^{}_3$ that
renders $m^{}_\mu \neq m^{}_\tau$. The heavy Majorana neutrino mass
matrix $M^{}_N$ breaks the lepton-number symmetries in a soft way
but maintains the other symmetries, so it also possesses the
$\mu$-$\tau$ permutation symmetry. Taking account of the seesaw
formula, one therefore arrives at the light Majorana neutrino mass
matrix $M^{}_\nu$ which shares the same symmetry.

Several comments on this simple but instructive model-building
approach are in order.

(a) In such a model the right-handed neutrinos necessarily suffer
the $\mu$-$\tau$ permutation symmetry. The reason is that $\rm
U(1)^{}_{\mu}$ and $\rm U(1)^{}_{\tau}$ do not commute with $\rm
Z^{\mu \tau}_2$. As a result of the $\rm U(1)^{}_{\mu}$ (or $\rm
U(1)^{}_{\tau}$) symmetry, $L^{}_{\mu}$ (or $L^{}_{\tau}$) is only
allowed to have couplings to $\mu^{}_{\rm R}$ and $N^{}_\mu$ (or
$\tau^{}_{\rm R}$ and $N^{}_\tau$). In order to keep $\mathcal
L^{}_{\rm mass}$ invariant under $\rm Z^{\mu \tau}_2$, $\mu^{}_{\rm
R}$ and $\tau^{}_{\rm R}$ have to transform into each other as
$L^{}_\mu$ and $L^{}_\tau$ do. So do $N^{}_\mu$ and $N^{}_\tau$.
Furthermore, $\rm U(1)^{}_{\mu}$, $\rm U(1)^{}_\tau$ and $\rm Z^{\mu
\tau}_2$ may be unified in a larger non-Abelian group which proves
to be the two-dimensional orthogonal group $O^{}_2$ \cite{GL5}.

(b) Another issue concerns the introduction of so many Higgs
doublets. Although the rates of lepton-flavor-changing processes can
be under control \cite{GL6}, one has to overcome some other
phenomenological problems brought by so many electroweak-scale
scalars (see, e.g., Refs. \cite{multi1,multi2,multi3} for the
discussions on some consequences of the multi-Higgs fields
introduced in a number of flavor-symmetry models). To isolate the
flavor symmetry issues from the electroweak issues, one can use the
flavons which emerge at a high energy scale $\Lambda$ to substitute
for the extra Higgs doublets. As an example, we only keep $H^{}_1$
while replacing $H^{}_{2}$ and $H^{}_3$ with two flavon fields
$\phi^{}_2$ and $\phi^{}_3$. And $\phi^{}_2$ (or $\phi^{}_3$) has
the same transformation properties as $H^{}_2$ (or $H^{}_3$) with
respect to $\rm Z^{\mu \tau}_2$ and $\rm Z^{\rm aux}_2$.
Correspondingly, the Lagrangian relevant for lepton masses can be
obtained from that in Eq. (5.3) with the replacements
\begin{eqnarray}
\left( \overline{L^{}_\mu} \mu^{}_{\rm R} +
\overline{L^{}_\tau} \tau^{}_{\rm R} \right)
H^{}_2 \hspace{0.2cm} \to \hspace{0.2cm}
\left( \overline{L^{}_\mu} \mu^{}_{\rm R} +
\overline{L^{}_\tau} \tau^{}_{\rm R} \right)
H^{}_1 \frac{\phi^{}_2}{\Lambda} \; ,
\nonumber \\
\left( \overline{L^{}_\mu} \mu^{}_{\rm R} -
\overline{L^{}_\tau} \tau^{}_{\rm R} \right)
H^{}_3 \hspace{0.2cm} \to \hspace{0.2cm}
\left( \overline{L^{}_\mu} \mu^{}_{\rm R} -
\overline{L^{}_\tau} \tau^{}_{\rm R} \right)
H^{}_1 \frac{\phi^{}_3}{\Lambda} \; .
\end{eqnarray}

(c) From the point of view of naturalness, the strong mass hierarchy
of charged leptons needs some explanations. As for the model under
discussion, the smallness of $m^{}_e$ can be easily explained with
the help of the Froggatt-Nielsen mechanism \cite{FN}. If one assigns
a $\rm U(1)^{}_{\rm FN}$ charge $n$ to $e^{}_{\rm R}$, then the
Yukawa interaction term for the electron will become
\begin{eqnarray}
y^{}_1 \overline{L^{}_e} e^{}_{\rm R} H^{}_1
\left( \frac{\langle \phi \rangle}{\Lambda} \right)^n \; ,
\end{eqnarray}
where $\phi$ is a flavon field carrying a $\rm U(1)^{}_{\rm FN}$
charge of $-1$. Given $\epsilon \equiv \langle \phi \rangle/\Lambda$
as a very small quantity, $m^{}_e$ is suppressed by $\epsilon^n$.
However, this model does not permit the same mechanism to explain
the smallness of $m^{}_\mu$ as compared with $m^{}_\tau$. One finds
that a relation $y^{}_2 \langle v^{}_2 \rangle \simeq -y^{}_3
\langle v^{}_3 \rangle$ with an accuracy of about $10\%$ is required
for obtaining the realistic values of $m^{}_\mu$ and $m^{}_\tau$.
Although a fine-tuning of this amount is not unacceptable, it is
better to find a more natural way to obtain this kind of approximate
relation. The idea is based on a new symmetry K which connects
$H^{}_2$ and $H^{}_3$ as follows \cite{GL4}:
\begin{eqnarray}
\mu^{}_{\rm R} \to -\mu^{}_{\rm R} \; ,
\hspace{0.5cm} H^{}_2 \leftrightarrow H^{}_3 \; .
\end{eqnarray}
Hence the coefficients $y^{}_2$ and $y^{}_3$ in Eq. (5.3) should be
opposite to each other. Furthermore, the K symmetry can constrain
the scalar potential to such a form that $v^{}_2 = v^{}_3$ is most
likely to take place \cite{GL4}, in which case $m^{}_\mu$ is exactly
vanishing. When this symmetry is softly broken in the scalar
potential, it is natural for us to expect a slight difference
between $v^{}_2$ and $v^{}_3$. And thus we arrive at a finite value
of $m^{}_\mu$ which is much smaller than that of $m^{}_\tau$ \cite{GL4}.

(d) The model under discussion can be slightly modified to
accommodate the $\mu$-$\tau$ reflection symmetry \cite{GL4}. For
this purpose, $\rm Z^{\mu \tau}_2$ is replaced by the $\mu$-$\tau$
reflection symmetry while the other symmetries keep unchanged. Note
that the $\mu$-$\tau$ reflection symmetry can be viewed as a GCP
symmetry described by
\begin{eqnarray}
F^{}_\alpha \hspace{0.2cm} \to \hspace{0.2cm} X^{}_{\alpha\beta}
F^c_\beta \; ,
\end{eqnarray}
where $F$ stands for all the fermion fields (i.e., $L$, $N$ and the
right-handed charged-lepton fields), $X = S^+$ as given by Eq. (3.6)
\footnote{A generalization of the $\mu$-$\tau$ reflection symmetry
can be found in Ref. \cite{CDGV}, where $X$ takes a form different
from $S^+$.},
and the Greek subscripts run over $e$, $\mu$ and $\tau$. On the
other hand, the actions of this symmetry on the three Higgs fields
are supposed to be
\begin{eqnarray}
H^{}_1 \to H^*_1 \; , \hspace{0.5cm} & H^{}_2 \to H^*_2 \; ,
\hspace{0.5cm} & H^{}_3 \to -H^*_3 \; .
\end{eqnarray}
The resulting $\mathcal L^{}_{\rm mass}$ appears as
\begin{eqnarray}
-\mathcal L^{}_{\rm mass} = & y^{}_1 \overline{L^{}_e} e^{}_{\rm R}
H^{}_1 + \left( y^{}_2 \overline{L^{}_\mu} \mu^{}_{\rm R} + y^{*}_2
\overline{L^{}_\tau} \tau^{}_{\rm R} \right) H^{}_2
\nonumber \\
& \hspace{-0.1cm} + \left( y^{}_3 \overline{L^{}_\mu} \mu^{}_{\rm R} -
y^{*}_3 \overline{L^{}_\tau} \tau^{}_{\rm R} \right) H^{}_3 +
y^{}_4 \overline{L^{}_e} N^{}_e \widetilde H^{}_1
\nonumber \\
& \hspace{-0.1cm} + \left( y^{}_5 \overline{L^{}_\mu} N^{}_\mu +
y^{*}_5 \overline{L^{}_\tau} N^{}_\tau \right) \widetilde H^{}_1 +
{\rm h.c.} \; ,
\end{eqnarray}
where $y^{}_1$ and $y^{}_4$ are real by definition. As a result, the
expressions of $M^{}_l$ and $M^{}_{\rm D}$ are changed to
\begin{eqnarray}
M^{}_l = \pmatrix{y^{}_1 v^{}_1 & 0 & 0 \cr 0 & y^{}_2 v^{}_2 +
y^{}_3 v^{}_3 & 0 \cr 0 & 0 & y^{*}_2 v^{}_2 - y^{*}_3 v^{}_3 \cr} \;
,
\nonumber \\
M^{}_{\rm D} = \pmatrix{y^{}_4 v^{}_1 & 0 & 0 \cr 0 & y^{}_5 v^{}_1
& 0 \cr 0 & 0 & y^{*}_5 v^{}_1 \cr} \; .
\end{eqnarray}
It should be noted that the phase difference between $v^{}_2$ and
$v^{}_3$ cannot be $\pm \pi/2$. Otherwise, one would be led to the
unrealistic result $m^{}_\mu = m^{}_\tau$. One may easily understand
this observation by taking account of the fact that the $\mu$-$\tau$
reflection symmetry still holds when $v^{}_2$ is real and $v^{}_3$
is purely imaginary. If $M^{}_N$ and $M^{}_{\rm D}$ take the forms
in Eqs. (3.24) and (5.11) respectively, then the seesaw formula
assures the texture of $M^{}_\nu$ to respect the $\mu$-$\tau$
reflection symmetry.

\subsection{Models with the $\mu$-$\tau$ reflection symmetry}

Now we turn to the model-building issues based on the $\mu$-$\tau$
reflection symmetry. First of all, let us consider an interesting
model developed by Mohapatra and Nishi for the purpose of
illustration \cite{MN}. In this model the $\mu$-$\tau$ reflection
symmetry is imposed on all the fermion fields in the manner defined
by Eq. (5.8). A $\rm U(1)^{}_{\mu-\tau}$ symmetry, which is helpful
to make $M^{}_l M^\dagger_l$ diagonal, is also introduced. As
suggested by the notation of $\rm U(1)^{}_{\mu-\tau}$ itself, the
$e$, $\mu$ and $\tau$ flavor fields carry the $\rm
U(1)^{}_{\mu-\tau}$ charges of 0, 1 and $-1$, respectively. So the
$\rm U(1)^{}_{\mu-\tau}$ symmetry transformation can be described by
\begin{eqnarray}
\mathcal T = \pmatrix{ 1 & 0 & 0 \cr 0 & e^{{\rm i}\theta} & 0 \cr
0 & 0 & e^{-{\rm i}\theta} } \; ,
\end{eqnarray}
where $\theta$ is an arbitrary constant. As compared with ${\rm
U(1)}^{}_e \times \rm U(1)^{}_\mu \times U(1)^{}_\tau$, $\rm
U(1)^{}_{\mu-\tau}$ has a quite appealing consequence: it commutes
with the $\mu$-$\tau$ reflection symmetry as indicated by $S^+
\mathcal T^* S^+ = \mathcal T$. This means that the whole flavor
symmetry may be simply a direct product of $\rm U(1)^{}_{\mu-\tau}$
and the $\mu$-$\tau$ reflection symmetry. In order to lift the
would-be degeneracy between $m^{}_\mu$ and $m^{}_\tau$, two extra
Higgs doublets $H^{}_{\pm 2}$ carrying the $\rm U(1)^{}_{\mu-\tau}$
charges of $\pm 2$ are introduced in addition to the SM Higgs
doublet $H^{}_0$. Given this framework, the Yukawa coupling terms
relevant for $M^{}_l$ include \cite{MN}
\begin{eqnarray}
\lambda^{}_0 \overline{L^{}_e} e^{}_{\rm R} H^{}_0
+ \lambda^{}_{+2} \overline{L^{}_\mu} \tau^{}_{\rm R} H^{}_{+2}
+ \lambda^{}_{-2} \overline{L^{}_\tau} \mu^{}_{\rm R}  H^{}_{-2} \; .
\end{eqnarray}
The invariance of these terms with respect to the $\mu$-$\tau$
reflection symmetry requires $\lambda^{}_0 = {\rm Re}
(\lambda^{}_0)$ and $\lambda^{}_{+2} = \lambda^*_{-2}$ as well as
the following transformation properties for the Higgs fields:
\begin{eqnarray}
H^{}_0 \to H^*_0 \; , \hspace{0.5cm} H^{}_{+2} \to H^*_{-2} \; .
\end{eqnarray}
Apparently, the $\mu$-$\tau$ reflection symmetry will be broken if
$|\langle H^{}_{+2} \rangle| \neq |\langle H^{}_{-2} \rangle|$
holds. In particular, the relation $|\langle H^{}_{+2} \rangle| \ll
|\langle H^{}_{-2} \rangle|$ is required in order to generate
$m^{}_\mu \ll m^{}_\tau$. Such a relation can be induced in the
scalar potential by a scalar field which changes its sign under the
$\mu$-$\tau$ reflection symmetry \cite{CMP}. It should be pointed
out that the unwanted terms $\overline{L^{}_\mu} \mu^{}_{\rm R}
H^{}_0$ and $\overline{L^{}_\tau} \tau^{}_{\rm R} H^{}_0$, which may
render $M^{}_l M^\dagger_l$ non-diagonal, have been forbidden by
some additional symmetries (e.g., a $\rm Z^{\rm aux}_2$ symmetry
under which only $\mu^{}_{\rm R}$, $\tau^{}_{\rm R}$ and $H^{}_{\pm
2}$ change signs). Note that the Yukawa coupling terms in Eq. (5.13)
will give rise to an accidental $\rm U(1)^{}_1 \times U(1)^{}_2$
symmetry under which the related fields transform as
\begin{eqnarray}
L^{}_\mu \to L^{}_\mu e^{{\rm i}\theta^{}_1} \; , \hspace{0.5cm}
\tau^{}_{\rm R} \to \tau^{}_{\rm R} e^{{\rm i}\theta^{}_1} \; ,
\nonumber \\
L^{}_\tau  \to L^{}_\tau e^{{\rm i}\theta^{}_2} \; , \hspace{0.5cm}
\mu^{}_{\rm R} \to \mu^{}_{\rm R} e^{{\rm i}\theta^{}_2} \; .
\end{eqnarray}
Although the $\rm U(1)^{}_{\mu-\tau}$ symmetry will be spontaneously
broken by $\langle H^{}_{\pm 2} \rangle$, the $\rm U(1)^{}_1 \times
U(1)^{}_2$ symmetry promises $M^{}_l M^\dagger_l$ to be diagonal. On
the other hand, the Yukawa coupling terms of the neutrinos are given
by \cite{MN}
\begin{eqnarray}
y^{}_0 \overline{L^{}_e} N^{}_e H^{}_0 + y^{}_2 \overline{L^{}_\mu}
N^{}_\mu H^{}_0 + y^{}_3 \overline{L^{}_\tau} N^{}_\tau H^{}_0  \; ,
\end{eqnarray}
where $y^{}_0 = {\rm Re} (y^{}_0)$ and $y^{}_2 = y^*_3$, as required
by the $\mu$-$\tau$ reflection symmetry. However, the resulting
$M^{}_\nu$ cannot be realistic unless $M^{}_N$ contains the $\rm
U(1)^{}_{\mu-\tau}$ symmetry breaking terms. The latter can be
achieved by including $\phi^{}_1$ and $\phi^{}_2$ which carry the
respective $\rm U(1)^{}_{\mu-\tau}$ charges of 1 and 2 but transform
trivially under the $\mu$-$\tau$ reflection symmetry. Consequently,
the mass or Yukawa-coupling terms contributing to $M^{}_{N}$ are
obtained as
\begin{eqnarray}
M^{}_{11} \overline{N^{c}_e} N^{}_e + M^{}_{23} \overline{N^{c}_\mu}
N^{}_\tau + y^{}_{12} \overline{N^c_e} N^{}_\mu \phi^{*}_1 +
y^{}_{13} \overline{N^c_e} N^{}_\tau \phi^{}_1
\nonumber \\
+ y^{}_{22} \overline{N^{c}_{\mu}} N^{}_\mu \phi^{*}_2 +
y^{}_{33} \overline{N^{c}_\tau} N^{}_\tau \phi^{}_2 \; ,
\end{eqnarray}
where $M^{}_{11}$ and $M^{}_{23}$ are real mass parameters, whereas
$y^{}_{12} = y^{*}_{13}$ and $y^{}_{22} = y^*_{33}$ hold. After
$\phi^{}_1$ and $\phi^{}_2$ develop their real VEVs, the texture of
$M^{}_N$ turns out to exhibit the $\mu$-$\tau$ reflection symmetry.
Given the form of $M^{}_{\rm D}$ in Eq. (5.16), the texture of
$M^{}_\nu$ resulting from the seesaw formula will also share the
$\mu$-$\tau$ reflection symmetry.

It has recently been pointed out that the $\mu$-$\tau$ reflection
symmetry is not a necessary condition for obtaining the equality
$|U^{}_{\mu i}| = |U^{}_{\tau i}|$ (for $i=1,2,3$) for the PMNS
matrix \cite{GMT1,GMT2}. In fact, this equality may arise in a more
general context to be discussed below. Such a statement is based on
the observation that an $O^\dagger_l$ of the form in Eq. (3.19)
together with a real orthogonal $O^{}_\nu$ always results in a PMNS
matrix featuring $U^{}_{\mu i} = U^{*}_{\tau i}$. Moreover, it is
found that $O^\dagger_l$ and $O^{}_\nu$ will satisfy the above
criteria when the residual symmetries ${\rm G}^{}_l$ and $\rm
G^{}_\nu$ are real and able to completely fix the neutrino mixing
pattern \cite{GMT1,GMT2}. Consider that ${\rm G}^{}_l$ is generated
by a real $\mathcal T$ in which case $O^{}_l$ can be identified as
the unitary matrix used to diagonalize $\mathcal T$. To obtain
$O^{}_l$, we first determine the eigenvalues of $\mathcal T$ with
the help of \cite{GMT1}
\begin{eqnarray}
\lambda^3_i - \chi \lambda^2_i + \chi \lambda^{}_i -1 = 0 \;,
\end{eqnarray}
where $\chi$ is the trace of $\mathcal T$, and $|\lambda^{}_i| = 1$
(for $i=1,2,3$) holds. In addition to the eigenvalue
$\lambda^{}_1=1$, one finds the other two eigenvalues
\begin{eqnarray}
\lambda^{}_{2,3} = \frac{1}{2} \left[ \chi-1 \pm
\sqrt{ \left(\chi-1\right)^2 - 4 } \right] \; .
\end{eqnarray}
Apart from the special case of $\chi = 3$ or $-1$, $\lambda^{}_{2}$
and $\lambda^{}_{3}$ are generally complex and conjugate to each
other. For the general case, one can choose the eigenvectors
corresponding to $\lambda^{}_{2}$ and $\lambda^{}_{3}$ to be complex
conjugate to each other and the eigenvector for $\lambda^{}_1$ to be
real, making $O^\dagger_l$ be of the form in Eq. (3.19). In
comparison, the generators $\mathcal S^{}_1$ and $\mathcal S^{}_2$
of $\rm G^{}_\nu = Z^{}_2 \times Z^{}_2$ are order 2 and thus only
have the eigenvalues 1 and $-1$. Since we have required $\mathcal
S^{}_1$ and $\mathcal S^{}_2$ themselves to be real, their common
eigenvectors can be chosen to be real as well. Then we arrive at
$O^{}_\nu = O^{}_{\mathcal S} P^{}_\nu$ with $O^{}_{\mathcal S}$
being the real orthogonal matrix diagonalizing $\mathcal S^{}_{1}$
and $\mathcal S^{}_{2}$ simultaneously. Note that the undetermined
diagonal phase matrix $P^{}_\nu$ arises from the fact that the
$\mathcal S^{}_{1}$ and $\mathcal S^{}_{2}$ symmetries can only
determine $O^{}_\nu$ up to the Majonara phases. Putting all these
pieces together, a PMNS matrix $U = O^\dagger_l O^{}_{\mathcal S}
P^{}_\nu$ fulfilling the relation $|U^{}_{\mu i}| = |U^{}_{\tau i}|$
is finally available. Some immediate comments are in order.

(1) If $\rm G^{}_\nu$ only contains one $\rm Z^{}_2$ factor, then
$O^{}_{\mathcal S}$ cannot be completely fixed --- an example of
this kind is the TM1 or TM2 mixing pattern given in Eq. (4.6). In
this case the reality of $O^{}_{\mathcal S}$ is lost and the
relation $|U^{}_{\mu i}| = |U^{}_{\tau i}|$ cannot hold any more.
Therefore, ${\rm G}^{}_\nu$ and ${\rm G}^{}_l$ are required to be
able to pin down the lepton flavor mixing pattern in a realistic
model-building exercise.

(2) When one considers unifying ${\rm G}^{}_l$ and $\rm G^{}_\nu$ in
a larger discrete symmetry $\rm G^{}_F$, a subgroup of $O(3)$ (e.g.,
$\rm A^{}_4$, $\rm S^{}_4$ and $\rm A^{}_5$) will be a promising
candidate in which ${\rm G}^{}_l$ and $\rm G^{}_\nu$ are inherently
real \cite{GMT1,GMT2}. Nevertheless, none of $\rm A^{}_4$, $\rm
S^{}_4$ or $\rm A^{}_5$ can give us a satisfactory $\theta^{}_{13}$
at the lowest order \cite{GMT1}. Hence higher-order contributions
have to be invoked in these symmetry-based models.

(3) This approach may serve as a complementarity to the $\mu$-$\tau$
reflection symmetry approach but cannot take over from it. Although
both of them can predict $\delta =\pm\pi/2$ and $\theta^{}_{23}
=\pi/4$, the former is also able to determine $\theta^{}_{12}$ and
$\theta^{}_{13}$ but the latter fails in this connection. On the
other hand, only the $\mu$-$\tau$ reflection symmetry dictates the
Majorana phases to take trivial values (i.e., $0$ or $\pm\pi/2$).

(4) In fact, it is possible to generalize the theorem formulated
here to recover the $\mu$-$\tau$ reflection symmetry such that the
Majorana phases can be fixed. Suppose that $M^{}_\nu$ itself,
instead of $\rm G^{}_\nu$, should be real in which case $O^{}_\nu$
is a real orthogonal matrix without being subject to an undetermined
$P^{}_\nu$. The resulting neutrino mixing matrix will have the form
in Eq. (3.19) with the Majorana phases being trivial. It is not
surprising to find that such results agree with those predicted by
the $\mu$-$\tau$ reflection symmetry, simply because the neutrino
mass matrix will take the form in Eq. (3.24) when one returns to the
$M^{}_l M^\dagger_l$-diagonal basis by an $O^{}_l$ transformation of
the form in Eq. (3.19) \cite{GMT3,GMT4,GMT5,GMT6}. This scenario
provides an alternative way of getting at some interesting
consequences of the $\mu$-$\tau$ reflection symmetry \cite{GMT2}.
\begin{figure*}
\centering
\includegraphics[width=0.95\textwidth]{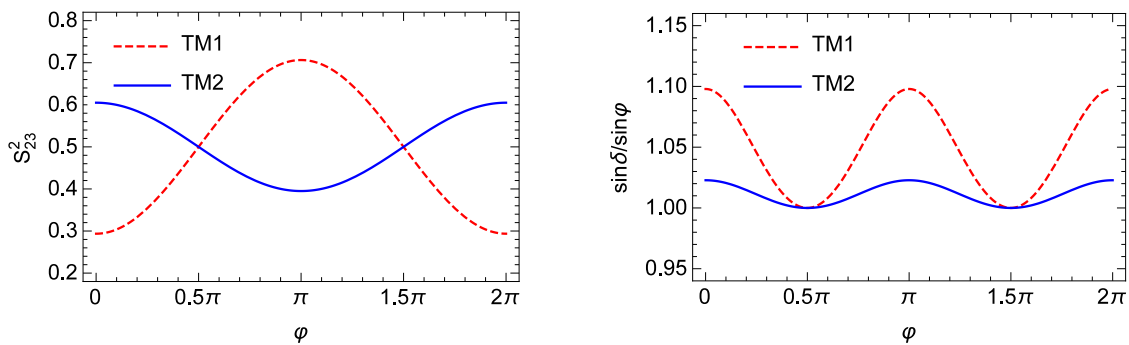}
\caption{The possible values of $s^{2}_{23}$ and
$\sin{\delta}/\sin{\varphi}$ against $\varphi$ in the TM1 (red
lines) or TM2 (blue lines) neutrino mixing case.}
\end{figure*}

\subsection{On the TM1 and TM2 neutrino mixing patterns}

As emphasized before, a general neutrino mass matrix with the
$\mu$-$\tau$ permutation symmetry is unable to give a definite
prediction for $\theta^{}_{12}$ unless some further conditions are
placed on it. The most popular example in this regard might be
$M^{}_{\mu\mu} + M^{}_{\mu\tau} = M^{}_{ee} + M^{}_{e\mu}$, a
condition that can lead us to the flavor mixing pattern $U^{}_{\rm
TB}$ \cite{TB1,TB2}. The latter yields $\theta^{}_{12} = \arctan
(1/\sqrt{2}) \simeq 35.3^\circ$, which is compatible with the
experimental value to a reasonably good degree of accuracy. Before
the size of $\theta^{}_{13}$ was measured in the Daya Bay experiment
\cite{DYB1}, the $U^{}_{\rm TB}$ pattern was widely believed to be a
good description of neutrino mixing and many discrete flavor
symmetry groups (most notably, $\rm A^{}_4$ \cite{A41,A42}) had been
employed to build the explicit models for deriving this special
flavor mixing matrix. However, the Daya Bay discovery of a
relatively large $\theta^{}_{13}$ requires a remarkable modification
of $U^{}_{\rm TB}$ \cite{X2012}. The simplest alternative turns out
to be the TM1 or TM2 flavor mixing pattern given in Eq. (4.6)
\cite{Lam1,MMTB1,TMa,TMb,TMc,MMTB2,TMd}. They owe their names to the
fact that they keep the first and second columns of the $U^{}_{\rm
TB}$ pattern, respectively. In fact, one has \cite{MMTB1,MMTB2}
\begin{eqnarray}
U^{}_{\rm TM1} = U^{}_{\rm TB} \pmatrix{ 1 & 0 & 0 \cr 0 & c & s
e^{-{\rm i} \varphi} \cr 0 & -s e^{{\rm i} \varphi} & c } \; ,
\nonumber \\
U^{}_{\rm TM2} = U^{}_{\rm TB} \pmatrix{ c & 0 & s e^{-{\rm i}
\varphi} \cr 0 & 1 & 0 \cr -s e^{{\rm i} \varphi} & 0 & c } \; ,
\end{eqnarray}
where $c \equiv \cos\theta$ and $s \equiv \sin\theta$. These two
patterns are attractive from a phenomenological point of view. First
of all, the observed $\theta^{}_{13}$ can be easily accommodated
thanks to the free parameter $\theta$. Second, the value of
$s^2_{12}$ predicted by $U^{}_{\rm TB}$ is only slightly
modified in the TM1 or TM2 pattern:
\begin{eqnarray}
s^2_{12} = \frac{1}{3} \cdot \frac{1-3s^2_{13}}{1-s^2_{13}} \simeq
0.318 \hspace{0.5cm} ({\rm for ~ TM1}) \; ,
\nonumber \\
s^2_{12} = \frac{1}{3} \cdot \frac{1}{1-s^2_{13}} \simeq 0.341
\hspace{0.68cm} ({\rm for ~ TM2}) \; .
\end{eqnarray}
Note that the value of $s^2_{12}$ given by the TM1 mixing matrix is
in better agreement with the experimental result than that given by
$U^{}_{\rm TB}$ or the TM2 mixing matrix. Third, $\theta^{}_{23}$ is
correlated with $\theta^{}_{13}$ and the phase parameter $\varphi$
as follows:
\begin{eqnarray}
s^2_{23} \simeq \frac{1}{2} \left(1 - 2\sqrt{2} s^{}_{13}
\cos{\varphi}\right) \hspace{0.5cm} ({\rm for ~ TM1}) \; ,
\nonumber \\
s^2_{23} \simeq \frac{1}{2} \left(1 + \sqrt{2} s^{}_{13}
\cos{\varphi}\right) \hspace{0.68cm} ({\rm for ~ TM2}) \; .
\end{eqnarray}
The dependence of $s^2_{23}$ on $\varphi$, with the value of
$\theta^{}_{13}$ as an input, is illustrated in Fig. 5.1. One can
see that $\theta^{}_{23}$ may stay close to $\pi/4$ if $\varphi$
takes a value around $\pi/2$ or $3\pi/2$. In addition, the
correlation between $\theta^{}_{23}$ and $\varphi$ is stronger in
the TM1 case and thus easier to be tested. Finally, the Jarlskog
invariant of CP violation in the lepton sector \cite{J} is found to
be
\begin{eqnarray}
{\cal J} = -\frac{1}{6} \sqrt{2 - 6 s^2_{13}} \ s^{}_{13} \sin{\varphi}
\hspace{0.5cm} ({\rm for ~ TM1}) \; ,
\nonumber \\
{\cal J} = -\frac{1}{6} \sqrt{2 - 3s^2_{13}} \ s^{}_{13} \sin{\varphi}
\hspace{0.5cm} ({\rm for ~ TM2}) \; ,
\end{eqnarray}
whose maximum magnitude is around $3.8\%$ at $\varphi = \pi/2$ or
$3\pi/2$. By comparing the above expression of $\cal J$ with ${\cal
J} = c^{}_{12} s^{}_{12} c^2_{13} s^{}_{13} c^{}_{23} s^{}_{23}
\sin{\delta}$ in the standard parametrization of $U$ given below Eq.
(2.13), we obtain the Dirac CP-violating phase $\delta$ as follows:
\begin{eqnarray}
\frac{\sin{\delta}}{\sin{\varphi}} = \frac{1 - s^2_{13}}
{\sqrt{\displaystyle 1 - 6s^2_{13} -4 s^2_{13} \cos{2\varphi}}}
\hspace{0.5cm} ({\rm for ~ TM1}) \; ,
\nonumber \\
\frac{\sin{\delta}}{\sin{\varphi}} = \frac{1 - s^2_{13}}
{\sqrt{\displaystyle 1 - 3s^2_{13} - s^2_{13} \cos{2\varphi}}}
\hspace{0.68cm} ({\rm for ~ TM2}) \; .
\end{eqnarray}
As shown in Fig. 5.1, $\delta$ is approximately equal to $\varphi$,
especially around $\varphi = \pi/2$ and $3\pi/2$.

Theoretically, exploring some appropriate physical contexts that can
justify these two particular mixing patterns makes sense. Besides
the residual-symmetry approach discussed in section 4.1
\cite{TMaRS1,TMaRS2,TMaRS3,TMaRS4,TMaRS5,TMaRS6,TMaRS7}, the
Friedberg-Lee (FL) symmetry
\cite{FL1,FL2,FL3,FL4,FL5,FL6,FL8,FL13,FL15} can also lead to the
TM1 and TM2 mixing patterns in a natural way. The FL symmetry in the
neutrino sector is defined in the sense that the neutrino mass
operators should be invariant under the following translational
transformations for the neutrino fields:
\begin{equation}
\nu^{}_\alpha \to \nu^{}_\alpha + \eta^{}_\alpha \xi \; ,
\end{equation}
in which $\alpha$ runs over $e$, $\mu$ and $\tau$,
$\xi$ is a spacetime-independent Grassmann-algebra element
which anti-commutes with the neutrino fields, and
$\eta^{}_{\alpha}$ are the complex coefficients.
The FL symmetry dictates the neutrino mass terms
to be of the form
\begin{eqnarray}
-{\mathcal L}^{}_{\rm mass} = & a \left(\eta^{*}_\tau \overline
\nu^{}_\mu - \eta^{*}_\mu \overline \nu^{}_\tau \right)
\left(\eta^{*}_\tau \nu^{c}_\mu - \eta^{*}_\mu \nu^{c}_\tau\right)
\nonumber \\
& \hspace{-0.12cm} + b \left(\eta^{*}_\mu \overline \nu^{}_e -
\eta^{*}_e \overline \nu^{}_\mu \right)
\left(\eta^{*}_\mu \nu^{c}_e - \eta^{*}_e \nu^{c}_\mu \right)
\nonumber \\
& \hspace{-0.12cm} + d \left(\eta^{*}_\tau \overline \nu^{}_e -
\eta^{*}_e \overline \nu^{}_\tau \right)
\left(\eta^{*}_\tau \nu^{c}_e - \eta^{*}_e \nu^{c}_\tau\right)
+ {\rm h.c.} \;
\end{eqnarray}
with $a$, $b$ and $d$ being the arbitrary complex numbers. When
proposing the original version of this effective flavor symmetry,
Friedberg and Lee assumed the neutrino fields to transform in a
universal way (i.e., $\eta^{}_{e} = \eta^{}_{\mu} = \eta^{}_{\tau}$)
\cite{FL1} in which case the neutrino mass matrix appears as
\cite{FL5}
\begin{equation}
M^{}_\nu = \pmatrix{
b + d & -b & -d \cr -b & a + b & -a \cr -d & -a & a + d } \; .
\end{equation}
This matrix can be transformed into the following form by using
the $U^{}_{\rm TB}$ matrix:
\begin{equation}
U^{\dagger}_{\rm TB} M^{}_\nu U^{*}_{\rm TB} = \frac{1}{2} \pmatrix{
3 \left(b + d\right) & 0 & \sqrt{3} \left(b - d\right) \cr 0 &  0 &
0 \cr \sqrt{3} \left(b - d\right) &  0 &  4a + b + d } \;
\end{equation}
which can be further diagonalized by a complex $(1,3)$ rotation
\footnote{Taking $b = d$, one is led to the $\mu$-$\tau$ permutation
symmetry and thus the flavor mixing pattern $U^{}_{\rm TB}$.}.
So the unitary matrix used to diagonalize $M^{}_\nu$ has the same
form as $U^{}_{\rm TM1}$ in Eq. (5.20). However, a potential problem
associated with $M^{}_\nu$ in Eq. (5.28) is its prediction $m^{}_2
=0$, which is in conflict with $\Delta m^2_{21} > 0$. A simple way
out is to add a universal term $m^{}_0 \left(\overline \nu^{}_e
\nu^c_e + \overline \nu^{}_\mu \nu^c_\mu + \overline \nu^{}_\tau
\nu^c_\tau\right)$ to the above neutrino mass operators \cite{FL1},
leading to $m^{}_2 = m^{}_0$. But this term explicitly breaks the FL
symmetry, and hence it needs a convincing explanation. Another way
out is to modify the FL symmetry by assuming $\eta^{}_e =
-2\eta^{}_\mu = -2\eta^{}_\tau$ \cite{ZZH}, and the associated
neutrino mass matrix reads
\begin{equation}
M^{\prime}_\nu = \pmatrix{
b + d & 2b & 2d \cr 2b & a + 4b & -a \cr 2d & -a & a + 4d } \; .
\end{equation}
Similarly, $U^{}_{\rm TB}$ can transform $M^\prime_\nu$ into the
form
\begin{equation}
U^{\dagger}_{\rm TB} M^\prime_\nu U^{*}_{\rm TB} = \pmatrix{ 0 & 0 &
0 \cr 0 & 3 \left(b + d\right) & \sqrt{6} \left(d - b\right) \cr 0 &
\sqrt{6} \left(d - b\right) & 2 \left(a + b + d\right) }
\end{equation}
which can be further diagonalized by a complex $(2,3)$ rotation.
Hence the resultant neutrino mixing matrix will be the TM2 pattern.
In particular, $m^{}_1 =0$ comes out from this ansatz, implying that
such a modified FL symmetry does not need to be broken. It is worth
pointing out that a combination of the modified FL symmetry and the
$\mu$-$\tau$ reflection symmetry can pin down all the physical
parameters of massive neutrinos. Note that $\eta^{}_\mu =
\eta^{}_\tau$ allows for this kind of operation which otherwise
would not make sense. In this specific scenario the neutrino mass
matrix maintains its form in Eq. (5.29) and is subject to some
further conditions such as $b = d^*$ and ${\rm Im}(a) = 0$. The
neutrino masses and flavor mixing quantities can be determined in
terms of three real parameters: ${\rm Re}(a)$, ${\rm Re}(b)$ and
${\rm Im}(b)$. While the above results for $(m^{}_1, m^{}_2,
m^{}_3)$ and $(\theta^{}_{12}, \theta^{}_{13}, \theta^{}_{23})$
still hold, the CP phases will take specific values ($\varphi=
\pi/2$ or $3\pi/2$ and $\rho, \sigma = 0$ or $\pi/2$). To summarize,
the modified FL symmetry can lead us to some definite predictions
which are compatible with current neutrino oscillation data, but the
FL symmetry itself remains puzzling and needs a further study.

Finally, let us emphasize that the TM1 and TM2 mixing patterns can
also be derived via the {\it indirect model} approach. To achieve
the TM1 or TM2 mixing matrix based on an indirect model which has
been introduced at the end of section 4.2 for realizing $U^{}_{\rm
TB}$, one has to modify the VEV alignments in Eq. (4.23) or
construct $M^{}_l$ and (or) $M^{}_N$ with off-diagonal elements. A
possible solution inspired by Eq. (5.29) is to change $\langle
\phi^{\bf 3}_1 \rangle$ and $\langle \phi^{\bf 3}_2 \rangle$ to the
following forms \cite{CSD2}:
\begin{eqnarray}
\langle \phi^{\bf 3}_1 \rangle = v^{}_1
\pmatrix{1 \cr 2 \cr 0 \cr} \; , \hspace{0.4cm}
\langle \phi^{\bf 3}_2 \rangle = v^{}_2
\pmatrix{1 \cr 0 \cr 2 \cr} \; .
\end{eqnarray}
It is obvious that the effective neutrino mass matrix obtained from
Eq. (4.22) with such VEV alignments will have the same form as that
given in Eq. (5.29). In this case one neutrino remains massless,
although the seesaw mechanism with three right-handed neutrinos has
been taken into account. The observation that the TM1 or TM2 mixing
pattern can be viewed as $U^{}_{\rm TB}$ multiplied by a complex
$(1,3)$ or $(2,3)$ rotation matrix from its right-hand side suggests
another possible solution: one may keep the regular VEV alignments
in Eq. (4.23) and ascribe the TM1 or TM2 mixing to the off-diagonal
terms in an approximately diagonal $M^{}_N$ \cite{ZZH14,KS}. To be
specific, $M^{}_N$ may have the form
\begin{eqnarray}
M^{}_N = \pmatrix{ M^{}_1 & 0  &0 \cr
0 & M^{}_{22} & M^{}_{23} \cr 0 & M^{}_{23}& M^{}_{33} }
\hspace{0.5cm} ({\rm for ~ TM1}) \; ,
\nonumber \\
M^{}_N = \pmatrix{ M^{}_{11} & 0  & M^{}_{13} \cr
0 & M^{}_{2} & 0 \cr M^{}_{13} & 0 & M^{}_{33} }
\hspace{0.5cm} ({\rm for ~ TM2}) \; ,
\end{eqnarray}
where the off-diagonal terms may be traced back to the
symmetry-breaking effects and thus relatively small. In this
situation one can go back to the mass basis of right-handed
neutrinos by means of a complex $(2,3)$ or $(1,3)$ rotation which
will in turn change the Yukawa coupling matrix of the neutrinos.
Then it is easy to check that the seesaw formula allows us to arrive
at a particular texture of $M^{}_\nu$, which finally leads us to the
TM1 or TM2 neutrino mixing matrix.

\subsection{When the sterile neutrinos are concerned}

A ``sterile" neutrino $\nu^{}_{s}$, as its name suggests, does not
carry any quantum number of the SM gauge symmetry and thus does not
directly take part in the standard weak interactions. Although there
has not been any convincing evidence for sterile neutrinos, their
existence is either theoretically motivated or experimentally
hinted. A good example of this kind is the {\it heavy} right-handed
neutrinos introduced for implementing the type-I seesaw mechanism.
On the other hand, the long-standing LSND anomaly \cite{LSND}
indicates the possible existence of an $\mathcal O(\rm eV)$ sterile
neutrino which can more or less mix with $\nu^{}_e$ and
$\nu^{}_\mu$. There are also a few other short-baseline
neutrino-oscillation anomalies, such as the reactor \cite{Reactor},
Gallium \cite{SAGE} and MiniBooNE \cite{MiniBooNE} anomalies, which
could be explained with the help of active-sterile neutrino mixing
\footnote{In addition, possible keV sterile neutrinos could be a good
candidate for warm dark matter in the Universe \cite{WDM1,WDM2,WDM3}.}.
It is therefore worthwhile to investigate an implementation of the
$\mu$-$\tau$ flavor symmetry in the presence of sterile neutrinos.
One interesting proposal in this respect is that the sterile
neutrino sector may be responsible for the $\mu$-$\tau$ symmetry
breaking \cite{GMN,GP,BRZ,MMW,Lixq,RP}. In this subsection we take
this possibility seriously in the context of {\it light} sterile
neutrinos.

According to Ref. \cite{GGLLZ}, there is little improvement of the
global-fit results in the 3+2 mixing scheme (i.e., the mixing of 3
active neutrinos and 2 sterile neutrinos) with respect to the 3+1
neutrino mixing scheme. For this reason, we restrict our discussion
to the 3+1 case although a generalization of our results to the 3+2
or 3+3 case is rather straightforward. In this context the Majorana
neutrino mass matrix for $\nu^{}_e$, $\nu^{}_\mu$, $\nu^{}_\tau$ and
$\nu^{}_s$ can be expressed as \cite{MMW}
\begin{eqnarray}
M^{}_s = \pmatrix{ m^{}_{ee} & m^{}_{e\mu} & m^{}_{e\mu} & m^{}_{e
s} \cr m^{}_{e\mu} & m^{}_{\mu\mu} & m^{}_{\mu\tau} & m^{}_{\mu s}
\cr m^{}_{e\mu} & m^{}_{\mu\tau} & m^{}_{\mu\mu} & m^{}_{\tau s} \cr
m^{}_{e s} & m^{}_{\mu s} & m^{}_{\tau s} & m^{}_{s s} } \; .
\end{eqnarray}
Here the $3\times 3$ submatrix of $M^{}_s$ for the active neutrinos
has already been assumed to have the $\mu$-$\tau$ permutation
symmetry. But it is unnecessary to assume $m^{}_{\mu s} = m^{}_{\tau
s}$ if the origin of sterile neutrinos is different from that of
active ones. For instance, the active neutrino sector can be
viewed as a consequence of the type-\Rmnum{2} seesaw mechanism
\cite{SS2a,SS2b,SS2c,SS2d,SS2e} while the other elements of $M^{}_s$
might originate from the canonical seesaw mechanism. A general $4
\times 4$ neutrino mass matrix can be diagonalized by a $4 \times 4$
unitary matrix $U^{}_s$ to give 4 real mass eigenvalues $m^{}_i$
(for $i = 1, 2, 3, 4$). And $U^{}_s$ contains 6 rotation angles
$\theta^{}_{ij}$ (for $ij = 12, 13, 23, 14, 24, 34$), 3 Dirac phases
and 3 Majorana phases. A quite popular parametrization of $U^{}_s$
reads as follows \cite{GK}:
\begin{eqnarray}
U^{}_s = O^{}_{34} U^{}_{24} U^{}_{14} O^{}_{23} U^{}_{13} O^{}_{12}
P \; ,
\end{eqnarray}
where $O^{}_{ij}$ is a real orthogonal rotation matrix with the
mixing angle $\theta^{}_{ij}$, and $U^{}_{ij}$ denotes a unitary
rotation matrix with both the mixing angle $\theta^{}_{ij}$ and the
phase parameter $\delta^{}_{ij}$ as illustrated in Eq. (3.90). In
addition, $P = {\rm Diag} \{ e^{{\rm i}\rho}, e^{{\rm i}\sigma}, 1,
e^{{\rm i}\gamma} \}$ denotes the Majorana phase matrix. This
parametrization has an advantage that the upper-left $3\times 3$
submatrix of $U^{}_s$ can simply reduce to the PMNS matrix $U$ of
the active neutrinos in the limit of vanishing active-sterile
mixing.
\begin{figure*}
\centering
\includegraphics[width=0.95\textwidth]{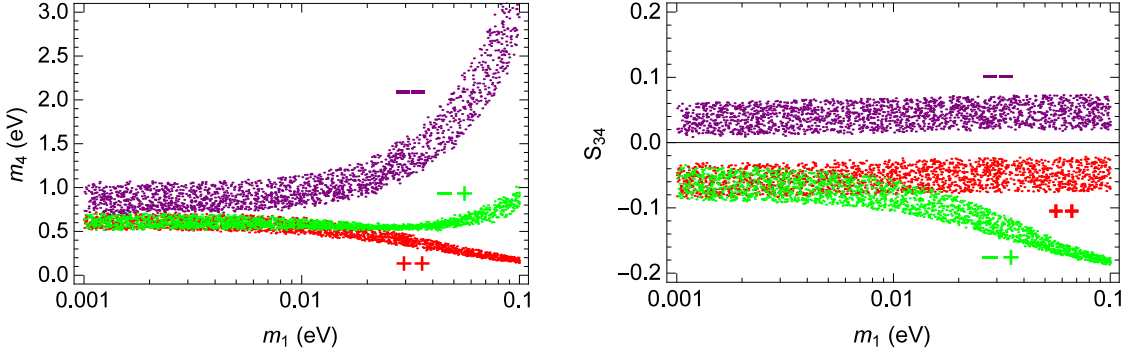}
\caption{The profiles of $m^{}_4$ and $s^{}_{34}$ versus $m^{}_{1}$
in the normal neutrino mass hierarchy. The results for ($\rho$,
$\sigma$) = (0, 0), ($\pi/2$, 0) and ($\pi/2$, $\pi/2$) are shown in
red, green and purple colors and denoted as ``$++$", ``$-+$"" and
``$--$", respectively. Note that we have presented $-s^{}_{34}$ in
the ($\rho$, $\sigma$) = ($\pi/2$, $\pi/2$) case for clarity. The
standard neutrino parameters are allowed to take values in their
1$\sigma$ intervals as listed in Table 2.1.}
\end{figure*}

Now we proceed to study the implications of the neutrino mass matrix
in Eq. (5.33) by assuming that it can explain the experimental data
concerning the active neutrinos. First of all, we intend to find out
the possible values of $m^{}_4$ and $\theta^{}_{i4}$. To serve this
purpose, let us reconstruct the elements of $M^{}_s$ in terms of
$U^{}_s$ and the neutrino masses \cite{Xing2011}:
\begin{eqnarray}
m^{}_{e\mu} \simeq & -\overline m^{}_{1} c^{}_{12} \left(s^{}_{12}
c^{}_{23} + c^{}_{12} \tilde s^{}_{13} s^{}_{23}\right) + \overline
m^{}_{2} s^{}_{12} \left(c^{}_{12} c^{}_{23} \right.
\nonumber  \\
& \left. -s^{}_{12} \tilde s^{}_{13} s^{}_{23}\right)
+ m^{}_{3} \tilde s^{*}_{13} s^{}_{23} + \overline m^{}_{4} s^{}_{14}
\tilde s^{*}_{24} \; ,
\nonumber  \\
m^{}_{e\tau} \simeq & \overline m^{}_{1} c^{}_{12}
\left(s^{}_{12}s^{}_{23} - c^{}_{12}\tilde s^{}_{13}
c^{}_{23}\right) - \overline m^{}_{2} s^{}_{12} \left(c^{}_{12}
s^{}_{23} \right.
\nonumber  \\
& \left. + s^{}_{12} \tilde s^{}_{13} c^{}_{23}\right)
+ m^{}_{3} \tilde s^{*}_{13} c^{}_{23} + \overline m^{}_{4} s^{}_{14}
\tilde s^{*}_{34} \; ,
\nonumber  \\
m^{}_{\mu \mu} \simeq & \overline  m^{}_1 \left(s^{}_{12} c^{}_{23} +
c^{}_{12} \tilde s^{}_{13} s^{}_{23}\right)^2+ \overline m^{}_2
\left(c^{}_{12}c^{}_{23} \right.
\nonumber  \\
& \left. -s^{}_{12}\tilde s^{}_{13} s^{}_{23}\right)^2 + m^{}_3
s^{2}_{23} + \overline  m^{}_4 \tilde s^{*2}_{24} \; ,
\nonumber \\
m^{}_{\tau \tau} \simeq & \overline  m^{}_1 \left(s^{}_{12} s^{}_{23} -
c^{}_{12} \tilde s^{}_{13} c^{}_{23}\right)^2+ \overline m^{}_2
\left(c^{}_{12} s^{}_{23} \right.
\nonumber  \\
& \left. + s^{}_{12} \tilde s^{}_{13} c^{}_{23}\right)^2 + m^{}_3
c^{2}_{23} +\overline m^{}_4 \tilde s^{*2}_{34} \; ,
\end{eqnarray}
where $\overline m^{}_1 \equiv m^{}_1 e^{2{\rm i} \rho}$, $\overline
m^{}_2 \equiv m^{}_2 e^{2{\rm i} \sigma}$ and $\overline m^{}_4
\equiv m^{}_4 e^{2{\rm i} \gamma}$. In obtaining these relations, we
have made use of the smallness of $\theta^{}_{i4}$ as implied by the
unitarity of $U$ which has been tested at the percent level
\cite{Antusch1,Antusch2}. And we are allowed to identify
$\theta^{}_{12}$, $\theta^{}_{13}$, $\theta^{}_{23}$ and
$\delta^{}_{13}$ in $U^{}_s$ with their counterparts in $U$ without
loss of much accuracy. Taking account of the $\mu$-$\tau$ symmetry
conditions $m^{}_{e\mu} = m^{}_{e\tau}$ and $m^{}_{\mu\mu} =
m^{}_{\tau\tau}$, we arrive at
\begin{eqnarray}
\tilde s^{*}_{34} & = & \sqrt{2} \frac{\left( m^{}_3 -
m^{}_{22}\right) \Delta \theta^{}_{23} - m^{}_{12} \tilde s^{}_{13}
} {m^{}_{12} \Delta \theta^{}_{23} + m^{}_{11} \tilde s^{}_{13} -
m^{}_3 \tilde s^{*}_{13}} s^{}_{14} - \tilde s^{*}_{24} \; ,
\nonumber \\
\overline m^{}_4 & = & \sqrt{2} \frac{m^{}_{12} \Delta
\theta^{}_{23} + m^{}_{11} \tilde s^{}_{13} - m^{}_3 \tilde
s^{*}_{13}} {s^{}_{14} \left(\tilde s^{*}_{34} - \tilde
s^{*}_{24}\right)} \; ,
\end{eqnarray}
where $m^{}_{11}$, $m^{}_{12}$ and $m^{}_{22}$ have already been
defined in Eq. (3.5), and $\Delta \theta^{}_{23} \equiv
\theta^{}_{23} - \pi/4$. There are so many free parameters that no
definite conclusions can be drawn. For the sake of simplicity and
illustration, let us consider the CP-conserving case by switching
off all the CP phases. Some discussions are in order.

(1) \underline{$m^{}_1 < m^{}_2 \ll m^{}_3$}.
In this case we obtain
\begin{eqnarray}
s^{}_{34} \sim -\sqrt{2} \ \frac{\Delta \theta^{}_{23}} {s^{}_{13}}
s^{}_{14} - s^{}_{24} \; ,
\nonumber \\
m^{}_{4} \sim \frac {s^2_{13}} { s^{}_{14} \left( s^{}_{14} \Delta
\theta^{}_{23}+ \sqrt{2} s^{}_{13} s^{}_{24} \right) } m^{}_3 \; ,
\end{eqnarray}
where $s^{}_{14}$ should not be too small, so as to make the
magnitude of $m^{}_4$ properly suppressed.

(2) \underline{$m^{}_1 \simeq m^{}_2 \gg m^{}_3$}. For ($\rho$,
$\sigma$) = (0, 0), $s^{}_{34}$ and $m^{}_{4}$ have the same
expressions as those in Eq. (5.37) if the replacement $m^{}_3
\Longrightarrow m^{}_{1}$ is made. In the ($\rho$, $\sigma$) = (0,
$\pi/2$) case, however, $s^{}_{34}$ and $m^{}_{4}$ approximate to
\begin{eqnarray}
s^{}_{34} \sim \sqrt{2} \ \frac{ \left(c^2_{12} - s^2_{12}\right)
\Delta \theta^{}_{23} - 2 c^{}_{12} s^{}_{12} s^{}_{13} } { 2
c^{}_{12} s^{}_{12} \Delta \theta^{}_{23} + \left(c^2_{12} -
s^2_{12}\right) s^{}_{13} } s^{}_{14} - s^{}_{24} \; ,
\nonumber \\
m^{}_{4} \sim \sqrt{2} \ \frac{ 2c^{}_{12} s^{}_{12} \Delta
\theta^{}_{23} + \left(c^2_{12} - s^2_{12}\right) s^{}_{13} } {
s^{}_{14} \left(s^{}_{34} - s^{}_{24}\right) } m^{}_2 \; ,
\end{eqnarray}
where the product of $s^{}_{14}$ and $s^{}_{34} - s^{}_{24}$ should
not be too small, such that $m^{}_4$ can be properly suppressed.

(3) \underline{$m^{}_1 \simeq m^{}_2 \simeq m^{}_3$}. For ($\rho$,
$\sigma$) = (0, 0), $s^{}_{34}$ and $m^{}_4$ are also the same as
those in Eq. (5.37) with the replacement $m^{}_3 \Longrightarrow
\Delta m^2_{31}/m^{}_1$. On the other hand, the results in the
($\rho$, $\sigma$) = (0, $\pi/2$) case are given by
\begin{eqnarray}
s^{}_{34} \sim \sqrt{2} \ \frac{ c^2_{12} \Delta \theta^{}_{23} -
c^{}_{12} s^{}_{12} s^{}_{13} } { c^{}_{12} s^{}_{12} \Delta
\theta^{}_{23} - s^2_{12} s^{}_{13} } s^{}_{14} - s^{}_{24} \; ,
\nonumber \\
m^{}_{4} \sim 2 \sqrt{2} \ \frac{ c^{}_{12} s^{}_{12} \Delta
\theta^{}_{23} - s^2_{12} s^{}_{13} } { s^{}_{14} \left(s^{}_{34} -
s^{}_{24}\right) } m^{}_1 \; .
\end{eqnarray}
These analytical approximations are consistent with the numerical
results illustrated in Fig. 5.2, in which the possible values of
$m^{}_4$ and $s^{}_{34}$ against $m^{}_1$ are shown in the normal
neutrino mass ordering case. In our numerical calculations we have
specified $s^{}_{14} \simeq s^{}_{24} \simeq 0.1$ as a compromise
between the unitarity test and short-baseline neutrino-oscillation
anomalies. One can see that the allowed range of $m^{}_{4}$ is
rather wide and covers the region favored by those preliminary
anomalies. In agreement with the unitarity test, the allowed range
of $|s^{}_{34}|$ is relatively small. Moreover, $|s^{}_{34}|$ is not
negligibly small in most of the parameter space, opening an
interesting possibility for the short-baseline neutrino oscillations
involving $\nu^{}_{\tau}$.

Given the above results, let us return to $M^{}_s$ itself to examine
its possible structures and the naturalness issue from a
model-building point of view. We focus on the cases that the flavor
mixing pattern and mass spectrum of three active neutrinos are
mainly determined by the $3 \times 3$ submatrix of $M^{}_s$,
although the contribution from the sterile neutrino sector might in
principle be dominant. To illustrate, we consider the following
three typical cases.

(a) In the $m^{}_1< m^{}_{2} \ll m^{}_{3}$ case, $M^{}_s$ can be
parameterized in the form
\begin{eqnarray}
M^{}_s = m \pmatrix{
e \epsilon^2 & d \epsilon^2 & d \epsilon^2 & \epsilon \cr
d \epsilon^2 & c \epsilon & -c \epsilon & a \epsilon \cr
d \epsilon^2 & -c \epsilon & c \epsilon & b \epsilon \cr
\epsilon & a \epsilon &  b \epsilon &  1 } \; ,
\end{eqnarray}
where $\epsilon \simeq 0.1$, and $a$, $b$, $c$, $d$ and $e$ are
all the ${\cal O}(1)$ coefficients. The relatively heavy $\nu^{}_4$
as compared with $\nu^{}_i$ (for $i = 1, 2, 3$) can be integrated
out, like the treatment for the heavy Majorana neutrinos in the
seesaw mechanism, leaving us a $3 \times 3$ neutrino mass matrix of
the form in Eq. (3.59). According to our previous analysis, we can
naturally obtain the phenomenologically acceptable results without
having to tune the relevant coefficients.

(b) The overall neutrino mass matrix that can result in
$m^{}_{1}\simeq m^{}_{2} \gg m^{}_{3}$ together with ($\rho$,
$\sigma$) = (0, 0) appears as
\begin{eqnarray}
M^{}_s = m \pmatrix{
2c \epsilon & d \epsilon^2 & d \epsilon^2 & \epsilon \cr
d \epsilon^2 & c \epsilon & c \epsilon & a \epsilon \cr
d \epsilon^2 & c \epsilon & c \epsilon & b \epsilon \cr
\epsilon & a \epsilon &  b \epsilon &  1 } \; .
\end{eqnarray}
The resulting active neutrino mass matrix is similar to that given
in Eq. (3.63) after the sterile neutrino is integrated out. As
discussed before, the fine-tuning conditions $a + b - 2d \simeq 0$
and $(a+b)^2 - 2 \simeq 0$ will be required to fit the experimental
data. It is therefore difficult to realize such a scenario from the
viewpoint of model building.

(c) The $m^{}_{1} \simeq m^{}_{2} \simeq m^{}_{3}$ case with
($\rho$, $\sigma$) = (0, 0), which allows a relatively light
$\nu^{}_s$, can be viewed as a consequence of the following mass
matrix:
\begin{eqnarray}
M^{}_s = m \pmatrix{
f \epsilon^2 + 1 & e \epsilon^2 & e \epsilon^2 & \epsilon \cr
e \epsilon^2 & d \epsilon + 1 & -d \epsilon & a \epsilon \cr
e \epsilon^2 & -d \epsilon & d \epsilon + 1 & b \epsilon \cr
\epsilon & a \epsilon &  b \epsilon &  c } \; .
\end{eqnarray}
Like the mass matrix given in Eq. (5.40), the present texture of
$M^{}_s$ can naturally fit the experimental results regarding the
active neutrino mixing and flavor oscillations. Of course, some
higher-order terms in the above three ans$\rm\ddot{a}$tze of the
neutrino mass matrix (e.g., those located at the $e\mu$ and $e\tau$
entries) can be neglected in the first place so as to get much more
and much simpler predictions.

\def\thefootnote{\arabic{footnote}}
\setcounter{footnote}{0}
\setcounter{equation}{0}
\setcounter{table}{0}
\setcounter{figure}{0}

\section{Some consequences of the $\mu$-$\tau$ symmetry}

The exact or approximate $\mu$-$\tau$ flavor symmetry may have some
important phenomenological consequences in neutrino physics and
cosmology. Here let us look at a few interesting examples of this
type, such as neutrino oscillations in matter, flavor distributions
of ultrahigh-energy (UHE) cosmic neutrinos, cosmic matter-antimatter
asymmetry via leptogenesis, and fermion mass matrices with a common
$\rm Z^{}_2$ symmetry.

\subsection{Neutrino oscillations in matter}

It is known that the behaviors of neutrino oscillations in matter
can be quite different from those in vacuum, simply because the
matter-induced {\it coherent forward scattering} effects modifies
the genuine neutrino masses and flavor mixing parameters when a
neutrino beam travels through a normal medium (e.g., the Sun or
Earth) \cite{MSW1,MSW2}. If the PMNS matrix $U$ in vacuum possesses
the $\mu$-$\tau$ flavor symmetry, one can show that the effective
PMNS flavor mixing matrix $\widetilde{U} = \widetilde{V} P^{}_\nu$
in matter must possess the same symmetry
\footnote{Note that the Majorana phase matrix $P^{}_\nu$ is not
affected by matter effects, as it has nothing
to do with neutrino oscillations no matter whether they occur in
vacuum or in matter.}.
In other words, the matter effects respect the $\mu$-$\tau$
permutation or reflection symmetry in neutrino oscillations
\cite{MT-R,XZ10}. Let us go into details of this
observation in two different ways in the following.

In the basis where the mass and flavor eigenstates of three charged
leptons are identical, we denote the effective neutrino mass matrix
in matter as $\widetilde{M}^{}_\nu$. No matter whether massive
neutrinos are of the Dirac or Majorana nature, the effective
Hamiltonian responsible for the propagation of a neutrino beam in
matter can be expressed in a similar way as that in Eq. (2.11):
\begin{eqnarray}
\widetilde{\cal H}^{}_{\rm eff} & = & \frac{1}{2 E}
\widetilde{M}^{}_\nu \widetilde{M}^\dagger_\nu = \frac{1}{2 E}
\widetilde{V} \widetilde{D}^2_\nu \widetilde{V}^\dagger
\nonumber \\
& = & \frac{1}{2 E} \left(V D^2_\nu V^\dagger + D^{}_{\rm m} \right)
\; ,
\end{eqnarray}
where $\widetilde{D}^{}_\nu = {\rm Diag}\{ \widetilde{m}^{}_1 ,
\widetilde{m}^{}_2 , \widetilde{m}^{}_3\}$ with $\widetilde{m}^{}_i$
(for $i=1,2,3$) being the effective neutrino masses in matter, and
$D^{}_{\rm m} = {\rm Diag}\{A , 0 , 0\}$ with $A = 2\sqrt{2} \
G^{}_{\rm F} N^{}_e E$ denoting the charged-current contribution to
the coherent $\nu^{}_e e^-$ forward scattering (here $N^{}_e$
stands for the background number density of electrons \cite{MSW1}).

Without loss of generality, it is always possible to redefine the
phases of three neutrino mass eigenstates such that $V^{}_{e i}$ and
$\widetilde{V}^{}_{e i}$ (for $i = 1, 2, 3$) are all real and
positive \cite{FX01}. To be explicit,
\begin{eqnarray}
V = O^{}_{23} O^\prime_\delta O^{}_{13} O^{}_{12}
= \pmatrix{ c^{}_{12} c^{}_{13} & s^{}_{12} c^{}_{13} & s^{}_{13} \cr
V^{}_{\mu 1} & V^{}_{\mu 2} & c^{}_{13} s^{}_{23} \cr V^{}_{\tau 1} &
V^{}_{\tau 2} & c^{}_{13} c^{}_{23} \cr} \;
\end{eqnarray}
with
\begin{eqnarray}
V^{}_{\mu 1} = -s^{}_{12} c^{}_{23} e^{-{\rm i} \delta} -
c^{}_{12} s^{}_{13} s^{}_{23} \; ,
\nonumber \\
V^{}_{\mu 2} = c^{}_{12} c^{}_{23} e^{-{\rm i} \delta} -
s^{}_{12} s^{}_{13} s^{}_{23} \; ,
\nonumber \\
V^{}_{\tau 1} = s^{}_{12} s^{}_{23} e^{-{\rm i} \delta} -
c^{}_{12} s^{}_{13} c^{}_{23} \; ,
\nonumber \\
V^{}_{\tau 2} = -c^{}_{12} s^{}_{23} e^{-{\rm i} \delta} -
s^{}_{12} s^{}_{13} c^{}_{23} \; ,
\end{eqnarray}
and $O^\prime_\delta = {\rm Diag} \left\{ 1, e^{-{\rm i} \delta}, 1
\right\}$. Such a phase convention is apparently different from that
taken in Eq. (2.8). In matter the effective matrix $\widetilde{V}$
takes the same phase convention as $V$ in Eq. (6.2). In this
particular basis $V^{}_{e i} = |V^{}_{e i}|$ and
$\widetilde{V}^{}_{e i} = |\widetilde{V}^{}_{e i}|$ (for $i=1, 2,
3$) hold, and thus it will be suitable for discussing the
$\mu$-$\tau$ flavor symmetry. Thanks to Eq. (6.1), we simply obtain
\begin{eqnarray}
\sum_i \widetilde{m}^2_i \widetilde{V}^{}_{\alpha i}
\widetilde{V}^*_{\beta i} = \sum_i m^2_i V^{}_{\alpha i}
V^*_{\beta i} + \delta^{}_{e \alpha} \delta^{}_{e \beta} A \; ,
\end{eqnarray}
where $\alpha$ and $\beta$ run over $e$, $\mu$ and $\tau$. If $V$
possesses the $\mu$-$\tau$ symmetry, then it is straightforward to
show that $\widetilde{V}$ must have the same symmetry. Let us
consider two distinct possibilities.

(A) {\it The $\mu$-$\tau$ permutation symmetry}. In this limit
$\theta^{}_{13} =0$ and $\theta^{}_{23} = \pi/4$ hold, and thus the
phase parameter $\delta$ in $V$ can be removed. We are then left
with a real $V$ with $V^{}_{e 3} =0$, $V^{}_{\mu 1} = -V^{}_{\tau
1}$, $V^{}_{\mu 2} = -V^{}_{\tau 2}$ and $V^{}_{\mu 3} = V^{}_{\tau
3}$ in the above parametrization. So the elements of $\widetilde{V}$
satisfy
\begin{eqnarray}
\sum_i \widetilde{m}^2_i |\widetilde{V}^{}_{e i}|
\widetilde{V}^*_{\mu i} = -\sum_i \widetilde{m}^2_i
|\widetilde{V}^{}_{e i}| \widetilde{V}^*_{\tau i} \; ,
\nonumber \\
\sum_i \widetilde{m}^2_i |\widetilde{V}^{}_{\mu i}|^2 = \sum_i
\widetilde{m}^2_i |\widetilde{V}^{}_{\tau i}|^2 \; ,
\end{eqnarray}
as one can see from Eq. (6.4). Because the matrix $\widetilde{V}$ is
actually real and $\widetilde{m}^2_i$ can be arbitrary for arbitrary
values of $E$, it is always possible to arrive at the results
$\widetilde{V}^{}_{e 3} =0$, $\widetilde{V}^{}_{\mu 1} =
-\widetilde{V}^{}_{\tau 1}$, $\widetilde{V}^{}_{\mu 2} =
-\widetilde{V}^{}_{\tau 2}$ and $\widetilde{V}^{}_{\mu 3} =
\widetilde{V}^{}_{\tau 3}$ from Eq. (6.5), corresponding to
$\widetilde{\theta}^{}_{13} =0$ and $\widetilde{\theta}^{}_{23} =
\pi/4$ in the parametrization of $\widetilde{V}$ similar to Eq.
(6.2). Note that it is impossible to simultaneously have
$\widetilde{V}^{}_{\mu 1} = \widetilde{V}^{}_{\tau 1}$,
$\widetilde{V}^{}_{\mu 2} = \widetilde{V}^{}_{\tau 2}$ and
$\widetilde{V}^{}_{\mu 3} = \widetilde{V}^{}_{\tau 3}$ (or
$\widetilde{V}^{}_{\mu 1} = -\widetilde{V}^{}_{\tau 1}$,
$\widetilde{V}^{}_{\mu 2} = -\widetilde{V}^{}_{\tau 2}$ and
$\widetilde{V}^{}_{\mu 3} = -\widetilde{V}^{}_{\tau 3}$), because
they violate the unitarity requirement of $\widetilde{V}$. We
conclude that $\widetilde{V}$ may have the same $\mu$-$\tau$
permutation symmetry as $V$ does.

(B) {\it The $\mu$-$\tau$ reflection symmetry}. In this more
interesting limit $\theta^{}_{23} = \pi/4$ and $\delta = \pm\pi/2$
hold, leading to $V^{}_{\mu 1} = V^*_{\tau 1}$, $V^{}_{\mu 2} =
V^*_{\tau 2}$ and $V^{}_{\mu 3} = V^{}_{\tau 3}$. As both $V^{}_{\mu
3}$ and $V^{}_{\tau 3}$ are real in the phase convention of Eq.
(6.2), one may also take $V^{}_{\mu 3} = V^*_{\tau 3}$ and thus
$V^{}_{\mu i} = V^*_{\tau i}$ (for $i=1, 2, 3$) in this case
\cite{MT-R}. As a result, Eq. (6.4) leads us to the relations
\begin{eqnarray}
\sum_i \widetilde{m}^2_i |\widetilde{V}^{}_{e i}|
\widetilde{V}^*_{\mu i} = \sum_i \widetilde{m}^2_i
|\widetilde{V}^{}_{e i}| \widetilde{V}^{}_{\tau i} \; ,
\nonumber \\
\sum_i \widetilde{m}^2_i |\widetilde{V}^{}_{\mu i}|^2 = \sum_i
\widetilde{m}^2_i |\widetilde{V}^{}_{\tau i}|^2 \; .
\end{eqnarray}
Hence it is straightforward to obtain $\widetilde{V}^{}_{\mu i} =
\widetilde{V}^*_{\tau i}$ from Eq. (6.6) for arbitrary
$\widetilde{m}^{2}_i$. Given the parametrization of $\widetilde{V}$
similar to Eq. (6.2), the above equalities give rise to
$\widetilde{\theta}^{}_{23} =\pi/4$ and $\widetilde{\delta} = \pm
\pi/2$. That is why $\widetilde{V}$ possesses the $\mu$-$\tau$
reflection symmetry as $V$ does.

Now we turn to another way to show that the matter effects respect
the $\mu$-$\tau$ reflection symmetry in neutrino oscillations.
Taking the parametrization of $V$ in Eq. (6.2), we find
\begin{eqnarray}
\widetilde{M}^{}_\nu \widetilde{M}^\dagger_\nu & = &
O^{}_{23} O^\prime_\delta \left( O^{}_{13} O^{}_{12}
D^2_\nu O^\dagger_{12} O^\dagger_{13} +
D^{}_{\rm m} \right) O^{\prime \dagger}_\delta O^\dagger_{23}
\nonumber \\
& = & O^{}_{23} O^\prime_\delta
\widehat{O}^{}_{23} \widehat{O}^{}_{13} \widehat{O}^{}_{12}
\widetilde{D}^2_\nu \widehat{O}^\dagger_{12}
\widehat{O}^\dagger_{13} \widehat{O}^\dagger_{23} O^{\prime
\dagger}_\delta O^\dagger_{23} \;
\end{eqnarray}
with $\widehat{O}^{}_{23} \widehat{O}^{}_{13} \widehat{O}^{}_{12}$
being a real orthogonal matrix which consists of three rotation
angles ($\widehat{\theta}^{}_{12}, \widehat{\theta}^{}_{13},
\widehat{\theta}^{}_{23}$) and has been defined to diagonalize the
real bracketed part in Eq. (6.7). Up to the freedom of diagonal
phase matrices $P^{}_{\rm L}$ and $P^{}_{\rm R}$ denoted below,
which are actually irrelevant to neutrino oscillations, the
effective neutrino mixing matrix in matter turns out to be
\begin{eqnarray}
\widetilde{V} & \equiv & \widetilde{O}^{}_{23}
\widetilde{O}^\prime_\delta \widetilde{O}^{}_{13}
\widetilde{O}^{}_{12} = P^{}_{\rm L} O^{}_{23} O^\prime_\delta
\widehat{O}^{}_{23} \widehat{O}^{}_{13} \widehat{O}^{}_{12}
P^{}_{\rm R} \nonumber \\
& = & P^{}_{\rm L} \left(O^{}_{23} O^\prime_\delta
\widehat{O}^{}_{23}\right) \widetilde{O}^{}_{13}
\widetilde{O}^{}_{12} P^{}_{\rm R} \; ,
\end{eqnarray}
where $\widetilde{O}^{}_{12}$ and $\widetilde{O}^{}_{13}$ have been
identified with $\widehat{O}^{}_{12}$ and $\widehat{O}^{}_{13}$,
respectively. In comparison, $\widetilde{\theta}^{}_{23}$ and
$\widetilde{\delta}$ must arise from a mixture of $\theta^{}_{23}$,
$\delta$ and $\widehat{\theta}^{}_{23}$. In a way analogous to Eq.
(2.13), we define the effective Jarlskog invariant $\widetilde{\cal
J}$ in matter and calculate it with the help of Eq. (6.8). After a
straightforward calculation, we arrive at
\begin{eqnarray}
\widetilde{\cal J} & = & \widetilde{c}^{}_{12} \widetilde{s}^{}_{12}
\widetilde{c}^2_{13} \widetilde{s}^{}_{13} \widetilde{c}^{}_{23}
\widetilde{s}^{}_{23} \sin\widetilde{\delta} \nonumber \\
& = & \widetilde{c}^{}_{12} \widetilde{s}^{}_{12}
\widetilde{c}^2_{13} \widetilde{s}^{}_{13} c^{}_{23} s^{}_{23}
\sin\delta \; ,
\end{eqnarray}
in which the formula proportional to $\sin\widetilde{\delta}$ is
derived from the definition of $\widetilde{V}$ in Eq. (6.8), while
the one proportional to $\sin\delta$ is obtained from the last
equality of Eq. (6.8). Note that the free parameter
$\widehat{\theta}^{}_{23}$ does not enter $\widetilde{\cal J}$,
although it mixes with $\theta^{}_{23}$ and $\delta$ in the
expressions of $\widetilde{\theta}^{}_{23}$ and
$\widetilde{\delta}$. Thus Eq. (6.9) leads us to the interesting
Toshev relation \cite{Toshev}
\begin{eqnarray}
\sin 2\widetilde{\theta}^{}_{23} \sin\widetilde{\delta} = \sin
2\theta^{}_{23} \sin\delta \; ,
\end{eqnarray}
which links $\widetilde{\theta}^{}_{23}$ and $\widetilde{\delta}$ in
matter to $\theta^{}_{23}$ and $\delta$ in vacuum. In addition to
the parametrization of $V$ in Eq. (2.8) or Eq. (6.2), one may also
derive the Toshev-like relation from two other parametrizations of
$V$ or $\widetilde{V}$ \cite{Zhou}.

Given $\theta^{}_{23} = \pi/4$ and $\delta =\pm \pi/2$ for the PMNS
flavor mixing matrix in vacuum, the Toshev relation in Eq. (6.10)
tells us that $\sin 2\widetilde{\theta}^{}_{23}
\sin\widetilde{\delta} =\pm 1$ must hold in matter. As a result, we
are left with $\widetilde{\theta}^{}_{23} = \theta^{}_{23} = \pi/4$
and $\widetilde{\delta} = \delta = \pm \pi/2$. We draw the
conclusion that the $\mu$-$\tau$ reflection symmetry keeps unchanged
for neutrino oscillations in matter. Finally, it is obvious that
these discussions are not applicable for the $\mu$-$\tau$
permutation symmetry case, in which $\widetilde{\delta}$ and
$\delta$ are irrelevant in physics because of
$\widetilde{\theta}^{}_{13} = \theta^{}_{13} = 0$.

\subsection{Flavor distributions of UHE cosmic neutrinos}

An extremely important topic in high-energy neutrino astronomy is to
search for the point-like neutrino sources that may help resolve a
long-standing problem --- the very origin of UHE cosmic rays. The
reason is simply that the UHE protons originating in a cosmic
accelerator (e.g., the gamma ray burst or active galactic nuclei
\cite{XZ}) unavoidably interact with their ambient photons via $p +
\gamma \to \Delta^+ \to \pi^+ + n$ or their ambient protons through
$p + p \to \pi^{\pm} + X$ with $X$ being other particles, producing
a large amount of energetic charged pions whose decays can therefore
produce copious UHE cosmic muon neutrinos and electron neutrinos or
their antiparticles. The so-called neutrino telescope is such a kind
of deep underground ice or water Cherenkov detector which can record
those rare events induced by the UHE cosmic neutrinos. Today the
most outstanding neutrino telescope is the km$^3$-volume IceCube
detector at the South Pole, which has successfully identified
thirty-seven extraterrestrial neutrino candidate events with
deposited energies ranging from 30 TeV to 2 PeV \cite{IC1,IC2}.
Among them, the three PeV events represent the highest-energy
neutrino interactions ever observed. But where such extraordinary
PeV neutrinos came from remains quite mysterious.

Given a distant astrophysical source of either $p\gamma$ or $pp$
collisions, it has the same $\nu^{}_\alpha +
\overline{\nu}^{}_\alpha$ flavor distribution $\Phi^{\rm S}_e :
\Phi^{\rm S}_\mu : \Phi^{\rm S}_\tau = 1 : 2 : 0$, where $\Phi^{\rm
S}_\alpha \equiv \Phi^{\rm S}_{\nu_\alpha} + \Phi^{\rm
S}_{\overline{\nu}_\alpha}$ with $\Phi^{\rm S}_{\nu_\alpha}$ and
$\Phi^{\rm S}_{\overline{\nu}_\alpha}$ being the fluxes of
$\nu^{}_\alpha$ and $\overline{\nu}^{}_\alpha$ (for $\alpha = e,
\mu, \tau$) at the source. Such an initial flavor distribution is
expected to change to $\Phi^{\rm T}_e : \Phi^{\rm T}_\mu : \Phi^{\rm
T}_\tau = 1 : 1 : 1$ at a neutrino telescope, where $\Phi^{\rm
T}_\alpha \equiv \Phi^{\rm T}_{\nu_\alpha} + \Phi^{\rm
T}_{\overline{\nu}_\alpha}$ is similarly defined, because the UHE
cosmic neutrinos may oscillate many times on the way to the Earth
and finally reach a flavor democracy \cite{Pakvasa} if the PMNS
neutrino mixing matrix $U$ satisfies the $|U^{}_{\mu i}| =
|U^{}_{\tau i}|$ condition (for $i=1,2,3$) \cite{XZ08}. As a
consequence, the effects of $\mu$-$\tau$ symmetry breaking can
modify the ratio $\Phi^{\rm T}_e : \Phi^{\rm T}_\mu : \Phi^{\rm
T}_\tau = 1 : 1 : 1$ in a nontrivial way.

To be more explicit, the flavor distribution of UHE cosmic neutrinos
at a terrestrial neutrino telescope is given by
\begin{eqnarray}
\Phi^{\rm T}_\alpha & = & \sum_\beta \left[\Phi^{\rm
S}_{\nu^{}_\beta} P(\nu^{}_\beta \to \nu^{}_\alpha) + \Phi^{\rm
S}_{\overline{\nu}^{}_\beta} P(\overline{\nu}^{}_\beta \to
\overline{\nu}^{}_\alpha) \right]
\nonumber \\
& = & \sum_i \sum_\beta |U^{}_{\alpha i}|^2 |U^{}_{\beta i}|^2
\Phi^{\rm S}_\beta \; ,
\end{eqnarray}
where we have used
\begin{eqnarray}
P(\nu^{}_\beta \to \nu^{}_\alpha) = P(\overline{\nu}^{}_\beta \to
\overline{\nu}^{}_\alpha) = \sum_i |U^{}_{\alpha i}|^2 |U^{}_{\beta
i}|^2 \; .
\end{eqnarray}
Note that Eq. (6.12) holds for a very simple reason: the galactic
distance that the non-monochromatic UHE cosmic neutrinos have
traveled far exceeds the observed neutrino oscillation lengths, and
thus $P(\nu^{}_\beta \to \nu^{}_\alpha)$ and
$P(\overline{\nu}^{}_\beta \to \overline{\nu}^{}_\alpha)$ are
averaged over many oscillations and finally become energy- and
distance-independent. Given $\Phi^{\rm S}_e : \Phi^{\rm S}_\mu :
\Phi^{\rm S}_\tau = 1 : 2 : 0$ and the unitarity of $U$, we
immediately obtain
\begin{eqnarray}
\Phi^{\rm T}_\alpha & = & \frac{\Phi^{}_0}{3} \sum_i |U^{}_{\alpha
i}|^2 \left(|U^{}_{e i}|^2 + 2 |U^{}_{\mu i}|^2 \right)
\nonumber \\
& = & \frac{\Phi^{}_0}{3} \left[1 + \sum_i |U^{}_{\alpha i}|^2
\left( |U^{}_{\mu i}|^2 - |U^{}_{\tau i}|^2 \right) \right] \; ,
\end{eqnarray}
where $\Phi^{}_0 \equiv \Phi^{\rm S}_e + \Phi^{\rm S}_\mu +
\Phi^{\rm S}_\tau$ denotes the total flux of neutrinos and
antineutrinos of all flavors. That is why $\Phi^{\rm T}_e :
\Phi^{\rm T}_\mu : \Phi^{\rm T}_\tau = 1 : 1 : 1$ can be achieved
provided the equalities $|U^{}_{\mu i}| = |U^{}_{\tau i}|$ (for
$i=1,2,3$) hold. In fact, the difference between $\Phi^{\rm T}_\mu$
and $\Phi^{\rm T}_\tau$ is a straightforward signature of the
$\mu$-$\tau$ symmetry breaking:
\begin{eqnarray}
\Phi^{\rm T}_\mu - \Phi^{\rm T}_\tau = \frac{\Phi^{}_0}{3} \sum_i
\left( |U^{}_{\mu i}|^2 - |U^{}_{\tau i}|^2 \right)^2 \; .
\end{eqnarray}
This exact and parametrization-independent result is very useful for
us to probe the leptonic flavor mixing structure via the detection
of UHE cosmic neutrinos at neutrino telescopes \cite{Xing2012}.

In the standard parametrization of $U$, let us define $\varepsilon
\equiv \theta^{}_{23} -\pi/4$ to measure a part of the $\mu$-$\tau$
symmetry-breaking effects. Then we arrive at
\begin{eqnarray}
\Phi^{\rm T}_e : \Phi^{\rm T}_\mu : \Phi^{\rm T}_\tau = \left(1 +
D^{}_e\right) : \left(1 + D^{}_\mu\right) : \left(1 +
D^{}_\tau\right) \;
\end{eqnarray}
with $D^{}_e = -2\Delta$, $D^{}_\mu = \Delta + \overline{\Delta}$
and $D^{}_\tau = \Delta - \overline{\Delta}$, where the first- and
second-order perturbation terms $\Delta$ \cite{Xing2006} and
$\overline{\Delta}$ \cite{Rodejohann2008} are expressed as
\begin{eqnarray}
\Delta & \simeq & \frac{1}{2} \sin^2 2\theta^{}_{12} \sin\varepsilon
-\frac{1}{4} \sin 4\theta^{}_{12} \sin\theta^{}_{13} \cos\delta \; ,
\nonumber \\
\overline{\Delta} & \simeq & \left(4 - \sin^2 2\theta^{}_{12}
\right) \sin^2\varepsilon + \sin^2 2\theta^{}_{12}
\sin^2\theta^{}_{13}\cos^2\delta
\nonumber \\
&& + \sin 4\theta^{}_{12} \sin\varepsilon \sin\theta^{}_{13}
\cos\delta \; ,
\end{eqnarray}
respectively. So a difference between $\Phi^{\rm T}_\mu$ and
$\Phi^{\rm T}_\tau$ (i.e., a difference between $D^{}_\mu$ and
$D^{}_\tau$) is actually an effect of the second-order $\mu$-$\tau$
symmetry breaking. It is easy to show $\overline{\Delta} \geq 0$ for
arbitrary values of $\delta$, but $\Delta$ may be either positive or
negative (or vanishing in the $\mu$-$\tau$ symmetry limit). When the
$3\sigma$ intervals of $\theta^{}_{12}$, $\theta^{}_{13}$,
$\theta^{}_{23}$ and $\delta$ in Table 2.1 are taken into account,
the upper bounds of $|\Delta|$ and $\overline{\Delta}$ can both
reach about $0.1$, implying that the flavor democracy of $\Phi^{\rm
T}_e$, $\Phi^{\rm T}_\mu$ and $\Phi^{\rm T}_\tau$ may maximally be
broken at the same level.

It is convenient to define three working observables at neutrino
telescopes and connect them to the $\mu$-$\tau$ symmetry-breaking
quantities:
\begin{eqnarray}
R^{}_e & \equiv & \frac{\Phi^{\rm T}_e}{\Phi^{\rm T}_\mu + \Phi^{\rm
T}_\tau} \simeq \frac{1}{2} - \frac{3}{2} \Delta \; ,
\nonumber \\
R^{}_\mu & \equiv & \frac{\Phi^{\rm T}_\mu}{\Phi^{\rm T}_\tau +
\Phi^{\rm T}_e} \simeq \frac{1}{2} + \frac{3}{4} \left(\Delta +
\overline{\Delta} \right) \; ,
\nonumber \\
R^{}_\tau & \equiv & \frac{\Phi^{\rm T}_\tau}{\Phi^{\rm T}_e +
\Phi^{\rm T}_\mu} \simeq \frac{1}{2} + \frac{3}{4} \left(\Delta -
\overline{\Delta} \right) \; .
\end{eqnarray}
The small departure of $R^{}_\alpha$ (for $\alpha =e, \mu, \tau$)
from $1/2$ is therefore a clear measure of the overall effects of
$\mu$-$\tau$ symmetry breaking.

Now we turn to the flavor distribution of UHE cosmic neutrinos at a
neutrino telescope by detecting the $\overline{\nu}^{}_e$ flux from
a very distant astrophysical source via the famous Glashow-resonance
(GR) channel $\overline{\nu}^{}_e e \rightarrow W^- \rightarrow ~
{\rm anything}$ \cite{Glashow}, which happens over a narrow energy
interval around $E^{\rm GR}_{\overline{\nu}^{}_e} \simeq
M^2_W/2m^{}_e \simeq 6.3 ~ {\rm PeV}$, and its cross section is
about two orders of magnitude larger than those of
$\overline{\nu}^{}_e N$ interactions of the same
$\overline{\nu}^{}_e$ energy \cite{Gandhi}. A measurement of the GR
phenomenon is also important for another reason: it may serve as a
sensitive discriminator of UHE cosmic neutrinos originating from
$p\gamma$ and $pp$ collisions \cite{G12,G22,XZ11,Lin}. Let us assume
that a neutrino telescope is able to observe both the GR-mediated
$\overline{\nu}^{}_e$ events and the $\nu^{}_\mu +
\overline{\nu}^{}_\mu$ events of charged-current interactions in the
vicinity of $E^{\rm GR}_{\overline{\nu}^{}_e}$. Then their ratio
$R^{}_{\rm GR} \equiv \Phi^{\rm T}_{\overline{\nu}^{}_e}/\Phi^{\rm
T}_\mu$ can tell us something about the lepton flavor mixing.

To see this point more clearly, we start from the initial flavor
distribution of UHE neutrinos at a cosmic accelerator: $\Phi^{\rm
S}_{\nu^{}_e} : \Phi^{\rm S}_{\overline{\nu}^{}_e} : \Phi^{\rm
S}_{\nu^{}_\mu} : \Phi^{\rm S}_{\overline{\nu}^{}_\mu} : \Phi^{\rm
S}_{\nu^{}_\tau} : \Phi^{\rm S}_{\overline{\nu}^{}_\tau} = 1 : 1 : 2
: 2 : 0 : 0$ ($pp$ collisions) or $1 : 0 : 1 : 1 : 0 : 0$ ($p\gamma$
collisions). Thanks to neutrino oscillations, the
$\overline{\nu}^{}_e$ flux at a neutrino telescope can be calculated
by using the following formula:
\begin{eqnarray}
\Phi^{\rm T}_{\overline{\nu}^{}_e} = \sum_i \sum_\beta |U^{}_{e
i}|^2 |U^{}_{\beta i}|^2 \Phi^{\rm S}_{\overline{\nu}^{}_\beta} \; .
\end{eqnarray}
Since the expression of $\Phi^{\rm T}_\mu$ has been given in
Eq. (6.13), it is straightforward to obtain $R^{}_{\rm GR}$
for two different astrophysical sources:
\begin{eqnarray}
R^{}_{\rm GR} (pp) & \simeq & \frac{1}{2} - \frac{3}{2} \Delta
- \frac{1}{2} \overline{\Delta} \; ,
\nonumber \\
R^{}_{\rm GR} (p\gamma) & \simeq & \frac{\sin^2 2\theta^{}_{12}}{4}
- \frac{4 + \sin^2 2\theta^{}_{12}}{4} \Delta -
\frac{\sin^2 2\theta^{}_{12}}{4} \overline{\Delta}
\nonumber \\
&& + \frac{1+\cos^2 2\theta^{}_{12}}{2} \sin^2 \theta^{}_{13} \; .
\end{eqnarray}
It is clear that the deviation of $R^{}_{\rm GR} (pp)$ from $1/2$
and that of $R^{}_{\rm GR} (p\gamma)$ from $\sin^2
2\theta^{}_{12}/4$ are both governed by the effects of $\mu$-$\tau$
symmetry breaking, which can maximally be of ${\cal O}(0.1)$. As
discussed in the literature \cite{G1,XZ11}, the IceCube detector
running at the South Pole has a discovery potential to probe
$R^{}_{\rm GR} (pp)$. In comparison, it is more challenging to
detect the GR-mediated UHE $\overline{\nu}^{}_e$ events originating
from the pure $p\gamma$ collisions at a cosmic accelerator.

In the above discussions we have neglected the uncertainties
associated with the initial neutrino fluxes at a given astrophysical
source. A careful analysis of the flavor ratios of UHE cosmic
neutrinos originating from $pp$ or $p\gamma$ collisions actually
leads us to the ratio $\Phi^{\rm S}_e : \Phi^{\rm S}_\mu : \Phi^{\rm
S}_\tau \simeq 1 : 2\left(1 -\eta\right) : 0$ with $\eta \simeq
0.08$ \cite{eta}. If this uncertainty is taken into account, then
Eq. (6.13) will accordingly change to
\begin{eqnarray}
\Phi^{\rm T}_\alpha & \simeq & \frac{\Phi^{}_0}{3} \left[1 + \sum_i
|U^{}_{\alpha i}|^2 \left(|U^{}_{\mu i}|^2 - |U^{}_{\tau i}|^2
\right) \right.
\nonumber \\
&& \left. - 2\eta \sum_i |U^{}_{\alpha i}|^2 |U^{}_{\mu i}|^2
\right] \;
\end{eqnarray}
for $\alpha = e$, $\mu$ and $\tau$. This result implies that the
$\eta$-induced correction is in general comparable with (or even
larger than) the effect of $\mu$-$\tau$ symmetry breaking. On the
other hand, there may exist the uncertainties associated with the
identification of different flavors at the neutrino telescope. Given
the IceCube detector for example, the experimental error for
determining the ratio of the muon track to the non-muon shower is
typically $\xi \sim 20\%$ \cite{Beacom}, depending on the event
numbers. Such an estimate means that the ratio $R^{}_\mu$ in Eq.
(6.17) may in practice be contaminated by $\xi$, and this
contamination is very likely to overwhelm the $\mu$-$\tau$ symmetry
breaking effect $(\Delta + \overline{\Delta})$ \cite{Xing2012}.

We have not considered other complexities and uncertainties
associated with the origin of UHE cosmic neutrinos, such as their
energy dependence, the effect of magnetic fields and possible new
physics \cite{Winter}. It remains too early to say that we have
correctly understood the production mechanism of UHE cosmic rays and
neutrinos from a given cosmic accelerator. But progress in
determining the neutrino mixing parameters is so encouraging that it
may finally allow us to well control the error bars from particle
physics (e.g., the effect of $\mu$-$\tau$ symmetry breaking) and
thus focus on the unknowns from astrophysics (e.g., the initial
flavor composition of UHE cosmic neutrinos \cite{XingZhou2006}).
Needless to say, any constraint on the flavor distribution of UHE
cosmic neutrinos to be achieved from a neutrino telescope will be
greatly useful in diagnosing the astrophysical sources and in
understanding the properties of neutrinos themselves.

\subsection{Matter-antimatter asymmetry via leptogenesis}

Besides interpreting why the masses of the three known neutrinos are
tiny in a qualitatively natural way, the canonical seesaw mechanism
has another attractive feature --- it can also provide a natural
explanation of the observed baryon-antibaryon asymmetry of the
Universe via the leptogenesis mechanism \cite{Leptogenesis}: the
CP-violating, lepton-number-violating and out-of-equilibrium decays
of the hypothetical heavy Majorana neutrinos $N^{}_i$ may give rise
to a lepton-antilepton asymmetry; and the latter is subsequently
converted to the baryon-antibaryon asymmetry thanks to the sphaleron
process \cite{Yanagida}. The amount of the lepton-antilepton
asymmetry depends on the CP-violating asymmetries $\epsilon^{}_i$
between the decays of $N^{}_i$ and their CP-conjugate processes. If
the masses of $N^{}_i$ have a strong hierarchy (i.e., $M^{}_1\ll
M^{}_{2} \ll M^{}_3$), then only $\epsilon^{}_1$ is expected to
really matter at a temperature lower than $M^{}_1$. As a good
approximation \cite{CPa},
\begin{eqnarray}
\epsilon^{}_1=-\frac{3}{16\pi(Y^\dagger_\nu Y^{}_\nu)^{}_{11}}
\sum^{}_{i\neq 1} {\rm Im}\left[(Y^\dagger_\nu
Y^{}_\nu)^2_{1i}\right] \frac{M^{}_1}{M^{}_i} \; ,
\end{eqnarray}
where $Y^{}_\nu$ denotes the Yukawa coupling matrix of the neutrinos
in the mass basis of $N^{}_i$. With the help of $M^{}_\nu = U
D^{}_\nu U^T$ in Eq. (2.10) and the seesaw formula $M^{}_\nu =
Y^{}_\nu D^{-1}_{N} Y^T_\nu \langle H\rangle^2$ in the chosen basis,
where $D^{}_{N} \equiv {\rm Diag}\{M^{}_1, M^{}_2, M^{}_3\}$ and
$\langle H\rangle = v/\sqrt{2} \simeq 174$ GeV, it is convenient to
parametrize $Y^{}_\nu$ in the following way \cite{CI}:
\begin{eqnarray}
Y^{}_\nu=\frac{1}{\langle H \rangle} U^{} \sqrt{D^{}_{\nu}} R
\sqrt{D^{}_{N}}  \; ,
\end{eqnarray}
in which $R$ is a complex orthogonal matrix satisfying
$RR^{T}=R^{T}R=1$. Inserting Eq. (6.22) into Eq. (6.21), we arrive
at
\begin{eqnarray}
\epsilon^{}_1 = -\frac{3}{16\pi}\frac{M^{}_1}{\langle H\rangle^2}
\frac{{\rm Im} \left( \Delta m^2_{21} R^{*2}_{21} +
\Delta m^2_{31} R^{*2}_{31} \right)}
{\displaystyle \sum^{}_i m^{}_i |R^{}_{i1}|^2} \; ,
\end{eqnarray}
where the orthogonality condition
$R^{2}_{11}+R^{2}_{21}+R^{2}_{31}=0$ has been used. Eq. (6.23) shows
that $\epsilon^{}_1$ is essentially independent of the PMNS matrix
$U$, implying that {\it in general} there is no direct link between
CP violation in $N^{}_i$ decays and that in low-energy neutrino
oscillations \cite{BP}. If a certain flavor symmetry is taken into
account to reduce the number of free parameters associated with the
heavy and (or) light neutrinos, then $R$ will get constrained to a
simpler form so that the above expression of $\epsilon^{}_1$ can be
simplified to some extent \cite{SV}. For instance, a model which can
predict a constant neutrino mixing pattern is expected to result in
a real diagonal $R$ and thus a vanishing $\epsilon^{}_1$
\cite{FSaLG1, FSaLG2,FSaLG3,FSaLG4,FSaLG5}. This point is
particularly transparent in the so-called ``form-dominance"
scenarios where $Y^{}_\nu$ is simply taken as $U$ multiplied by a
diagonal matrix \cite{FD}.

In fact, it is easy to see that the $\mu$-$\tau$ permutation
symmetry may have some interesting implications on the leptogenesis
picture \cite{RNMN,RNMNY,MTaLG1,MTaLG2,MTaLG3,MTaLG,MTaLG4}. In the
limit of such a flavor symmetry the Yukawa coupling matrix of the
neutrinos appears as \cite{RNMN,RNMNY}
\begin{eqnarray}
\widetilde Y^{}_\nu = \pmatrix{ y^{}_{11} &  y^{}_{12} & y^{}_{12}
\cr y^{}_{21} & y^{}_{22} & y^{}_{23} \cr  y^{}_{21} & y^{}_{23} &
y^{}_{22}} \; ,
\end{eqnarray}
and the heavy Majorana neutrino mass matrix $M^{}_{N}$ has the same
form as the light Majorana neutrino mass matrix $M^{}_\nu$ given in
Eq. (3.7). One can make the transformation $U^T_{N} M^{}_{N}
U^{}_{N} = D^{}_{N}$, where $U^{}_{N} = O^{}_{\rm M} U^{N}_{12}$
with $O^{}_{\rm M}$ being the maximal mixing matrix:
\begin{eqnarray}
O^{}_{\rm M} = \frac{1}{\sqrt{2}} \pmatrix{ \sqrt{2} & 0 & 0 \cr
0 & 1 & -1 \cr 0 & 1 & 1 \cr} \; .
\end{eqnarray}
In this mass basis $\widetilde Y^{}_\nu$ becomes
\begin{eqnarray}
Y^{}_\nu = \widetilde Y^{}_\nu U^{}_{N} = U U^{\dagger}
\widetilde Y^{}_\nu U^{}_{N} = U U^{\dagger}_{12} O^{T}_{\rm M}
\widetilde Y^{}_\nu O^{}_{\rm M} U^{N}_{12}
\nonumber \\
= U U^\dagger_{12} \pmatrix{y^{}_{11} &\sqrt{2} y^{}_{12} & 0 \cr
\sqrt{2} y^{}_{21} & y^{}_{22}+y^{}_{23} & 0 \cr
0 & 0 & y^{}_{22}-y^{}_{23} } U^{N}_{12} \;,
\end{eqnarray}
in which $U=O^{}_{\rm M}U^{}_{12}$ is just the PMNS matrix of the
three light Majorana neutrinos consisting of $\theta^{}_{23}=\pi/4$.
Comparing Eq. (6.22) with Eq. (6.26), we see that $R$ is now subject
to the $(1,2)$ rotation subspace \cite{RNMN,RNMNY}. In this case
$\epsilon^{}_1$ is given by
\begin{eqnarray}
\epsilon^{}_1 = -\frac{3}{16\pi} \cdot\frac{M^{}_1}{\langle
H\rangle^2} \cdot\frac{{\rm Im}\left( \Delta m^2_{21} R^{*2}_{21}
\right)} {\displaystyle m^{}_1 |R^{}_{11}|^2+m^{}_2 |R^{}_{21}|^2} \; .
\end{eqnarray}
With the help of this result and $R^2_{11}+R^2_{21}=1$, it is easy
to show that $|\epsilon^{}_1|$ has an upper bound
\begin{eqnarray}
|\epsilon^{}_1| \leq \frac{3}{16\pi} \cdot \frac{M^{}_1} {\langle
H\rangle^2} \left|m^{}_2-m^{}_1 \right| \; .
\end{eqnarray}
Note that such an upper bound of $\epsilon^{}_1$ can be used to set
a lower bound on $M^{}_1$ if the leptogenesis mechanism works well
in explaining the cosmological matter-antimatter asymmetry. In
comparison with the upper bound of $\epsilon^{}_1$ obtained in a
more general case, where $m^{}_2$ in Eq. (6.28) ought to be replaced
by $m^{}_3$ \cite{Ubound}, the present upper bound can be lowered by
at least a factor $\sqrt{\Delta m^2_{21}/\Delta m^2_{31}}~$. When
the effect of $\mu$-$\tau$ flavor symmetry breaking is concerned,
the situation will become more complicated and thus a careful
analysis has to be done \cite{RNMNY}.

In the literature the so-called minimal seesaw model \cite{FGY} has
received some special attention due to its simplicity and stronger
predictive power, so its connection to the leptogenesis mechanism
deserves some discussions. Because there are only two heavy Majorana
neutrinos in this model, $M^{}_{N}$ and $\widetilde Y^{}_\nu$ may
read as follows \cite{RNMN,RNMNY}:
\begin{eqnarray}
M^{}_{N} = \pmatrix{M^{}_{22} & M^{}_{23} \cr M^{}_{23} & M^{}_{22}}
\; , \hspace{0.2cm}
\widetilde Y^{}_\nu = \pmatrix{ y^{}_{12} &
y^{}_{12} \cr y^{}_{22} & y^{}_{23} \cr  y^{}_{23} & y^{}_{22}} \; .
\end{eqnarray}
The transformation $Q^T_{\rm M} M^{}_{N} Q^{}_{\rm M} = D^{}_{N}$,
where
\begin{eqnarray}
Q^{}_{\rm M} = \frac{1}{\sqrt{2}} \pmatrix{ 1 & -1 \cr
1 & 1 \cr} \; ,
\end{eqnarray}
leads us to the mass eigenvalues $M^{}_2 = M^{}_{22} + M^{}_{23}$
and $M^{}_3 = M^{}_{22} - M^{}_{23}$. In the mass basis the Yukawa
coupling matrix $\widetilde{Y}^{}_\nu$ becomes
\begin{eqnarray}
Y^{}_\nu = \widetilde Y^{}_\nu Q^{}_{\rm M} = \frac{1}{\sqrt{2}}
\pmatrix{ 2 y^{}_{12} &  0 \cr y^{}_{22}+ y^{}_{23} &  y^{}_{23}-
y^{}_{22} \cr  y^{}_{22}+y^{}_{23} & y^{}_{22}- y^{}_{23}} \; .
\end{eqnarray}
It is interesting to see that the two columns of $Y^{}_\nu$ are
orthogonal to each other, and thus $Y^{\dagger}_\nu Y^{}_\nu$ must
be diagonal and $\epsilon^{}_1$ must be vanishing. So a finite value
of $\epsilon^{}_1$ has to come from the $\mu$-$\tau$ permutation
symmetry breaking. Let us take a simple example to illustrate the
effect of $\mu$-$\tau$ symmetry breaking and its connection to CP
violation in the decays of two heavy Majorana neutrinos. To be
explicit, we assume $M^{}_{N}$ and $\widetilde Y^{}_\nu$ to be real
and introduce the symmetry breaking term as
\begin{eqnarray}
\widetilde Y^{\prime}_\nu = \widetilde{Y}^{}_\nu + \pmatrix{
-y^{\prime}_{12} & y^{\prime}_{12} \cr 0 & 0 \cr  0 & 0 } \; ,
\end{eqnarray}
in which $y^\prime_{12}$ is complex so as to accommodate CP
violation. In the mass basis we obtain
\begin{eqnarray}
Y^{\prime}_\nu = \widetilde Y^{\prime}_\nu Q^{}_{\rm M} =
\pmatrix{ r^{}_1 a &  r^{}_2 b \cr a & -b \cr a & b} \; ,
\end{eqnarray}
where $a \equiv \left(y^{}_{22} + y^{}_{23}\right)/\sqrt{2}$, $b
\equiv \left(y^{}_{22} - y^{}_{23}\right)/\sqrt{2}$, $r^{}_1 \equiv
\sqrt{2} y^{}_{12}/a$ and $r^{}_2 \equiv \sqrt{2} y^\prime_{12}/b$
are defined. The parameter $r^{}_2$ is therefore responsible for the
generation of $\theta^{}_{13}$ and CP violation. It is found that
the experimental data at low energies can be reproduced when one
assumes $r^{}_1\simeq 1$, $|r^{}_2|\simeq \sqrt{2}\theta^{}_{13}$,
$3a^2\langle H\rangle^2/M^{}_2=m^{}_2$ and $2b^2\langle
H\rangle^2/M^{}_3=m^{}_3$ \cite{SFKL}. In this case the CP-violating
asymmetry between the decay of the lighter heavy Majorana neutrino
(with mass $M^{}_3$) and its CP-conjugate process can approximate to
\begin{eqnarray}
\epsilon^{} = \frac{b^2}{16\pi} \cdot\frac{m^{}_2}{m^{}_3} {\rm Im}
\left(r^2_2\right) \sim 10^{-4} \sin{2\phi} \; ,
\end{eqnarray}
where $\phi \equiv \arg\left(r^{}_2\right)$ and $b \sim \mathcal
O(1)$ has been taken into account. This estimate implies that the
strength of $\mu$-$\tau$ symmetry breaking required by the observed
value of $\theta^{}_{13}$ is sufficient for a successful
leptogenesis mechanism to work in such a minimal seesaw scenario
\cite{RNMNY}.

When the $\mu$-$\tau$ reflection symmetry is imposed on both the
Yukawa coupling matrix $\widetilde{Y}^{}_\nu$ and the heavy Majorana
neutrino matrix $M^{}_N$, one has
\begin{eqnarray}
\widetilde Y^{}_\nu = \pmatrix{ y^{}_{11} &  y^{}_{12} & y^{*}_{12}
\cr  y^{}_{21} & y^{}_{22} & y^{}_{23} \cr  y^{*}_{21} & y^{*}_{23}
& y^{*}_{22}} \;
\end{eqnarray}
with $y^{}_{11}$ being real, and the form of $M^{}_{N}$ is the same
as that given in Eq. (3.24). In this case $M^{}_N$ can be
diagonalized by a unitary matrix $U^{}_{N}$, whose complex conjugate
is of the form shown in Eq. (3.19). Given the Yukawa coupling matrix
$Y^{}_\nu=\widetilde Y^{}_\nu U^{}_{N}$ in the mass basis of
$N^{}_i$, a straightforward calculation shows that $Y^\dagger_\nu
Y^{}_\nu$ is actually a real matrix which prohibits CP violation
\cite{GL}. To make the leptogenesis idea viable, one may choose to
softly break the $\mu$-$\tau$ reflection symmetry \cite{AKKN}.
Another way out is to keep the $\mu$-$\tau$ reflection symmetry but
take into account the so-called flavor effects to make the
leptogenesis mechanism work \cite{MN}. If the mass of the
lightest heavy Majorana neutrino $N^{}_1$ lies in the range $10^{9}\
{\rm GeV} < M^{}_1 <10^{12}\ {\rm GeV}$, the $\tau$ leptons can be
in thermal equilibrium, making them distinguishable from the $e$ and
$\mu$ flavors. In this situation the decay of $N^{}_1$ to the $\tau$
flavor should be treated separately from the decays of $N^{}_1$ to
the other two flavors \cite{Feffect1,Feffect2}. Accordingly, the
{\it flavored} CP-violating asymmetries associated with the decays
of $N^{}_1$ are given by
\begin{eqnarray}
\epsilon^{\alpha}_1 & = & -\frac{1}{8\pi} \cdot\frac{M^{}_1}{\langle
H\rangle^2} \sum^{}_{i\neq 1} {\rm Im} \left[ \frac{3}{2}
(Y^{*}_{\nu})^{}_{\alpha 1} (Y^{}_{\nu})^{}_{\alpha i}
(Y^\dagger_\nu Y^{}_\nu)^{}_{1i} \frac{M^{}_1}{M^{}_i} \right.
\nonumber \\
&& \left. -(Y^{*}_{\nu})^{}_{\alpha 1} (Y^{}_{\nu})^{}_{\alpha i}
(Y^\dagger_\nu Y^{}_\nu)^{*}_{1i} \frac{M^2_1}{\displaystyle M^2_i}
\right] \; ,
\end{eqnarray}
in which $\alpha$ runs over $e$, $\mu$ and $\tau$. After a
straightforward calculation, one can obtain $\epsilon^e_1=0$ and
$\epsilon^\mu_1=-\epsilon^\tau_1$ as a direct consequence of the
$\mu$-$\tau$ reflection symmetry. This result confirms the vanishing
of the overall CP-violating asymmetry (i.e.,
$\epsilon^{}_1=\epsilon^e_1+\epsilon^\mu_1+\epsilon^\tau_1 =0$). In
spite of $\epsilon^{}_1=0$, the lepton-antilepton asymmetry produced
from the decays of $N^{}_1$ is still likely to survive if such
processes took place in the temperature range $10^{9}\ {\rm GeV} < T
= M^{}_1 <10^{12}\ {\rm GeV}$ in which the flavor effects should
take effect \cite{MN}. When the temperature of the Universe
dropped below $10^9$ GeV, however, the $\mu$ leptons would also be
in thermal equilibrium. In this case the simple leptogenesis
scenario under discussion would not work anymore if $M^{}_1 < 10^9$
GeV held.

\subsection{Fermion mass matrices with the $\rm Z^{}_2$ symmetry}

Being capable of relating the smallness of quark flavor mixing
angles to the smallness of quark mass ratios in a way similar to the
GST relation, the Fritzsch texture-zero ansatz is a simple and
instructive example for studying the possible underlying flavor
structures of three quark families \cite{Fritzsch1,Fritzsch2}:
\begin{equation}
M^{}_{\rm q}= \pmatrix{ 0& c^{}_{\rm q} &0 \cr c^*_{\rm q} & 0 &
b^{}_{\rm q} \cr 0& b^*_{\rm q} & a^{}_{\rm q} } \; ,
\end{equation}
where q = u (up) or d (down), and the Hermiticity has been assumed
to reduce the number of free parameters. Given the strong quark mass
hierarchies as shown in Fig. 2.2, $M^{}_{\rm q}$ is expected to have
a strongly hierarchical structure (i.e., $|c^{}_q| \ll |b^{}_q| \ll
|a^{}_q|$). Unfortunately, such a simple texture has been ruled out
mainly because its phenomenological consequences cannot
simultaneously fit the small size of $V^{}_{cb}$ and the large value
of $m^{}_t$. One way out is to introduce an additional free
parameter to make the $(2,2)$ entry of $M^{}_{\rm q}$ nonzero
\cite{4zero1,4zero2,4zero3}:
\begin{equation}
M^{}_{\rm q}= \pmatrix{ 0& c^{}_{\rm q} &0 \cr c^*_{\rm q} &
d^{}_{\rm q} &  b^{}_{\rm q} \cr 0& b^*_{\rm q} & a^{}_{\rm q} } \; .
\end{equation}
Then it is easy to check that this new ansatz, which may have a more
or less weaker hierarchy than its original counterpart, can be in
good agreement with the experimental data on quark flavor mixing and
CP violation \cite{4zero4,XZhao}. Although the Fritzsch-like texture
in Eq. (6.38) is apparently different from the pattern of $M^{}_\nu$
with a $\mu$-$\tau$ flavor symmetry (i.e., the former contains a few
texture zeros while the latter is characterized by some linear
relations or equalities of different entries), it is possible to
find a flavor basis where $M^{}_{\rm q}$ may essentially have the
same form as $M^{}_\nu$. Here we consider a special but interesting
possibility that the lepton and quark sectors share a permutation
symmetry between their respective second and third families. For
simplicity, we assume that the relevant mass matrices $M^{}_{\rm f}$
(for $\rm f = u, d$, $l, \nu$) take the following form
\cite{2-31,2-32,2-33,2-34,2-35,2-36}
\begin{equation}
M^{}_{\rm f}= P^{\dagger}_{\rm f} \widehat{M}^{}_{\rm f} P^{}_{\rm
f}= P^{\dagger}_{\rm f} \pmatrix{ A^{}_{\rm f} & B^{}_{\rm f} &
B^{}_{\rm f} \cr B^{}_{\rm f} &  C^{}_{\rm f} &  D^{}_{\rm f} \cr
B^{}_{\rm f} &  D^{}_{\rm f} &  C^{}_{\rm f} } P^{}_{\rm f} \;
\end{equation}
with $\widehat M^{}_{\rm f}$ being real and $P^{}_{\rm f} \equiv
{\rm Diag}\{e^{{\rm i}\phi^{\rm f}_1}, e^{{\rm i}\phi^{\rm
f}_2},e^{{\rm i}\phi^{\rm f}_3}\}$. Obviously, the $(2,3)$
permutation symmetry requires $\phi^{\rm f}_2=\phi^{\rm f}_3$. If
the massive neutrinos are the Majorana particles, however, the
$P^{}_\nu$ matrix on the right-hand side of $\widehat M^{}_\nu$ in
Eq. (6.39) should be replaced by $P^*_\nu$. The unitary matrices
used to diagonalize $M^{}_{\rm f}$ can be universally expressed as
\begin{eqnarray}
O^{}_{\rm f} = P^{\dagger}_{\rm f} O^{\rm f}_{23} O^{\rm f}_{12} \; ,
\end{eqnarray}
where $O^{\rm f}_{23}$ has the same form as $O^{}_{\rm M}$ in Eq.
(6.25), and $O^{\rm f}_{12}$ contains an unspecified rotation angle
$\theta^{\rm f}_{12}$.

Now let us pay particular attention to the $A^{}_{\rm q}=0$ case,
where $\theta^{\rm q}_{12}$ can be connected to the quark mass ratio
via $\tan{\theta^{\rm q}_{12}}=\sqrt{m^{\rm q}_1/m^{\rm q}_2}$.
Hence it is easy to obtain the GST relation. In view of Eq. (6.40),
we find that the CKM quark mixing matrix $V^{}_{\rm
CKM}=O^{\dagger}_{\rm u} O^{}_{\rm d}$ only contains one nonzero
mixing angle (i.e., the Cabibbo angle $\theta^{\rm CKM}_{12} \equiv
\theta^{}_{\rm C}$) in the $(2,3)$ permutation symmetry limit. The
vanishing of $\theta^{\rm CKM}_{13}$ and $\theta^{\rm CKM}_{23}$ in
this scenario qualitatively agrees with the experimental fact that
these two angles are very small: $\theta^{\rm CKM}_{13} \simeq
0.2^\circ$ and $\theta^{\rm CKM}_{23} \sim 2^\circ$ \cite{PDG}. So
their finite values can be ascribed to the small $(2,3)$ permutation
symmetry breaking \cite{JAS}. Similar to the $\mu$-$\tau$
permutation symmetry, the $(2,3)$ permutation symmetry can be broken
by allowing
\begin{eqnarray}
\widehat M^{\rm f}_{12} \neq \widehat{M}^{\rm f}_{13} \; ,
\hspace{0.4cm} \widehat{M}^{\rm f}_{22} \neq \widehat{M}^{\rm
f}_{33} \; ,  \hspace{0.4cm} \phi^{\rm f}_2 \neq \phi^{\rm f}_3 \; .
\end{eqnarray}
If the symmetry breaking only comes from the phase parameters (i.e.,
$\phi^{\rm f}_2 \neq \phi^{\rm f}_3$), then $\theta^{\rm CKM}_{23}$
can fit its experimental value by taking $\phi^{\rm u}_{32}
-\phi^{\rm d}_{32} \sim 0.08$, where $\phi^{\rm q}_{32} \equiv
\phi^{\rm q}_{3}-\phi^{\rm q}_{2}$ is defined. On the other hand,
$\theta^{\rm CKM}_{13}$ can be calculated via the correlation
$\theta^{\rm CKM}_{13}\sim \theta^{\rm u}_{12}\theta^{\rm
CKM}_{23}$, but the result is somewhat smaller than the observed
value \cite{XZhao}. In this case one may either give up $A^{}_{\rm
q} =0$ or break the $(2,3)$ permutation symmetry of $\widehat
M^{}_{\rm q}$ in a slightly stronger way. The former approach may
not make a realization of the GST relation self-evident. Following
the latter approach, we can obtain nonzero $\theta^{\rm q}_{13}$ by
allowing $\widehat M^{\rm q}_{12} \neq \widehat M^{\rm q}_{13}$ but
keeping $\widehat M^{\rm q}_{22} = \widehat M^{\rm q}_{33}$. Note
that one may simply take $\widehat M^{q}_{13} =0$ and arrive at the
following zero texture of fermion mass matrices:
\begin{equation}
M^{}_{\rm f}= P^{\dagger}_{\rm f} \widehat M^{}_{\rm f} P^{}_{\rm f}
= P^{\dagger}_{\rm f} \pmatrix{ 0 & B^{}_{\rm f} &  0 \cr B^{}_{\rm
f} &  C^{}_{\rm f} &  D^{}_{\rm f} \cr 0 &  D^{}_{\rm f} & C^{}_{\rm
f} } P^{}_{\rm f} \; .
\end{equation}
Then it is easy to show that $\theta^{\rm u}_{13}$ is vanishingly
small and $\theta^{\rm d}_{13}$ can be given by $\theta^{\rm d}_{13}
\simeq \sqrt{m^{}_d m^{}_s/m^{2}_b}$, which is large enough to
generate a phenomenologically acceptable value of $\theta^{\rm
CKM}_{13}$ \cite{XZhao}. Since the fermion mass matrices in Eq.
(6.42) have the same structure as those quark mass matrices in Eq.
(6.38) if $a^{}_{\rm q} = d^{}_{\rm q}$ is taken, a universal
treatment of lepton and quark flavor mixing issues seems quite
natural in this sense.

We proceed to discuss some phenomenological consequences of Eq.
(6.39) in the lepton sector. Above all, the experimental result
$\theta^{}_{23} \simeq \pi/4$ signifies the strong $(2,3)$
permutation symmetry breaking. If the symmetry is only broken by
$\phi^{\rm f}_2 \neq \phi^{\rm f}_3$, then the phase difference
$\phi^{l}_{32}-\phi^{\nu}_{32}$ has to be very close to $\pm \pi/2$
in order to avoid a significant cancellation between the
charged-lepton and neutrino sectors. To be explicit, we focus on the
texture in Eq. (6.42) to obtain more definite predictions. We have
$\theta^{l}_{12} \simeq \sqrt{m^{}_e/m^{}_\mu} \simeq 0.07$, and
$\theta^{l}_{13}$ is highly suppressed due to $m^{}_e \ll m^{}_\mu
\ll m^{}_\tau$. So $\theta^{\nu}_{12}$ should take the dominant
responsibility for $\theta^{}_{12} \simeq 34^\circ$, implying
$\theta^{\nu}_{12} \simeq \theta^{}_{12}$ as a good approximation.
Taking $\tan{\theta^{\nu}_{12}}\simeq \sqrt{m^{}_1/m^{}_2}$ in a
parallel way, one can roughly fix the three neutrino masses: $m^{}_1
\sim 0.008 ~{\rm eV}$, $m^{}_2 \sim 0.011 ~{\rm eV}$ and $m^{}_3
\sim 0.051 ~{\rm eV}$. In addition, $\theta^{\nu}_{13} \sim
\sqrt{m^{}_1 m^{}_2/m^2_3} \simeq 0.18$, and hence the correct value
of $\theta^{}_{13}$ can be achieved after the contribution from
$\theta^{l}_{12}$ (ranging between $-0.05$ and 0.05, as a function
of $\phi^{l}_{21}-\phi^{\nu}_{21}$) is properly included. Finally,
it is also possible to obtain $\delta \simeq \pm \pi/2$.

At this point it is worth mentioning the idea of ``flavor democracy"
\cite{FX96,FX98} or ``universal strength of Yukawa couplings"
\cite{Branco} for the mass matrices of leptons and quarks, which can
be regarded as an extreme case of $M^{}_{\rm f}$ shown in Eq. (6.39)
after they receive some proper perturbations respecting the $\rm
Z^{}_2$ permutation symmetry. While the charged-lepton mass
hierarchy, the quark mass hierarchy and small quark flavor mixing
effects are easily understood in such an approach, one has to
constrain the neutrino mass matrix $M^{}_\nu$ to be nearly diagonal
so as to obtain significant flavor mixing effects in the lepton
sector \cite{Barr}. It is also found that a combination of the
flavor democracy idea and the seesaw mechanism allows one to build a
phenomenologically viable model \cite{RX2004}.

To summarize, the universal texture of fermion mass matrices shown
in Eq. (6.39), which possesses the $(2,3)$ or $Z^{}_2$ permutation
symmetry --- an analogue of the $\mu$-$\tau$ permutation symmetry,
can serve as a starting point to describe the flavor structures of
quarks and leptons and to identify their similarities and
differences. Its variation with some symmetry-breaking effects in
Eq. (6.42) contains both the texture zeros and the linear equality
of different entries \cite{AGR1,AGR2}. Such an ansatz is predictive
and can essentially fit current experimental data. For simplicity,
here we have only discussed the possibility that all the fermion
mass matrices share the same texture but their respective parameters
are independent of one another. The latter can get correlated if the
GUT framework is taken into account (see, e.g., Refs.
\cite{MNY,JKP}, where the $(2,3)$ permutation symmetry is combined
with the SU(5) and SO(10) GUT models, respectively).

\def\thefootnote{\arabic{footnote}}
\setcounter{footnote}{0}
\setcounter{equation}{0}
\setcounter{table}{0}
\setcounter{figure}{0}

\section{Summary and outlook}

In the past two decades we have witnessed a booming period in
neutrino physics --- an extremely important part of particle physics
and cosmology. Especially since the first discovery of atmospheric
neutrino oscillations at the Super-Kamiokande detector in 1998,
quite a lot of significant breakthroughs have been made in
experimental neutrino physics, as recognized by both the 2015 Nobel
Prize in Physics and the 2016 Breakthrough Prize in Fundamental
Physics. On the one hand, the striking and appealing phenomena of
atmospheric, solar, reactor and accelerator neutrino (or
antineutrino) oscillations have all been observed in a convincing
way, and the oscillation parameters $\Delta m^2_{21}$, $|\Delta
m^2_{31}|$, $\theta^{}_{12}$, $\theta^{}_{13}$ and $\theta^{}_{23}$
have been determined to an impressive degree of accuracy. On the
other hand, the unusual geo-antineutrino events and extraterrestrial
PeV neutrino events have also been observed, and the sensitivities
to neutrino masses in the beta decays, $0\nu 2\beta$ decays and
cosmological observations have been improved to a great extent. On
the theoretical side, a lot of efforts have been made towards a
deeper understanding of the origin of tiny neutrino masses, the
secret of flavor mixing and CP violation and a unified picture of
leptons and quarks at a much larger framework beyond the SM.
Moreover, one has explored various implications of massive neutrinos
on the cosmological matter-antimatter asymmetry, warm dark matter
and many violent astrophysical or astronomical processes. All these
have demonstrated neutrino physics to be one of the most important
and exciting frontiers in modern science.

What the present review article has concentrated on is the ``minimal
flavor symmetry" behind the observed lepton flavor mixing pattern
--- a partial (or approximate) $\mu$-$\tau$ symmetry which is
definitely favored by current experimental data, and its various
phenomenological implications in neutrino physics. We have discussed
both the $\mu$-$\tau$ {\it permutation} symmetry and the
$\mu$-$\tau$ {\it reflection} symmetry, and pointed out a few
typical ways to slightly break such a symmetry. Some larger discrete
flavor symmetry groups, in which the intriguing $\mu$-$\tau$
symmetry can naturally arise as a residual symmetry, have been
briefly described. Both the bottom-up approach and the top-down
approach have been illustrated in this connection, in order to
bridge the gap between model building attempts and neutrino
oscillation data. To be more explicit, we have summarized the basic
strategies of model building with the help of the $\mu$-$\tau$
symmetry, either in the presence or in the absence of the seesaw
mechanism. The phenomenological consequences of the $\mu$-$\tau$
flavor symmetry on some interesting and important topics, such as
neutrino oscillations in matter, radiative corrections to the
equalities $|U^{}_{\mu i}| = |U^{}_{\tau i}|$ from a
superhigh-energy scale down to the electroweak scale, flavor
distributions of the UHE cosmic neutrinos at a neutrino telescope, a
possible connection between the leptogenesis and low-energy CP
violation through the seesaw mechanism and a unified flavor
structure of leptons and quarks, have also been discussed.
Therefore, we hope that this review could serve as a new milestone
in the development of neutrino phenomenology, in order to make so
many theoretical ideas of ours much deeper and more convergent.
\begin{figure}
\vspace{0.2cm}
\centerline{\includegraphics[width=6cm]{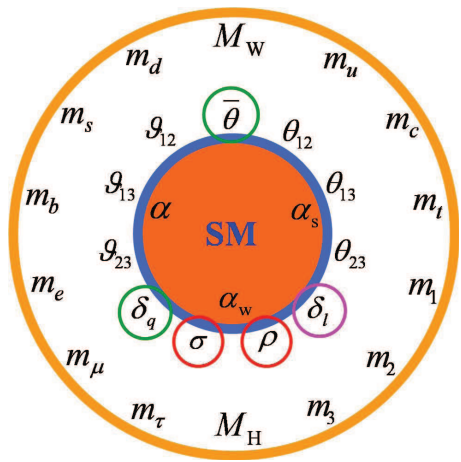}}
\caption{The Fritzsch-Xing ``pizza" plot of 28 parameters associated
with the SM itself and neutrino masses, lepton flavor mixing angles
and CP-violating phases.}
\end{figure}

There are still quite a number of open fundamental questions about
massive neutrinos and lepton flavor issues. The immediate ones
include how small the absolute neutrino mass scale is, whether the
neutrino mass spectrum is normal as those of the charged leptons and
quarks, whether the antiparticles of massive neutrinos are just
themselves, how large the CP-violating phase $\delta$ can really be,
which octant the largest flavor mixing angle $\theta^{}_{23}$
belongs to, whether there are light and (or) heavy sterile
neutrinos, what the role of neutrinos is in dark matter, whether the
observed matter-antimatter asymmetry of the Universe is directly
related to CP violation in neutrino oscillations, etc. Motivated by
so many challenging and exciting questions, we are on our way to
discovering a new physics world in the coming decades
\footnote{See some recent reviews on the future prospects of
leptonic CP violation \cite{Branco2}, neutrino oscillation
experiments \cite{LBNE}, searches for the $0\nu 2\beta$ decays
\cite{Carlo}, electromagnetic properties of massive neutrinos
\cite{Carlo2} and sterile neutrino physics \cite{WDM3}.}.

Last but not least, let us make some concluding remarks with the
help of the Fritzsch-Xing ``pizza" plot shown in Fig. 7.1. This
picture provides a brief summary of 28 fundamental parameters
associated with the SM itself and neutrino masses, lepton flavor
mixing angles and CP-violating phases. Among them, the five
parameters of strong and weak CP violation deserve some special
attention. In the quark sector the strong CP-violating phase
$\overline{\theta}$ remains a mystery, but the weak CP-violating
phase $\delta^{}_q$ has been determined to a good degree of
accuracy. In the lepton sector none of the CP-violating phases has
been directly measured, although a very preliminary hint
$\delta^{}_l \sim -\pi/2$ has been achieved from a comparison
between the present T2K and Daya Bay data
\footnote{Here we use $\delta^{}_l$ to denote the Dirac CP phase in
the standard parametrization of the PMNS matrix $U$, so as to
distinguish it from its analogue $\delta^{}_q$ in the CKM quark
flavor mixing matrix.}.
The true value of $\delta^{}_l$ is expected to be determined in the
future long-baseline neutrino oscillation experiments. If massive
neutrinos are really the Majorana particles, however, it will be
extremely challenging to probe or constrain the Majorana
CP-violating phases $\rho$ and $\sigma$ which can only show up in
some extremely rare processes involving lepton number
nonconservation.

Perhaps some of the flavor puzzles cannot be resolved unless we
finally find out the fundamental flavor theory \cite{Q50}. But the
latter cannot be formulated without a lot of phenomenological and
experimental inputs. As Leonardo da Vinci once stressed, ``Although
nature commences with reason and ends in experience, it is necessary
for us to do the opposite. That is, to commence with experience and
from this to proceed to investigate the reason."

\vspace{0.2cm}

We would like to thank Leszek Roszkowski for inviting us to write
this review article. One of us (Z.Z.X.) is indebted to Harald Fritzsch,
Shu Luo, He Zhang, Shun Zhou and Ye-Ling Zhou for fruitful collaboration
on the $\mu$-$\tau$ flavor symmetry issues. We are also grateful to
Wan-lei Guo, Yu-Feng Li, Jue Zhang and Jing-yu Zhu for their kind
helps in plotting several figures. This work is supported in part by
the National Natural Science Foundation of China under grant No.
11135009 and grant No. 11375207 (Z.Z.X.); by the National Basic
Research Program of China under grant No. 2013CB834300 (Z.Z.X.);
by the China Postdoctoral Science Foundation under grant No.
2015M570150 (Z.H.Z.); and by the CAS Center for Excellence in Particle
Physics.


\section{Appendix}

As discussed in section 3.4, the delicate effects of $\mu$-$\tau$
reflection symmetry breaking are characterized by the quantities
$\Delta \theta^{}_{23}$, $\Delta \delta$, $\Delta \rho$ and $\Delta
\sigma$. Their expressions in our analytical approximations can be
parametrized as
\begin{eqnarray}
\Delta \theta^{}_{23} & = c^{\theta}_{r1} {\rm R}^{}_1 +
c^{\theta}_{i1} {\rm I}^{}_1 + c^{\theta}_{r2} {\rm R}^{}_2
+ c^{\theta}_{i2} {\rm I}^{}_2 \; ,
\nonumber \\
\Delta \delta & = c^{\delta}_{r1} {\rm R}^{}_1 +
c^{\delta}_{i1} {\rm I}^{}_1 + c^{\delta}_{r2} {\rm R}^{}_2
+ c^{\delta}_{i2} {\rm I}^{}_2 \; ,
\nonumber \\
\Delta \rho & = c^{\rho}_{r1} {\rm R}^{}_1 +
c^{\rho}_{i1} {\rm I}^{}_1 + c^{\rho}_{r2} {\rm R}^{}_2
+ c^{\rho}_{i2} {\rm I}^{}_2 \; ,
\nonumber \\
\Delta \sigma & = c^{\sigma}_{r1} {\rm R}^{}_1 +
c^{\sigma}_{i1} {\rm I}^{}_1 + c^{\sigma}_{r2} {\rm R}^{}_2
+ c^{\sigma}_{i2} {\rm I}^{}_2 \; ,
\end{eqnarray}
where the relevant coefficients are given by
\begin{eqnarray}
c^{\theta}_{r1} & = & 2 \left(m^2_{12} - m^2_{11} s^2_{13} +
m^2_3 s^2_{13}\right) \Omega^{}_\theta \; ,
\nonumber \\
c^{\theta}_{i1} & = & 4 m^{}_{11} m^{}_{12} \tilde
s^{}_{13} \Omega^{}_\theta \; ,
\nonumber \\
c^{\theta}_{r2} & = & - m^{}_{11-3} m^{}_{22+3}
\Omega^{}_\theta \; ,
\nonumber \\
c^{\theta}_{i2} & = & 4 m^{}_{11} m^{}_{12} \tilde
s^{}_{13} \Omega^{}_\theta \; ;
\end{eqnarray}
and
\begin{eqnarray}
c^{\delta}_{r1} & = &
2 \left[ m^{}_{12} \left(\overline m^{}_1 + \overline m^{}_2 \right)
m^{}_{22-3} T \right.
\nonumber \\
& & +  2 m^{}_{11-3}  m^{}_{11+3} m^{}_{22-3} s^2_{13}
\nonumber \\
& & \left. - 2 m^2_{12} (2 m^{}_{11+3} + m^{}_{12} T) s^2_{13}
\right] \Omega^{}_\delta \; ,
\nonumber \\
c^{\delta}_{i1} & = &
2 \left[ \left(\overline m^{}_1 + \overline m^{}_2\right)
m^{}_{11+3} m^{}_{22-3} T \right.
\nonumber \\
& & \left. - 2 m^{}_{12} \left(\overline m^{}_1 - m^{}_3\right)
\left(\overline m^{}_2 - m^{}_3\right) \right]
\tilde s^{}_{13} \Omega^{}_\delta \; ,
\nonumber \\
c^{\delta}_{r2} & = &
\left[- m^{}_{12} \left(\overline m^{}_1 + \overline m^{}_2\right)
m^{}_{22+3} T \right.
\nonumber \\
& & - 2m^{}_{12} \left( m^{2}_{12} - 2m^{}_{22} m^{}_{11-3}\right)
T s^2_{13}
\nonumber \\
& & \left. + 2m^{}_{11-3} m^{}_{11+3} m^{}_{22+3} s^2_{13} \right]
\Omega^{}_\delta \; ,
\nonumber \\
c^{\delta}_{i2} & = & m^{}_{22+3} \left[ m^2_{12} - m^{}_{22-3}
\left(2m^{}_{11} \right. \right.
\nonumber \\
& & \left. \left. + m^{}_{22-3}\right) \right] T
\tilde s^{}_{13} \Omega^{}_\delta \; ;
\end{eqnarray}
and
\begin{eqnarray}
c^{\rho}_{r1} & = & 8 m^{}_{11} m^{}_3 \left[ m^2_{12}
\left(m^{}_{11+3} + m^{}_{12} T\right) \right.
\nonumber \\
& & \left. - \left(\overline m^{}_1 - m^{}_3\right) \left(\overline
m^{}_2 - m^{}_3\right) m^{}_{11+3} \right] c^2_{12} \tilde s^{}_{13}
\Omega^{}_\rho \; ,
\nonumber \\
c^{\rho}_{i1} & = & - 8 m^{}_{11} m^{}_{12} m^{}_3 \left[
\left(\overline m^{}_1 - m^{}_3\right) \left(\overline m^{}_2 -
m^{}_3\right) \right.
\nonumber \\
& & \left. - 2 m^{}_{11}  m^{}_{11+3} s^2_{13} \right]
c^2_{12} \Omega^{}_\rho \; ,
\nonumber \\
c^{\rho}_{r2} & = & 4 m^{}_{11} m^{}_3 \left[ m^{}_{12}
\left(m^2_{12} - 2m^{}_{22} m^{}_{11-3} \right) T  \right.
\nonumber \\
& & \left. -m^{}_{11-3} m^{}_{11+3} m^{}_{22+3} \right] c^2_{12}
\tilde s^{}_{13} \Omega^{}_\rho \; ,
\nonumber \\
c^{\rho}_{i2} & = & \left(\overline m^{}_1 - m^{}_3\right)
\left(\overline m^{}_2 - m^{}_3\right) m^{}_{22+3} \left[  2
m^{}_{11} m^{}_3 c^2_{12} \right.
\nonumber \\
& & \left. + \left(\overline m^{}_1 + \overline m^{}_2\right)
\left(\overline m^{}_1 s^2_{12} - \overline m^{}_1 c^2_{12} \right.
\right.
\nonumber \\
& & \left. \left. -m^{}_3 s^2_{12}\right) \right] T \Omega^{}_\rho
\; ;
\end{eqnarray}
and
\begin{eqnarray}
c^{\sigma}_{r1} & = & -8 m^{}_{11} m^{}_3 \left[ m^2_{12}
\left(m^{}_{11+3} + m^{}_{12} T\right)
\right. \nonumber \\
& & \left. - \left(\overline m^{}_1 - m^{}_3\right) \left(\overline
m^{}_2 - m^{}_3\right) m^{}_{11+3} \right] s^2_{12} \tilde s^{}_{13}
\Omega^{}_\sigma \; ,
\nonumber \\
c^{\sigma}_{i1} & = & 8 m^{}_{11} m^{}_{12} m^{}_3 \left[
\left(\overline m^{}_1 - m^{}_3\right) \left(\overline m^{}_2 -
m^{}_3\right) \right.
\nonumber \\
& & \left. - 2 m^{}_{11}  m^{}_{11+3} s^2_{13} \right]
s^2_{12} \Omega^{}_\sigma \; ,
\nonumber \\
c^{\sigma}_{r2} & = &
-4 m^{}_{11} m^{}_3 \left[ m^{}_{12} \left(m^2_{12} - 2m^{}_{22}
m^{}_{11-3}\right) T  \right.
\nonumber \\
& & \left. -m^{}_{11-3} m^{}_{11+3} m^{}_{22+3} \right ] s^2_{12}
\tilde s^{}_{13} \Omega^{}_\sigma \; ,
\nonumber \\
c^{\sigma}_{i2} & = & -\left(\overline m^{}_1 - m^{}_3\right)
\left(\overline m^{}_2 - m^{}_3\right) m^{}_{22+3} \left[  2
m^{}_{11} m^{}_3 s^2_{12} \right.
\nonumber \\
& & \left. + \left(\overline m^{}_1 + \overline m^{}_2\right)
\left(\overline m^{}_1
c^2_{12} - \overline m^{}_1 s^2_{12} \right. \right.
\nonumber \\
& & \left. \left. -m^{}_3 c^2_{12}\right) \right] T
\Omega^{}_\sigma \; .
\end{eqnarray}
In Eqs. (8.2)---(8.5), $T = \tan 2\theta^{}_{12}$, and
$\Omega^{}_\theta$, $\Omega^{}_\delta$, $\Omega^{}_\rho$ and
$\Omega^{}_\sigma$ are defined through the equations
\begin{eqnarray}
\Omega^{-1}_\theta & = & 2 \left(\overline m^{}_1 - m^{}_3 \right)
\left( \overline m^{}_2 - m^{}_3 \right) \; ,
\nonumber \\
\Omega^{-1}_\delta & = & 2 \left(\overline m^{}_1 + \overline
m^{}_2\right) \left(\overline m^{}_1 - m^{}_3\right) \left(\overline
m^{}_2 - m^{}_3\right) T \tilde s^{}_{13} \; ,
\nonumber \\
\Omega^{-1}_\rho & = & 4 \overline m^{}_1 m^{}_3 c^{}_{12} s^{}_{12}
\left(\Omega^{}_\delta T \tilde{s}^{}_{13}\right)^{-1} \; ,
\nonumber \\
\Omega^{-1}_\sigma & = & 4 \overline m^{}_2 m^{}_3 c^{}_{12} s^{}_{12}
\left(\Omega^{}_\delta T \tilde{s}^{}_{13}\right)^{-1} \; .
\end{eqnarray}

\vspace{1cm}


\end{document}